\documentclass[%
 reprint,
 amsmath,amssymb,
 aps,
 showkeys,
]{revtex4-2}

\usepackage{hyperref}
\usepackage{graphicx}
\usepackage{dcolumn}
\usepackage{bm}
\usepackage{lipsum}
\usepackage{booktabs}
\usepackage{tabularx} 
\usepackage{aasmacros} 
\usepackage[dvipsnames]{xcolor}
\usepackage[export]{adjustbox}

\begin{document}

\preprint{APS/123-QED}

\newcommand{\etaIA}{\ensuremath{{\eta_{A_{\mathrm{IA}}}}}}
\newcommand{\etab}{\ensuremath{{\eta_{b_{g}}}}}
\newcommand{\etabnl}{\ensuremath{{\eta_{b_{g,2}}}}}
\newcommand{\AIA}{\ensuremath{A_{\mathrm{IA}}}}
\newcommand{\omatter}{\ensuremath{\Omega_{\mathrm{m}}}}
\newcommand{\sigeight}{\ensuremath{\sigma_8}}
\newcommand{\biasg}{\ensuremath{b_{g}}}
\newcommand{\rstoch}{\ensuremath{r_{g}}}
\newcommand{\biasnl}{\ensuremath{b_{g,2}}}
\newcommand{\Seight}{\ensuremath{S_8}}
\newcommand{\x}{\ensuremath{\times}}
\defcitealias{Fluri2019kids}{F19}

\title[DeepLSS: LSS with deep learning]{DeepLSS: breaking parameter degeneracies in large scale structure\\with deep learning analysis of combined probes}


\author{Tomasz Kacprzak}
\email{tomaszk@phys.ethz.ch}
\affiliation{Institute for Particle Physics and Astrophysics, ETH Zurich, 8093 Zurich, Switzerland}
\affiliation{Swiss Data Science Center, Paul Scherrer Institute, 5232 Villigen, Switzerland}

\author{Janis Fluri}%
\affiliation{Institute for Particle Physics and Astrophysics, ETH Zurich, 8093 Zurich, Switzerland}

\date{\today}

\begin{abstract}
In classical cosmological analysis of large scale structure surveys with 2-pt functions, the parameter measurement precision is limited by several key degeneracies within the cosmology and astrophysics sectors. For cosmic shear, clustering amplitude $\sigma_8$ and matter density $\Omega_m$ roughly follow the \mbox{$S_8=\sigma_8(\Omega_m/0.3)^{0.5}$} relation. In turn, $S_8$ is highly correlated with the intrinsic galaxy alignment amplitude $A_{\mathrm{IA}}$. For galaxy clustering, the bias $b_g$ is degenerate with both $\sigma_8$ and $\Omega_m$, as well as the stochasticity $r_g$. Moreover, the redshift evolution of IA and bias can cause further parameter confusion. A tomographic 2-pt probe combination can partially lift these degeneracies. In this work we demonstrate that a deep learning analysis of combined probes of weak gravitational lensing and galaxy clustering, which we call \textsc{DeepLSS}, can effectively break these degeneracies and yield significantly more precise constraints on $\sigma_8$, $\Omega_m$, $A_{\mathrm{IA}}$, $b_g$, $r_g$, and IA redshift evolution parameter $\eta_{\mathrm{IA}}$. The most significant gains are in the IA sector: the precision of $A_{\mathrm{IA}}$ is increased by approximately 8$\times$ and is almost perfectly decorrelated from $S_8$. Galaxy bias $b_g$ is improved by 1.5$\times$, stochasticity $r_g$ by 3$\times$, and the redshift evolution $\eta_{\mathrm{IA}}$ and $\eta_b$ by 1.6$\times$. Breaking these degeneracies leads to a significant gain in constraining power for $\sigma_8$ and $\Omega_m$, with the figure of merit improved by 15$\times$. We give an intuitive explanation for the origin of this information gain using sensitivity maps. These results indicate that the fully numerical, map-based forward modeling approach to cosmological inference with machine learning may play an important role in upcoming LSS surveys. We discuss perspectives and challenges in its practical deployment for a full survey analysis.
\end{abstract}

\maketitle

\section{Introduction}

Combined probes of large scale structure (LSS) contain information about the late-time evolution of the universe and thus are a unique laboratory for testing cosmological models.
The structure of the matter density field, observed through weak gravitational lensing and galaxy clustering, is used to constrain the present day matter and dark energy densities, \omatter\ and $\Omega_\Lambda$, as well as matter clustering strength \sigeight\ and the dark energy equation of state $w$, and other parameters  (see \cite{Kilbinger2015review,Zhan2018lsst,Albrecht2006darkenergy} for reviews).

In recent years, dedicated LSS observing programs, such as 
the Dark Energy Survey \footnote{\url{darkenergysurvey.org}} (DES), 
the Kilo Degree Survey \footnote{\url{kids.strw.leidenuniv.nl}} (KiDS) and 
the Hyper-Suprime Cam Survey \footnote{\url{hsc.mtk.nao.ac.jp/ssp/survey}} (HSC) 
measured cosmological parameters with less than 5\% precision \cite{Des2022combined,Heymans2020multiprobe,Hikage2019cosmicshear}.
Upcoming surveys, such as 
Euclid \footnote{\url{https://www.euclid-ec.org}} and 
Vera Rubin Observatory \footnote{\url{lsstdesc.org}} (LSST) 
will provide orders of magnitude richer datasets and enable sub-percent measurements of these parameters.

In a LSS analysis, cosmology is typically constrained jointly with a large number of other parameters, corresponding to both astrophysical uncertainties and measurement systematics.
These uncertainties can constitute a significant part of the cosmology error budget.
In particular, the key degeneracies that limit our ability to constrain \sigeight\ and \omatter\ are galaxy intrinsic alignments and galaxy bias \cite{Kirk2015intrinsic,Eriksen2018bias}. 

Intrinsic alignments (IA) of galaxy shape and its large scale environment arise due to tidal fields acting on them during their formation and evolution \cite{Kirk2015intrinsic,Blazek2019beyond}.
IA effects can perfectly mimic weak gravitational lensing and thus cause a degeneracy between their amplitude \AIA\ and $S_8$ \cite{Secco2022cosmicshear}. 
Galaxy biasing describes how galaxies trace the underlying dark matter density field.
Changes in biasing it can have a similar effect on clustering maps as changes in cosmology, causing a sharp degeneracy between linear bias parameter $b_g$ and \sigeight\ and \omatter\ \cite{Porredon2021clustering}.
Varying galaxy stochasticity parameter $r_g$ can also mimic the effects of varying cosmology.
These degeneracies are even stronger for models that include redshift evolution of IA and biasing, or their higher order properties.
Finally, for WL, the \sigeight\ and \omatter\ are itself degenerate, roughly following the $S_8=\sigma_8(\Omega_m/0.3)^{0.5}$ relation.

The tomographic probe combination can alleviate these degeneracies due to the use of joint information about galaxy positions and shapes.
For the scales considered, 2-pt functions -based analyses (3$\times$2) managed to reduce these degeneracies, but did not remove them completely \cite{Eifler2014cosmolike,Krause2021des,Des2022combined,Heymans2020multiprobe}.
Moreover, this type of analysis has a very limited ability to constrain higher order properties and redshift evolution of IA and biasing.
This can cause further challenges for accurate cosmology measurement, it is now known that the use of wrong IA or biasing models can cause significant errors in inferred cosmology \cite{Kirk2012intrinsic,Secco2022cosmicshear}, due to limited information on these effects that can be extracted from the data.
In that case, even the full Bayesian marginalization of these unconstrained parameters can lead to biases due to prior volume effects \cite{Fischbacher2022ia}.
Addressing these challenges is particularly important in the light of recent hints of tensions between the \Seight\ measurements between LSS and the early universe, as extrapolated from CMB measurements \cite{Amon2022consistent,Leauthaud2017lensinglow,Leauthaud2022withoutborders,Des2022combined,Heymans2020multiprobe,cosmology2022intertwined}.

At intermediate and small scales, the late-time LSS density field contains non-Gaussian information, which can also be extracted from lensing and clustering data.
Recently, peak statistics \cite{Zuercher2022despeaks} and deep learning \cite{Fluri2019kids,Fluri2022kids} have been demonstrated to achieve significant measurement precision gain of \sigeight\ and \omatter\ from weak lensing maps.
The convolutional neural network (CNN) results \cite{Fluri2019kids} and the DES Y3 peaks analysis \cite{Zuercher2022despeaks} hinted at the possibility of delivering improved IA constraints compared to the power spectrum.
However, a map-level probe combination has not been extensively explored to date.

In this paper, we propose a deep learning analysis of combined probes of weak lensing and galaxy clustering and investigate its potential to break degeneracies between \sigeight, \omatter, \rstoch, \AIA, \biasg, as well as their redshift evolution, parameterized by a simple power law with \etaIA\ and \etab.
We use a fully forward-modeling approach, where we train CNNs on a stacked tomographic weak lensing convergence $\kappa_g$ and galaxy counts $\delta_g$.
This is similar a to the 2D ``counts-in-cells'' technique \cite{Gruen2018densitysplit,Salvador2019sv}, where the galaxy catalogs are binned into pixel or voxel ``cells'', which are then analyzed directly without using 2-pt functions.

We compare the results of the CNN analysis with the equivalent tomographic power spectrum (PSD) analysis, which uses data vectors similar to the 3$\times$2 method.
We use a Stage-III simulated survey configuration with 900 deg$^2$, 10 galaxies/arcmin$^2$, and four broad redshift bins.
The analysis is limited to intermediate scales, corresponding to smoothing FWHM of 8 Mpc/h for galaxy clustering and 4 Mpc/h for weak lensing.
We also explore a small scales configuration with non-linear bias, with smoothing at 4 Mpc/h for both probes.
We report the forecasted cosmology constraints expected from the CNN and PSD, and describe the information gain for all parameters considered.
Finally, we investigate the sensitivity maps for CNNs to gain more intuition about the origin of the information used by CNN.

This paper is structured as follows. 
We describe the simulated lensing and clustering data in Section~\ref{sec:theory}.
The measurement methods for deep learning and power spectrum are presented in Section~\ref{sec:methods}.
We describe our results in Section~\ref{sec:results}.
We investigate the origin of the information used by our methods in Section~\ref{sec:sensitivity} and conclude in Section~\ref{sec:conclusions}.


\section{Theory modelling}
\label{sec:theory} 


\begin{table}
    \centering
    \begin{tabularx}{0.475\textwidth}{lXrr}
    \ & description & prior & fiducial \\
    \toprule
    \omatter & matter density today          & [0.15, 0.45]  & 0.29 \\ 
    \sigeight& clustering amplitude          & [0.5, 1.2]    & 0.71 \\ 
    \AIA     & intrinsic alignment amplitude & [-6, 6]       & 0.5 \\
    \biasg   & linear galaxy bias            & [0.5, 2.5]    & 1.5 \\
    \rstoch  & galaxy stochasticity          & [0.4, 1]      & 0.7 \\
    \etaIA   & IA redshift evolution         & [-4, 6]       & 1.6 \\
    \etab    & linear bias evolution         & [-3, 3]       & 0.5 \\
    \midrule
    \biasnl  & non-linear galaxy bias        & [-3, 1]       & 0.5 \\
    \etabnl  & non-linear bias evolution     & [-2, 2]       & 0.0 \\
    \toprule
    \end{tabularx}
    \caption{Summary of parameters used in the data models. Parameters \biasnl\ and \etabnl\ are used only in the non-linear galaxy bias model. Parameters \omatter\ and \sigeight\ are additionally restricted by convex hull prior defined by the simulation grid.}
    \label{tab:params}
\end{table}

We create a set of consistent tomographic maps of weak lensing convergence $\kappa_{g}$ and galaxy clustering $\delta_g$.
We use a flat $\Lambda\mathrm{CDM}$ model for cosmology with fixed dark energy equation of state $w=-1$.
For intrinsic alignments, we employ the Non-linear - Linear Alignment model (NLA) \cite{Hirata2004intrinsic,Bridle2007constraints,Joachimi2011ia}. 
The galaxy number counts follow a biasing model, with optional non-linear terms.
The redshift evolution of intrinsic alignments and galaxy biasing is also included in this model.
We do not include the Redshift Space Distortions (RSD) or magnification contributions to clustering in this work, as it is subdominant to other effects included.
The full model has the parameters described below, and summarized in Table~\ref{tab:params}.

\begin{itemize}
    \item 
    We vary cosmology parameters \omatter, \sigeight, with other parameters fixed at $\Omega_b = 0.0493$, $H_0 = 67.36$ and $n_s = 0.9649$, which corresponds to the baseline results ($\Lambda$CDM,TT,TE,EE+lowE+lensing) of Planck 2018 \cite{Planck2018parameters}.

    \item 
    Intrinsic alignment amplitude is controlled by 
    \AIA, while its redshift dependence by the power law \etaIA\ (see Equation~\ref{eqn:resdhift_evol_ia} below).
    We do not include the luminosity or color dependence of intrinsic alignments.

    \item
    Galaxy density field is controlled by linear bias \biasg\ and its redshift evolution \etab, which is also a power law parameter (Equation~\ref{eqn:resdhift_evol_b}).
    The galaxy stochasticity \rstoch\ captures the degree of correlation between the galaxy and matter density field.
    We reduce the correlation between these fields by adding uniform noise to phases of the galaxy density field.
    In our extended model, we also add non-linear galaxy bias \biasnl\ and its redshift evolution \etabnl.

\end{itemize}

\begin{figure*}
    \centering
        \includegraphics[width=1.00\textwidth]{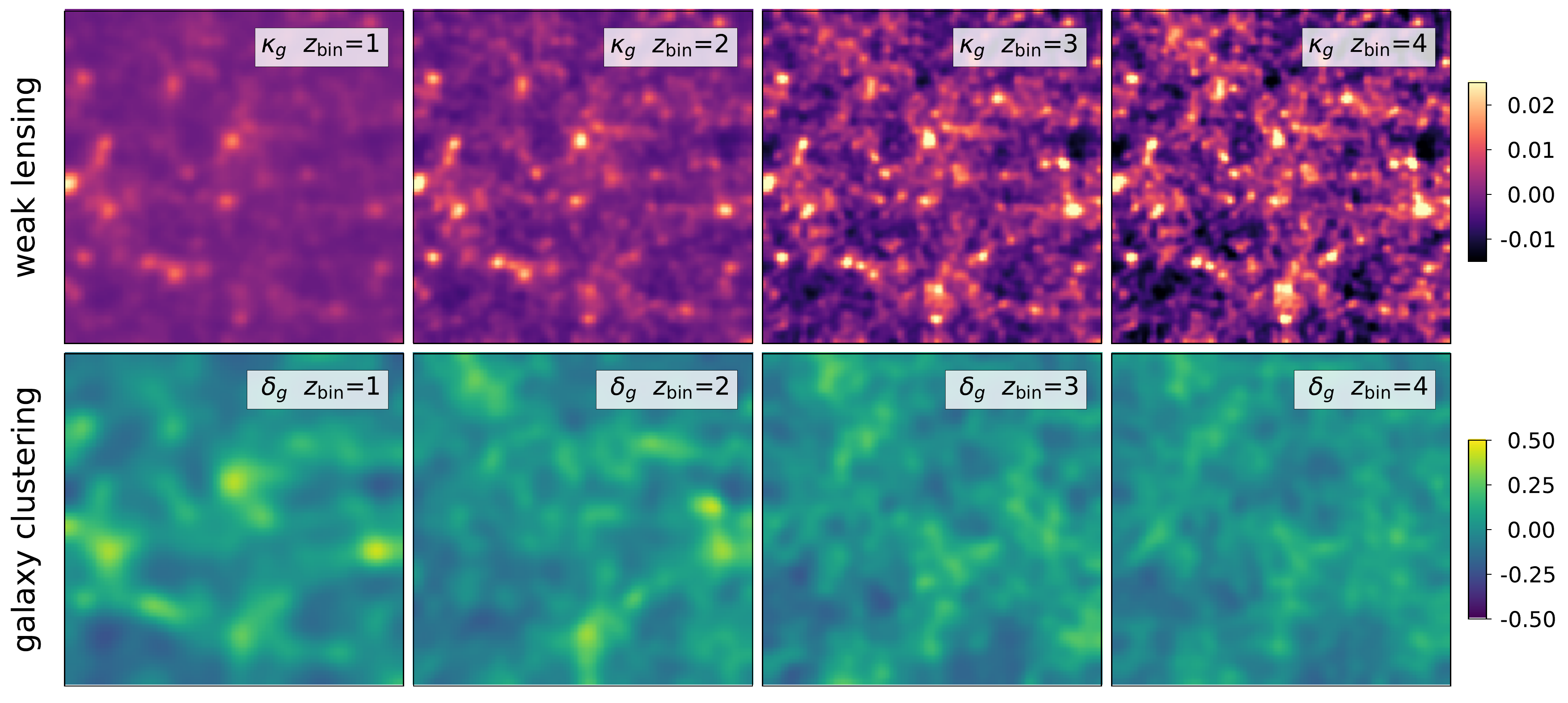}
    \caption{
    Example maps for weak lensing convergence $\kappa_m$ and galaxy clustering $\delta_m$, before the addition of noise. The size of the maps is 5$\times$5 degrees. The maps were smoothed with the redshift bin -dependent Gaussian kernel, with FWHM corresponding to $\sim$4 Mpc/h at the mean redshift of the bin for weak lensing and $\sim$8 Mpc/h for galaxy clustering (see Section~\ref{sec:smoothing}).}
    \label{fig:maps}
\end{figure*}

We consider a Stage-III -like survey configuration with 900 deg$^2$ with 10 galaxies/arcmin$^2$ distributed evenly in 4 broad redshift bins.
The redshift bins are shown in Appendix~\ref{sec:reshift_bins}, and have the following mean redshifts $\langle z \rangle=0.31, 0.48, 0.75, 0.94$.
We use the same galaxy selection for both lensing and clustering; we do not create a separate ``lens'' sample, as often done \cite{RodriguezMonroy2022clustering,Porredon2021optimizing,Porredon2021clustering}.
We also do not include uncertainties in measurement systematic biases, such as redshift errors, shear calibration, or selection function uncertainty for clustering.
These may play an important role amplifying degeneracies in the parameter measurements, especially in the redshift error and IA sector \cite{Fischbacher2022ia}.
We leave the optimization of the lens sample and the investigation of the systematics effects to future work.


\subsection{Multi-probe maps} 


We closely follow the method we introduced in \cite[][hereafter \citetalias{Fluri2019kids}]{Fluri2019kids} for calculating convergence $\kappa_g$ maps and create the corresponding clustering $\delta_g$ maps using the same pencil-beam simulations.
The simulations in \citetalias{Fluri2019kids} consist of 57 unique cosmologies spanning the \omatter -- \sigeight\ plane.
In that work we used the \textsc{PkdGrav3} code \cite{potter2016pkdgrav3} to run a total of 12 simulations at each cosmology.
Each simulation used 256$^3$ particles in a volume of 500$^3$ Mpc$^3$, and the initial conditions were generated at redshift $z_\mathrm{init} = 50$, using the the \textsc{Music} code \cite{MUSIC2011}. 
All simulations were run with 500 time steps, writing snapshots at the interval of $\Delta z=0.1$ from z = 3.45 to z = 1.55 and $\Delta z = 0.05$ down to redshift z = 0. 
See \citetalias{Fluri2019kids} for more details of these simulations.

The 2D map of these fields $m_{\mathrm{2D}}$ is projected from the 3D simulated overdensity $\delta_{\mathrm{3D}}$ using the Born approximation.
Maps are calculated using the \textsc{UFalcon} code \cite{Sgier2018}, in a very similar way as \cite{Fluri2019kids,zuercher2020forecast,Zuercher2022despeaks,Fluri2022kids}.
The 2D maps are calculated using the following equation:
\begin{align}
\label{eqn:projection}
m_{\mathrm{2D}}^{\mathrm{pix}} &\approx \sum_bW^{m}\int_{\Delta z_b}\frac{\mathrm{d}z}{E(z)}\delta_{\mathrm{3D}}\left[\frac{c}{H_0}\mathcal{D}(z)\hat{n}^\mathrm{pix},z\right],
\end{align}
where $W_m$ is the weight kernel corresponding to the considered field, $\mathcal{D}(z)$ is the dimensionless comoving distance, $\hat{n}^{\mathrm{pix}}$ is a unit vector pointing to the pixels center and $E(z)$ is given by $\mathrm{d}\mathcal{D} = \mathrm{d}z/E(z).$
The sum runs over all redshift shells and $\Delta z_b$ is the thickness of shell $b$.

The kernels corresponding to weak lensing $W^{\mathrm{WL}}$, intrinsic alignments $W^\mathrm{IA}$, and galaxy clustering $W^\mathrm{G}$ are:
\begin{align}
W^{\mathrm{WL}} & = \frac{3}{2}\Omega_\mathrm{m}  \frac{\int_{\Delta z_b}\frac{\mathrm{d}z}{E(z)}\int_z^{z_s}\mathrm{d}z'n(z')\frac{\mathcal{D}(z)\mathcal{D}(z,z')}{\mathcal{D}(z')}\frac{1}{a(z)}}{\int_{\Delta z_b}\frac{\mathrm{d}z}{E(z)}\int_{z_0}^{z_s}\mathrm{d}z'n(z')} \\[1em]
W^\mathrm{IA} &= \frac{\int_{\Delta z_b}\mathrm{d}zF(z)n(z)}{\int_{\Delta z_b} \frac{\mathrm{d}z}{E(z)}\int_{z_0}^{z_s}\mathrm{d}z'n(z')} \\[1em]
W^\mathrm{G} &= \frac{\int_{\Delta z_b}\mathrm{d}z \ n(z)}{\int_{\Delta z_b} \frac{\mathrm{d}z}{E(z)}\int_{z_0}^{z_s}\mathrm{d}z'n(z')} 
\end{align}
where $n(z)$ is the redshift distribution of galaxies in a given bin and $F(z)$ is a cosmology and redshift dependent term:
\begin{equation}
F(z) = -C_1\rho_\mathrm{crit}\frac{\Omega_\mathrm{m}}{D_+(z)},
\end{equation}
where ${C_1 = 5 \times 10^{-14} h^{-2}M_\odot\mathrm{Mpc}^3}$ is a normalization constant, $\rho_\mathrm{crit}$ is the critical density at $z=0$ and $D_+(z)$ normalized linear growth factor, so that $D_+(0) = 1$. 

We use kernels $W^{\mathrm{WL}}$, $W^\mathrm{IA}$, and $W^\mathrm{G}$ with Equation~\ref{eqn:projection} to create convergence maps of lensing $\kappa_m$, intrinsic alignment $\kappa_{\rm{IA}}$, and galaxy density contrast $\delta_m$.
Then, we subtract the mean of the convergence fields and normalize the galaxy density contrast:
\begin{align}
    \kappa &\leftarrow \kappa - \langle \kappa \rangle \\
    \delta &\leftarrow (\delta - \langle \delta \rangle)/\langle \delta \rangle.
\end{align}

The redshift dependence of IA and biasing is calculated for a given redshift bin $n^i(z)$ by using a power law. 
A similar method was previously used in \cite{Zuercher2022despeaks}.
We use a simplified formulation of this model, where we calculate a single effective scaling value per redshift bin $i$ using the $n^i(z)$ only:
\begin{align}
\label{eqn:resdhift_evol_ia}
\AIA^{i} &= \AIA \int_z dz \ n^i(z) \left( \frac{1 + z}{1 + z_0} \right) ^ \etaIA \\
\label{eqn:resdhift_evol_b}
\biasg^{i} &= \biasg \int_z dz \ n^i(z) \left( \frac{1 + z}{1 + z_0} \right) ^ {\etab}
\end{align}
where $z_0=0.7$ is the fixed pivot redshift.
This allows us to remove the redshift dependence in the process of summing the shell maps to create projected maps.
By doing this, we can pre-compute a set of field maps and apply the IA and biasing variation on-the-fly during training and predictions steps.
We verified that this approximation gives similar results to the full formulation. 
The exact implementation of this dependence is not crucial in this work, as its main focus is a constraining power forecast.
To make the interpretation of the values of \etab\ parameter easier, we calculated the values corresponding to uncertainty of current measurements from the results in DES Y3 combined probes \cite{Des2022combined}.
We used the linear bias measurements (their Table V) and fitted them with the $\eta$-evolution model in Equation~\ref{eqn:resdhift_evol_b}.
The uncertainty in these measurements translate to the uncertainty on bias evolution $\sigma\left[ \etab \right] = 0.21$. 

To create the forward-modeled probe maps, we add the noise to the pixels directly. 
The observed lensing map $\kappa_g$ and galaxy counts maps $\delta_g$ are:
\begin{align}
    \delta_g &= \mathrm{Poisson} \Big[ \bar n_{\mathrm{gal}} (1 + b_g   \delta_m^r ) \Big] \\
    \kappa_g &= \mathrm{Normal}\Big[ \kappa_{\mathrm{WL}} + \AIA \kappa_{\mathrm{IA}}, \frac{\sigma_e}{\sqrt{\delta_{g}}} \Big] 
\end{align}
where $\bar n_{\mathrm{gal}}$ is the mean number of galaxies per map pixel, $\sigma_e=0.4$ is the galaxy shape noise, which does not change across redshift bins as we assume the same number of galaxies from each bin. 
We create an auxiliary map $\delta_m^r$ to be partially decorrelated form the density contrast $\delta_m$, such that the correlation coefficient is \rstoch.
This is calculated from Fourier transform $\tilde \delta_m$, such that their phase angles $\angle$ are related by
\begin{equation}
    \label{eqn:stochasticity}
    \angle \tilde \delta_m^r  = \angle \tilde \delta_m + (1-\rstoch)^{2/3} \cdot \mathrm{Uniform}[-\pi, \pi],
\end{equation}
which reduces to $\delta_m^r=\delta_m$ for \rstoch=1.
We create this relation empirically for the $\delta_m$ maps we considered.
We find that this relation gives Pearson's correlation coefficient of pixel values roughly $\mathrm{corr}[\delta_m^r, \delta_m] = r_g$.
This method is described in more detail in Appendix~\ref{sec:stochasticity}.
While this implementation may differ from others used in previous work \cite{Gruen2018densitysplit,Friedrich2018density,Eriksen2018bias}, it gives a full degree of variation and should be sufficient for the purpose of this study.

For a model that includes non-linear galaxy bias, the galaxy counts map is calculated as
\begin{align}
    \delta_g &= \mathrm{Poisson}\left[ \bar n_{\mathrm{gal}} (1 + b_g \delta_m^{r} + \biasnl (\delta_m^{r})^2 ) \right] 
\end{align}
where \biasnl\ controls the strength of non-linear biasing and follows the same redshift evolution the other fields (Equation~\ref{eqn:resdhift_evol_b}), controlled by the parameter $\etabnl$.

We use these maps consistently for both individual and combined probes analyses; for the lensing maps, the noise level is calculated using the actual galaxy density map in each bin.
While lensing-only inference does not aim to constrain \biasg, \rstoch, \etab\ parameters, they are still used to make the maps.
Their values are always taken from the same prior, shown in Table~\ref{tab:params}.
This way the lensing-only networks effectively marginalize over the uncertainty on these parameters without aiming to constrain them.

As the Poisson random generator does not have GPU kernels available in \textsc{Tensorflow}, we use the inverse Anscombe transform \cite{Makitalo2011anscombe} to approximate it, as described in Appendix~\ref{sec:anscombe}.
We perform a number of additional scalings on the map to ensure numerical stability of the training for all parameters in the prior range, see Appendix~\ref{sec:numerics}.


\subsection{Scale cuts and smoothing} 
\label{sec:smoothing}


In this work we do not attempt to model baryonic effects, which can strongly modify the matter distribution at small scales \cite{Schneider2020baryons1,Weiss2019peakbaryons}. 
Similarly, the galaxy biasing on small scales can significantly deviate from the linear bias model \cite{Porredon2021clustering}.
To avoid using small scales, we smooth the maps with a redshift-dependent kernel.
Following choices in \cite{Porredon2021clustering}, we use the smoothing scales of 
$R$=4 Mpc/h for lensing, $R$=8 Mpc/h for clustering maps for the linear bias model, and $R$=4 Mpc/h for the non-linear bias model.
Including the pixel kernel, which is of size $d=4.68$ arcmin, we calculate that the additional Gaussian smoothing to apply is:
\begin{align*}
\label{eqn:smoothing_scales}
  R=4~\mathrm{Mpc/h} \ \ &\Rightarrow \ \ \sigma = [4.8, \ 3.5, \ 2.8, \ 2.5] \ \ \mathrm{arcmin} \\
  R=8~\mathrm{Mpc/h} \ \ &\Rightarrow \ \ \sigma = [9.8, \ 7.4, \ 6.0, \ 5.6] \ \ \mathrm{arcmin}
\end{align*}
for four redshift bins used in our analysis. 
An example of the maps is shown in Figure~\ref{fig:maps}.

In the original simulations from \citepalias{Fluri2019kids}, the field size is 5$\times$5 degrees and 128$\times$128 pixels.
As we use larger scales in this work, we down-sample the original maps to the size of 64$\times$64, which results in pixel size of $d=4.68$ \mbox{arcmin}.
In \citetalias{Fluri2019kids}, the survey consisted of 20 fields, which were passed to the networks as channels.
We construct random simulated surveys of 900 deg$^2$ by using 36 fields.
We employ, however, a different method of including these fields: we create a mosaic of size $6 \times 6$ fields. 
We add noise on-the-fly at each realization during training and prediction.
We avoid repeating the same mosaic by placing each field randomly within it, as well as performing a random flip.
The smoothing is done before creating the mosaic to avoid blurring over sharp edges.


\section{Deep learning and power spectra}
\label{sec:methods}


We analyze the maps using two approaches: a convolutional neural network (CNN) and a power spectrum (PSD).
The CNN maps from pixel maps to summary statistics corresponding to the model parameters.
For the PSD, we also use the summary statistics compression.
We calculate all auto and cross-spectra from the maps, and then employ a simple neural network (NN) to compress these PSD vectors into summary statistics.
We use this approach for operational reasons, as it removes the necessity of creating a dedicated likelihood analysis for PSDs only.
A similar approach was used in \cite{Fluri2022kids}.
The PSD NNs can be trained simultaneously with the CNN, with minimal time overhead.
Both CNN and PSD NNs give then the same form of summary output that can be interpreted in a likelihood analysis the same way.

We create a separate network for combined probes \mbox{(CP-CNN)}, which takes a stack of 4 $\kappa_g$ and 4 $\delta_g$ maps, as well as separate networks for lensing \mbox{(KG-CNN)} and clustering \mbox{(DG-CNN)}.
For the CNN, we use a ResNet architecture \cite{He2015resnet} with 4 convolutional layers, 10 residual layers, and kernel size of 5 with stride of 2, and the Relu activation function.
The final residual layer is flattened and fully-connected to the output, which contains the summaries and their covariances.
That gives a network with $\sim$10m -- 12m trainable parameters, depending on the model.
This configuration was inspired by \citetalias{Fluri2019kids} and we did not optimize it further.

As the equivalent NN for the PSD case, we also create separate networks for combined probes \mbox{(CP-PSD-NN)}, lensing \mbox{(KG-PSD-NN)} and clustering \mbox{(DG-PSD-NN)}.
The power spectra were calculated using FFT and averaged in 20 $\ell$-bins in the range of $\ell \in [36,4536]$.
The bins are spaced in logarithmic way, with the minimum interval of $\delta \ell=36$, which is the resolution of the FFT given the pixel sizes used.
We first concatenate the auto and cross -spectra from the maps, which gives 36 and 10 spectra for combined and individual probes, respectively.
We do not cut the scales, as the smoothing already removes most small scale power.
Then we pass the PSDs through 2 fully connected layers with 1024 hidden units, also with the Relu activation.
Finally, the output layer predicts the summaries and their covariances, as for the CNN.
This gives $\sim$ 1.2m - 1.8m trainable parameters, depending on the output size. 
For the PSD-NNs, we test a number of architectures and model sizes. 
We find that they all give very similar results and that choice does not change in comparison with the CNNs.
This test is described in Appendix~\ref{sec:psd_nets}
Further details of the implementation of these networks can be found in the public \href{https://github.com/tomaszkacprzak/DeepLSS}{\textsc{DeepLSS} code repository}.

We employ the same training strategy as \citetalias{Fluri2019kids}.
We use the negative log-likelihood loss function:
\begin{equation}
\label{eqn:likelihoodloss}
L = \ln\left(\left\vert\Sigma\right\vert\right) + \left(\theta_p - \theta_t\right)^{\top}\Sigma_p^{-1}\left(\theta_p - \theta_t\right),
\end{equation}
where $\theta_t$ is the true parameter vector, $\theta_p$ is the summary vector and $\Sigma_p$ is the predicted covariance matrix.
We train the networks using stochastic gradient descent with the \textsc{Adam} optimizer \cite{Kingma2014} with batch size of 32 mosaic maps and learning rates of 0.00005 and 0.0025 for CNNs and PSD-NNs, respectively. 
We additionally applied gradient clipping using the \cite{Seetharaman2020autoclip} method, using 50\% percentile.
The training took 885k batches and the loss did not improve over the last 200k batches, which indicates reasonable convergence.
The networks were created in \textsc{Tensorflow} \cite{tensorflow2015-whitepaper} and trained on \mbox{NVIDIA A100-SXM4-40GB GPUs}.

To avoid overfitting the training set, we calculate the loss for the test set, which is not used during training. 
The test set consisted of $\approx$8\% of the full simulation set.
We did not notice significant differences between the training and testing loss at any point in the training process.
This indicates that the networks have a good generalization ability and do not overfit the training set.
The reason for this is the on-the-fly addition of noise, which is a very efficient regularizer.


\section{Likelihood analysis}
\label{sec:likelihood}


Once the networks are trained, we create the conditional probability distribution $p(\theta_p|\theta_t)$ of the network output summary statistic $\theta_p$ given the true parameter value $\theta_t$.
We run a prediction for a set of 7615200 samples from $p(\theta_p|\theta_t)$, with 228000 unique parameter combinations.
While values of $\Omega_m$ and $\sigma_8$ were fixed at all 57 simulation points, the values for the remaining parameters were drawn from flat priors using a Sobol sampling.
We then create a model of $p(\theta_p|\theta_t)$ using a Mixture Density Network (MDN), which uses a mixture of Gaussians at each $\theta_t$;
it predicts the relative weights $w_j(\theta_t)$ of components, their means $\mu_j(\theta_t)$ and covariances $\Psi_j(\theta_t)$, where $j=1,\dots,K$, where $K$ is the number of Gaussians in the mixture model.
We train this model using stochastic gradient descent and monitor its validation loss to prevent overfitting, see Appendix~\ref{sec:mdn_tests} for details.
We confirmed that this network predicts the right means and variances for our choice of $K=4$, for all $\theta_t$, in Appendix~\ref{sec:mdn_tests}.

The final constraints are calculated using Markov Chain Monte Carlo (MCMC) using the \textsc{Emcee} algorithm \cite{ForemanMackey2013emcee}.
We use flat priors on parameters, with ranges shown in Table~\ref{tab:params}, and obtain a chain with 128k samples (1.28m for plotted chains).
As the \omatter\ and \sigeight\ parameters were sampled on a grid, we use the convex hull of this grid as the prior for this parameter combination.
\begin{figure*}
    \includegraphics[width=1\textwidth]{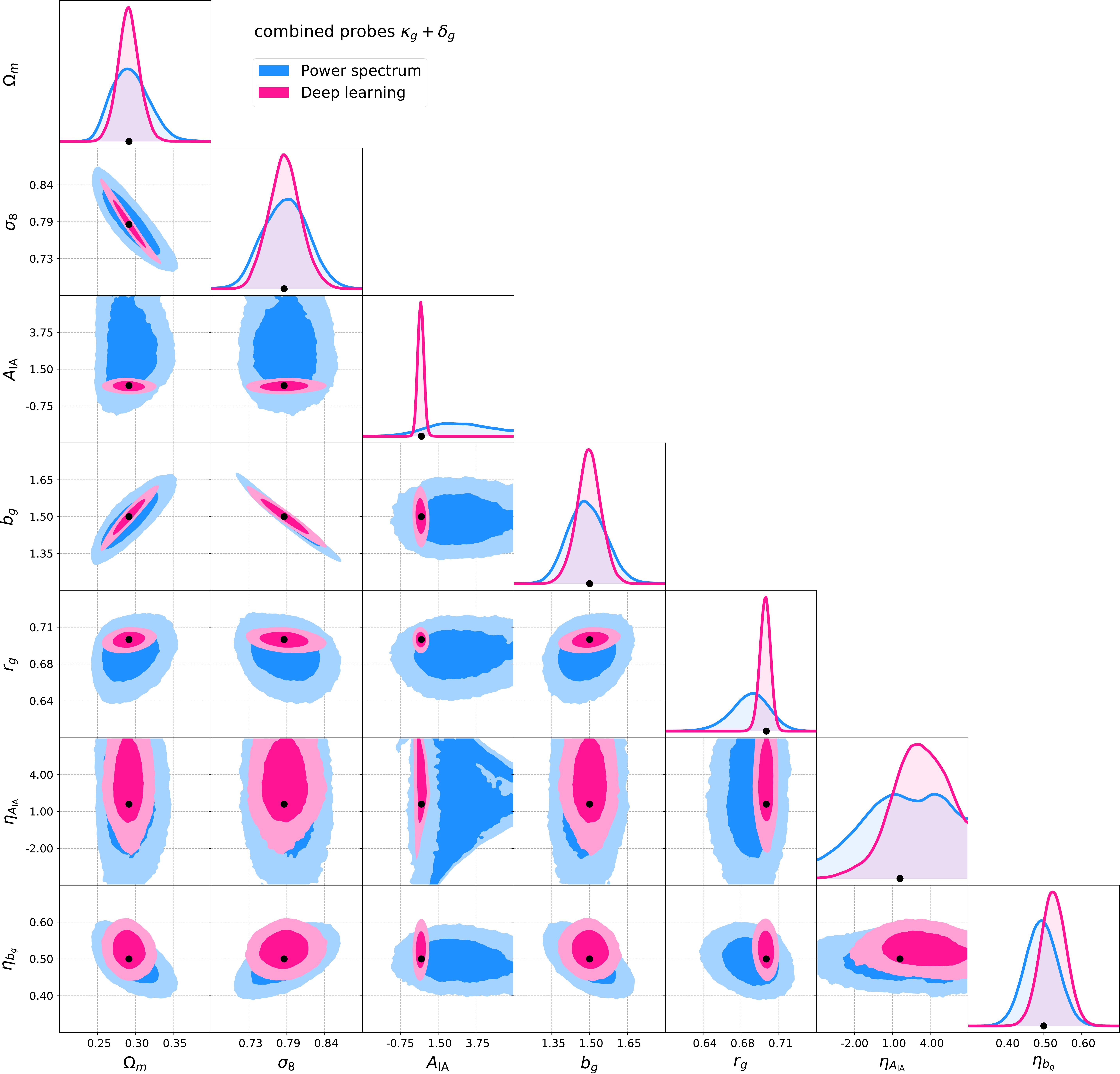} 
    \caption{
    Constraints for combined probes with deep learning (pink), compared to power spectra (blue), for the main model with linear galaxy bias.
    The black dots mark the true value of parameters used here, from the fiducial model summarized in Table~\ref{tab:params}.
    The mock observation used here was taken as the most likely model prediction for the fiducial model.
    }
    \label{fig:constraints}
\end{figure*}


\section{Results}
\label{sec:results}


We calculate constraints for CNNs and PSDs for the fiducial true parameter set $\theta_t^{\mathrm{fid}}$ given in Table~\ref{tab:params}. 
We create a mock observation $\theta_p^{\mathrm{obs}}$ for CNNs and PSDs by taking the most likely prediction for $\theta_t^{\mathrm{fid}}$.
Figure~\ref{fig:constraints} shows the comparison of the constraints for the combined probes analysis, with CNN result in pink and PSD in blue. 
The regions correspond to 68\% and 95\% confidence intervals.

\begin{figure*}
    \includegraphics[width=0.43\textwidth,valign=t]{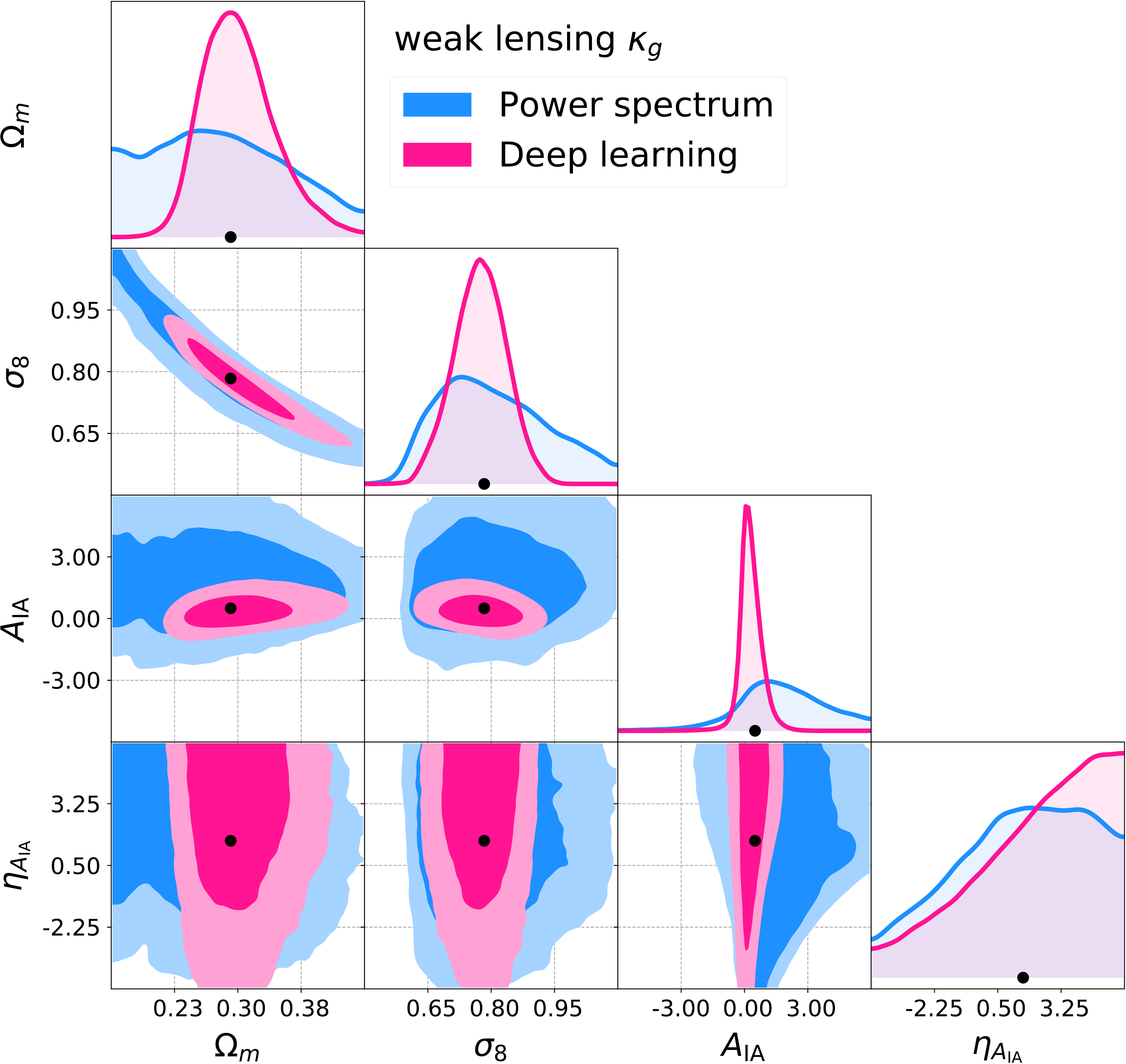} 
    \hspace{1em}
    \includegraphics[width=0.53\textwidth,valign=t]{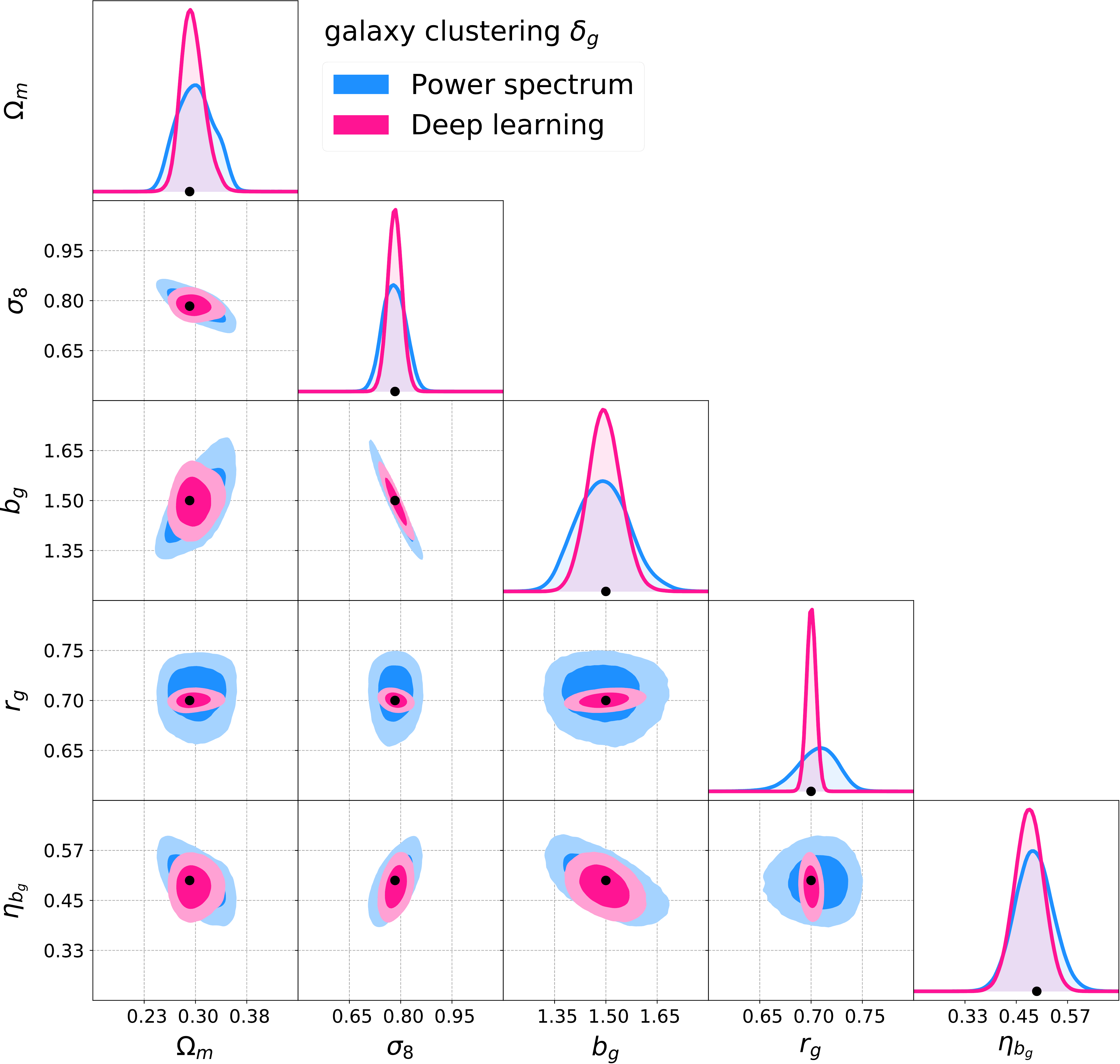} 
    \caption{
    Constraints for weak lensing (left) and galaxy clustering (right), with deep learning (pink) and power spectra (blue), for the main model with linear galaxy bias.
    As in Figure~\ref{fig:constraints}, the black dots show the true parameters.
    }
    \label{fig:constraints_individual}
\end{figure*}

The CNN yields more precise measurement than the PSD for all parameters considered.
The strongest constraining power gain is for the intrinsic alignment amplitude \AIA, which is of the order of 8\x. 
CNN measurement effectively breaks the degeneracy between the IA and other parameters.
We explore this further in following paragraphs.
Another significant gain comes from breaking the degeneracy between stochasticity \rstoch\ and \sigeight\ and \biasg, with \rstoch\ improved by 4\x.
The measurement of galaxy bias \biasg\ and its evolution \etab\ is also decorrelated, with respective improvements of 1.5\x\ and 1.2\x.
Moreover, for CNN, the cosmology parameters \omatter\ and \sigeight\  are also significantly less dependent on the bias evolution, with constraints improved by 1.6\x\ and 1.3\x, respectively.
The redshift evolution of intrinsic alignments is weakly constrained, compared to no constraint by PSD.
Overall, breaking of these degeneracies contributes to a very substantial improvement of the \omatter --\sigeight\ figure of merit: the FoM is 15\x\ higher for the deep learning analysis.

Figure~\ref{fig:constraints_individual} shows the equivalent comparison for individual probes: weak lensing and galaxy clustering.
For weak lensing, the CNNs are able to break the \sigeight\ -- \omatter\ ``banana'' degeneracy, as previously shown by \citetalias{Fluri2019kids}.
There also is an improvement for the \AIA\ parameter, as also found in \citetalias{Fluri2019kids}.
It is, however, significantly smaller than for the combined probes analysis.
Moreover, the IA redshift evolution \etaIA\ is unconstrained for both CNNs and PSDs.
This suggests that the \etaIA\ information from combined probes come purely from a better understanding of the $\kappa_g$ and $\delta_g$ maps jointly, since the clustering maps are independent of this IA.

The constraints from galaxy clustering are also significantly improved. 
The measurement of stochasticity benefits the most from using CNNs, with 4\x\ better precision.
Again, the CNNs break the degeneracy between \biasg\ and its evolution \etab, which in turn helps with constraining \sigeight\ and \omatter.
Generally, these parameters are improved by a factor of around 1.3\x\ - 1.8\x, which is also a significant information gain.
The degeneracy between \sigeight\ and \biasg\ is slightly reduced, with a precision gains on the level of 1.6\x.

The comparison for all parameters and probes is shown in Table~\ref{tab:gains}. 
The gain value was calculated using 200 mock observations selected at random from all the predictions for the fiducial parameter set $\theta_t^{\mathrm{fid}}$.
The value cited is the median of the ratios of parameter standard deviations.

\begin{figure}
    \vspace{0.5cm}
    \includegraphics[width=0.475\textwidth]{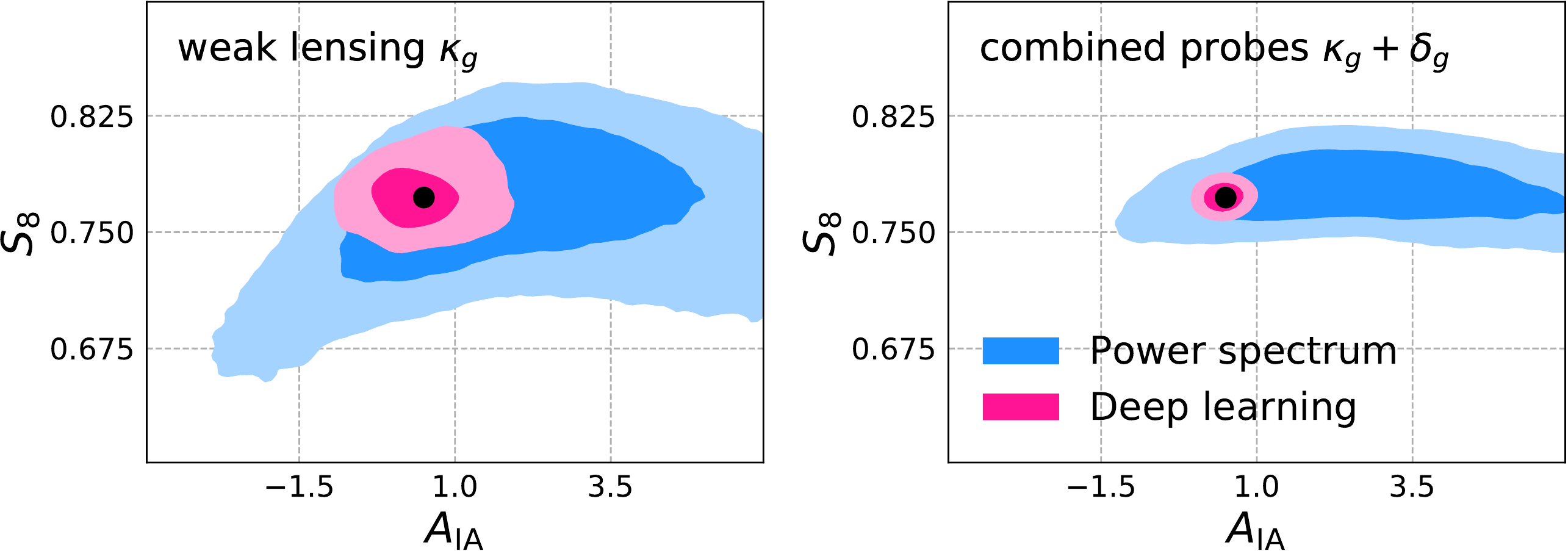} 
    \caption{
    Constraints on \Seight, intrinsic alignments \AIA, and IA evolution \etaIA, for the combined probes (left) and lensing only (right), using the linear galaxy bias model.
    }
    \label{fig:constraints_IA}
\end{figure}

We further investigate the powerful IA constraints obtained by the CNN.
Figure~\ref{fig:constraints_IA} shows the uncertainty on \Seight\ versus \AIA, similarly to \cite{Fischbacher2022ia}.
For the PSD, there is a clear degeneracy between these parameters of both lensing and combined probes.
For lensing, the degeneracy is still present also for the CNN.
It is, however, efficiently broken by CNNs for the probe combination, yielding a 2.3\x\ improvement in \Seight, on average.

\begin{figure}
    \includegraphics[width=0.4755\textwidth]{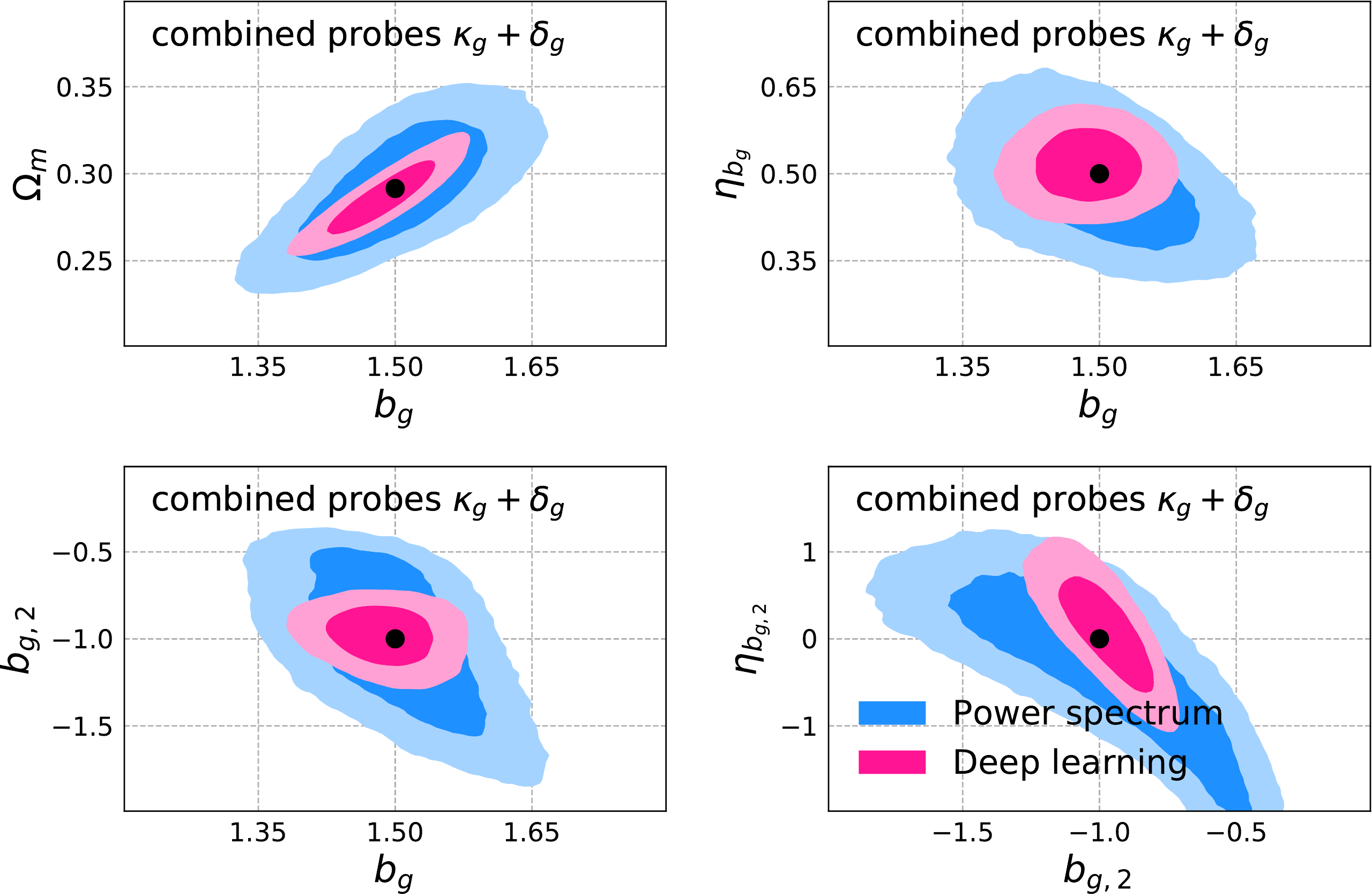} 
    \caption{
    Galaxy bias constraints for the non-linear bias model for combined probes.
    Deep learning results are shown in pink, while power spectra in blue.
    To give some intuition about the meaning of the \etab\ parameter, we calculated its uncertainty using bias measurements from DES Y3 \cite{Des2022combined} and obtained $\sigma\left[ \etab \right] = 0.21$. 
    }
    \label{fig:constraints_nonlin}
\end{figure}

We present the results for the non-linear galaxy biasing model analysis in Figure~\ref{fig:constraints_nonlin}. 
For the sake of clarity, we limit the panels to \omatter, \biasg, the non-linear bias strength \biasnl, as well as its evolution \etabnl.
For the PSD, a clear degeneracy appears between the linear and non-linear galaxy bias parameters, \biasg\ and \biasnl, as well as between the \biasnl\ and its redshift evolution \etabnl.
We also notice that the PSD constraints on linear galaxy bias evolution \etab\ are much worse than for the linear bias case, despite using smaller smoothing scales.
Deep learning is able to break these degeneracies and constrain the \biasg\ and \biasnl\ with 1.4\x\ and 2.7\x\ precision increase, respectively.
The clustering-only analysis of the non-linear model gives similar gains, as shown in Table~\ref{tab:gains}.

\section{Where is the information coming from?}
\label{sec:sensitivity}


\begin{figure*}
    \includegraphics[width=0.45\textwidth,valign=t]{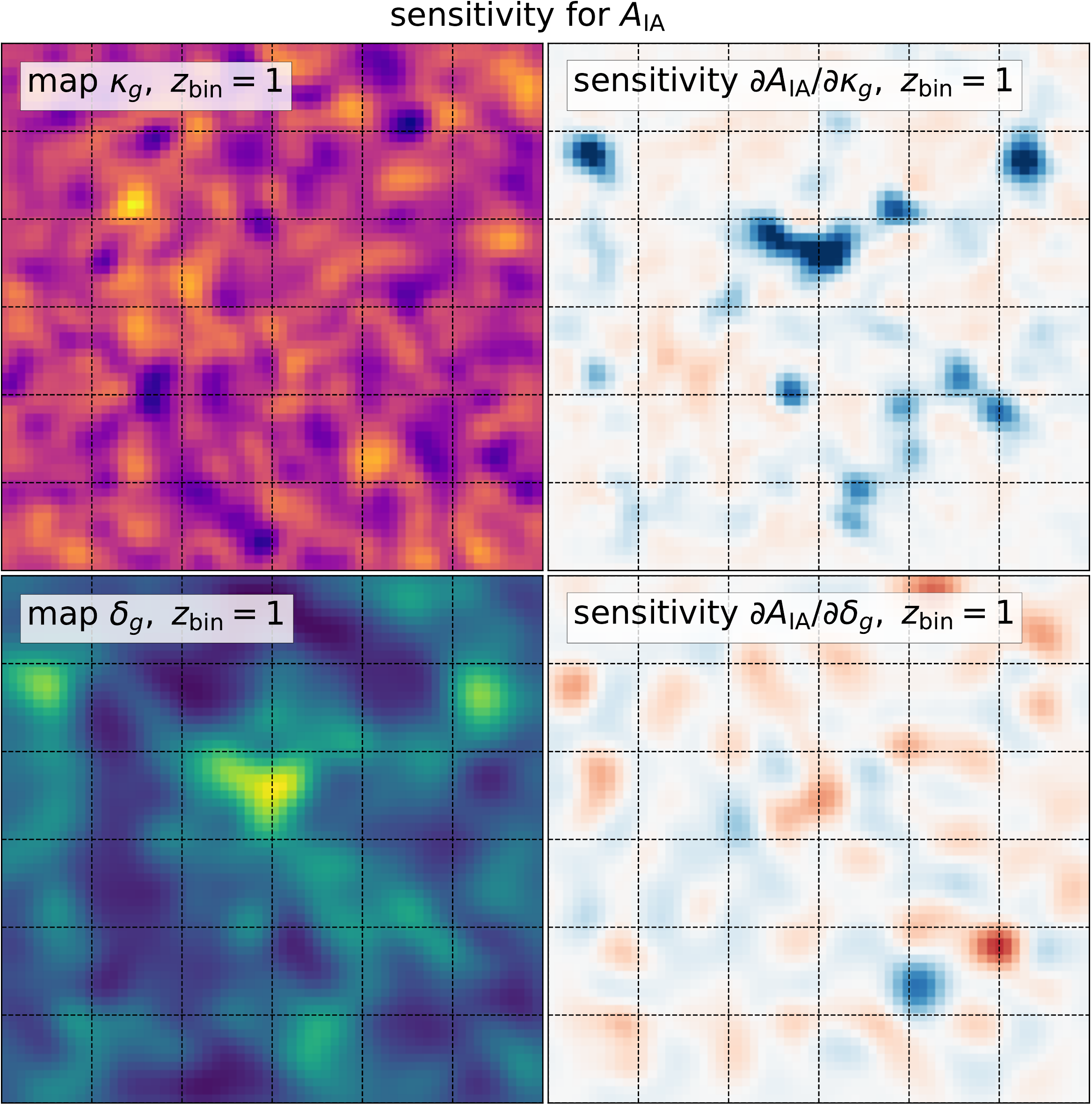} 
    \hspace{1em}
    \includegraphics[width=0.45\textwidth,valign=t]{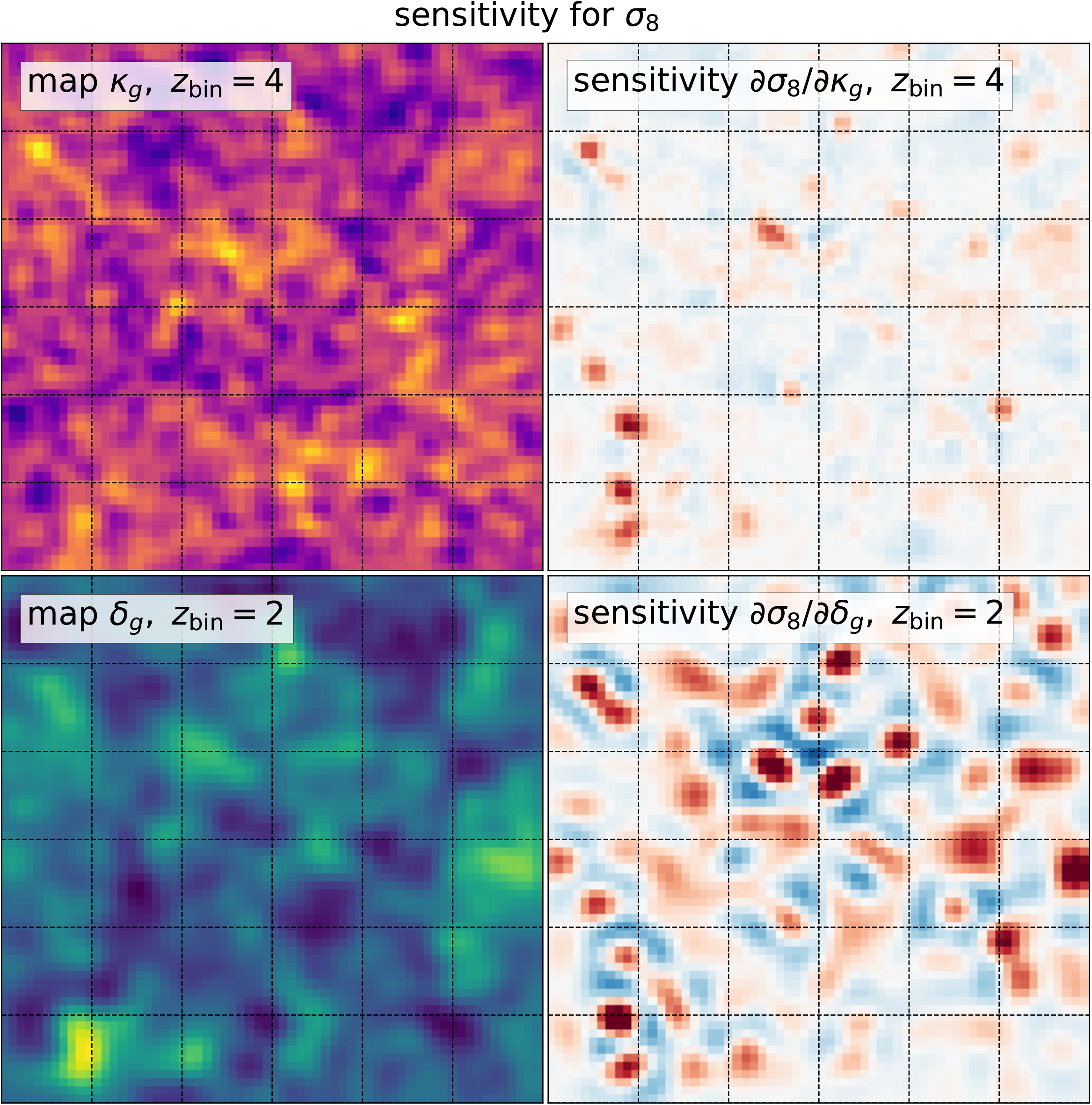} 
    \caption{
    Sensitivity analysis for the \AIA\ (left panels) and \sigeight\ (right penels).
    We show the $\kappa_m$ and $\delta_g$ maps on the left, and the corresponding sensitivity maps on the right.
    To suppress the noise, the lensing maps are additionally smoothed using a Gaussian kernel with $\sigma=1$ pixel.
    Regions with higher absolute sensitivity contribute more to the decision of the network about the predicted value of \AIA\ and \sigeight.
    Note the correspondence between the galaxy overdensity map and the lensing sensitivity map for bin 1; the areas of large overdensities are weighted higher in the $\kappa_m$ map.
    } 
    \label{fig:sensitivity}
\end{figure*}

Given the significant improvement in constraining power for the CNNs, it is important to gain an intuitive understanding about where is the additional information coming from.
Firstly, the non-linear part of the matter density field contains significant information, even at intermediate scales \cite{Fluri2022kids,Zuercher2022despeaks}.
Multiple approaches have been designed to extract it from the lensing convergence field \cite{kratochvil2011probing,zuercher2020forecast,HarnoisDeraps2020peaks}, most recently tomographic shear peak statistics.
Moreover, the full forward analysis on map level enables us to include the phase information of the field; the 2-pt functions are effectively discarding it.

But this may not be the only mechanism employed by the CNNs to increase constraining power.
To investigate this further, we look at \emph{sensitivity} maps.
For a trained network, one can create such maps by calculating gradients the network output with respect to the image pixels. 
In our configuration, where the network outputs are summaries that correspond directly to a given parameter, such a map is easy to create and interpret.
Given a set of input maps $m$ and output summaries $\theta_p$, the sensitivity map $s$ is defined as $s = \partial \theta_p/\partial m$.
Changes in the regions of the map with highest gradient absolute value will have the most impact on the final prediction, while regions with sensitivity close to zero have almost no impact.
However, as all pixels are considered, it does not directly indicate what the value of the predicted parameter will be. 
Intuitively, one may also interpret these maps as a weight function, which shows which parts of the map are used by the network, and which are ignored.
This is a first order analysis, i.e. it ignores the correlations between pixels and only focuses on the leading term.
An approach similar to this has been introduced for lensing maps by  \cite{ZorillaMatilla2020interpreting}, and more broadly in the field of machine learning called \emph{interpretability} \cite{olah2018building,Samek2021explaining}.

We focus on the \AIA\ and \sigeight\ parameter constrains, as these display the strongest gain and the most benefit from the combined probes analysis.
Figure~\ref{fig:sensitivity} shows input maps and the corresponding sensitivity maps.
For the purpose of this figure, additional Gaussian smoothing was applied to the $\kappa_g$ maps to suppress the noise and make them easier to read.
For \AIA, we show the maps for redshift bin 1, which contain the largest sensitivity signal. 
We notice that there is a large overdensity in the middle of the $\delta_g^1$ map, as well as in the upper left and right corners.
The sensitivity to the $\kappa_g^1$ is focused on precisely these regions, while the rest of the map is practically ignored.
This makes intuitive sense: the IA signal is expected to be present in the convergence maps at the positions of overdensities.
For \sigeight, the sensitivity is the strongest for $\kappa_g^{2}$ and $\delta_g^{4}$.
The mean redshifts of these bins are $\langle z^{2} \rangle \approx 0.5$ and $\langle z^{4} \rangle\approx1$, which is exactly the configuration that leads to the strongest lensing signal.
The $\kappa_g$ sensitivity map seem to be focused at positions of peaks, additionally matching their sizes.
The $\delta_g$ sensitivity is also focused on position of peaks;
it seems that the CNNs detect overdensities in the $\kappa_g$ map and look for corresponding signal in the $\delta_g$ map.

The fact that the network only takes information from specific regions of the map seems quite simple, but has profound consequences, especially for the  \AIA\ constraint.
While the power spectrum analysis also takes advantage of the $\kappa_g \times \delta_g$ cross-correlation, it does not employ such weighting.
There, the total cross-correlation signal is an average of regions that are important with regions that are not informative at all.
This leads to the dilution of the signal, as the rest of the field contributes only random noise.
This in turn lowers the signal-to-noise on the measured parameter and consequently makes the constraints less precise.
This can explain the $8\times$ information gain for \AIA, as it draws information mostly from the lowest redshift bin.
For that bin, the maps are usually sparsely populated by overdensities, as seen the bottom left panel of Figure~\ref{fig:sensitivity}, the dilution of the signal is the most significant.
The CNNs automatically select the important regions and ignore the rest, which avoids signal dilution by unimportant regions.

While this interpretation is qualitative and subjective, it can already shed some initial light on the origin of the information used by CNNs.
The impact of non-Gaussian information, access to phases of the map, and optimal map weighting, could all be quantitatively investigated further.
However, we leave it to the dedicated future work.


\section{Conclusions}
\label{sec:conclusions}


\begin{table*}
    \centering
     \begin{tabularx}{0.8\textwidth}{lXrXrrXrr}
                         \toprule
                           & & \multicolumn{1}{c}{\textbf{Weak Lensing}} &  & \multicolumn{2}{c}{\textbf{Galaxy Clustering}} &  & \multicolumn{2}{c}{\textbf{Combined Probes}} \\
                         \cmidrule(lr){5-6} 
                         \cmidrule(lr){8-9} 
                              & & \ & & large scales & small scales & & large scales &small scales\\
                             \midrule
                   
\omatter                                 & &                            1.7$\times$                                       &  &                             1.5$\times$                                       &                             1.3$\times$                                       &  &                             1.6$\times$                                       &                             1.5$\times$  \\
\sigeight                                & &                            2.1$\times$                                       &  &                             1.6$\times$                                       &                             1.2$\times$                                       &  &                             1.3$\times$                                       &                             1.2$\times$  \\
\Seight                                  & &                            3.0$\times$                                       &  &                             0.9$\times$                                       &                             1.3$\times$                                       &  &                             2.3$\times$                                       &                             1.5$\times$  \\
\rm{FoM} \omatter -- \sigeight           & &                           11.0$\times$                                       &  &                             1.7$\times$                                       &                             6.2$\times$                                       &  &                            14.9$\times$                                       &                            13.2$\times$  \\
\AIA                                     & &                            3.9$\times$                                       &  &                                     --                                        &                                     --                                        &  &                             8.3$\times$                                       &                             8.1$\times$  \\
\biasg                                   & &                                    --                                        &  &                             1.6$\times$                                       &                             1.5$\times$                                       &  &                             1.5$\times$                                       &                             1.4$\times$  \\
\rstoch                                  & &                                    --                                        &  &                             4.4$\times$                                       &                             4.3$\times$                                       &  &                             3.6$\times$                                       &                             4.8$\times$  \\
\etaIA                                   & &                            1.0$\times$                                       &  &                                     --                                        &                                     --                                        &  &                             1.4$\times$                                       &                             1.6$\times$  \\
\etab                                    & &                                    --                                        &  &                             1.3$\times$                                       &                             1.7$\times$                                       &  &                             1.2$\times$                                       &                             1.7$\times$  \\
\biasnl                                  & &                                    --                                        &  &                                     --                                        &                             2.6$\times$                                       &  &                                     --                                        &                             2.7$\times$  \\
\etabnl                                  & &                                    --                                        &  &                                     --                                        &                             1.4$\times$                                       &  &                                     --                                        &                             1.7$\times$  \\

 \bottomrule
                     \end{tabularx}
                    
    \caption{
    Measurement precision gain using with DeepLSS over the power spectrum analysis. 
    Results for galaxy clustering and combined probes are given for two redshift-dependent smoothing scales of galaxy position maps, with FWHM of 8 Mpc/h and 4 Mpc/h for ``large scales'' and ``small scales'', respectively. 
    Weak lensing maps were always smoothed with 4 Mpc/h kernel.
    The difference is expressed as $\mathrm{std}(\theta_{\mathrm{PSD}})/\mathrm{std}(\theta_{\mathrm{CNN}})$, where value of $1\times$ means no precision gain.
    It was calculated as a median of 200 noisy mock observations taken at the fiducial cosmology.  
    }
    \label{tab:gains}
\end{table*}

We present a novel approach to analysis of large scale structure data by using full forward modelling on map level and its interpretation with deep learning.
In a method dubbed \textsc{DeepLSS}, we create a combined probes analysis of weak lensing convergence $\kappa_g$ and galaxy clustering maps $\delta_g$.
We focus on improving the constraints on the \omatter\ and \sigeight\ parameters through breaking their degeneracies with intrinsic galaxy alignments and galaxy biasing; specifically between \Seight\ and intrinsic alignment amplitude \AIA, linear bias \biasg\ and matter density \omatter\, and linear \biasg\ and non-linear bias \biasnl.
Internal degeneracies between these parameters and their redshift evolution cause further increase in marginalized parameter uncertainty.

We create consistent sets of simulations of galaxy clustering and weak lensing within the space of two models: 
(i) large smoothing and linear bias, and
(ii) small smoothing and non-linear bias.
These models have 7 and 9 free parameters, respectively.

As the most common way to analyze combined probes of LSS is to use 2-pt functions, we focus on a fair comparison between the deep learning and the power spectrum methods.
We create a set of residual convolutional neural networks on the individual probes $\kappa_g$ and $\delta_g$, and the probe combination $\kappa_g+\delta_g$.
We pass the sets of tomographic maps to the networks as channels: 4 for individual probes and 8 for probe combination.
The networks are trained using likelihood loss between outputs $\theta_p$ and the true input parameters $\theta_t$.
This way, the network creates informative output summaries corresponding to the model parameters.
We interpret these summaries by performing conditional density estimation on the likelihood $p(\theta_p|\theta_t)$.
The constraints are the obtained using Bayesian analysis with a Markov Chain Monte Carlo sampler.

We report a remarkable ability of deep learning to break degeneracies between the LSS parameters and cosmology, as compared to the power spectrum method.
The most significant gain is for intrinsic galaxy alignments, where the \Seight\ -- \AIA\ correlation is effectively broken and the \AIA\ constraint improves by a factor of 8\x.
Galaxy stochasticity constraints improve by a factor of 3\x.
The improvement in the galaxy biasing sector is around 1.3\x.
For the non-linear model, the CNN effectively breaks the degeneracy between the linear and non-linear bias parameters, \biasg\ and \biasnl.
We also observe a significant gain in the constraining power of the redshift evolution of intrinsic alignments \etaIA, where the gain is 1.4\x.
Similarly, for $z$-evolution of galaxy biasing, the CNN achieves 1.2\x\ improvement for \etab\ together with 1.7\x\ improvement in \etabnl.
Overall, breaking these degeneracies leads to a very significant gain in constraints along the \sigeight\ and \omatter\ degeneracy, with the figure of merit improved by 15\x.
 
We investigate the source of the information gained in the CNN analysis by looking at sensitivity maps.
These maps show which regions of the input $\kappa_g$ and $\delta_g$ maps have the most impact on the network's prediction.
We focus on the sensitivity to \AIA\ and \sigeight, which gain the most from the \textsc{DeepLSS} analysis.
We notice that, for the \AIA\ parameter, the network draws most information from $\kappa_g$ maps in the regions corresponding to high overdensities in the $\delta_g$ maps, while other pixels are heavily down-weighted. 
This picture suggests the following intuitive explanation.
The highlighted regions are exactly where the IA signal comes from: galaxy shape alignment around overdense regions at the same redshift.
Ignoring the rest of the map decreases the dilution of the signal by the noise coming from uninformative regions.
The PSD method, on the other hand, does not perform such weighting and considers all pixels in the field equally, which leads to increased impact of the noise from regions that have no signal.
As high density regions in low redshift maps are rare, the gain of the networks due to this effect becomes very significant.
This interpretation, however, is probably not the full story: the PSD also ignores the phases of the maps, as well as all non-Gaussian information.
Further work would be needed to gain more insights about relative importance of these effects.

This work serves as a demonstration of the constraining power of the \textsc{DeepLSS} method and the exact gains are specific to the analysis configuration used here.
Our theory and data models, while simpler than for a typical survey analysis, are highly realistic and contain most of the needed degrees of freedom.
This suggests that the gains presented here should be also recovered by a LSS analysis with more precise theory and data modeling.

More development on the forward modeling side is needed before practical deployment of this method.
Firstly, the galaxy biasing prescriptions could be compared to the more advanced modeling using Halo Occupation Distributions (HOD), Subhalo Abundance Matching (SHAM) models, or using rapid Halo and Subhalo simulations with PINOCCHIO \cite{Berner2021rapid,Taffoni2002pinnochio}, and others \cite{DeRose2019buzzard,Fosalba2015mice,Friedrich2022pdf,HarnoisDeraps2018simulations}.
Secondly, the baryonic effects \cite{Schneider2020xray} should be included in the forward model, similarly to \cite{Fluri2022kids}.
Finally, given the large constraining power gain for intrinsic alignments, it may be feasible to constrain much more complex models with AI, such as Tidal Alignment Tidal Torque (TATT) \cite{Blazek2019beyond}, or color-dependent IAs \cite{Samuroff2019desia}.
Conversely, one may imagine using the clustering maps with even larger smoothing scales for simple AI models. 
This would reduce the dependence of the forward model on details of galaxy clustering modeling, non-linear effects, and systematics, while sill providing a large improvement on \AIA.
Such large scale-only model would be much easier to implement in practice.

A careful study of the impact of survey systematic effects, such as galaxy selection function uncertainty, redshift measurement errors, and shear calibration, should be performed before the practical application of \textsc{DeepLSS}.
Their requirements may prove to be different for the AI analysis compared to the 2-pt.
For shear peak statistics, the DES Y3 analysis \cite{Zuercher2022despeaks} revisited the shear calibration requirements; similar strategies can be used here.

The machine learning methods used in this work were very simple and we did not optimize them for better information gain.
We did, however, optimize the architectures of the networks interpreting the PSD vectors.
It is therefore conceivable that the gain from CNNs can be further increased by using larger models and better optimization techniques available now in the field of machine learning.

In the near future, large lightcone simulation grids will become more available \cite{Fluri2022kids,Kacprzak2022cosmogrid}.
Along with the growth in processing capacities of hardware, the full forward model will become more practical and reproducible.
Given that the advantage of this analysis is substantial and able to break degeneracies in the model, it is therefore possible that a map-level probe combination using deep learning may play an important role in future large scale structure surveys.


\section*{Acknowledgements}


TK thanks Javier Francisco Castander, Pablo Fosalba, Martin Crocce and Enrique Enrique Gazta\~naga for very helpful discussions.
TK also thanks Marc Caubet Serrabou for excellent HPC support for the Merlin cluster at PSI, as well as Andreas Adelmann for access to the Gwendolen machine.
We would like to thank Aurelien Lucchi for ongoing collaboration on deep learning for cosmology and providing machine learning expertise and advice.
TK thanks David Bacon for support and useful discussions.


\section*{Data Availability}

 
The code is made publicly available at \href{https://github.com/tomaszkacprzak/DeepLSS}{\url{https://github.com/tomaszkacprzak/DeepLSS}}.
The data will be made public as a part of larger data release of \textsc{CosmoGrid}.
In the meantime, the data can be made available upon request.


\bibliography{bibliography}

\begin{thebibliography}{69}%
\makeatletter
\providecommand \@ifxundefined [1]{%
 \@ifx{#1\undefined}
}%
\providecommand \@ifnum [1]{%
 \ifnum #1\expandafter \@firstoftwo
 \else \expandafter \@secondoftwo
 \fi
}%
\providecommand \@ifx [1]{%
 \ifx #1\expandafter \@firstoftwo
 \else \expandafter \@secondoftwo
 \fi
}%
\providecommand \natexlab [1]{#1}%
\providecommand \enquote  [1]{``#1''}%
\providecommand \bibnamefont  [1]{#1}%
\providecommand \bibfnamefont [1]{#1}%
\providecommand \citenamefont [1]{#1}%
\providecommand \href@noop [0]{\@secondoftwo}%
\providecommand \href [0]{\begingroup \@sanitize@url \@href}%
\providecommand \@href[1]{\@@startlink{#1}\@@href}%
\providecommand \@@href[1]{\endgroup#1\@@endlink}%
\providecommand \@sanitize@url [0]{\catcode `\\12\catcode `\$12\catcode
  `\&12\catcode `\#12\catcode `\^12\catcode `\_12\catcode `\%12\relax}%
\providecommand \@@startlink[1]{}%
\providecommand \@@endlink[0]{}%
\providecommand \url  [0]{\begingroup\@sanitize@url \@url }%
\providecommand \@url [1]{\endgroup\@href {#1}{\urlprefix }}%
\providecommand \urlprefix  [0]{URL }%
\providecommand \Eprint [0]{\href }%
\providecommand \doibase [0]{https://doi.org/}%
\providecommand \selectlanguage [0]{\@gobble}%
\providecommand \bibinfo  [0]{\@secondoftwo}%
\providecommand \bibfield  [0]{\@secondoftwo}%
\providecommand \translation [1]{[#1]}%
\providecommand \BibitemOpen [0]{}%
\providecommand \bibitemStop [0]{}%
\providecommand \bibitemNoStop [0]{.\EOS\space}%
\providecommand \EOS [0]{\spacefactor3000\relax}%
\providecommand \BibitemShut  [1]{\csname bibitem#1\endcsname}%
\let\auto@bib@innerbib\@empty
\bibitem [{\citenamefont {{Kilbinger}}(2015)}]{Kilbinger2015review}%
  \BibitemOpen
  \bibfield  {author} {\bibinfo {author} {\bibfnamefont {M.}~\bibnamefont
  {{Kilbinger}}},\ }\bibfield  {title} {\bibinfo {title} {{Cosmology with
  cosmic shear observations: a review}},\ }\href
  {https://doi.org/10.1088/0034-4885/78/8/086901} {\bibfield  {journal}
  {\bibinfo  {journal} {Reports on Progress in Physics}\ }\textbf {\bibinfo
  {volume} {78}},\ \bibinfo {eid} {086901} (\bibinfo {year}
  {2015})}\BibitemShut {NoStop}%
\bibitem [{\citenamefont {{Zhan}}\ and\ \citenamefont
  {{Tyson}}(2018)}]{Zhan2018lsst}%
  \BibitemOpen
  \bibfield  {author} {\bibinfo {author} {\bibfnamefont {H.}~\bibnamefont
  {{Zhan}}}\ and\ \bibinfo {author} {\bibfnamefont {J.~A.}\ \bibnamefont
  {{Tyson}}},\ }\bibfield  {title} {\bibinfo {title} {{Cosmology with the Large
  Synoptic Survey Telescope: an overview}},\ }\href
  {https://doi.org/10.1088/1361-6633/aab1bd} {\bibfield  {journal} {\bibinfo
  {journal} {Reports on Progress in Physics}\ }\textbf {\bibinfo {volume}
  {81}},\ \bibinfo {eid} {066901} (\bibinfo {year} {2018})},\ \Eprint
  {https://arxiv.org/abs/1707.06948} {arXiv:1707.06948 [astro-ph.CO]}
  \BibitemShut {NoStop}%
\bibitem [{\citenamefont {{Albrecht}}\ \emph {et~al.}(2006)\citenamefont
  {{Albrecht}}, \citenamefont {{Bernstein}}, \citenamefont {{Cahn}},
  \citenamefont {{Freedman}}, \citenamefont {{Hewitt}}, \citenamefont {{Hu}},
  \citenamefont {{Huth}}, \citenamefont {{Kamionkowski}}, \citenamefont
  {{Kolb}}, \citenamefont {{Knox}}, \citenamefont {{Mather}}, \citenamefont
  {{Staggs}},\ and\ \citenamefont {{Suntzeff}}}]{Albrecht2006darkenergy}%
  \BibitemOpen
  \bibfield  {author} {\bibinfo {author} {\bibfnamefont {A.}~\bibnamefont
  {{Albrecht}}}, \bibinfo {author} {\bibfnamefont {G.}~\bibnamefont
  {{Bernstein}}}, \bibinfo {author} {\bibfnamefont {R.}~\bibnamefont {{Cahn}}},
  \bibinfo {author} {\bibfnamefont {W.~L.}\ \bibnamefont {{Freedman}}},
  \bibinfo {author} {\bibfnamefont {J.}~\bibnamefont {{Hewitt}}}, \bibinfo
  {author} {\bibfnamefont {W.}~\bibnamefont {{Hu}}}, \bibinfo {author}
  {\bibfnamefont {J.}~\bibnamefont {{Huth}}}, \bibinfo {author} {\bibfnamefont
  {M.}~\bibnamefont {{Kamionkowski}}}, \bibinfo {author} {\bibfnamefont
  {E.~W.}\ \bibnamefont {{Kolb}}}, \bibinfo {author} {\bibfnamefont
  {L.}~\bibnamefont {{Knox}}}, \bibinfo {author} {\bibfnamefont {J.~C.}\
  \bibnamefont {{Mather}}}, \bibinfo {author} {\bibfnamefont {S.}~\bibnamefont
  {{Staggs}}},\ and\ \bibinfo {author} {\bibfnamefont {N.~B.}\ \bibnamefont
  {{Suntzeff}}},\ }\bibfield  {title} {\bibinfo {title} {{Report of the Dark
  Energy Task Force}},\ }\href@noop {} {\bibfield  {journal} {\bibinfo
  {journal} {ArXiv Astrophysics}\ } (\bibinfo {year} {2006})}\BibitemShut
  {NoStop}%
\bibitem [{Note1()}]{Note1}%
  \BibitemOpen
  \bibinfo {note} {\protect \url {darkenergysurvey.org}}\BibitemShut {NoStop}%
\bibitem [{Note2()}]{Note2}%
  \BibitemOpen
  \bibinfo {note} {\protect \url {kids.strw.leidenuniv.nl}}\BibitemShut
  {NoStop}%
\bibitem [{Note3()}]{Note3}%
  \BibitemOpen
  \bibinfo {note} {\protect \url {hsc.mtk.nao.ac.jp/ssp/survey}}\BibitemShut
  {NoStop}%
\bibitem [{\citenamefont {Abbott}\ \emph {et~al.}(2022)\citenamefont {Abbott},
  \citenamefont {Aguena}, \citenamefont {Alarcon}, \citenamefont {Allam},
  \citenamefont {Alves}, \citenamefont {Amon}, \citenamefont
  {Andrade-Oliveira}, \citenamefont {Annis}, \citenamefont {Avila},
  \citenamefont {Bacon}, \citenamefont {Baxter}, \citenamefont {Bechtol},
  \citenamefont {Becker}, \citenamefont {Bernstein}, \citenamefont {Bhargava},
  \citenamefont {Birrer}, \citenamefont {Blazek}, \citenamefont
  {Brandao-Souza}, \citenamefont {Bridle}, \citenamefont {Brooks},
  \citenamefont {Buckley-Geer}, \citenamefont {Burke}, \citenamefont {Camacho},
  \citenamefont {Campos}, \citenamefont {Carnero~Rosell}, \citenamefont
  {Carrasco~Kind}, \citenamefont {Carretero}, \citenamefont {Castander},
  \citenamefont {Cawthon}, \citenamefont {Chang}, \citenamefont {Chen},
  \citenamefont {Chen}, \citenamefont {Choi}, \citenamefont {Conselice},
  \citenamefont {Cordero}, \citenamefont {Costanzi}, \citenamefont {Crocce},
  \citenamefont {da~Costa}, \citenamefont {da~Silva~Pereira}, \citenamefont
  {Davis}, \citenamefont {Davis}, \citenamefont {De~Vicente}, \citenamefont
  {DeRose}, \citenamefont {Desai}, \citenamefont {Di~Valentino}, \citenamefont
  {Diehl}, \citenamefont {Dietrich}, \citenamefont {Dodelson}, \citenamefont
  {Doel}, \citenamefont {Doux}, \citenamefont {Drlica-Wagner}, \citenamefont
  {Eckert}, \citenamefont {Eifler}, \citenamefont {Elsner}, \citenamefont
  {Elvin-Poole}, \citenamefont {Everett}, \citenamefont {Evrard}, \citenamefont
  {Fang}, \citenamefont {Farahi}, \citenamefont {Fernandez}, \citenamefont
  {Ferrero}, \citenamefont {Fert\'e}, \citenamefont {Fosalba}, \citenamefont
  {Friedrich}, \citenamefont {Frieman}, \citenamefont {Garc\'{\i}a-Bellido},
  \citenamefont {Gatti}, \citenamefont {Gaztanaga}, \citenamefont {Gerdes},
  \citenamefont {Giannantonio}, \citenamefont {Giannini}, \citenamefont
  {Gruen}, \citenamefont {Gruendl}, \citenamefont {Gschwend}, \citenamefont
  {Gutierrez}, \citenamefont {Harrison}, \citenamefont {Hartley}, \citenamefont
  {Herner}, \citenamefont {Hinton}, \citenamefont {Hollowood}, \citenamefont
  {Honscheid}, \citenamefont {Hoyle}, \citenamefont {Huff}, \citenamefont
  {Huterer}, \citenamefont {Jain}, \citenamefont {James}, \citenamefont
  {Jarvis}, \citenamefont {Jeffrey}, \citenamefont {Jeltema}, \citenamefont
  {Kovacs}, \citenamefont {Krause}, \citenamefont {Kron}, \citenamefont
  {Kuehn}, \citenamefont {Kuropatkin}, \citenamefont {Lahav}, \citenamefont
  {Leget}, \citenamefont {Lemos}, \citenamefont {Liddle}, \citenamefont
  {Lidman}, \citenamefont {Lima}, \citenamefont {Lin}, \citenamefont
  {MacCrann}, \citenamefont {Maia}, \citenamefont {Marshall}, \citenamefont
  {Martini}, \citenamefont {McCullough}, \citenamefont {Melchior},
  \citenamefont {Mena-Fern\'andez}, \citenamefont {Menanteau}, \citenamefont
  {Miquel}, \citenamefont {Mohr}, \citenamefont {Morgan}, \citenamefont {Muir},
  \citenamefont {Myles}, \citenamefont {Nadathur}, \citenamefont
  {Navarro-Alsina}, \citenamefont {Nichol}, \citenamefont {Ogando},
  \citenamefont {Omori}, \citenamefont {Palmese}, \citenamefont {Pandey},
  \citenamefont {Park}, \citenamefont {Paz-Chinch\'on}, \citenamefont
  {Petravick}, \citenamefont {Pieres}, \citenamefont {Plazas~Malag\'on},
  \citenamefont {Porredon}, \citenamefont {Prat}, \citenamefont {Raveri},
  \citenamefont {Rodriguez-Monroy}, \citenamefont {Rollins}, \citenamefont
  {Romer}, \citenamefont {Roodman}, \citenamefont {Rosenfeld}, \citenamefont
  {Ross}, \citenamefont {Rykoff}, \citenamefont {Samuroff}, \citenamefont
  {S\'anchez}, \citenamefont {Sanchez}, \citenamefont {Sanchez}, \citenamefont
  {Sanchez~Cid}, \citenamefont {Scarpine}, \citenamefont {Schubnell},
  \citenamefont {Scolnic}, \citenamefont {Secco}, \citenamefont {Serrano},
  \citenamefont {Sevilla-Noarbe}, \citenamefont {Sheldon}, \citenamefont
  {Shin}, \citenamefont {Smith}, \citenamefont {Soares-Santos}, \citenamefont
  {Suchyta}, \citenamefont {Swanson}, \citenamefont {Tabbutt}, \citenamefont
  {Tarle}, \citenamefont {Thomas}, \citenamefont {To}, \citenamefont {Troja},
  \citenamefont {Troxel}, \citenamefont {Tucker}, \citenamefont {Tutusaus},
  \citenamefont {Varga}, \citenamefont {Walker}, \citenamefont {Weaverdyck},
  \citenamefont {Wechsler}, \citenamefont {Weller}, \citenamefont {Yanny},
  \citenamefont {Yin}, \citenamefont {Zhang},\ and\ \citenamefont
  {Zuntz}}]{Des2022combined}%
  \BibitemOpen
  \bibfield  {author} {\bibinfo {author} {\bibfnamefont {T.~M.~C.}\
  \bibnamefont {Abbott}}, \bibinfo {author} {\bibfnamefont {M.}~\bibnamefont
  {Aguena}}, \bibinfo {author} {\bibfnamefont {A.}~\bibnamefont {Alarcon}},
  \bibinfo {author} {\bibfnamefont {S.}~\bibnamefont {Allam}}, \bibinfo
  {author} {\bibfnamefont {O.}~\bibnamefont {Alves}}, \bibinfo {author}
  {\bibfnamefont {A.}~\bibnamefont {Amon}}, \bibinfo {author} {\bibfnamefont
  {F.}~\bibnamefont {Andrade-Oliveira}}, \bibinfo {author} {\bibfnamefont
  {J.}~\bibnamefont {Annis}}, \bibinfo {author} {\bibfnamefont
  {S.}~\bibnamefont {Avila}}, \bibinfo {author} {\bibfnamefont
  {D.}~\bibnamefont {Bacon}}, \bibinfo {author} {\bibfnamefont
  {E.}~\bibnamefont {Baxter}}, \bibinfo {author} {\bibfnamefont
  {K.}~\bibnamefont {Bechtol}}, \bibinfo {author} {\bibfnamefont {M.~R.}\
  \bibnamefont {Becker}}, \bibinfo {author} {\bibfnamefont {G.~M.}\
  \bibnamefont {Bernstein}}, \bibinfo {author} {\bibfnamefont {S.}~\bibnamefont
  {Bhargava}}, \bibinfo {author} {\bibfnamefont {S.}~\bibnamefont {Birrer}},
  \bibinfo {author} {\bibfnamefont {J.}~\bibnamefont {Blazek}}, \bibinfo
  {author} {\bibfnamefont {A.}~\bibnamefont {Brandao-Souza}}, \bibinfo {author}
  {\bibfnamefont {S.~L.}\ \bibnamefont {Bridle}}, \bibinfo {author}
  {\bibfnamefont {D.}~\bibnamefont {Brooks}}, \bibinfo {author} {\bibfnamefont
  {E.}~\bibnamefont {Buckley-Geer}}, \bibinfo {author} {\bibfnamefont {D.~L.}\
  \bibnamefont {Burke}}, \bibinfo {author} {\bibfnamefont {H.}~\bibnamefont
  {Camacho}}, \bibinfo {author} {\bibfnamefont {A.}~\bibnamefont {Campos}},
  \bibinfo {author} {\bibfnamefont {A.}~\bibnamefont {Carnero~Rosell}},
  \bibinfo {author} {\bibfnamefont {M.}~\bibnamefont {Carrasco~Kind}}, \bibinfo
  {author} {\bibfnamefont {J.}~\bibnamefont {Carretero}}, \bibinfo {author}
  {\bibfnamefont {F.~J.}\ \bibnamefont {Castander}}, \bibinfo {author}
  {\bibfnamefont {R.}~\bibnamefont {Cawthon}}, \bibinfo {author} {\bibfnamefont
  {C.}~\bibnamefont {Chang}}, \bibinfo {author} {\bibfnamefont
  {A.}~\bibnamefont {Chen}}, \bibinfo {author} {\bibfnamefont {R.}~\bibnamefont
  {Chen}}, \bibinfo {author} {\bibfnamefont {A.}~\bibnamefont {Choi}}, \bibinfo
  {author} {\bibfnamefont {C.}~\bibnamefont {Conselice}}, \bibinfo {author}
  {\bibfnamefont {J.}~\bibnamefont {Cordero}}, \bibinfo {author} {\bibfnamefont
  {M.}~\bibnamefont {Costanzi}}, \bibinfo {author} {\bibfnamefont
  {M.}~\bibnamefont {Crocce}}, \bibinfo {author} {\bibfnamefont {L.~N.}\
  \bibnamefont {da~Costa}}, \bibinfo {author} {\bibfnamefont {M.~E.}\
  \bibnamefont {da~Silva~Pereira}}, \bibinfo {author} {\bibfnamefont
  {C.}~\bibnamefont {Davis}}, \bibinfo {author} {\bibfnamefont {T.~M.}\
  \bibnamefont {Davis}}, \bibinfo {author} {\bibfnamefont {J.}~\bibnamefont
  {De~Vicente}}, \bibinfo {author} {\bibfnamefont {J.}~\bibnamefont {DeRose}},
  \bibinfo {author} {\bibfnamefont {S.}~\bibnamefont {Desai}}, \bibinfo
  {author} {\bibfnamefont {E.}~\bibnamefont {Di~Valentino}}, \bibinfo {author}
  {\bibfnamefont {H.~T.}\ \bibnamefont {Diehl}}, \bibinfo {author}
  {\bibfnamefont {J.~P.}\ \bibnamefont {Dietrich}}, \bibinfo {author}
  {\bibfnamefont {S.}~\bibnamefont {Dodelson}}, \bibinfo {author}
  {\bibfnamefont {P.}~\bibnamefont {Doel}}, \bibinfo {author} {\bibfnamefont
  {C.}~\bibnamefont {Doux}}, \bibinfo {author} {\bibfnamefont {A.}~\bibnamefont
  {Drlica-Wagner}}, \bibinfo {author} {\bibfnamefont {K.}~\bibnamefont
  {Eckert}}, \bibinfo {author} {\bibfnamefont {T.~F.}\ \bibnamefont {Eifler}},
  \bibinfo {author} {\bibfnamefont {F.}~\bibnamefont {Elsner}}, \bibinfo
  {author} {\bibfnamefont {J.}~\bibnamefont {Elvin-Poole}}, \bibinfo {author}
  {\bibfnamefont {S.}~\bibnamefont {Everett}}, \bibinfo {author} {\bibfnamefont
  {A.~E.}\ \bibnamefont {Evrard}}, \bibinfo {author} {\bibfnamefont
  {X.}~\bibnamefont {Fang}}, \bibinfo {author} {\bibfnamefont {A.}~\bibnamefont
  {Farahi}}, \bibinfo {author} {\bibfnamefont {E.}~\bibnamefont {Fernandez}},
  \bibinfo {author} {\bibfnamefont {I.}~\bibnamefont {Ferrero}}, \bibinfo
  {author} {\bibfnamefont {A.}~\bibnamefont {Fert\'e}}, \bibinfo {author}
  {\bibfnamefont {P.}~\bibnamefont {Fosalba}}, \bibinfo {author} {\bibfnamefont
  {O.}~\bibnamefont {Friedrich}}, \bibinfo {author} {\bibfnamefont
  {J.}~\bibnamefont {Frieman}}, \bibinfo {author} {\bibfnamefont
  {J.}~\bibnamefont {Garc\'{\i}a-Bellido}}, \bibinfo {author} {\bibfnamefont
  {M.}~\bibnamefont {Gatti}}, \bibinfo {author} {\bibfnamefont
  {E.}~\bibnamefont {Gaztanaga}}, \bibinfo {author} {\bibfnamefont {D.~W.}\
  \bibnamefont {Gerdes}}, \bibinfo {author} {\bibfnamefont {T.}~\bibnamefont
  {Giannantonio}}, \bibinfo {author} {\bibfnamefont {G.}~\bibnamefont
  {Giannini}}, \bibinfo {author} {\bibfnamefont {D.}~\bibnamefont {Gruen}},
  \bibinfo {author} {\bibfnamefont {R.~A.}\ \bibnamefont {Gruendl}}, \bibinfo
  {author} {\bibfnamefont {J.}~\bibnamefont {Gschwend}}, \bibinfo {author}
  {\bibfnamefont {G.}~\bibnamefont {Gutierrez}}, \bibinfo {author}
  {\bibfnamefont {I.}~\bibnamefont {Harrison}}, \bibinfo {author}
  {\bibfnamefont {W.~G.}\ \bibnamefont {Hartley}}, \bibinfo {author}
  {\bibfnamefont {K.}~\bibnamefont {Herner}}, \bibinfo {author} {\bibfnamefont
  {S.~R.}\ \bibnamefont {Hinton}}, \bibinfo {author} {\bibfnamefont {D.~L.}\
  \bibnamefont {Hollowood}}, \bibinfo {author} {\bibfnamefont {K.}~\bibnamefont
  {Honscheid}}, \bibinfo {author} {\bibfnamefont {B.}~\bibnamefont {Hoyle}},
  \bibinfo {author} {\bibfnamefont {E.~M.}\ \bibnamefont {Huff}}, \bibinfo
  {author} {\bibfnamefont {D.}~\bibnamefont {Huterer}}, \bibinfo {author}
  {\bibfnamefont {B.}~\bibnamefont {Jain}}, \bibinfo {author} {\bibfnamefont
  {D.~J.}\ \bibnamefont {James}}, \bibinfo {author} {\bibfnamefont
  {M.}~\bibnamefont {Jarvis}}, \bibinfo {author} {\bibfnamefont
  {N.}~\bibnamefont {Jeffrey}}, \bibinfo {author} {\bibfnamefont
  {T.}~\bibnamefont {Jeltema}}, \bibinfo {author} {\bibfnamefont
  {A.}~\bibnamefont {Kovacs}}, \bibinfo {author} {\bibfnamefont
  {E.}~\bibnamefont {Krause}}, \bibinfo {author} {\bibfnamefont
  {R.}~\bibnamefont {Kron}}, \bibinfo {author} {\bibfnamefont {K.}~\bibnamefont
  {Kuehn}}, \bibinfo {author} {\bibfnamefont {N.}~\bibnamefont {Kuropatkin}},
  \bibinfo {author} {\bibfnamefont {O.}~\bibnamefont {Lahav}}, \bibinfo
  {author} {\bibfnamefont {P.-F.}\ \bibnamefont {Leget}}, \bibinfo {author}
  {\bibfnamefont {P.}~\bibnamefont {Lemos}}, \bibinfo {author} {\bibfnamefont
  {A.~R.}\ \bibnamefont {Liddle}}, \bibinfo {author} {\bibfnamefont
  {C.}~\bibnamefont {Lidman}}, \bibinfo {author} {\bibfnamefont
  {M.}~\bibnamefont {Lima}}, \bibinfo {author} {\bibfnamefont {H.}~\bibnamefont
  {Lin}}, \bibinfo {author} {\bibfnamefont {N.}~\bibnamefont {MacCrann}},
  \bibinfo {author} {\bibfnamefont {M.~A.~G.}\ \bibnamefont {Maia}}, \bibinfo
  {author} {\bibfnamefont {J.~L.}\ \bibnamefont {Marshall}}, \bibinfo {author}
  {\bibfnamefont {P.}~\bibnamefont {Martini}}, \bibinfo {author} {\bibfnamefont
  {J.}~\bibnamefont {McCullough}}, \bibinfo {author} {\bibfnamefont
  {P.}~\bibnamefont {Melchior}}, \bibinfo {author} {\bibfnamefont
  {J.}~\bibnamefont {Mena-Fern\'andez}}, \bibinfo {author} {\bibfnamefont
  {F.}~\bibnamefont {Menanteau}}, \bibinfo {author} {\bibfnamefont
  {R.}~\bibnamefont {Miquel}}, \bibinfo {author} {\bibfnamefont {J.~J.}\
  \bibnamefont {Mohr}}, \bibinfo {author} {\bibfnamefont {R.}~\bibnamefont
  {Morgan}}, \bibinfo {author} {\bibfnamefont {J.}~\bibnamefont {Muir}},
  \bibinfo {author} {\bibfnamefont {J.}~\bibnamefont {Myles}}, \bibinfo
  {author} {\bibfnamefont {S.}~\bibnamefont {Nadathur}}, \bibinfo {author}
  {\bibfnamefont {A.}~\bibnamefont {Navarro-Alsina}}, \bibinfo {author}
  {\bibfnamefont {R.~C.}\ \bibnamefont {Nichol}}, \bibinfo {author}
  {\bibfnamefont {R.~L.~C.}\ \bibnamefont {Ogando}}, \bibinfo {author}
  {\bibfnamefont {Y.}~\bibnamefont {Omori}}, \bibinfo {author} {\bibfnamefont
  {A.}~\bibnamefont {Palmese}}, \bibinfo {author} {\bibfnamefont
  {S.}~\bibnamefont {Pandey}}, \bibinfo {author} {\bibfnamefont
  {Y.}~\bibnamefont {Park}}, \bibinfo {author} {\bibfnamefont {F.}~\bibnamefont
  {Paz-Chinch\'on}}, \bibinfo {author} {\bibfnamefont {D.}~\bibnamefont
  {Petravick}}, \bibinfo {author} {\bibfnamefont {A.}~\bibnamefont {Pieres}},
  \bibinfo {author} {\bibfnamefont {A.~A.}\ \bibnamefont {Plazas~Malag\'on}},
  \bibinfo {author} {\bibfnamefont {A.}~\bibnamefont {Porredon}}, \bibinfo
  {author} {\bibfnamefont {J.}~\bibnamefont {Prat}}, \bibinfo {author}
  {\bibfnamefont {M.}~\bibnamefont {Raveri}}, \bibinfo {author} {\bibfnamefont
  {M.}~\bibnamefont {Rodriguez-Monroy}}, \bibinfo {author} {\bibfnamefont
  {R.~P.}\ \bibnamefont {Rollins}}, \bibinfo {author} {\bibfnamefont {A.~K.}\
  \bibnamefont {Romer}}, \bibinfo {author} {\bibfnamefont {A.}~\bibnamefont
  {Roodman}}, \bibinfo {author} {\bibfnamefont {R.}~\bibnamefont {Rosenfeld}},
  \bibinfo {author} {\bibfnamefont {A.~J.}\ \bibnamefont {Ross}}, \bibinfo
  {author} {\bibfnamefont {E.~S.}\ \bibnamefont {Rykoff}}, \bibinfo {author}
  {\bibfnamefont {S.}~\bibnamefont {Samuroff}}, \bibinfo {author}
  {\bibfnamefont {C.}~\bibnamefont {S\'anchez}}, \bibinfo {author}
  {\bibfnamefont {E.}~\bibnamefont {Sanchez}}, \bibinfo {author} {\bibfnamefont
  {J.}~\bibnamefont {Sanchez}}, \bibinfo {author} {\bibfnamefont
  {D.}~\bibnamefont {Sanchez~Cid}}, \bibinfo {author} {\bibfnamefont
  {V.}~\bibnamefont {Scarpine}}, \bibinfo {author} {\bibfnamefont
  {M.}~\bibnamefont {Schubnell}}, \bibinfo {author} {\bibfnamefont
  {D.}~\bibnamefont {Scolnic}}, \bibinfo {author} {\bibfnamefont {L.~F.}\
  \bibnamefont {Secco}}, \bibinfo {author} {\bibfnamefont {S.}~\bibnamefont
  {Serrano}}, \bibinfo {author} {\bibfnamefont {I.}~\bibnamefont
  {Sevilla-Noarbe}}, \bibinfo {author} {\bibfnamefont {E.}~\bibnamefont
  {Sheldon}}, \bibinfo {author} {\bibfnamefont {T.}~\bibnamefont {Shin}},
  \bibinfo {author} {\bibfnamefont {M.}~\bibnamefont {Smith}}, \bibinfo
  {author} {\bibfnamefont {M.}~\bibnamefont {Soares-Santos}}, \bibinfo {author}
  {\bibfnamefont {E.}~\bibnamefont {Suchyta}}, \bibinfo {author} {\bibfnamefont
  {M.~E.~C.}\ \bibnamefont {Swanson}}, \bibinfo {author} {\bibfnamefont
  {M.}~\bibnamefont {Tabbutt}}, \bibinfo {author} {\bibfnamefont
  {G.}~\bibnamefont {Tarle}}, \bibinfo {author} {\bibfnamefont
  {D.}~\bibnamefont {Thomas}}, \bibinfo {author} {\bibfnamefont
  {C.}~\bibnamefont {To}}, \bibinfo {author} {\bibfnamefont {A.}~\bibnamefont
  {Troja}}, \bibinfo {author} {\bibfnamefont {M.~A.}\ \bibnamefont {Troxel}},
  \bibinfo {author} {\bibfnamefont {D.~L.}\ \bibnamefont {Tucker}}, \bibinfo
  {author} {\bibfnamefont {I.}~\bibnamefont {Tutusaus}}, \bibinfo {author}
  {\bibfnamefont {T.~N.}\ \bibnamefont {Varga}}, \bibinfo {author}
  {\bibfnamefont {A.~R.}\ \bibnamefont {Walker}}, \bibinfo {author}
  {\bibfnamefont {N.}~\bibnamefont {Weaverdyck}}, \bibinfo {author}
  {\bibfnamefont {R.}~\bibnamefont {Wechsler}}, \bibinfo {author}
  {\bibfnamefont {J.}~\bibnamefont {Weller}}, \bibinfo {author} {\bibfnamefont
  {B.}~\bibnamefont {Yanny}}, \bibinfo {author} {\bibfnamefont
  {B.}~\bibnamefont {Yin}}, \bibinfo {author} {\bibfnamefont {Y.}~\bibnamefont
  {Zhang}},\ and\ \bibinfo {author} {\bibfnamefont {J.}~\bibnamefont {Zuntz}}
  (\bibinfo {collaboration} {DES Collaboration}),\ }\bibfield  {title}
  {\bibinfo {title} {Dark energy survey year 3 results: Cosmological
  constraints from galaxy clustering and weak lensing},\ }\href
  {https://doi.org/10.1103/PhysRevD.105.023520} {\bibfield  {journal} {\bibinfo
   {journal} {Phys. Rev. D}\ }\textbf {\bibinfo {volume} {105}},\ \bibinfo
  {pages} {023520} (\bibinfo {year} {2022})}\BibitemShut {NoStop}%
\bibitem [{\citenamefont {{Heymans}}\ \emph {et~al.}(2021)\citenamefont
  {{Heymans}}, \citenamefont {{Tr{\"o}ster}}, \citenamefont {{Asgari}},
  \citenamefont {{Blake}}, \citenamefont {{Hildebrandt}}, \citenamefont
  {{Joachimi}}, \citenamefont {{Kuijken}}, \citenamefont {{Lin}}, \citenamefont
  {{S{\'a}nchez}}, \citenamefont {{van den Busch}}, \citenamefont {{Wright}},
  \citenamefont {{Amon}}, \citenamefont {{Bilicki}}, \citenamefont {{de Jong}},
  \citenamefont {{Crocce}}, \citenamefont {{Dvornik}}, \citenamefont {{Erben}},
  \citenamefont {{Fortuna}}, \citenamefont {{Getman}}, \citenamefont
  {{Giblin}}, \citenamefont {{Glazebrook}}, \citenamefont {{Hoekstra}},
  \citenamefont {{Joudaki}}, \citenamefont {{Kannawadi}}, \citenamefont
  {{K{\"o}hlinger}}, \citenamefont {{Lidman}}, \citenamefont {{Miller}},
  \citenamefont {{Napolitano}}, \citenamefont {{Parkinson}}, \citenamefont
  {{Schneider}}, \citenamefont {{Shan}}, \citenamefont {{Valentijn}},
  \citenamefont {{Verdoes Kleijn}},\ and\ \citenamefont
  {{Wolf}}}]{Heymans2020multiprobe}%
  \BibitemOpen
  \bibfield  {author} {\bibinfo {author} {\bibfnamefont {C.}~\bibnamefont
  {{Heymans}}}, \bibinfo {author} {\bibfnamefont {T.}~\bibnamefont
  {{Tr{\"o}ster}}}, \bibinfo {author} {\bibfnamefont {M.}~\bibnamefont
  {{Asgari}}}, \bibinfo {author} {\bibfnamefont {C.}~\bibnamefont {{Blake}}},
  \bibinfo {author} {\bibfnamefont {H.}~\bibnamefont {{Hildebrandt}}}, \bibinfo
  {author} {\bibfnamefont {B.}~\bibnamefont {{Joachimi}}}, \bibinfo {author}
  {\bibfnamefont {K.}~\bibnamefont {{Kuijken}}}, \bibinfo {author}
  {\bibfnamefont {C.-A.}\ \bibnamefont {{Lin}}}, \bibinfo {author}
  {\bibfnamefont {A.~G.}\ \bibnamefont {{S{\'a}nchez}}}, \bibinfo {author}
  {\bibfnamefont {J.~L.}\ \bibnamefont {{van den Busch}}}, \bibinfo {author}
  {\bibfnamefont {A.~H.}\ \bibnamefont {{Wright}}}, \bibinfo {author}
  {\bibfnamefont {A.}~\bibnamefont {{Amon}}}, \bibinfo {author} {\bibfnamefont
  {M.}~\bibnamefont {{Bilicki}}}, \bibinfo {author} {\bibfnamefont
  {J.}~\bibnamefont {{de Jong}}}, \bibinfo {author} {\bibfnamefont
  {M.}~\bibnamefont {{Crocce}}}, \bibinfo {author} {\bibfnamefont
  {A.}~\bibnamefont {{Dvornik}}}, \bibinfo {author} {\bibfnamefont
  {T.}~\bibnamefont {{Erben}}}, \bibinfo {author} {\bibfnamefont {M.~C.}\
  \bibnamefont {{Fortuna}}}, \bibinfo {author} {\bibfnamefont {F.}~\bibnamefont
  {{Getman}}}, \bibinfo {author} {\bibfnamefont {B.}~\bibnamefont {{Giblin}}},
  \bibinfo {author} {\bibfnamefont {K.}~\bibnamefont {{Glazebrook}}}, \bibinfo
  {author} {\bibfnamefont {H.}~\bibnamefont {{Hoekstra}}}, \bibinfo {author}
  {\bibfnamefont {S.}~\bibnamefont {{Joudaki}}}, \bibinfo {author}
  {\bibfnamefont {A.}~\bibnamefont {{Kannawadi}}}, \bibinfo {author}
  {\bibfnamefont {F.}~\bibnamefont {{K{\"o}hlinger}}}, \bibinfo {author}
  {\bibfnamefont {C.}~\bibnamefont {{Lidman}}}, \bibinfo {author}
  {\bibfnamefont {L.}~\bibnamefont {{Miller}}}, \bibinfo {author}
  {\bibfnamefont {N.~R.}\ \bibnamefont {{Napolitano}}}, \bibinfo {author}
  {\bibfnamefont {D.}~\bibnamefont {{Parkinson}}}, \bibinfo {author}
  {\bibfnamefont {P.}~\bibnamefont {{Schneider}}}, \bibinfo {author}
  {\bibfnamefont {H.}~\bibnamefont {{Shan}}}, \bibinfo {author} {\bibfnamefont
  {E.~A.}\ \bibnamefont {{Valentijn}}}, \bibinfo {author} {\bibfnamefont
  {G.}~\bibnamefont {{Verdoes Kleijn}}},\ and\ \bibinfo {author} {\bibfnamefont
  {C.}~\bibnamefont {{Wolf}}},\ }\bibfield  {title} {\bibinfo {title}
  {{KiDS-1000 Cosmology: Multi-probe weak gravitational lensing and
  spectroscopic galaxy clustering constraints}},\ }\href
  {https://doi.org/10.1051/0004-6361/202039063} {\bibfield  {journal} {\bibinfo
   {journal} {\aap}\ }\textbf {\bibinfo {volume} {646}},\ \bibinfo {eid} {A140}
  (\bibinfo {year} {2021})}\BibitemShut {NoStop}%
\bibitem [{\citenamefont {{Hikage}}\ \emph {et~al.}(2019)\citenamefont
  {{Hikage}}, \citenamefont {{Oguri}}, \citenamefont {{Hamana}}, \citenamefont
  {{More}}, \citenamefont {{Mandelbaum}}, \citenamefont {{Takada}},
  \citenamefont {{K{\"o}hlinger}}, \citenamefont {{Miyatake}}, \citenamefont
  {{Nishizawa}}, \citenamefont {{Aihara}}, \citenamefont {{Armstrong}},
  \citenamefont {{Bosch}}, \citenamefont {{Coupon}}, \citenamefont {{Ducout}},
  \citenamefont {{Ho}}, \citenamefont {{Hsieh}}, \citenamefont {{Komiyama}},
  \citenamefont {{Lanusse}}, \citenamefont {{Leauthaud}}, \citenamefont
  {{Lupton}}, \citenamefont {{Medezinski}}, \citenamefont {{Mineo}},
  \citenamefont {{Miyama}}, \citenamefont {{Miyazaki}}, \citenamefont
  {{Murata}}, \citenamefont {{Murayama}}, \citenamefont {{Shirasaki}},
  \citenamefont {{Sif{\'o}n}}, \citenamefont {{Simet}}, \citenamefont
  {{Speagle}}, \citenamefont {{Spergel}}, \citenamefont {{Strauss}},
  \citenamefont {{Sugiyama}}, \citenamefont {{Tanaka}}, \citenamefont
  {{Utsumi}}, \citenamefont {{Wang}},\ and\ \citenamefont
  {{Yamada}}}]{Hikage2019cosmicshear}%
  \BibitemOpen
  \bibfield  {author} {\bibinfo {author} {\bibfnamefont {C.}~\bibnamefont
  {{Hikage}}}, \bibinfo {author} {\bibfnamefont {M.}~\bibnamefont {{Oguri}}},
  \bibinfo {author} {\bibfnamefont {T.}~\bibnamefont {{Hamana}}}, \bibinfo
  {author} {\bibfnamefont {S.}~\bibnamefont {{More}}}, \bibinfo {author}
  {\bibfnamefont {R.}~\bibnamefont {{Mandelbaum}}}, \bibinfo {author}
  {\bibfnamefont {M.}~\bibnamefont {{Takada}}}, \bibinfo {author}
  {\bibfnamefont {F.}~\bibnamefont {{K{\"o}hlinger}}}, \bibinfo {author}
  {\bibfnamefont {H.}~\bibnamefont {{Miyatake}}}, \bibinfo {author}
  {\bibfnamefont {A.~J.}\ \bibnamefont {{Nishizawa}}}, \bibinfo {author}
  {\bibfnamefont {H.}~\bibnamefont {{Aihara}}}, \bibinfo {author}
  {\bibfnamefont {R.}~\bibnamefont {{Armstrong}}}, \bibinfo {author}
  {\bibfnamefont {J.}~\bibnamefont {{Bosch}}}, \bibinfo {author} {\bibfnamefont
  {J.}~\bibnamefont {{Coupon}}}, \bibinfo {author} {\bibfnamefont
  {A.}~\bibnamefont {{Ducout}}}, \bibinfo {author} {\bibfnamefont
  {P.}~\bibnamefont {{Ho}}}, \bibinfo {author} {\bibfnamefont {B.-C.}\
  \bibnamefont {{Hsieh}}}, \bibinfo {author} {\bibfnamefont {Y.}~\bibnamefont
  {{Komiyama}}}, \bibinfo {author} {\bibfnamefont {F.}~\bibnamefont
  {{Lanusse}}}, \bibinfo {author} {\bibfnamefont {A.}~\bibnamefont
  {{Leauthaud}}}, \bibinfo {author} {\bibfnamefont {R.~H.}\ \bibnamefont
  {{Lupton}}}, \bibinfo {author} {\bibfnamefont {E.}~\bibnamefont
  {{Medezinski}}}, \bibinfo {author} {\bibfnamefont {S.}~\bibnamefont
  {{Mineo}}}, \bibinfo {author} {\bibfnamefont {S.}~\bibnamefont {{Miyama}}},
  \bibinfo {author} {\bibfnamefont {S.}~\bibnamefont {{Miyazaki}}}, \bibinfo
  {author} {\bibfnamefont {R.}~\bibnamefont {{Murata}}}, \bibinfo {author}
  {\bibfnamefont {H.}~\bibnamefont {{Murayama}}}, \bibinfo {author}
  {\bibfnamefont {M.}~\bibnamefont {{Shirasaki}}}, \bibinfo {author}
  {\bibfnamefont {C.}~\bibnamefont {{Sif{\'o}n}}}, \bibinfo {author}
  {\bibfnamefont {M.}~\bibnamefont {{Simet}}}, \bibinfo {author} {\bibfnamefont
  {J.}~\bibnamefont {{Speagle}}}, \bibinfo {author} {\bibfnamefont {D.~N.}\
  \bibnamefont {{Spergel}}}, \bibinfo {author} {\bibfnamefont {M.~A.}\
  \bibnamefont {{Strauss}}}, \bibinfo {author} {\bibfnamefont {N.}~\bibnamefont
  {{Sugiyama}}}, \bibinfo {author} {\bibfnamefont {M.}~\bibnamefont
  {{Tanaka}}}, \bibinfo {author} {\bibfnamefont {Y.}~\bibnamefont {{Utsumi}}},
  \bibinfo {author} {\bibfnamefont {S.-Y.}\ \bibnamefont {{Wang}}},\ and\
  \bibinfo {author} {\bibfnamefont {Y.}~\bibnamefont {{Yamada}}},\ }\href@noop
  {} {\bibfield  {journal} {\bibinfo  {journal} {\pasj}\ }\textbf {\bibinfo
  {volume} {71}},\ \bibinfo {eid} {43} (\bibinfo {year} {2019})}\BibitemShut
  {NoStop}%
\bibitem [{Note4()}]{Note4}%
  \BibitemOpen
  \bibinfo {note} {\protect \url {https://www.euclid-ec.org}}\BibitemShut
  {NoStop}%
\bibitem [{Note5()}]{Note5}%
  \BibitemOpen
  \bibinfo {note} {\protect \url {lsstdesc.org}}\BibitemShut {NoStop}%
\bibitem [{\citenamefont {{Kirk}}\ \emph {et~al.}(2015)\citenamefont {{Kirk}},
  \citenamefont {{Brown}}, \citenamefont {{Hoekstra}}, \citenamefont
  {{Joachimi}}, \citenamefont {{Kitching}}, \citenamefont {{Mandelbaum}},
  \citenamefont {{Sif{\'o}n}}, \citenamefont {{Cacciato}}, \citenamefont
  {{Choi}}, \citenamefont {{Kiessling}}, \citenamefont {{Leonard}},
  \citenamefont {{Rassat}},\ and\ \citenamefont
  {{Sch{\"a}fer}}}]{Kirk2015intrinsic}%
  \BibitemOpen
  \bibfield  {author} {\bibinfo {author} {\bibfnamefont {D.}~\bibnamefont
  {{Kirk}}}, \bibinfo {author} {\bibfnamefont {M.~L.}\ \bibnamefont {{Brown}}},
  \bibinfo {author} {\bibfnamefont {H.}~\bibnamefont {{Hoekstra}}}, \bibinfo
  {author} {\bibfnamefont {B.}~\bibnamefont {{Joachimi}}}, \bibinfo {author}
  {\bibfnamefont {T.~D.}\ \bibnamefont {{Kitching}}}, \bibinfo {author}
  {\bibfnamefont {R.}~\bibnamefont {{Mandelbaum}}}, \bibinfo {author}
  {\bibfnamefont {C.}~\bibnamefont {{Sif{\'o}n}}}, \bibinfo {author}
  {\bibfnamefont {M.}~\bibnamefont {{Cacciato}}}, \bibinfo {author}
  {\bibfnamefont {A.}~\bibnamefont {{Choi}}}, \bibinfo {author} {\bibfnamefont
  {A.}~\bibnamefont {{Kiessling}}}, \bibinfo {author} {\bibfnamefont
  {A.}~\bibnamefont {{Leonard}}}, \bibinfo {author} {\bibfnamefont
  {A.}~\bibnamefont {{Rassat}}},\ and\ \bibinfo {author} {\bibfnamefont
  {B.~M.}\ \bibnamefont {{Sch{\"a}fer}}},\ }\bibfield  {title} {\bibinfo
  {title} {{Galaxy Alignments: Observations and Impact on Cosmology}},\ }\href
  {https://doi.org/10.1007/s11214-015-0213-4} {\bibfield  {journal} {\bibinfo
  {journal} {\ssr}\ }\textbf {\bibinfo {volume} {193}},\ \bibinfo {pages} {139}
  (\bibinfo {year} {2015})}\BibitemShut {NoStop}%
\bibitem [{\citenamefont {{Eriksen}}\ and\ \citenamefont
  {{Gazta{\~n}aga}}(2018)}]{Eriksen2018bias}%
  \BibitemOpen
  \bibfield  {author} {\bibinfo {author} {\bibfnamefont {M.}~\bibnamefont
  {{Eriksen}}}\ and\ \bibinfo {author} {\bibfnamefont {E.}~\bibnamefont
  {{Gazta{\~n}aga}}},\ }\bibfield  {title} {\bibinfo {title} {{Combining
  spectroscopic and photometric surveys using angular cross-correlations - III.
  Galaxy bias and stochasticity}},\ }\href
  {https://doi.org/10.1093/mnras/sty2168} {\bibfield  {journal} {\bibinfo
  {journal} {\mnras}\ }\textbf {\bibinfo {volume} {480}},\ \bibinfo {pages}
  {5226} (\bibinfo {year} {2018})},\ \Eprint {https://arxiv.org/abs/1508.00035}
  {arXiv:1508.00035 [astro-ph.CO]} \BibitemShut {NoStop}%
\bibitem [{\citenamefont {{Blazek}}\ \emph {et~al.}(2019)\citenamefont
  {{Blazek}}, \citenamefont {{MacCrann}}, \citenamefont {{Troxel}},\ and\
  \citenamefont {{Fang}}}]{Blazek2019beyond}%
  \BibitemOpen
  \bibfield  {author} {\bibinfo {author} {\bibfnamefont {J.~A.}\ \bibnamefont
  {{Blazek}}}, \bibinfo {author} {\bibfnamefont {N.}~\bibnamefont
  {{MacCrann}}}, \bibinfo {author} {\bibfnamefont {M.~A.}\ \bibnamefont
  {{Troxel}}},\ and\ \bibinfo {author} {\bibfnamefont {X.}~\bibnamefont
  {{Fang}}},\ }\bibfield  {title} {\bibinfo {title} {{Beyond linear galaxy
  alignments}},\ }\href {https://doi.org/10.1103/PhysRevD.100.103506}
  {\bibfield  {journal} {\bibinfo  {journal} {\prd}\ }\textbf {\bibinfo
  {volume} {100}},\ \bibinfo {eid} {103506} (\bibinfo {year}
  {2019})}\BibitemShut {NoStop}%
\bibitem [{\citenamefont {{Secco}}\ \emph {et~al.}(2022)\citenamefont
  {{Secco}}, \citenamefont {{Samuroff}}, \citenamefont {{Krause}},
  \citenamefont {{Jain}}, \citenamefont {{Blazek}}, \citenamefont {{Raveri}},
  \citenamefont {{Campos}}, \citenamefont {{Amon}}, \citenamefont {{Chen}},
  \citenamefont {{Doux}}, \citenamefont {{Choi}}, \citenamefont {{Gruen}},
  \citenamefont {{Bernstein}}, \citenamefont {{Chang}}, \citenamefont
  {{DeRose}}, \citenamefont {{Myles}}, \citenamefont {{Fert{\'e}}},
  \citenamefont {{Lemos}}, \citenamefont {{Huterer}}, \citenamefont {{Prat}},
  \citenamefont {{Troxel}}, \citenamefont {{MacCrann}}, \citenamefont
  {{Liddle}}, \citenamefont {{Kacprzak}}, \citenamefont {{Fang}}, \citenamefont
  {{S{\'a}nchez}}, \citenamefont {{Pandey}}, \citenamefont {{Dodelson}},
  \citenamefont {{Chintalapati}}, \citenamefont {{Hoffmann}}, \citenamefont
  {{Alarcon}}, \citenamefont {{Alves}}, \citenamefont {{Andrade-Oliveira}},
  \citenamefont {{Baxter}}, \citenamefont {{Bechtol}}, \citenamefont
  {{Becker}}, \citenamefont {{Brandao-Souza}}, \citenamefont {{Camacho}},
  \citenamefont {{Carnero Rosell}}, \citenamefont {{Carrasco Kind}},
  \citenamefont {{Cawthon}}, \citenamefont {{Cordero}}, \citenamefont
  {{Crocce}}, \citenamefont {{Davis}}, \citenamefont {{Di Valentino}},
  \citenamefont {{Drlica-Wagner}}, \citenamefont {{Eckert}}, \citenamefont
  {{Eifler}}, \citenamefont {{Elidaiana}}, \citenamefont {{Elsner}},
  \citenamefont {{Elvin-Poole}}, \citenamefont {{Everett}}, \citenamefont
  {{Fosalba}}, \citenamefont {{Friedrich}}, \citenamefont {{Gatti}},
  \citenamefont {{Giannini}}, \citenamefont {{Gruendl}}, \citenamefont
  {{Harrison}}, \citenamefont {{Hartley}}, \citenamefont {{Herner}},
  \citenamefont {{Huang}}, \citenamefont {{Huff}}, \citenamefont {{Jarvis}},
  \citenamefont {{Jeffrey}}, \citenamefont {{Kuropatkin}}, \citenamefont
  {{Leget}}, \citenamefont {{Muir}}, \citenamefont {{Mccullough}},
  \citenamefont {{Navarro Alsina}}, \citenamefont {{Omori}}, \citenamefont
  {{Park}}, \citenamefont {{Porredon}}, \citenamefont {{Rollins}},
  \citenamefont {{Roodman}}, \citenamefont {{Rosenfeld}}, \citenamefont
  {{Ross}}, \citenamefont {{Rykoff}}, \citenamefont {{Sanchez}}, \citenamefont
  {{Sevilla-Noarbe}}, \citenamefont {{Sheldon}}, \citenamefont {{Shin}},
  \citenamefont {{Tutusaus}}, \citenamefont {{Varga}}, \citenamefont
  {{Weaverdyck}}, \citenamefont {{Wechsler}}, \citenamefont {{Yanny}},
  \citenamefont {{Yin}}, \citenamefont {{Zhang}}, \citenamefont {{Zuntz}},
  \citenamefont {{Abbott}}, \citenamefont {{Aguena}}, \citenamefont {{Allam}},
  \citenamefont {{Annis}}, \citenamefont {{Bacon}}, \citenamefont {{Bertin}},
  \citenamefont {{Bhargava}}, \citenamefont {{Bridle}}, \citenamefont
  {{Brooks}}, \citenamefont {{Buckley-Geer}}, \citenamefont {{Burke}},
  \citenamefont {{Carretero}}, \citenamefont {{Costanzi}}, \citenamefont {{da
  Costa}}, \citenamefont {{De Vicente}}, \citenamefont {{Diehl}}, \citenamefont
  {{Dietrich}}, \citenamefont {{Doel}}, \citenamefont {{Ferrero}},
  \citenamefont {{Flaugher}}, \citenamefont {{Frieman}}, \citenamefont
  {{Garc{\'\i}a-Bellido}}, \citenamefont {{Gaztanaga}}, \citenamefont
  {{Gerdes}}, \citenamefont {{Giannantonio}}, \citenamefont {{Gschwend}},
  \citenamefont {{Gutierrez}}, \citenamefont {{Hinton}}, \citenamefont
  {{Hollowood}}, \citenamefont {{Honscheid}}, \citenamefont {{Hoyle}},
  \citenamefont {{James}}, \citenamefont {{Jeltema}}, \citenamefont {{Kuehn}},
  \citenamefont {{Lahav}}, \citenamefont {{Lima}}, \citenamefont {{Lin}},
  \citenamefont {{Maia}}, \citenamefont {{Marshall}}, \citenamefont
  {{Martini}}, \citenamefont {{Melchior}}, \citenamefont {{Menanteau}},
  \citenamefont {{Miquel}}, \citenamefont {{Mohr}}, \citenamefont {{Morgan}},
  \citenamefont {{Ogando}}, \citenamefont {{Palmese}}, \citenamefont
  {{Paz-Chinch{\'o}n}}, \citenamefont {{Petravick}}, \citenamefont {{Pieres}},
  \citenamefont {{Plazas Malag{\'o}n}}, \citenamefont {{Rodriguez-Monroy}},
  \citenamefont {{Romer}}, \citenamefont {{Sanchez}}, \citenamefont
  {{Scarpine}}, \citenamefont {{Schubnell}}, \citenamefont {{Scolnic}},
  \citenamefont {{Serrano}}, \citenamefont {{Smith}}, \citenamefont
  {{Soares-Santos}}, \citenamefont {{Suchyta}}, \citenamefont {{Swanson}},
  \citenamefont {{Tarle}}, \citenamefont {{Thomas}}, \citenamefont {{To}},\
  and\ \citenamefont {{DES Collaboration}}}]{Secco2022cosmicshear}%
  \BibitemOpen
  \bibfield  {author} {\bibinfo {author} {\bibfnamefont {L.~F.}\ \bibnamefont
  {{Secco}}}, \bibinfo {author} {\bibfnamefont {S.}~\bibnamefont {{Samuroff}}},
  \bibinfo {author} {\bibfnamefont {E.}~\bibnamefont {{Krause}}}, \bibinfo
  {author} {\bibfnamefont {B.}~\bibnamefont {{Jain}}}, \bibinfo {author}
  {\bibfnamefont {J.}~\bibnamefont {{Blazek}}}, \bibinfo {author}
  {\bibfnamefont {M.}~\bibnamefont {{Raveri}}}, \bibinfo {author}
  {\bibfnamefont {A.}~\bibnamefont {{Campos}}}, \bibinfo {author}
  {\bibfnamefont {A.}~\bibnamefont {{Amon}}}, \bibinfo {author} {\bibfnamefont
  {A.}~\bibnamefont {{Chen}}}, \bibinfo {author} {\bibfnamefont
  {C.}~\bibnamefont {{Doux}}}, \bibinfo {author} {\bibfnamefont
  {A.}~\bibnamefont {{Choi}}}, \bibinfo {author} {\bibfnamefont
  {D.}~\bibnamefont {{Gruen}}}, \bibinfo {author} {\bibfnamefont {G.~M.}\
  \bibnamefont {{Bernstein}}}, \bibinfo {author} {\bibfnamefont
  {C.}~\bibnamefont {{Chang}}}, \bibinfo {author} {\bibfnamefont
  {J.}~\bibnamefont {{DeRose}}}, \bibinfo {author} {\bibfnamefont
  {J.}~\bibnamefont {{Myles}}}, \bibinfo {author} {\bibfnamefont
  {A.}~\bibnamefont {{Fert{\'e}}}}, \bibinfo {author} {\bibfnamefont
  {P.}~\bibnamefont {{Lemos}}}, \bibinfo {author} {\bibfnamefont
  {D.}~\bibnamefont {{Huterer}}}, \bibinfo {author} {\bibfnamefont
  {J.}~\bibnamefont {{Prat}}}, \bibinfo {author} {\bibfnamefont {M.~A.}\
  \bibnamefont {{Troxel}}}, \bibinfo {author} {\bibfnamefont {N.}~\bibnamefont
  {{MacCrann}}}, \bibinfo {author} {\bibfnamefont {A.~R.}\ \bibnamefont
  {{Liddle}}}, \bibinfo {author} {\bibfnamefont {T.}~\bibnamefont
  {{Kacprzak}}}, \bibinfo {author} {\bibfnamefont {X.}~\bibnamefont {{Fang}}},
  \bibinfo {author} {\bibfnamefont {C.}~\bibnamefont {{S{\'a}nchez}}}, \bibinfo
  {author} {\bibfnamefont {S.}~\bibnamefont {{Pandey}}}, \bibinfo {author}
  {\bibfnamefont {S.}~\bibnamefont {{Dodelson}}}, \bibinfo {author}
  {\bibfnamefont {P.}~\bibnamefont {{Chintalapati}}}, \bibinfo {author}
  {\bibfnamefont {K.}~\bibnamefont {{Hoffmann}}}, \bibinfo {author}
  {\bibfnamefont {A.}~\bibnamefont {{Alarcon}}}, \bibinfo {author}
  {\bibfnamefont {O.}~\bibnamefont {{Alves}}}, \bibinfo {author} {\bibfnamefont
  {F.}~\bibnamefont {{Andrade-Oliveira}}}, \bibinfo {author} {\bibfnamefont
  {E.~J.}\ \bibnamefont {{Baxter}}}, \bibinfo {author} {\bibfnamefont
  {K.}~\bibnamefont {{Bechtol}}}, \bibinfo {author} {\bibfnamefont {M.~R.}\
  \bibnamefont {{Becker}}}, \bibinfo {author} {\bibfnamefont {A.}~\bibnamefont
  {{Brandao-Souza}}}, \bibinfo {author} {\bibfnamefont {H.}~\bibnamefont
  {{Camacho}}}, \bibinfo {author} {\bibfnamefont {A.}~\bibnamefont {{Carnero
  Rosell}}}, \bibinfo {author} {\bibfnamefont {M.}~\bibnamefont {{Carrasco
  Kind}}}, \bibinfo {author} {\bibfnamefont {R.}~\bibnamefont {{Cawthon}}},
  \bibinfo {author} {\bibfnamefont {J.~P.}\ \bibnamefont {{Cordero}}}, \bibinfo
  {author} {\bibfnamefont {M.}~\bibnamefont {{Crocce}}}, \bibinfo {author}
  {\bibfnamefont {C.}~\bibnamefont {{Davis}}}, \bibinfo {author} {\bibfnamefont
  {E.}~\bibnamefont {{Di Valentino}}}, \bibinfo {author} {\bibfnamefont
  {A.}~\bibnamefont {{Drlica-Wagner}}}, \bibinfo {author} {\bibfnamefont
  {K.}~\bibnamefont {{Eckert}}}, \bibinfo {author} {\bibfnamefont {T.~F.}\
  \bibnamefont {{Eifler}}}, \bibinfo {author} {\bibfnamefont {M.}~\bibnamefont
  {{Elidaiana}}}, \bibinfo {author} {\bibfnamefont {F.}~\bibnamefont
  {{Elsner}}}, \bibinfo {author} {\bibfnamefont {J.}~\bibnamefont
  {{Elvin-Poole}}}, \bibinfo {author} {\bibfnamefont {S.}~\bibnamefont
  {{Everett}}}, \bibinfo {author} {\bibfnamefont {P.}~\bibnamefont
  {{Fosalba}}}, \bibinfo {author} {\bibfnamefont {O.}~\bibnamefont
  {{Friedrich}}}, \bibinfo {author} {\bibfnamefont {M.}~\bibnamefont
  {{Gatti}}}, \bibinfo {author} {\bibfnamefont {G.}~\bibnamefont {{Giannini}}},
  \bibinfo {author} {\bibfnamefont {R.~A.}\ \bibnamefont {{Gruendl}}}, \bibinfo
  {author} {\bibfnamefont {I.}~\bibnamefont {{Harrison}}}, \bibinfo {author}
  {\bibfnamefont {W.~G.}\ \bibnamefont {{Hartley}}}, \bibinfo {author}
  {\bibfnamefont {K.}~\bibnamefont {{Herner}}}, \bibinfo {author}
  {\bibfnamefont {H.}~\bibnamefont {{Huang}}}, \bibinfo {author} {\bibfnamefont
  {E.~M.}\ \bibnamefont {{Huff}}}, \bibinfo {author} {\bibfnamefont
  {M.}~\bibnamefont {{Jarvis}}}, \bibinfo {author} {\bibfnamefont
  {N.}~\bibnamefont {{Jeffrey}}}, \bibinfo {author} {\bibfnamefont
  {N.}~\bibnamefont {{Kuropatkin}}}, \bibinfo {author} {\bibfnamefont {P.~F.}\
  \bibnamefont {{Leget}}}, \bibinfo {author} {\bibfnamefont {J.}~\bibnamefont
  {{Muir}}}, \bibinfo {author} {\bibfnamefont {J.}~\bibnamefont
  {{Mccullough}}}, \bibinfo {author} {\bibfnamefont {A.}~\bibnamefont {{Navarro
  Alsina}}}, \bibinfo {author} {\bibfnamefont {Y.}~\bibnamefont {{Omori}}},
  \bibinfo {author} {\bibfnamefont {Y.}~\bibnamefont {{Park}}}, \bibinfo
  {author} {\bibfnamefont {A.}~\bibnamefont {{Porredon}}}, \bibinfo {author}
  {\bibfnamefont {R.}~\bibnamefont {{Rollins}}}, \bibinfo {author}
  {\bibfnamefont {A.}~\bibnamefont {{Roodman}}}, \bibinfo {author}
  {\bibfnamefont {R.}~\bibnamefont {{Rosenfeld}}}, \bibinfo {author}
  {\bibfnamefont {A.~J.}\ \bibnamefont {{Ross}}}, \bibinfo {author}
  {\bibfnamefont {E.~S.}\ \bibnamefont {{Rykoff}}}, \bibinfo {author}
  {\bibfnamefont {J.}~\bibnamefont {{Sanchez}}}, \bibinfo {author}
  {\bibfnamefont {I.}~\bibnamefont {{Sevilla-Noarbe}}}, \bibinfo {author}
  {\bibfnamefont {E.~S.}\ \bibnamefont {{Sheldon}}}, \bibinfo {author}
  {\bibfnamefont {T.}~\bibnamefont {{Shin}}}, \bibinfo {author} {\bibfnamefont
  {I.}~\bibnamefont {{Tutusaus}}}, \bibinfo {author} {\bibfnamefont {T.~N.}\
  \bibnamefont {{Varga}}}, \bibinfo {author} {\bibfnamefont {N.}~\bibnamefont
  {{Weaverdyck}}}, \bibinfo {author} {\bibfnamefont {R.~H.}\ \bibnamefont
  {{Wechsler}}}, \bibinfo {author} {\bibfnamefont {B.}~\bibnamefont {{Yanny}}},
  \bibinfo {author} {\bibfnamefont {B.}~\bibnamefont {{Yin}}}, \bibinfo
  {author} {\bibfnamefont {Y.}~\bibnamefont {{Zhang}}}, \bibinfo {author}
  {\bibfnamefont {J.}~\bibnamefont {{Zuntz}}}, \bibinfo {author} {\bibfnamefont
  {T.~M.~C.}\ \bibnamefont {{Abbott}}}, \bibinfo {author} {\bibfnamefont
  {M.}~\bibnamefont {{Aguena}}}, \bibinfo {author} {\bibfnamefont
  {S.}~\bibnamefont {{Allam}}}, \bibinfo {author} {\bibfnamefont
  {J.}~\bibnamefont {{Annis}}}, \bibinfo {author} {\bibfnamefont
  {D.}~\bibnamefont {{Bacon}}}, \bibinfo {author} {\bibfnamefont
  {E.}~\bibnamefont {{Bertin}}}, \bibinfo {author} {\bibfnamefont
  {S.}~\bibnamefont {{Bhargava}}}, \bibinfo {author} {\bibfnamefont {S.~L.}\
  \bibnamefont {{Bridle}}}, \bibinfo {author} {\bibfnamefont {D.}~\bibnamefont
  {{Brooks}}}, \bibinfo {author} {\bibfnamefont {E.}~\bibnamefont
  {{Buckley-Geer}}}, \bibinfo {author} {\bibfnamefont {D.~L.}\ \bibnamefont
  {{Burke}}}, \bibinfo {author} {\bibfnamefont {J.}~\bibnamefont
  {{Carretero}}}, \bibinfo {author} {\bibfnamefont {M.}~\bibnamefont
  {{Costanzi}}}, \bibinfo {author} {\bibfnamefont {L.~N.}\ \bibnamefont {{da
  Costa}}}, \bibinfo {author} {\bibfnamefont {J.}~\bibnamefont {{De Vicente}}},
  \bibinfo {author} {\bibfnamefont {H.~T.}\ \bibnamefont {{Diehl}}}, \bibinfo
  {author} {\bibfnamefont {J.~P.}\ \bibnamefont {{Dietrich}}}, \bibinfo
  {author} {\bibfnamefont {P.}~\bibnamefont {{Doel}}}, \bibinfo {author}
  {\bibfnamefont {I.}~\bibnamefont {{Ferrero}}}, \bibinfo {author}
  {\bibfnamefont {B.}~\bibnamefont {{Flaugher}}}, \bibinfo {author}
  {\bibfnamefont {J.}~\bibnamefont {{Frieman}}}, \bibinfo {author}
  {\bibfnamefont {J.}~\bibnamefont {{Garc{\'\i}a-Bellido}}}, \bibinfo {author}
  {\bibfnamefont {E.}~\bibnamefont {{Gaztanaga}}}, \bibinfo {author}
  {\bibfnamefont {D.~W.}\ \bibnamefont {{Gerdes}}}, \bibinfo {author}
  {\bibfnamefont {T.}~\bibnamefont {{Giannantonio}}}, \bibinfo {author}
  {\bibfnamefont {J.}~\bibnamefont {{Gschwend}}}, \bibinfo {author}
  {\bibfnamefont {G.}~\bibnamefont {{Gutierrez}}}, \bibinfo {author}
  {\bibfnamefont {S.~R.}\ \bibnamefont {{Hinton}}}, \bibinfo {author}
  {\bibfnamefont {D.~L.}\ \bibnamefont {{Hollowood}}}, \bibinfo {author}
  {\bibfnamefont {K.}~\bibnamefont {{Honscheid}}}, \bibinfo {author}
  {\bibfnamefont {B.}~\bibnamefont {{Hoyle}}}, \bibinfo {author} {\bibfnamefont
  {D.~J.}\ \bibnamefont {{James}}}, \bibinfo {author} {\bibfnamefont
  {T.}~\bibnamefont {{Jeltema}}}, \bibinfo {author} {\bibfnamefont
  {K.}~\bibnamefont {{Kuehn}}}, \bibinfo {author} {\bibfnamefont
  {O.}~\bibnamefont {{Lahav}}}, \bibinfo {author} {\bibfnamefont
  {M.}~\bibnamefont {{Lima}}}, \bibinfo {author} {\bibfnamefont
  {H.}~\bibnamefont {{Lin}}}, \bibinfo {author} {\bibfnamefont {M.~A.~G.}\
  \bibnamefont {{Maia}}}, \bibinfo {author} {\bibfnamefont {J.~L.}\
  \bibnamefont {{Marshall}}}, \bibinfo {author} {\bibfnamefont
  {P.}~\bibnamefont {{Martini}}}, \bibinfo {author} {\bibfnamefont
  {P.}~\bibnamefont {{Melchior}}}, \bibinfo {author} {\bibfnamefont
  {F.}~\bibnamefont {{Menanteau}}}, \bibinfo {author} {\bibfnamefont
  {R.}~\bibnamefont {{Miquel}}}, \bibinfo {author} {\bibfnamefont {J.~J.}\
  \bibnamefont {{Mohr}}}, \bibinfo {author} {\bibfnamefont {R.}~\bibnamefont
  {{Morgan}}}, \bibinfo {author} {\bibfnamefont {R.~L.~C.}\ \bibnamefont
  {{Ogando}}}, \bibinfo {author} {\bibfnamefont {A.}~\bibnamefont {{Palmese}}},
  \bibinfo {author} {\bibfnamefont {F.}~\bibnamefont {{Paz-Chinch{\'o}n}}},
  \bibinfo {author} {\bibfnamefont {D.}~\bibnamefont {{Petravick}}}, \bibinfo
  {author} {\bibfnamefont {A.}~\bibnamefont {{Pieres}}}, \bibinfo {author}
  {\bibfnamefont {A.~A.}\ \bibnamefont {{Plazas Malag{\'o}n}}}, \bibinfo
  {author} {\bibfnamefont {M.}~\bibnamefont {{Rodriguez-Monroy}}}, \bibinfo
  {author} {\bibfnamefont {A.~K.}\ \bibnamefont {{Romer}}}, \bibinfo {author}
  {\bibfnamefont {E.}~\bibnamefont {{Sanchez}}}, \bibinfo {author}
  {\bibfnamefont {V.}~\bibnamefont {{Scarpine}}}, \bibinfo {author}
  {\bibfnamefont {M.}~\bibnamefont {{Schubnell}}}, \bibinfo {author}
  {\bibfnamefont {D.}~\bibnamefont {{Scolnic}}}, \bibinfo {author}
  {\bibfnamefont {S.}~\bibnamefont {{Serrano}}}, \bibinfo {author}
  {\bibfnamefont {M.}~\bibnamefont {{Smith}}}, \bibinfo {author} {\bibfnamefont
  {M.}~\bibnamefont {{Soares-Santos}}}, \bibinfo {author} {\bibfnamefont
  {E.}~\bibnamefont {{Suchyta}}}, \bibinfo {author} {\bibfnamefont {M.~E.~C.}\
  \bibnamefont {{Swanson}}}, \bibinfo {author} {\bibfnamefont {G.}~\bibnamefont
  {{Tarle}}}, \bibinfo {author} {\bibfnamefont {D.}~\bibnamefont {{Thomas}}},
  \bibinfo {author} {\bibfnamefont {C.}~\bibnamefont {{To}}},\ and\ \bibinfo
  {author} {\bibnamefont {{DES Collaboration}}},\ }\bibfield  {title} {\bibinfo
  {title} {{Dark Energy Survey Year 3 results: Cosmology from cosmic shear and
  robustness to modeling uncertainty}},\ }\href
  {https://doi.org/10.1103/PhysRevD.105.023515} {\bibfield  {journal} {\bibinfo
   {journal} {\prd}\ }\textbf {\bibinfo {volume} {105}},\ \bibinfo {eid}
  {023515} (\bibinfo {year} {2022})},\ \Eprint
  {https://arxiv.org/abs/2105.13544} {arXiv:2105.13544 [astro-ph.CO]}
  \BibitemShut {NoStop}%
\bibitem [{\citenamefont {{Porredon}}\ \emph
  {et~al.}(2021{\natexlab{a}})\citenamefont {{Porredon}}, \citenamefont
  {{Crocce}}, \citenamefont {{Elvin-Poole}}, \citenamefont {{Cawthon}},
  \citenamefont {{Giannini}}, \citenamefont {{De Vicente}}, \citenamefont
  {{Carnero Rosell}}, \citenamefont {{Ferrero}}, \citenamefont {{Krause}},
  \citenamefont {{Fang}}, \citenamefont {{Prat}}, \citenamefont
  {{Rodriguez-Monroy}}, \citenamefont {{Pandey}}, \citenamefont {{Pocino}},
  \citenamefont {{Castander}}, \citenamefont {{Choi}}, \citenamefont {{Amon}},
  \citenamefont {{Tutusaus}}, \citenamefont {{Dodelson}}, \citenamefont
  {{Sevilla-Noarbe}}, \citenamefont {{Fosalba}}, \citenamefont {{Gaztanaga}},
  \citenamefont {{Alarcon}}, \citenamefont {{Alves}}, \citenamefont
  {{Andrade-Oliveira}}, \citenamefont {{Baxter}}, \citenamefont {{Bechtol}},
  \citenamefont {{Becker}}, \citenamefont {{Bernstein}}, \citenamefont
  {{Blazek}}, \citenamefont {{Camacho}}, \citenamefont {{Campos}},
  \citenamefont {{Carrasco Kind}}, \citenamefont {{Chintalapati}},
  \citenamefont {{Cordero}}, \citenamefont {{DeRose}}, \citenamefont {{Di
  Valentino}}, \citenamefont {{Doux}}, \citenamefont {{Eifler}}, \citenamefont
  {{Everett}}, \citenamefont {{Fert{\'e}}}, \citenamefont {{Friedrich}},
  \citenamefont {{Gatti}}, \citenamefont {{Gruen}}, \citenamefont {{Harrison}},
  \citenamefont {{Hartley}}, \citenamefont {{Herner}}, \citenamefont {{Huff}},
  \citenamefont {{Huterer}}, \citenamefont {{Jain}}, \citenamefont {{Jarvis}},
  \citenamefont {{Lee}}, \citenamefont {{Lemos}}, \citenamefont {{MacCrann}},
  \citenamefont {{Mena-Fern{\'a}ndez}}, \citenamefont {{Muir}}, \citenamefont
  {{Myles}}, \citenamefont {{Park}}, \citenamefont {{Raveri}}, \citenamefont
  {{Rosenfeld}}, \citenamefont {{Ross}}, \citenamefont {{Rykoff}},
  \citenamefont {{Samuroff}}, \citenamefont {{S{\'a}nchez}}, \citenamefont
  {{Sanchez}}, \citenamefont {{Sanchez}}, \citenamefont {{Sanchez Cid}},
  \citenamefont {{Scolnic}}, \citenamefont {{Secco}}, \citenamefont
  {{Sheldon}}, \citenamefont {{Troja}}, \citenamefont {{Troxel}}, \citenamefont
  {{Weaverdyck}}, \citenamefont {{Yanny}}, \citenamefont {{Zuntz}},
  \citenamefont {{Abbott}}, \citenamefont {{Aguena}}, \citenamefont {{Allam}},
  \citenamefont {{Annis}}, \citenamefont {{Avila}}, \citenamefont {{Bacon}},
  \citenamefont {{Bertin}}, \citenamefont {{Bhargava}}, \citenamefont
  {{Brooks}}, \citenamefont {{Buckley-Geer}}, \citenamefont {{Burke}},
  \citenamefont {{Carretero}}, \citenamefont {{Costanzi}}, \citenamefont {{da
  Costa}}, \citenamefont {{Pereira}}, \citenamefont {{Davis}}, \citenamefont
  {{Desai}}, \citenamefont {{Diehl}}, \citenamefont {{Dietrich}}, \citenamefont
  {{Doel}}, \citenamefont {{Drlica-Wagner}}, \citenamefont {{Eckert}},
  \citenamefont {{Evrard}}, \citenamefont {{Flaugher}}, \citenamefont
  {{Frieman}}, \citenamefont {{Garc{\'\i}a-Bellido}}, \citenamefont {{Gerdes}},
  \citenamefont {{Giannantonio}}, \citenamefont {{Gruendl}}, \citenamefont
  {{Gschwend}}, \citenamefont {{Gutierrez}}, \citenamefont {{Hinton}},
  \citenamefont {{Hollowood}}, \citenamefont {{Honscheid}}, \citenamefont
  {{Hoyle}}, \citenamefont {{James}}, \citenamefont {{Kuehn}}, \citenamefont
  {{Kuropatkin}}, \citenamefont {{Lahav}}, \citenamefont {{Lidman}},
  \citenamefont {{Lima}}, \citenamefont {{Lin}}, \citenamefont {{Maia}},
  \citenamefont {{Marshall}}, \citenamefont {{Martini}}, \citenamefont
  {{Melchior}}, \citenamefont {{Menanteau}}, \citenamefont {{Miquel}},
  \citenamefont {{Mohr}}, \citenamefont {{Morgan}}, \citenamefont {{Ogando}},
  \citenamefont {{Palmese}}, \citenamefont {{Paz-Chinch{\'o}n}}, \citenamefont
  {{Petravick}}, \citenamefont {{Pieres}}, \citenamefont {{Plazas
  Malag{\'o}n}}, \citenamefont {{Romer}}, \citenamefont {{Santiago}},
  \citenamefont {{Scarpine}}, \citenamefont {{Schubnell}}, \citenamefont
  {{Serrano}}, \citenamefont {{Smith}}, \citenamefont {{Soares-Santos}},
  \citenamefont {{Suchyta}}, \citenamefont {{Tarle}}, \citenamefont {{Thomas}},
  \citenamefont {{To}}, \citenamefont {{Varga}},\ and\ \citenamefont
  {{Weller}}}]{Porredon2021clustering}%
  \BibitemOpen
  \bibfield  {author} {\bibinfo {author} {\bibfnamefont {A.}~\bibnamefont
  {{Porredon}}}, \bibinfo {author} {\bibfnamefont {M.}~\bibnamefont
  {{Crocce}}}, \bibinfo {author} {\bibfnamefont {J.}~\bibnamefont
  {{Elvin-Poole}}}, \bibinfo {author} {\bibfnamefont {R.}~\bibnamefont
  {{Cawthon}}}, \bibinfo {author} {\bibfnamefont {G.}~\bibnamefont
  {{Giannini}}}, \bibinfo {author} {\bibfnamefont {J.}~\bibnamefont {{De
  Vicente}}}, \bibinfo {author} {\bibfnamefont {A.}~\bibnamefont {{Carnero
  Rosell}}}, \bibinfo {author} {\bibfnamefont {I.}~\bibnamefont {{Ferrero}}},
  \bibinfo {author} {\bibfnamefont {E.}~\bibnamefont {{Krause}}}, \bibinfo
  {author} {\bibfnamefont {X.}~\bibnamefont {{Fang}}}, \bibinfo {author}
  {\bibfnamefont {J.}~\bibnamefont {{Prat}}}, \bibinfo {author} {\bibfnamefont
  {M.}~\bibnamefont {{Rodriguez-Monroy}}}, \bibinfo {author} {\bibfnamefont
  {S.}~\bibnamefont {{Pandey}}}, \bibinfo {author} {\bibfnamefont
  {A.}~\bibnamefont {{Pocino}}}, \bibinfo {author} {\bibfnamefont {F.~J.}\
  \bibnamefont {{Castander}}}, \bibinfo {author} {\bibfnamefont
  {A.}~\bibnamefont {{Choi}}}, \bibinfo {author} {\bibfnamefont
  {A.}~\bibnamefont {{Amon}}}, \bibinfo {author} {\bibfnamefont
  {I.}~\bibnamefont {{Tutusaus}}}, \bibinfo {author} {\bibfnamefont
  {S.}~\bibnamefont {{Dodelson}}}, \bibinfo {author} {\bibfnamefont
  {I.}~\bibnamefont {{Sevilla-Noarbe}}}, \bibinfo {author} {\bibfnamefont
  {P.}~\bibnamefont {{Fosalba}}}, \bibinfo {author} {\bibfnamefont
  {E.}~\bibnamefont {{Gaztanaga}}}, \bibinfo {author} {\bibfnamefont
  {A.}~\bibnamefont {{Alarcon}}}, \bibinfo {author} {\bibfnamefont
  {O.}~\bibnamefont {{Alves}}}, \bibinfo {author} {\bibfnamefont
  {F.}~\bibnamefont {{Andrade-Oliveira}}}, \bibinfo {author} {\bibfnamefont
  {E.}~\bibnamefont {{Baxter}}}, \bibinfo {author} {\bibfnamefont
  {K.}~\bibnamefont {{Bechtol}}}, \bibinfo {author} {\bibfnamefont {M.~R.}\
  \bibnamefont {{Becker}}}, \bibinfo {author} {\bibfnamefont {G.~M.}\
  \bibnamefont {{Bernstein}}}, \bibinfo {author} {\bibfnamefont
  {J.}~\bibnamefont {{Blazek}}}, \bibinfo {author} {\bibfnamefont
  {H.}~\bibnamefont {{Camacho}}}, \bibinfo {author} {\bibfnamefont
  {A.}~\bibnamefont {{Campos}}}, \bibinfo {author} {\bibfnamefont
  {M.}~\bibnamefont {{Carrasco Kind}}}, \bibinfo {author} {\bibfnamefont
  {P.}~\bibnamefont {{Chintalapati}}}, \bibinfo {author} {\bibfnamefont
  {J.}~\bibnamefont {{Cordero}}}, \bibinfo {author} {\bibfnamefont
  {J.}~\bibnamefont {{DeRose}}}, \bibinfo {author} {\bibfnamefont
  {E.}~\bibnamefont {{Di Valentino}}}, \bibinfo {author} {\bibfnamefont
  {C.}~\bibnamefont {{Doux}}}, \bibinfo {author} {\bibfnamefont {T.~F.}\
  \bibnamefont {{Eifler}}}, \bibinfo {author} {\bibfnamefont {S.}~\bibnamefont
  {{Everett}}}, \bibinfo {author} {\bibfnamefont {A.}~\bibnamefont
  {{Fert{\'e}}}}, \bibinfo {author} {\bibfnamefont {O.}~\bibnamefont
  {{Friedrich}}}, \bibinfo {author} {\bibfnamefont {M.}~\bibnamefont
  {{Gatti}}}, \bibinfo {author} {\bibfnamefont {D.}~\bibnamefont {{Gruen}}},
  \bibinfo {author} {\bibfnamefont {I.}~\bibnamefont {{Harrison}}}, \bibinfo
  {author} {\bibfnamefont {W.~G.}\ \bibnamefont {{Hartley}}}, \bibinfo {author}
  {\bibfnamefont {K.}~\bibnamefont {{Herner}}}, \bibinfo {author}
  {\bibfnamefont {E.~M.}\ \bibnamefont {{Huff}}}, \bibinfo {author}
  {\bibfnamefont {D.}~\bibnamefont {{Huterer}}}, \bibinfo {author}
  {\bibfnamefont {B.}~\bibnamefont {{Jain}}}, \bibinfo {author} {\bibfnamefont
  {M.}~\bibnamefont {{Jarvis}}}, \bibinfo {author} {\bibfnamefont
  {S.}~\bibnamefont {{Lee}}}, \bibinfo {author} {\bibfnamefont
  {P.}~\bibnamefont {{Lemos}}}, \bibinfo {author} {\bibfnamefont
  {N.}~\bibnamefont {{MacCrann}}}, \bibinfo {author} {\bibfnamefont
  {J.}~\bibnamefont {{Mena-Fern{\'a}ndez}}}, \bibinfo {author} {\bibfnamefont
  {J.}~\bibnamefont {{Muir}}}, \bibinfo {author} {\bibfnamefont
  {J.}~\bibnamefont {{Myles}}}, \bibinfo {author} {\bibfnamefont
  {Y.}~\bibnamefont {{Park}}}, \bibinfo {author} {\bibfnamefont
  {M.}~\bibnamefont {{Raveri}}}, \bibinfo {author} {\bibfnamefont
  {R.}~\bibnamefont {{Rosenfeld}}}, \bibinfo {author} {\bibfnamefont {A.~J.}\
  \bibnamefont {{Ross}}}, \bibinfo {author} {\bibfnamefont {E.~S.}\
  \bibnamefont {{Rykoff}}}, \bibinfo {author} {\bibfnamefont {S.}~\bibnamefont
  {{Samuroff}}}, \bibinfo {author} {\bibfnamefont {C.}~\bibnamefont
  {{S{\'a}nchez}}}, \bibinfo {author} {\bibfnamefont {E.}~\bibnamefont
  {{Sanchez}}}, \bibinfo {author} {\bibfnamefont {J.}~\bibnamefont
  {{Sanchez}}}, \bibinfo {author} {\bibfnamefont {D.}~\bibnamefont {{Sanchez
  Cid}}}, \bibinfo {author} {\bibfnamefont {D.}~\bibnamefont {{Scolnic}}},
  \bibinfo {author} {\bibfnamefont {L.~F.}\ \bibnamefont {{Secco}}}, \bibinfo
  {author} {\bibfnamefont {E.}~\bibnamefont {{Sheldon}}}, \bibinfo {author}
  {\bibfnamefont {A.}~\bibnamefont {{Troja}}}, \bibinfo {author} {\bibfnamefont
  {M.~A.}\ \bibnamefont {{Troxel}}}, \bibinfo {author} {\bibfnamefont
  {N.}~\bibnamefont {{Weaverdyck}}}, \bibinfo {author} {\bibfnamefont
  {B.}~\bibnamefont {{Yanny}}}, \bibinfo {author} {\bibfnamefont
  {J.}~\bibnamefont {{Zuntz}}}, \bibinfo {author} {\bibfnamefont {T.~M.~C.}\
  \bibnamefont {{Abbott}}}, \bibinfo {author} {\bibfnamefont {M.}~\bibnamefont
  {{Aguena}}}, \bibinfo {author} {\bibfnamefont {S.}~\bibnamefont {{Allam}}},
  \bibinfo {author} {\bibfnamefont {J.}~\bibnamefont {{Annis}}}, \bibinfo
  {author} {\bibfnamefont {S.}~\bibnamefont {{Avila}}}, \bibinfo {author}
  {\bibfnamefont {D.}~\bibnamefont {{Bacon}}}, \bibinfo {author} {\bibfnamefont
  {E.}~\bibnamefont {{Bertin}}}, \bibinfo {author} {\bibfnamefont
  {S.}~\bibnamefont {{Bhargava}}}, \bibinfo {author} {\bibfnamefont
  {D.}~\bibnamefont {{Brooks}}}, \bibinfo {author} {\bibfnamefont
  {E.}~\bibnamefont {{Buckley-Geer}}}, \bibinfo {author} {\bibfnamefont
  {D.~L.}\ \bibnamefont {{Burke}}}, \bibinfo {author} {\bibfnamefont
  {J.}~\bibnamefont {{Carretero}}}, \bibinfo {author} {\bibfnamefont
  {M.}~\bibnamefont {{Costanzi}}}, \bibinfo {author} {\bibfnamefont {L.~N.}\
  \bibnamefont {{da Costa}}}, \bibinfo {author} {\bibfnamefont {M.~E.~S.}\
  \bibnamefont {{Pereira}}}, \bibinfo {author} {\bibfnamefont {T.~M.}\
  \bibnamefont {{Davis}}}, \bibinfo {author} {\bibfnamefont {S.}~\bibnamefont
  {{Desai}}}, \bibinfo {author} {\bibfnamefont {H.~T.}\ \bibnamefont
  {{Diehl}}}, \bibinfo {author} {\bibfnamefont {J.~P.}\ \bibnamefont
  {{Dietrich}}}, \bibinfo {author} {\bibfnamefont {P.}~\bibnamefont {{Doel}}},
  \bibinfo {author} {\bibfnamefont {A.}~\bibnamefont {{Drlica-Wagner}}},
  \bibinfo {author} {\bibfnamefont {K.}~\bibnamefont {{Eckert}}}, \bibinfo
  {author} {\bibfnamefont {A.~E.}\ \bibnamefont {{Evrard}}}, \bibinfo {author}
  {\bibfnamefont {B.}~\bibnamefont {{Flaugher}}}, \bibinfo {author}
  {\bibfnamefont {J.}~\bibnamefont {{Frieman}}}, \bibinfo {author}
  {\bibfnamefont {J.}~\bibnamefont {{Garc{\'\i}a-Bellido}}}, \bibinfo {author}
  {\bibfnamefont {D.~W.}\ \bibnamefont {{Gerdes}}}, \bibinfo {author}
  {\bibfnamefont {T.}~\bibnamefont {{Giannantonio}}}, \bibinfo {author}
  {\bibfnamefont {R.~A.}\ \bibnamefont {{Gruendl}}}, \bibinfo {author}
  {\bibfnamefont {J.}~\bibnamefont {{Gschwend}}}, \bibinfo {author}
  {\bibfnamefont {G.}~\bibnamefont {{Gutierrez}}}, \bibinfo {author}
  {\bibfnamefont {S.~R.}\ \bibnamefont {{Hinton}}}, \bibinfo {author}
  {\bibfnamefont {D.~L.}\ \bibnamefont {{Hollowood}}}, \bibinfo {author}
  {\bibfnamefont {K.}~\bibnamefont {{Honscheid}}}, \bibinfo {author}
  {\bibfnamefont {B.}~\bibnamefont {{Hoyle}}}, \bibinfo {author} {\bibfnamefont
  {D.~J.}\ \bibnamefont {{James}}}, \bibinfo {author} {\bibfnamefont
  {K.}~\bibnamefont {{Kuehn}}}, \bibinfo {author} {\bibfnamefont
  {N.}~\bibnamefont {{Kuropatkin}}}, \bibinfo {author} {\bibfnamefont
  {O.}~\bibnamefont {{Lahav}}}, \bibinfo {author} {\bibfnamefont
  {C.}~\bibnamefont {{Lidman}}}, \bibinfo {author} {\bibfnamefont
  {M.}~\bibnamefont {{Lima}}}, \bibinfo {author} {\bibfnamefont
  {H.}~\bibnamefont {{Lin}}}, \bibinfo {author} {\bibfnamefont {M.~A.~G.}\
  \bibnamefont {{Maia}}}, \bibinfo {author} {\bibfnamefont {J.~L.}\
  \bibnamefont {{Marshall}}}, \bibinfo {author} {\bibfnamefont
  {P.}~\bibnamefont {{Martini}}}, \bibinfo {author} {\bibfnamefont
  {P.}~\bibnamefont {{Melchior}}}, \bibinfo {author} {\bibfnamefont
  {F.}~\bibnamefont {{Menanteau}}}, \bibinfo {author} {\bibfnamefont
  {R.}~\bibnamefont {{Miquel}}}, \bibinfo {author} {\bibfnamefont {J.~J.}\
  \bibnamefont {{Mohr}}}, \bibinfo {author} {\bibfnamefont {R.}~\bibnamefont
  {{Morgan}}}, \bibinfo {author} {\bibfnamefont {R.~L.~C.}\ \bibnamefont
  {{Ogando}}}, \bibinfo {author} {\bibfnamefont {A.}~\bibnamefont {{Palmese}}},
  \bibinfo {author} {\bibfnamefont {F.}~\bibnamefont {{Paz-Chinch{\'o}n}}},
  \bibinfo {author} {\bibfnamefont {D.}~\bibnamefont {{Petravick}}}, \bibinfo
  {author} {\bibfnamefont {A.}~\bibnamefont {{Pieres}}}, \bibinfo {author}
  {\bibfnamefont {A.~A.}\ \bibnamefont {{Plazas Malag{\'o}n}}}, \bibinfo
  {author} {\bibfnamefont {A.~K.}\ \bibnamefont {{Romer}}}, \bibinfo {author}
  {\bibfnamefont {B.}~\bibnamefont {{Santiago}}}, \bibinfo {author}
  {\bibfnamefont {V.}~\bibnamefont {{Scarpine}}}, \bibinfo {author}
  {\bibfnamefont {M.}~\bibnamefont {{Schubnell}}}, \bibinfo {author}
  {\bibfnamefont {S.}~\bibnamefont {{Serrano}}}, \bibinfo {author}
  {\bibfnamefont {M.}~\bibnamefont {{Smith}}}, \bibinfo {author} {\bibfnamefont
  {M.}~\bibnamefont {{Soares-Santos}}}, \bibinfo {author} {\bibfnamefont
  {E.}~\bibnamefont {{Suchyta}}}, \bibinfo {author} {\bibfnamefont
  {G.}~\bibnamefont {{Tarle}}}, \bibinfo {author} {\bibfnamefont
  {D.}~\bibnamefont {{Thomas}}}, \bibinfo {author} {\bibfnamefont
  {C.}~\bibnamefont {{To}}}, \bibinfo {author} {\bibfnamefont {T.~N.}\
  \bibnamefont {{Varga}}},\ and\ \bibinfo {author} {\bibfnamefont
  {J.}~\bibnamefont {{Weller}}},\ }\bibfield  {title} {\bibinfo {title} {{Dark
  Energy Survey Year 3 results: Cosmological constraints from galaxy clustering
  and galaxy-galaxy lensing using the MagLim lens sample}},\ }\href@noop {}
  {\bibfield  {journal} {\bibinfo  {journal} {arXiv e-prints}\ ,\ \bibinfo
  {eid} {arXiv:2105.13546}} (\bibinfo {year} {2021}{\natexlab{a}})},\ \Eprint
  {https://arxiv.org/abs/2105.13546} {arXiv:2105.13546 [astro-ph.CO]}
  \BibitemShut {NoStop}%
\bibitem [{\citenamefont {{Eifler}}\ \emph {et~al.}(2014)\citenamefont
  {{Eifler}}, \citenamefont {{Krause}}, \citenamefont {{Schneider}},\ and\
  \citenamefont {{Honscheid}}}]{Eifler2014cosmolike}%
  \BibitemOpen
  \bibfield  {author} {\bibinfo {author} {\bibfnamefont {T.}~\bibnamefont
  {{Eifler}}}, \bibinfo {author} {\bibfnamefont {E.}~\bibnamefont {{Krause}}},
  \bibinfo {author} {\bibfnamefont {P.}~\bibnamefont {{Schneider}}},\ and\
  \bibinfo {author} {\bibfnamefont {K.}~\bibnamefont {{Honscheid}}},\
  }\bibfield  {title} {\bibinfo {title} {{Combining probes of large-scale
  structure with COSMOLIKE}},\ }\href {https://doi.org/10.1093/mnras/stu251}
  {\bibfield  {journal} {\bibinfo  {journal} {\mnras}\ }\textbf {\bibinfo
  {volume} {440}},\ \bibinfo {pages} {1379} (\bibinfo {year} {2014})},\ \Eprint
  {https://arxiv.org/abs/1302.2401} {arXiv:1302.2401 [astro-ph.CO]}
  \BibitemShut {NoStop}%
\bibitem [{\citenamefont {{Krause}}\ \emph {et~al.}(2021)\citenamefont
  {{Krause}}, \citenamefont {{Fang}}, \citenamefont {{Pandey}}, \citenamefont
  {{Secco}}, \citenamefont {{Alves}}, \citenamefont {{Huang}}, \citenamefont
  {{Blazek}}, \citenamefont {{Prat}}, \citenamefont {{Zuntz}}, \citenamefont
  {{Eifler}}, \citenamefont {{MacCrann}}, \citenamefont {{DeRose}},
  \citenamefont {{Crocce}}, \citenamefont {{Porredon}}, \citenamefont {{Jain}},
  \citenamefont {{Troxel}}, \citenamefont {{Dodelson}}, \citenamefont
  {{Huterer}}, \citenamefont {{Liddle}}, \citenamefont {{Leonard}},
  \citenamefont {{Amon}}, \citenamefont {{Chen}}, \citenamefont
  {{Elvin-Poole}}, \citenamefont {{Fert{\'e}}}, \citenamefont {{Muir}},
  \citenamefont {{Park}}, \citenamefont {{Samuroff}}, \citenamefont
  {{Brandao-Souza}}, \citenamefont {{Weaverdyck}}, \citenamefont
  {{Zacharegkas}}, \citenamefont {{Rosenfeld}}, \citenamefont {{Campos}},
  \citenamefont {{Chintalapati}}, \citenamefont {{Choi}}, \citenamefont {{Di
  Valentino}}, \citenamefont {{Doux}}, \citenamefont {{Herner}}, \citenamefont
  {{Lemos}}, \citenamefont {{Mena-Fern{\'a}ndez}}, \citenamefont {{Omori}},
  \citenamefont {{Paterno}}, \citenamefont {{Rodriguez-Monroy}}, \citenamefont
  {{Rogozenski}}, \citenamefont {{Rollins}}, \citenamefont {{Troja}},
  \citenamefont {{Tutusaus}}, \citenamefont {{Wechsler}}, \citenamefont
  {{Abbott}}, \citenamefont {{Aguena}}, \citenamefont {{Allam}}, \citenamefont
  {{Andrade-Oliveira}}, \citenamefont {{Annis}}, \citenamefont {{Bacon}},
  \citenamefont {{Baxter}}, \citenamefont {{Bechtol}}, \citenamefont
  {{Bernstein}}, \citenamefont {{Brooks}}, \citenamefont {{Buckley-Geer}},
  \citenamefont {{Burke}}, \citenamefont {{Carnero Rosell}}, \citenamefont
  {{Carrasco Kind}}, \citenamefont {{Carretero}}, \citenamefont {{Castander}},
  \citenamefont {{Cawthon}}, \citenamefont {{Chang}}, \citenamefont
  {{Costanzi}}, \citenamefont {{da Costa}}, \citenamefont {{Pereira}},
  \citenamefont {{De Vicente}}, \citenamefont {{Desai}}, \citenamefont
  {{Diehl}}, \citenamefont {{Doel}}, \citenamefont {{Everett}}, \citenamefont
  {{Evrard}}, \citenamefont {{Ferrero}}, \citenamefont {{Flaugher}},
  \citenamefont {{Fosalba}}, \citenamefont {{Frieman}}, \citenamefont
  {{Garc{\'\i}a-Bellido}}, \citenamefont {{Gaztanaga}}, \citenamefont
  {{Gerdes}}, \citenamefont {{Giannantonio}}, \citenamefont {{Gruen}},
  \citenamefont {{Gruendl}}, \citenamefont {{Gschwend}}, \citenamefont
  {{Gutierrez}}, \citenamefont {{Hartley}}, \citenamefont {{Hinton}},
  \citenamefont {{Hollowood}}, \citenamefont {{Honscheid}}, \citenamefont
  {{Hoyle}}, \citenamefont {{Huff}}, \citenamefont {{James}}, \citenamefont
  {{Kuehn}}, \citenamefont {{Kuropatkin}}, \citenamefont {{Lahav}},
  \citenamefont {{Lima}}, \citenamefont {{Maia}}, \citenamefont {{Marshall}},
  \citenamefont {{Martini}}, \citenamefont {{Melchior}}, \citenamefont
  {{Menanteau}}, \citenamefont {{Miquel}}, \citenamefont {{Mohr}},
  \citenamefont {{Morgan}}, \citenamefont {{Myles}}, \citenamefont {{Palmese}},
  \citenamefont {{Paz-Chinch{\'o}n}}, \citenamefont {{Petravick}},
  \citenamefont {{Pieres}}, \citenamefont {{Plazas Malag{\'o}n}}, \citenamefont
  {{Sanchez}}, \citenamefont {{Scarpine}}, \citenamefont {{Schubnell}},
  \citenamefont {{Serrano}}, \citenamefont {{Sevilla-Noarbe}}, \citenamefont
  {{Smith}}, \citenamefont {{Soares-Santos}}, \citenamefont {{Suchyta}},
  \citenamefont {{Tarle}}, \citenamefont {{Thomas}}, \citenamefont {{To}},
  \citenamefont {{Varga}},\ and\ \citenamefont {{Weller}}}]{Krause2021des}%
  \BibitemOpen
  \bibfield  {author} {\bibinfo {author} {\bibfnamefont {E.}~\bibnamefont
  {{Krause}}}, \bibinfo {author} {\bibfnamefont {X.}~\bibnamefont {{Fang}}},
  \bibinfo {author} {\bibfnamefont {S.}~\bibnamefont {{Pandey}}}, \bibinfo
  {author} {\bibfnamefont {L.~F.}\ \bibnamefont {{Secco}}}, \bibinfo {author}
  {\bibfnamefont {O.}~\bibnamefont {{Alves}}}, \bibinfo {author} {\bibfnamefont
  {H.}~\bibnamefont {{Huang}}}, \bibinfo {author} {\bibfnamefont
  {J.}~\bibnamefont {{Blazek}}}, \bibinfo {author} {\bibfnamefont
  {J.}~\bibnamefont {{Prat}}}, \bibinfo {author} {\bibfnamefont
  {J.}~\bibnamefont {{Zuntz}}}, \bibinfo {author} {\bibfnamefont {T.~F.}\
  \bibnamefont {{Eifler}}}, \bibinfo {author} {\bibfnamefont {N.}~\bibnamefont
  {{MacCrann}}}, \bibinfo {author} {\bibfnamefont {J.}~\bibnamefont
  {{DeRose}}}, \bibinfo {author} {\bibfnamefont {M.}~\bibnamefont {{Crocce}}},
  \bibinfo {author} {\bibfnamefont {A.}~\bibnamefont {{Porredon}}}, \bibinfo
  {author} {\bibfnamefont {B.}~\bibnamefont {{Jain}}}, \bibinfo {author}
  {\bibfnamefont {M.~A.}\ \bibnamefont {{Troxel}}}, \bibinfo {author}
  {\bibfnamefont {S.}~\bibnamefont {{Dodelson}}}, \bibinfo {author}
  {\bibfnamefont {D.}~\bibnamefont {{Huterer}}}, \bibinfo {author}
  {\bibfnamefont {A.~R.}\ \bibnamefont {{Liddle}}}, \bibinfo {author}
  {\bibfnamefont {C.~D.}\ \bibnamefont {{Leonard}}}, \bibinfo {author}
  {\bibfnamefont {A.}~\bibnamefont {{Amon}}}, \bibinfo {author} {\bibfnamefont
  {A.}~\bibnamefont {{Chen}}}, \bibinfo {author} {\bibfnamefont
  {J.}~\bibnamefont {{Elvin-Poole}}}, \bibinfo {author} {\bibfnamefont
  {A.}~\bibnamefont {{Fert{\'e}}}}, \bibinfo {author} {\bibfnamefont
  {J.}~\bibnamefont {{Muir}}}, \bibinfo {author} {\bibfnamefont
  {Y.}~\bibnamefont {{Park}}}, \bibinfo {author} {\bibfnamefont
  {S.}~\bibnamefont {{Samuroff}}}, \bibinfo {author} {\bibfnamefont
  {A.}~\bibnamefont {{Brandao-Souza}}}, \bibinfo {author} {\bibfnamefont
  {N.}~\bibnamefont {{Weaverdyck}}}, \bibinfo {author} {\bibfnamefont
  {G.}~\bibnamefont {{Zacharegkas}}}, \bibinfo {author} {\bibfnamefont
  {R.}~\bibnamefont {{Rosenfeld}}}, \bibinfo {author} {\bibfnamefont
  {A.}~\bibnamefont {{Campos}}}, \bibinfo {author} {\bibfnamefont
  {P.}~\bibnamefont {{Chintalapati}}}, \bibinfo {author} {\bibfnamefont
  {A.}~\bibnamefont {{Choi}}}, \bibinfo {author} {\bibfnamefont
  {E.}~\bibnamefont {{Di Valentino}}}, \bibinfo {author} {\bibfnamefont
  {C.}~\bibnamefont {{Doux}}}, \bibinfo {author} {\bibfnamefont
  {K.}~\bibnamefont {{Herner}}}, \bibinfo {author} {\bibfnamefont
  {P.}~\bibnamefont {{Lemos}}}, \bibinfo {author} {\bibfnamefont
  {J.}~\bibnamefont {{Mena-Fern{\'a}ndez}}}, \bibinfo {author} {\bibfnamefont
  {Y.}~\bibnamefont {{Omori}}}, \bibinfo {author} {\bibfnamefont
  {M.}~\bibnamefont {{Paterno}}}, \bibinfo {author} {\bibfnamefont
  {M.}~\bibnamefont {{Rodriguez-Monroy}}}, \bibinfo {author} {\bibfnamefont
  {P.}~\bibnamefont {{Rogozenski}}}, \bibinfo {author} {\bibfnamefont {R.~P.}\
  \bibnamefont {{Rollins}}}, \bibinfo {author} {\bibfnamefont {A.}~\bibnamefont
  {{Troja}}}, \bibinfo {author} {\bibfnamefont {I.}~\bibnamefont {{Tutusaus}}},
  \bibinfo {author} {\bibfnamefont {R.~H.}\ \bibnamefont {{Wechsler}}},
  \bibinfo {author} {\bibfnamefont {T.~M.~C.}\ \bibnamefont {{Abbott}}},
  \bibinfo {author} {\bibfnamefont {M.}~\bibnamefont {{Aguena}}}, \bibinfo
  {author} {\bibfnamefont {S.}~\bibnamefont {{Allam}}}, \bibinfo {author}
  {\bibfnamefont {F.}~\bibnamefont {{Andrade-Oliveira}}}, \bibinfo {author}
  {\bibfnamefont {J.}~\bibnamefont {{Annis}}}, \bibinfo {author} {\bibfnamefont
  {D.}~\bibnamefont {{Bacon}}}, \bibinfo {author} {\bibfnamefont
  {E.}~\bibnamefont {{Baxter}}}, \bibinfo {author} {\bibfnamefont
  {K.}~\bibnamefont {{Bechtol}}}, \bibinfo {author} {\bibfnamefont {G.~M.}\
  \bibnamefont {{Bernstein}}}, \bibinfo {author} {\bibfnamefont
  {D.}~\bibnamefont {{Brooks}}}, \bibinfo {author} {\bibfnamefont
  {E.}~\bibnamefont {{Buckley-Geer}}}, \bibinfo {author} {\bibfnamefont
  {D.~L.}\ \bibnamefont {{Burke}}}, \bibinfo {author} {\bibfnamefont
  {A.}~\bibnamefont {{Carnero Rosell}}}, \bibinfo {author} {\bibfnamefont
  {M.}~\bibnamefont {{Carrasco Kind}}}, \bibinfo {author} {\bibfnamefont
  {J.}~\bibnamefont {{Carretero}}}, \bibinfo {author} {\bibfnamefont {F.~J.}\
  \bibnamefont {{Castander}}}, \bibinfo {author} {\bibfnamefont
  {R.}~\bibnamefont {{Cawthon}}}, \bibinfo {author} {\bibfnamefont
  {C.}~\bibnamefont {{Chang}}}, \bibinfo {author} {\bibfnamefont
  {M.}~\bibnamefont {{Costanzi}}}, \bibinfo {author} {\bibfnamefont {L.~N.}\
  \bibnamefont {{da Costa}}}, \bibinfo {author} {\bibfnamefont {M.~E.~S.}\
  \bibnamefont {{Pereira}}}, \bibinfo {author} {\bibfnamefont {J.}~\bibnamefont
  {{De Vicente}}}, \bibinfo {author} {\bibfnamefont {S.}~\bibnamefont
  {{Desai}}}, \bibinfo {author} {\bibfnamefont {H.~T.}\ \bibnamefont
  {{Diehl}}}, \bibinfo {author} {\bibfnamefont {P.}~\bibnamefont {{Doel}}},
  \bibinfo {author} {\bibfnamefont {S.}~\bibnamefont {{Everett}}}, \bibinfo
  {author} {\bibfnamefont {A.~E.}\ \bibnamefont {{Evrard}}}, \bibinfo {author}
  {\bibfnamefont {I.}~\bibnamefont {{Ferrero}}}, \bibinfo {author}
  {\bibfnamefont {B.}~\bibnamefont {{Flaugher}}}, \bibinfo {author}
  {\bibfnamefont {P.}~\bibnamefont {{Fosalba}}}, \bibinfo {author}
  {\bibfnamefont {J.}~\bibnamefont {{Frieman}}}, \bibinfo {author}
  {\bibfnamefont {J.}~\bibnamefont {{Garc{\'\i}a-Bellido}}}, \bibinfo {author}
  {\bibfnamefont {E.}~\bibnamefont {{Gaztanaga}}}, \bibinfo {author}
  {\bibfnamefont {D.~W.}\ \bibnamefont {{Gerdes}}}, \bibinfo {author}
  {\bibfnamefont {T.}~\bibnamefont {{Giannantonio}}}, \bibinfo {author}
  {\bibfnamefont {D.}~\bibnamefont {{Gruen}}}, \bibinfo {author} {\bibfnamefont
  {R.~A.}\ \bibnamefont {{Gruendl}}}, \bibinfo {author} {\bibfnamefont
  {J.}~\bibnamefont {{Gschwend}}}, \bibinfo {author} {\bibfnamefont
  {G.}~\bibnamefont {{Gutierrez}}}, \bibinfo {author} {\bibfnamefont {W.~G.}\
  \bibnamefont {{Hartley}}}, \bibinfo {author} {\bibfnamefont {S.~R.}\
  \bibnamefont {{Hinton}}}, \bibinfo {author} {\bibfnamefont {D.~L.}\
  \bibnamefont {{Hollowood}}}, \bibinfo {author} {\bibfnamefont
  {K.}~\bibnamefont {{Honscheid}}}, \bibinfo {author} {\bibfnamefont
  {B.}~\bibnamefont {{Hoyle}}}, \bibinfo {author} {\bibfnamefont {E.~M.}\
  \bibnamefont {{Huff}}}, \bibinfo {author} {\bibfnamefont {D.~J.}\
  \bibnamefont {{James}}}, \bibinfo {author} {\bibfnamefont {K.}~\bibnamefont
  {{Kuehn}}}, \bibinfo {author} {\bibfnamefont {N.}~\bibnamefont
  {{Kuropatkin}}}, \bibinfo {author} {\bibfnamefont {O.}~\bibnamefont
  {{Lahav}}}, \bibinfo {author} {\bibfnamefont {M.}~\bibnamefont {{Lima}}},
  \bibinfo {author} {\bibfnamefont {M.~A.~G.}\ \bibnamefont {{Maia}}}, \bibinfo
  {author} {\bibfnamefont {J.~L.}\ \bibnamefont {{Marshall}}}, \bibinfo
  {author} {\bibfnamefont {P.}~\bibnamefont {{Martini}}}, \bibinfo {author}
  {\bibfnamefont {P.}~\bibnamefont {{Melchior}}}, \bibinfo {author}
  {\bibfnamefont {F.}~\bibnamefont {{Menanteau}}}, \bibinfo {author}
  {\bibfnamefont {R.}~\bibnamefont {{Miquel}}}, \bibinfo {author}
  {\bibfnamefont {J.~J.}\ \bibnamefont {{Mohr}}}, \bibinfo {author}
  {\bibfnamefont {R.}~\bibnamefont {{Morgan}}}, \bibinfo {author}
  {\bibfnamefont {J.}~\bibnamefont {{Myles}}}, \bibinfo {author} {\bibfnamefont
  {A.}~\bibnamefont {{Palmese}}}, \bibinfo {author} {\bibfnamefont
  {F.}~\bibnamefont {{Paz-Chinch{\'o}n}}}, \bibinfo {author} {\bibfnamefont
  {D.}~\bibnamefont {{Petravick}}}, \bibinfo {author} {\bibfnamefont
  {A.}~\bibnamefont {{Pieres}}}, \bibinfo {author} {\bibfnamefont {A.~A.}\
  \bibnamefont {{Plazas Malag{\'o}n}}}, \bibinfo {author} {\bibfnamefont
  {E.}~\bibnamefont {{Sanchez}}}, \bibinfo {author} {\bibfnamefont
  {V.}~\bibnamefont {{Scarpine}}}, \bibinfo {author} {\bibfnamefont
  {M.}~\bibnamefont {{Schubnell}}}, \bibinfo {author} {\bibfnamefont
  {S.}~\bibnamefont {{Serrano}}}, \bibinfo {author} {\bibfnamefont
  {I.}~\bibnamefont {{Sevilla-Noarbe}}}, \bibinfo {author} {\bibfnamefont
  {M.}~\bibnamefont {{Smith}}}, \bibinfo {author} {\bibfnamefont
  {M.}~\bibnamefont {{Soares-Santos}}}, \bibinfo {author} {\bibfnamefont
  {E.}~\bibnamefont {{Suchyta}}}, \bibinfo {author} {\bibfnamefont
  {G.}~\bibnamefont {{Tarle}}}, \bibinfo {author} {\bibfnamefont
  {D.}~\bibnamefont {{Thomas}}}, \bibinfo {author} {\bibfnamefont
  {C.}~\bibnamefont {{To}}}, \bibinfo {author} {\bibfnamefont {T.~N.}\
  \bibnamefont {{Varga}}},\ and\ \bibinfo {author} {\bibfnamefont
  {J.}~\bibnamefont {{Weller}}},\ }\bibfield  {title} {\bibinfo {title} {{Dark
  Energy Survey Year 3 Results: Multi-Probe Modeling Strategy and
  Validation}},\ }\href@noop {} {\bibfield  {journal} {\bibinfo  {journal}
  {arXiv e-prints}\ ,\ \bibinfo {eid} {arXiv:2105.13548}} (\bibinfo {year}
  {2021})},\ \Eprint {https://arxiv.org/abs/2105.13548} {arXiv:2105.13548
  [astro-ph.CO]} \BibitemShut {NoStop}%
\bibitem [{\citenamefont {{Kirk}}\ \emph {et~al.}(2012)\citenamefont {{Kirk}},
  \citenamefont {{Rassat}}, \citenamefont {{Host}},\ and\ \citenamefont
  {{Bridle}}}]{Kirk2012intrinsic}%
  \BibitemOpen
  \bibfield  {author} {\bibinfo {author} {\bibfnamefont {D.}~\bibnamefont
  {{Kirk}}}, \bibinfo {author} {\bibfnamefont {A.}~\bibnamefont {{Rassat}}},
  \bibinfo {author} {\bibfnamefont {O.}~\bibnamefont {{Host}}},\ and\ \bibinfo
  {author} {\bibfnamefont {S.}~\bibnamefont {{Bridle}}},\ }\bibfield  {title}
  {\bibinfo {title} {{The cosmological impact of intrinsic alignment model
  choice for cosmic shear}},\ }\href
  {https://doi.org/10.1111/j.1365-2966.2012.21099.x} {\bibfield  {journal}
  {\bibinfo  {journal} {\mnras}\ }\textbf {\bibinfo {volume} {424}},\ \bibinfo
  {pages} {1647} (\bibinfo {year} {2012})}\BibitemShut {NoStop}%
\bibitem [{\citenamefont {et~al.}({\natexlab{a}})}]{Fischbacher2022ia}%
  \BibitemOpen
  \bibfield  {author} {\bibinfo {author} {\bibfnamefont {F.}~\bibnamefont
  {et~al.}},\ }\bibfield  {title} {\bibinfo {title} {Requirements for
  ucertaingy on reshfit distribution width measurement for complex intrinsic
  alignment models},\ }\href@noop {} {\bibfield  {journal} {\bibinfo  {journal}
  {in prep}\ } ({\natexlab{a}})}\BibitemShut {NoStop}%
\bibitem [{\citenamefont {{Amon}}\ \emph {et~al.}(2022)\citenamefont {{Amon}},
  \citenamefont {{Robertson}}, \citenamefont {{Miyatake}}, \citenamefont
  {{Heymans}}, \citenamefont {{White}}, \citenamefont {{DeRose}}, \citenamefont
  {{Yuan}}, \citenamefont {{Wechsler}}, \citenamefont {{Varga}}, \citenamefont
  {{Bocquet}}, \citenamefont {{Dvornik}}, \citenamefont {{More}}, \citenamefont
  {{Ross}}, \citenamefont {{Hoekstra}}, \citenamefont {{Alarcon}},
  \citenamefont {{Asgari}}, \citenamefont {{Blazek}}, \citenamefont {{Campos}},
  \citenamefont {{Chen}}, \citenamefont {{Choi}}, \citenamefont {{Crocce}},
  \citenamefont {{Diehl}}, \citenamefont {{Doux}}, \citenamefont {{Eckert}},
  \citenamefont {{Elvin-Poole}}, \citenamefont {{Everett}}, \citenamefont
  {{Fert{\'e}}}, \citenamefont {{Gatti}}, \citenamefont {{Giannini}},
  \citenamefont {{Gruen}}, \citenamefont {{Gruendl}}, \citenamefont
  {{Hartley}}, \citenamefont {{Herner}}, \citenamefont {{Hildebrandt}},
  \citenamefont {{Huang}}, \citenamefont {{Huff}}, \citenamefont {{Joachimi}},
  \citenamefont {{Lee}}, \citenamefont {{MacCrann}}, \citenamefont {{Myles}},
  \citenamefont {{Navarro- Alsina}}, \citenamefont {{Nishimichi}},
  \citenamefont {{Prat}}, \citenamefont {{Secco}}, \citenamefont
  {{Sevilla-Noarbe}}, \citenamefont {{Sheldon}}, \citenamefont {{Shin}},
  \citenamefont {{Trster}}, \citenamefont {{Troxel}}, \citenamefont
  {{Tutusaus}}, \citenamefont {{Wright}}, \citenamefont {{Yin}}, \citenamefont
  {{Aguena}}, \citenamefont {{Allam}}, \citenamefont {{Annis}}, \citenamefont
  {{Bacon}}, \citenamefont {{Bilicki}}, \citenamefont {{Brooks}}, \citenamefont
  {{Burke}}, \citenamefont {{Carnero Rosell}}, \citenamefont {{Carretero}},
  \citenamefont {{Castander}}, \citenamefont {{Cawthon}}, \citenamefont
  {{Costanzi}}, \citenamefont {{da Costa}}, \citenamefont {{Pereira}},
  \citenamefont {{de Jong}}, \citenamefont {{De Vicente}}, \citenamefont
  {{Desai}}, \citenamefont {{Dietrich}}, \citenamefont {{Doel}}, \citenamefont
  {{Ferrero}}, \citenamefont {{Frieman}}, \citenamefont
  {{Garc{\'\i}a-Bellido}}, \citenamefont {{Gerdes}}, \citenamefont
  {{Gschwend}}, \citenamefont {{Gutierrez}}, \citenamefont {{Hinton}},
  \citenamefont {{Hollowood}}, \citenamefont {{Honscheid}}, \citenamefont
  {{Huterer}}, \citenamefont {{Kannawadi}}, \citenamefont {{Kuehn}},
  \citenamefont {{Kuropatkin}}, \citenamefont {{Lahav}}, \citenamefont
  {{Lima}}, \citenamefont {{Maia}}, \citenamefont {{Marshall}}, \citenamefont
  {{Menanteau}}, \citenamefont {{Miquel}}, \citenamefont {{Mohr}},
  \citenamefont {{Morgan}}, \citenamefont {{Muir}}, \citenamefont
  {{Paz-Chinchon}}, \citenamefont {{Pieres}}, \citenamefont {{Plazas
  Malag{\'o}n}}, \citenamefont {{Porredon}}, \citenamefont
  {{Rodriguez-Monroy}}, \citenamefont {{Roodman}}, \citenamefont {{Sanchez}},
  \citenamefont {{Serrano}}, \citenamefont {{Shan}}, \citenamefont {{Suchyta}},
  \citenamefont {{Swanson}}, \citenamefont {{Tarle}}, \citenamefont {{Thomas}},
  \citenamefont {{To}},\ and\ \citenamefont {{Zhang}}}]{Amon2022consistent}%
  \BibitemOpen
  \bibfield  {author} {\bibinfo {author} {\bibfnamefont {A.}~\bibnamefont
  {{Amon}}}, \bibinfo {author} {\bibfnamefont {N.~C.}\ \bibnamefont
  {{Robertson}}}, \bibinfo {author} {\bibfnamefont {H.}~\bibnamefont
  {{Miyatake}}}, \bibinfo {author} {\bibfnamefont {C.}~\bibnamefont
  {{Heymans}}}, \bibinfo {author} {\bibfnamefont {M.}~\bibnamefont {{White}}},
  \bibinfo {author} {\bibfnamefont {J.}~\bibnamefont {{DeRose}}}, \bibinfo
  {author} {\bibfnamefont {S.}~\bibnamefont {{Yuan}}}, \bibinfo {author}
  {\bibfnamefont {R.~H.}\ \bibnamefont {{Wechsler}}}, \bibinfo {author}
  {\bibfnamefont {T.~N.}\ \bibnamefont {{Varga}}}, \bibinfo {author}
  {\bibfnamefont {S.}~\bibnamefont {{Bocquet}}}, \bibinfo {author}
  {\bibfnamefont {A.}~\bibnamefont {{Dvornik}}}, \bibinfo {author}
  {\bibfnamefont {S.}~\bibnamefont {{More}}}, \bibinfo {author} {\bibfnamefont
  {A.~J.}\ \bibnamefont {{Ross}}}, \bibinfo {author} {\bibfnamefont
  {H.}~\bibnamefont {{Hoekstra}}}, \bibinfo {author} {\bibfnamefont
  {A.}~\bibnamefont {{Alarcon}}}, \bibinfo {author} {\bibfnamefont
  {M.}~\bibnamefont {{Asgari}}}, \bibinfo {author} {\bibfnamefont
  {J.}~\bibnamefont {{Blazek}}}, \bibinfo {author} {\bibfnamefont
  {A.}~\bibnamefont {{Campos}}}, \bibinfo {author} {\bibfnamefont
  {R.}~\bibnamefont {{Chen}}}, \bibinfo {author} {\bibfnamefont
  {A.}~\bibnamefont {{Choi}}}, \bibinfo {author} {\bibfnamefont
  {M.}~\bibnamefont {{Crocce}}}, \bibinfo {author} {\bibfnamefont {H.~T.}\
  \bibnamefont {{Diehl}}}, \bibinfo {author} {\bibfnamefont {C.}~\bibnamefont
  {{Doux}}}, \bibinfo {author} {\bibfnamefont {K.}~\bibnamefont {{Eckert}}},
  \bibinfo {author} {\bibfnamefont {J.}~\bibnamefont {{Elvin-Poole}}}, \bibinfo
  {author} {\bibfnamefont {S.}~\bibnamefont {{Everett}}}, \bibinfo {author}
  {\bibfnamefont {A.}~\bibnamefont {{Fert{\'e}}}}, \bibinfo {author}
  {\bibfnamefont {M.}~\bibnamefont {{Gatti}}}, \bibinfo {author} {\bibfnamefont
  {G.}~\bibnamefont {{Giannini}}}, \bibinfo {author} {\bibfnamefont
  {D.}~\bibnamefont {{Gruen}}}, \bibinfo {author} {\bibfnamefont {R.~A.}\
  \bibnamefont {{Gruendl}}}, \bibinfo {author} {\bibfnamefont {W.~G.}\
  \bibnamefont {{Hartley}}}, \bibinfo {author} {\bibfnamefont {K.}~\bibnamefont
  {{Herner}}}, \bibinfo {author} {\bibfnamefont {H.}~\bibnamefont
  {{Hildebrandt}}}, \bibinfo {author} {\bibfnamefont {S.}~\bibnamefont
  {{Huang}}}, \bibinfo {author} {\bibfnamefont {E.~M.}\ \bibnamefont {{Huff}}},
  \bibinfo {author} {\bibfnamefont {B.}~\bibnamefont {{Joachimi}}}, \bibinfo
  {author} {\bibfnamefont {S.}~\bibnamefont {{Lee}}}, \bibinfo {author}
  {\bibfnamefont {N.}~\bibnamefont {{MacCrann}}}, \bibinfo {author}
  {\bibfnamefont {J.}~\bibnamefont {{Myles}}}, \bibinfo {author} {\bibfnamefont
  {A.}~\bibnamefont {{Navarro- Alsina}}}, \bibinfo {author} {\bibfnamefont
  {T.}~\bibnamefont {{Nishimichi}}}, \bibinfo {author} {\bibfnamefont
  {J.}~\bibnamefont {{Prat}}}, \bibinfo {author} {\bibfnamefont {L.~F.}\
  \bibnamefont {{Secco}}}, \bibinfo {author} {\bibfnamefont {I.}~\bibnamefont
  {{Sevilla-Noarbe}}}, \bibinfo {author} {\bibfnamefont {E.}~\bibnamefont
  {{Sheldon}}}, \bibinfo {author} {\bibfnamefont {T.}~\bibnamefont {{Shin}}},
  \bibinfo {author} {\bibfnamefont {T.}~\bibnamefont {{Trster}}}, \bibinfo
  {author} {\bibfnamefont {M.~A.}\ \bibnamefont {{Troxel}}}, \bibinfo {author}
  {\bibfnamefont {I.}~\bibnamefont {{Tutusaus}}}, \bibinfo {author}
  {\bibfnamefont {A.~H.}\ \bibnamefont {{Wright}}}, \bibinfo {author}
  {\bibfnamefont {B.}~\bibnamefont {{Yin}}}, \bibinfo {author} {\bibfnamefont
  {M.}~\bibnamefont {{Aguena}}}, \bibinfo {author} {\bibfnamefont
  {S.}~\bibnamefont {{Allam}}}, \bibinfo {author} {\bibfnamefont
  {J.}~\bibnamefont {{Annis}}}, \bibinfo {author} {\bibfnamefont
  {D.}~\bibnamefont {{Bacon}}}, \bibinfo {author} {\bibfnamefont
  {M.}~\bibnamefont {{Bilicki}}}, \bibinfo {author} {\bibfnamefont
  {D.}~\bibnamefont {{Brooks}}}, \bibinfo {author} {\bibfnamefont {D.~L.}\
  \bibnamefont {{Burke}}}, \bibinfo {author} {\bibfnamefont {A.}~\bibnamefont
  {{Carnero Rosell}}}, \bibinfo {author} {\bibfnamefont {J.}~\bibnamefont
  {{Carretero}}}, \bibinfo {author} {\bibfnamefont {F.~J.}\ \bibnamefont
  {{Castander}}}, \bibinfo {author} {\bibfnamefont {R.}~\bibnamefont
  {{Cawthon}}}, \bibinfo {author} {\bibfnamefont {M.}~\bibnamefont
  {{Costanzi}}}, \bibinfo {author} {\bibfnamefont {L.~N.}\ \bibnamefont {{da
  Costa}}}, \bibinfo {author} {\bibfnamefont {M.~E.~S.}\ \bibnamefont
  {{Pereira}}}, \bibinfo {author} {\bibfnamefont {J.}~\bibnamefont {{de
  Jong}}}, \bibinfo {author} {\bibfnamefont {J.}~\bibnamefont {{De Vicente}}},
  \bibinfo {author} {\bibfnamefont {S.}~\bibnamefont {{Desai}}}, \bibinfo
  {author} {\bibfnamefont {J.~P.}\ \bibnamefont {{Dietrich}}}, \bibinfo
  {author} {\bibfnamefont {P.}~\bibnamefont {{Doel}}}, \bibinfo {author}
  {\bibfnamefont {I.}~\bibnamefont {{Ferrero}}}, \bibinfo {author}
  {\bibfnamefont {J.}~\bibnamefont {{Frieman}}}, \bibinfo {author}
  {\bibfnamefont {J.}~\bibnamefont {{Garc{\'\i}a-Bellido}}}, \bibinfo {author}
  {\bibfnamefont {D.~W.}\ \bibnamefont {{Gerdes}}}, \bibinfo {author}
  {\bibfnamefont {J.}~\bibnamefont {{Gschwend}}}, \bibinfo {author}
  {\bibfnamefont {G.}~\bibnamefont {{Gutierrez}}}, \bibinfo {author}
  {\bibfnamefont {S.~R.}\ \bibnamefont {{Hinton}}}, \bibinfo {author}
  {\bibfnamefont {D.~L.}\ \bibnamefont {{Hollowood}}}, \bibinfo {author}
  {\bibfnamefont {K.}~\bibnamefont {{Honscheid}}}, \bibinfo {author}
  {\bibfnamefont {D.}~\bibnamefont {{Huterer}}}, \bibinfo {author}
  {\bibfnamefont {A.}~\bibnamefont {{Kannawadi}}}, \bibinfo {author}
  {\bibfnamefont {K.}~\bibnamefont {{Kuehn}}}, \bibinfo {author} {\bibfnamefont
  {N.}~\bibnamefont {{Kuropatkin}}}, \bibinfo {author} {\bibfnamefont
  {O.}~\bibnamefont {{Lahav}}}, \bibinfo {author} {\bibfnamefont
  {M.}~\bibnamefont {{Lima}}}, \bibinfo {author} {\bibfnamefont {M.~A.~G.}\
  \bibnamefont {{Maia}}}, \bibinfo {author} {\bibfnamefont {J.~L.}\
  \bibnamefont {{Marshall}}}, \bibinfo {author} {\bibfnamefont
  {F.}~\bibnamefont {{Menanteau}}}, \bibinfo {author} {\bibfnamefont
  {R.}~\bibnamefont {{Miquel}}}, \bibinfo {author} {\bibfnamefont {J.~J.}\
  \bibnamefont {{Mohr}}}, \bibinfo {author} {\bibfnamefont {R.}~\bibnamefont
  {{Morgan}}}, \bibinfo {author} {\bibfnamefont {J.}~\bibnamefont {{Muir}}},
  \bibinfo {author} {\bibfnamefont {F.}~\bibnamefont {{Paz-Chinchon}}},
  \bibinfo {author} {\bibfnamefont {A.}~\bibnamefont {{Pieres}}}, \bibinfo
  {author} {\bibfnamefont {A.~A.}\ \bibnamefont {{Plazas Malag{\'o}n}}},
  \bibinfo {author} {\bibfnamefont {A.}~\bibnamefont {{Porredon}}}, \bibinfo
  {author} {\bibfnamefont {M.}~\bibnamefont {{Rodriguez-Monroy}}}, \bibinfo
  {author} {\bibfnamefont {A.}~\bibnamefont {{Roodman}}}, \bibinfo {author}
  {\bibfnamefont {E.}~\bibnamefont {{Sanchez}}}, \bibinfo {author}
  {\bibfnamefont {S.}~\bibnamefont {{Serrano}}}, \bibinfo {author}
  {\bibfnamefont {H.}~\bibnamefont {{Shan}}}, \bibinfo {author} {\bibfnamefont
  {E.}~\bibnamefont {{Suchyta}}}, \bibinfo {author} {\bibfnamefont {M.~E.~C.}\
  \bibnamefont {{Swanson}}}, \bibinfo {author} {\bibfnamefont {G.}~\bibnamefont
  {{Tarle}}}, \bibinfo {author} {\bibfnamefont {D.}~\bibnamefont {{Thomas}}},
  \bibinfo {author} {\bibfnamefont {C.}~\bibnamefont {{To}}},\ and\ \bibinfo
  {author} {\bibfnamefont {Y.}~\bibnamefont {{Zhang}}},\ }\bibfield  {title}
  {\bibinfo {title} {{Consistent lensing and clustering in a low-$S_8$ Universe
  with BOSS, DES Year 3, HSC Year 1 and KiDS-1000}},\ }\href@noop {} {\bibfield
   {journal} {\bibinfo  {journal} {arXiv e-prints}\ ,\ \bibinfo {eid}
  {arXiv:2202.07440}} (\bibinfo {year} {2022})},\ \Eprint
  {https://arxiv.org/abs/2202.07440} {arXiv:2202.07440 [astro-ph.CO]}
  \BibitemShut {NoStop}%
\bibitem [{\citenamefont {Leauthaud}\ \emph {et~al.}(2017)\citenamefont
  {Leauthaud}, \citenamefont {Saito}, \citenamefont {Hilbert}, \citenamefont
  {Barreira}, \citenamefont {More}, \citenamefont {White}, \citenamefont
  {Alam}, \citenamefont {Behroozi}, \citenamefont {Bundy}, \citenamefont
  {Coupon}, \citenamefont {Erben}, \citenamefont {Heymans}, \citenamefont
  {Hildebrandt}, \citenamefont {Mandelbaum}, \citenamefont {Miller},
  \citenamefont {Moraes}, \citenamefont {Pereira}, \citenamefont
  {Rodríguez-Torres}, \citenamefont {Schmidt}, \citenamefont {Shan},
  \citenamefont {Viel},\ and\ \citenamefont
  {Villaescusa-Navarro}}]{Leauthaud2017lensinglow}%
  \BibitemOpen
  \bibfield  {author} {\bibinfo {author} {\bibfnamefont {A.}~\bibnamefont
  {Leauthaud}}, \bibinfo {author} {\bibfnamefont {S.}~\bibnamefont {Saito}},
  \bibinfo {author} {\bibfnamefont {S.}~\bibnamefont {Hilbert}}, \bibinfo
  {author} {\bibfnamefont {A.}~\bibnamefont {Barreira}}, \bibinfo {author}
  {\bibfnamefont {S.}~\bibnamefont {More}}, \bibinfo {author} {\bibfnamefont
  {M.}~\bibnamefont {White}}, \bibinfo {author} {\bibfnamefont
  {S.}~\bibnamefont {Alam}}, \bibinfo {author} {\bibfnamefont {P.}~\bibnamefont
  {Behroozi}}, \bibinfo {author} {\bibfnamefont {K.}~\bibnamefont {Bundy}},
  \bibinfo {author} {\bibfnamefont {J.}~\bibnamefont {Coupon}}, \bibinfo
  {author} {\bibfnamefont {T.}~\bibnamefont {Erben}}, \bibinfo {author}
  {\bibfnamefont {C.}~\bibnamefont {Heymans}}, \bibinfo {author} {\bibfnamefont
  {H.}~\bibnamefont {Hildebrandt}}, \bibinfo {author} {\bibfnamefont
  {R.}~\bibnamefont {Mandelbaum}}, \bibinfo {author} {\bibfnamefont
  {L.}~\bibnamefont {Miller}}, \bibinfo {author} {\bibfnamefont
  {B.}~\bibnamefont {Moraes}}, \bibinfo {author} {\bibfnamefont {M.~E.~S.}\
  \bibnamefont {Pereira}}, \bibinfo {author} {\bibfnamefont {S.~A.}\
  \bibnamefont {Rodríguez-Torres}}, \bibinfo {author} {\bibfnamefont
  {F.}~\bibnamefont {Schmidt}}, \bibinfo {author} {\bibfnamefont {H.-Y.}\
  \bibnamefont {Shan}}, \bibinfo {author} {\bibfnamefont {M.}~\bibnamefont
  {Viel}},\ and\ \bibinfo {author} {\bibfnamefont {F.}~\bibnamefont
  {Villaescusa-Navarro}},\ }\bibfield  {title} {\bibinfo {title} {{Lensing is
  low: cosmology, galaxy formation or new physics?}},\ }\href
  {https://doi.org/10.1093/mnras/stx258} {\bibfield  {journal} {\bibinfo
  {journal} {Monthly Notices of the Royal Astronomical Society}\ }\textbf
  {\bibinfo {volume} {467}},\ \bibinfo {pages} {3024} (\bibinfo {year}
  {2017})},\ \Eprint
  {https://arxiv.org/abs/https://academic.oup.com/mnras/article-pdf/467/3/3024/10875307/stx258.pdf}
  {https://academic.oup.com/mnras/article-pdf/467/3/3024/10875307/stx258.pdf}
  \BibitemShut {NoStop}%
\bibitem [{\citenamefont {{Leauthaud}}\ \emph {et~al.}(2022)\citenamefont
  {{Leauthaud}}, \citenamefont {{Amon}}, \citenamefont {{Singh}}, \citenamefont
  {{Gruen}}, \citenamefont {{Lange}}, \citenamefont {{Huang}}, \citenamefont
  {{Robertson}}, \citenamefont {{Varga}}, \citenamefont {{Luo}}, \citenamefont
  {{Heymans}}, \citenamefont {{Hildebrandt}}, \citenamefont {{Blake}},
  \citenamefont {{Aguena}}, \citenamefont {{Allam}}, \citenamefont
  {{Andrade-Oliveira}}, \citenamefont {{Annis}}, \citenamefont {{Bertin}},
  \citenamefont {{Bhargava}}, \citenamefont {{Blazek}}, \citenamefont
  {{Bridle}}, \citenamefont {{Brooks}}, \citenamefont {{Burke}}, \citenamefont
  {{Rosell}}, \citenamefont {{Carrasco Kind}}, \citenamefont {{Carretero}},
  \citenamefont {{Castander}}, \citenamefont {{Cawthon}}, \citenamefont
  {{Choi}}, \citenamefont {{Costanzi}}, \citenamefont {{da Costa}},
  \citenamefont {{Pereira}}, \citenamefont {{Davis}}, \citenamefont {{De
  Vicente}}, \citenamefont {{DeRose}}, \citenamefont {{Diehl}}, \citenamefont
  {{Dietrich}}, \citenamefont {{Doel}}, \citenamefont {{Eckert}}, \citenamefont
  {{Everett}}, \citenamefont {{Evrard}}, \citenamefont {{Ferrero}},
  \citenamefont {{Flaugher}}, \citenamefont {{Fosalba}}, \citenamefont
  {{Garc{\'\i}a-Bellido}}, \citenamefont {{Gatti}}, \citenamefont
  {{Gaztanaga}}, \citenamefont {{Gruendl}}, \citenamefont {{Gschwend}},
  \citenamefont {{Hartley}}, \citenamefont {{Hollowood}}, \citenamefont
  {{Honscheid}}, \citenamefont {{Jain}}, \citenamefont {{James}}, \citenamefont
  {{Jarvis}}, \citenamefont {{Joachimi}}, \citenamefont {{Kannawadi}},
  \citenamefont {{Kim}}, \citenamefont {{Krause}}, \citenamefont {{Kuehn}},
  \citenamefont {{Kuijken}}, \citenamefont {{Kuropatkin}}, \citenamefont
  {{Lima}}, \citenamefont {{MacCrann}}, \citenamefont {{Maia}}, \citenamefont
  {{Makler}}, \citenamefont {{March}}, \citenamefont {{Marshall}},
  \citenamefont {{Melchior}}, \citenamefont {{Menanteau}}, \citenamefont
  {{Miquel}}, \citenamefont {{Miyatake}}, \citenamefont {{Mohr}}, \citenamefont
  {{Moraes}}, \citenamefont {{More}}, \citenamefont {{Surhud}}, \citenamefont
  {{Morgan}}, \citenamefont {{Myles}}, \citenamefont {{Ogando}}, \citenamefont
  {{Palmese}}, \citenamefont {{Paz-Chinch{\'o}n}}, \citenamefont
  {{Malag{\'o}n}}, \citenamefont {{Prat}}, \citenamefont {{Rau}}, \citenamefont
  {{Rhodes}}, \citenamefont {{Rodriguez-Monroy}}, \citenamefont {{Roodman}},
  \citenamefont {{Ross}}, \citenamefont {{Samuroff}}, \citenamefont
  {{S{\'a}nchez}}, \citenamefont {{Sanchez}}, \citenamefont {{Scarpine}},
  \citenamefont {{Schlegel}}, \citenamefont {{Schubnell}}, \citenamefont
  {{Serrano}}, \citenamefont {{Sevilla-Noarbe}}, \citenamefont {{Sif{\'o}n}},
  \citenamefont {{Smith}}, \citenamefont {{Speagle}}, \citenamefont
  {{Suchyta}}, \citenamefont {{Tarle}}, \citenamefont {{Thomas}}, \citenamefont
  {{Tinker}}, \citenamefont {{To}}, \citenamefont {{Troxel}}, \citenamefont
  {{Van Waerbeke}}, \citenamefont {{Vielzeuf}},\ and\ \citenamefont
  {{Wright}}}]{Leauthaud2022withoutborders}%
  \BibitemOpen
  \bibfield  {author} {\bibinfo {author} {\bibfnamefont {A.}~\bibnamefont
  {{Leauthaud}}}, \bibinfo {author} {\bibfnamefont {A.}~\bibnamefont {{Amon}}},
  \bibinfo {author} {\bibfnamefont {S.}~\bibnamefont {{Singh}}}, \bibinfo
  {author} {\bibfnamefont {D.}~\bibnamefont {{Gruen}}}, \bibinfo {author}
  {\bibfnamefont {J.~U.}\ \bibnamefont {{Lange}}}, \bibinfo {author}
  {\bibfnamefont {S.}~\bibnamefont {{Huang}}}, \bibinfo {author} {\bibfnamefont
  {N.~C.}\ \bibnamefont {{Robertson}}}, \bibinfo {author} {\bibfnamefont
  {T.~N.}\ \bibnamefont {{Varga}}}, \bibinfo {author} {\bibfnamefont
  {Y.}~\bibnamefont {{Luo}}}, \bibinfo {author} {\bibfnamefont
  {C.}~\bibnamefont {{Heymans}}}, \bibinfo {author} {\bibfnamefont
  {H.}~\bibnamefont {{Hildebrandt}}}, \bibinfo {author} {\bibfnamefont
  {C.}~\bibnamefont {{Blake}}}, \bibinfo {author} {\bibfnamefont
  {M.}~\bibnamefont {{Aguena}}}, \bibinfo {author} {\bibfnamefont
  {S.}~\bibnamefont {{Allam}}}, \bibinfo {author} {\bibfnamefont
  {F.}~\bibnamefont {{Andrade-Oliveira}}}, \bibinfo {author} {\bibfnamefont
  {J.}~\bibnamefont {{Annis}}}, \bibinfo {author} {\bibfnamefont
  {E.}~\bibnamefont {{Bertin}}}, \bibinfo {author} {\bibfnamefont
  {S.}~\bibnamefont {{Bhargava}}}, \bibinfo {author} {\bibfnamefont
  {J.}~\bibnamefont {{Blazek}}}, \bibinfo {author} {\bibfnamefont {S.~L.}\
  \bibnamefont {{Bridle}}}, \bibinfo {author} {\bibfnamefont {D.}~\bibnamefont
  {{Brooks}}}, \bibinfo {author} {\bibfnamefont {D.~L.}\ \bibnamefont
  {{Burke}}}, \bibinfo {author} {\bibfnamefont {A.~C.}\ \bibnamefont
  {{Rosell}}}, \bibinfo {author} {\bibfnamefont {M.}~\bibnamefont {{Carrasco
  Kind}}}, \bibinfo {author} {\bibfnamefont {J.}~\bibnamefont {{Carretero}}},
  \bibinfo {author} {\bibfnamefont {F.~J.}\ \bibnamefont {{Castander}}},
  \bibinfo {author} {\bibfnamefont {R.}~\bibnamefont {{Cawthon}}}, \bibinfo
  {author} {\bibfnamefont {A.}~\bibnamefont {{Choi}}}, \bibinfo {author}
  {\bibfnamefont {M.}~\bibnamefont {{Costanzi}}}, \bibinfo {author}
  {\bibfnamefont {L.~N.}\ \bibnamefont {{da Costa}}}, \bibinfo {author}
  {\bibfnamefont {M.~E.~S.}\ \bibnamefont {{Pereira}}}, \bibinfo {author}
  {\bibfnamefont {C.}~\bibnamefont {{Davis}}}, \bibinfo {author} {\bibfnamefont
  {J.}~\bibnamefont {{De Vicente}}}, \bibinfo {author} {\bibfnamefont
  {J.}~\bibnamefont {{DeRose}}}, \bibinfo {author} {\bibfnamefont {H.~T.}\
  \bibnamefont {{Diehl}}}, \bibinfo {author} {\bibfnamefont {J.~P.}\
  \bibnamefont {{Dietrich}}}, \bibinfo {author} {\bibfnamefont
  {P.}~\bibnamefont {{Doel}}}, \bibinfo {author} {\bibfnamefont
  {K.}~\bibnamefont {{Eckert}}}, \bibinfo {author} {\bibfnamefont
  {S.}~\bibnamefont {{Everett}}}, \bibinfo {author} {\bibfnamefont {A.~E.}\
  \bibnamefont {{Evrard}}}, \bibinfo {author} {\bibfnamefont {I.}~\bibnamefont
  {{Ferrero}}}, \bibinfo {author} {\bibfnamefont {B.}~\bibnamefont
  {{Flaugher}}}, \bibinfo {author} {\bibfnamefont {P.}~\bibnamefont
  {{Fosalba}}}, \bibinfo {author} {\bibfnamefont {J.}~\bibnamefont
  {{Garc{\'\i}a-Bellido}}}, \bibinfo {author} {\bibfnamefont {M.}~\bibnamefont
  {{Gatti}}}, \bibinfo {author} {\bibfnamefont {E.}~\bibnamefont
  {{Gaztanaga}}}, \bibinfo {author} {\bibfnamefont {R.~A.}\ \bibnamefont
  {{Gruendl}}}, \bibinfo {author} {\bibfnamefont {J.}~\bibnamefont
  {{Gschwend}}}, \bibinfo {author} {\bibfnamefont {W.~G.}\ \bibnamefont
  {{Hartley}}}, \bibinfo {author} {\bibfnamefont {D.~L.}\ \bibnamefont
  {{Hollowood}}}, \bibinfo {author} {\bibfnamefont {K.}~\bibnamefont
  {{Honscheid}}}, \bibinfo {author} {\bibfnamefont {B.}~\bibnamefont {{Jain}}},
  \bibinfo {author} {\bibfnamefont {D.~J.}\ \bibnamefont {{James}}}, \bibinfo
  {author} {\bibfnamefont {M.}~\bibnamefont {{Jarvis}}}, \bibinfo {author}
  {\bibfnamefont {B.}~\bibnamefont {{Joachimi}}}, \bibinfo {author}
  {\bibfnamefont {A.}~\bibnamefont {{Kannawadi}}}, \bibinfo {author}
  {\bibfnamefont {A.~G.}\ \bibnamefont {{Kim}}}, \bibinfo {author}
  {\bibfnamefont {E.}~\bibnamefont {{Krause}}}, \bibinfo {author}
  {\bibfnamefont {K.}~\bibnamefont {{Kuehn}}}, \bibinfo {author} {\bibfnamefont
  {K.}~\bibnamefont {{Kuijken}}}, \bibinfo {author} {\bibfnamefont
  {N.}~\bibnamefont {{Kuropatkin}}}, \bibinfo {author} {\bibfnamefont
  {M.}~\bibnamefont {{Lima}}}, \bibinfo {author} {\bibfnamefont
  {N.}~\bibnamefont {{MacCrann}}}, \bibinfo {author} {\bibfnamefont {M.~A.~G.}\
  \bibnamefont {{Maia}}}, \bibinfo {author} {\bibfnamefont {M.}~\bibnamefont
  {{Makler}}}, \bibinfo {author} {\bibfnamefont {M.}~\bibnamefont {{March}}},
  \bibinfo {author} {\bibfnamefont {J.~L.}\ \bibnamefont {{Marshall}}},
  \bibinfo {author} {\bibfnamefont {P.}~\bibnamefont {{Melchior}}}, \bibinfo
  {author} {\bibfnamefont {F.}~\bibnamefont {{Menanteau}}}, \bibinfo {author}
  {\bibfnamefont {R.}~\bibnamefont {{Miquel}}}, \bibinfo {author}
  {\bibfnamefont {H.}~\bibnamefont {{Miyatake}}}, \bibinfo {author}
  {\bibfnamefont {J.~J.}\ \bibnamefont {{Mohr}}}, \bibinfo {author}
  {\bibfnamefont {B.}~\bibnamefont {{Moraes}}}, \bibinfo {author}
  {\bibfnamefont {S.}~\bibnamefont {{More}}}, \bibinfo {author} {\bibfnamefont
  {M.}~\bibnamefont {{Surhud}}}, \bibinfo {author} {\bibfnamefont
  {R.}~\bibnamefont {{Morgan}}}, \bibinfo {author} {\bibfnamefont
  {J.}~\bibnamefont {{Myles}}}, \bibinfo {author} {\bibfnamefont {R.~L.~C.}\
  \bibnamefont {{Ogando}}}, \bibinfo {author} {\bibfnamefont {A.}~\bibnamefont
  {{Palmese}}}, \bibinfo {author} {\bibfnamefont {F.}~\bibnamefont
  {{Paz-Chinch{\'o}n}}}, \bibinfo {author} {\bibfnamefont {A.~A.~P.}\
  \bibnamefont {{Malag{\'o}n}}}, \bibinfo {author} {\bibfnamefont
  {J.}~\bibnamefont {{Prat}}}, \bibinfo {author} {\bibfnamefont {M.~M.}\
  \bibnamefont {{Rau}}}, \bibinfo {author} {\bibfnamefont {J.}~\bibnamefont
  {{Rhodes}}}, \bibinfo {author} {\bibfnamefont {M.}~\bibnamefont
  {{Rodriguez-Monroy}}}, \bibinfo {author} {\bibfnamefont {A.}~\bibnamefont
  {{Roodman}}}, \bibinfo {author} {\bibfnamefont {A.~J.}\ \bibnamefont
  {{Ross}}}, \bibinfo {author} {\bibfnamefont {S.}~\bibnamefont {{Samuroff}}},
  \bibinfo {author} {\bibfnamefont {C.}~\bibnamefont {{S{\'a}nchez}}}, \bibinfo
  {author} {\bibfnamefont {E.}~\bibnamefont {{Sanchez}}}, \bibinfo {author}
  {\bibfnamefont {V.}~\bibnamefont {{Scarpine}}}, \bibinfo {author}
  {\bibfnamefont {D.~J.}\ \bibnamefont {{Schlegel}}}, \bibinfo {author}
  {\bibfnamefont {M.}~\bibnamefont {{Schubnell}}}, \bibinfo {author}
  {\bibfnamefont {S.}~\bibnamefont {{Serrano}}}, \bibinfo {author}
  {\bibfnamefont {I.}~\bibnamefont {{Sevilla-Noarbe}}}, \bibinfo {author}
  {\bibfnamefont {C.}~\bibnamefont {{Sif{\'o}n}}}, \bibinfo {author}
  {\bibfnamefont {M.}~\bibnamefont {{Smith}}}, \bibinfo {author} {\bibfnamefont
  {J.~S.}\ \bibnamefont {{Speagle}}}, \bibinfo {author} {\bibfnamefont
  {E.}~\bibnamefont {{Suchyta}}}, \bibinfo {author} {\bibfnamefont
  {G.}~\bibnamefont {{Tarle}}}, \bibinfo {author} {\bibfnamefont
  {D.}~\bibnamefont {{Thomas}}}, \bibinfo {author} {\bibfnamefont
  {J.}~\bibnamefont {{Tinker}}}, \bibinfo {author} {\bibfnamefont
  {C.}~\bibnamefont {{To}}}, \bibinfo {author} {\bibfnamefont {M.~A.}\
  \bibnamefont {{Troxel}}}, \bibinfo {author} {\bibfnamefont {L.}~\bibnamefont
  {{Van Waerbeke}}}, \bibinfo {author} {\bibfnamefont {P.}~\bibnamefont
  {{Vielzeuf}}},\ and\ \bibinfo {author} {\bibfnamefont {A.~H.}\ \bibnamefont
  {{Wright}}},\ }\bibfield  {title} {\bibinfo {title} {{Lensing without borders
  - I. A blind comparison of the amplitude of galaxy-galaxy lensing between
  independent imaging surveys}},\ }\href
  {https://doi.org/10.1093/mnras/stab3586} {\bibfield  {journal} {\bibinfo
  {journal} {\mnras}\ }\textbf {\bibinfo {volume} {510}},\ \bibinfo {pages}
  {6150} (\bibinfo {year} {2022})},\ \Eprint {https://arxiv.org/abs/2111.13805}
  {arXiv:2111.13805 [astro-ph.CO]} \BibitemShut {NoStop}%
\bibitem [{\citenamefont {{Snowmass Collaboration
  2021}}(2022)}]{cosmology2022intertwined}%
  \BibitemOpen
  \bibfield  {author} {\bibinfo {author} {\bibnamefont {{Snowmass Collaboration
  2021}}},\ }\bibfield  {title} {\bibinfo {title} {{Cosmology Intertwined: A
  Review of the Particle Physics, Astrophysics, and Cosmology Associated with
  the Cosmological Tensions and Anomalies}},\ }\href@noop {} {\bibfield
  {journal} {\bibinfo  {journal} {arXiv e-prints}\ ,\ \bibinfo {eid}
  {arXiv:2203.06142}} (\bibinfo {year} {2022})},\ \Eprint
  {https://arxiv.org/abs/2203.06142} {arXiv:2203.06142 [astro-ph.CO]}
  \BibitemShut {NoStop}%
\bibitem [{\citenamefont {{Z{\"u}rcher}}\ \emph {et~al.}(2022)\citenamefont
  {{Z{\"u}rcher}}, \citenamefont {{Fluri}}, \citenamefont {{Sgier}},
  \citenamefont {{Kacprzak}}, \citenamefont {{Gatti}}, \citenamefont {{Doux}},
  \citenamefont {{Whiteway}}, \citenamefont {{R{\'e}fr{\'e}gier}},
  \citenamefont {{Chang}}, \citenamefont {{Jeffrey}}, \citenamefont {{Jain}},
  \citenamefont {{Lemos}}, \citenamefont {{Bacon}}, \citenamefont {{Alarcon}},
  \citenamefont {{Amon}}, \citenamefont {{Bechtol}}, \citenamefont {{Becker}},
  \citenamefont {{Bernstein}}, \citenamefont {{Campos}}, \citenamefont
  {{Chen}}, \citenamefont {{Choi}}, \citenamefont {{Davis}}, \citenamefont
  {{Derose}}, \citenamefont {{Dodelson}}, \citenamefont {{Elsner}},
  \citenamefont {{Elvin-Poole}}, \citenamefont {{Everett}}, \citenamefont
  {{Ferte}}, \citenamefont {{Gruen}}, \citenamefont {{Harrison}}, \citenamefont
  {{Huterer}}, \citenamefont {{Jarvis}}, \citenamefont {{Leget}}, \citenamefont
  {{Maccrann}}, \citenamefont {{Mccullough}}, \citenamefont {{Muir}},
  \citenamefont {{Myles}}, \citenamefont {{Navarro Alsina}}, \citenamefont
  {{Pandey}}, \citenamefont {{Prat}}, \citenamefont {{Raveri}}, \citenamefont
  {{Rollins}}, \citenamefont {{Roodman}}, \citenamefont {{Sanchez}},
  \citenamefont {{Secco}}, \citenamefont {{Sheldon}}, \citenamefont {{Shin}},
  \citenamefont {{Troxel}}, \citenamefont {{Tutusaus}}, \citenamefont {{Yin}},
  \citenamefont {{Aguena}}, \citenamefont {{Allam}}, \citenamefont
  {{Andrade-Oliveira}}, \citenamefont {{Annis}}, \citenamefont {{Bertin}},
  \citenamefont {{Brooks}}, \citenamefont {{Burke}}, \citenamefont {{Carnero
  Rosell}}, \citenamefont {{Carrasco Kind}}, \citenamefont {{Carretero}},
  \citenamefont {{Castander}}, \citenamefont {{Cawthon}}, \citenamefont
  {{Conselice}}, \citenamefont {{Costanzi}}, \citenamefont {{da Costa}},
  \citenamefont {{da Silva Pereira}}, \citenamefont {{Davis}}, \citenamefont
  {{De Vicente}}, \citenamefont {{Desai}}, \citenamefont {{Diehl}},
  \citenamefont {{Dietrich}}, \citenamefont {{Doel}}, \citenamefont {{Eckert}},
  \citenamefont {{Evrard}}, \citenamefont {{Ferrero}}, \citenamefont
  {{Flaugher}}, \citenamefont {{Fosalba}}, \citenamefont {{Friedel}},
  \citenamefont {{Frieman}}, \citenamefont {{Garcia-Bellido}}, \citenamefont
  {{Gaztanaga}}, \citenamefont {{Gerdes}}, \citenamefont {{Giannantonio}},
  \citenamefont {{Gruendl}}, \citenamefont {{Gschwend}}, \citenamefont
  {{Gutierrez}}, \citenamefont {{Hinton}}, \citenamefont {{Hollowood}},
  \citenamefont {{Honscheid}}, \citenamefont {{Hoyle}}, \citenamefont
  {{James}}, \citenamefont {{Kuehn}}, \citenamefont {{Kuropatkin}},
  \citenamefont {{Lahav}}, \citenamefont {{Lidman}}, \citenamefont {{Lima}},
  \citenamefont {{Maia}}, \citenamefont {{Marshall}}, \citenamefont
  {{Melchior}}, \citenamefont {{Menanteau}}, \citenamefont {{Miquel}},
  \citenamefont {{Morgan}}, \citenamefont {{Palmese}}, \citenamefont
  {{Paz-Chinchon}}, \citenamefont {{Pieres}}, \citenamefont {{Plazas
  Malag{\'o}n}}, \citenamefont {{Reil}}, \citenamefont {{Rodriguez Monroy}},
  \citenamefont {{Romer}}, \citenamefont {{Sanchez}}, \citenamefont
  {{Scarpine}}, \citenamefont {{Schubnell}}, \citenamefont {{Serrano}},
  \citenamefont {{Sevilla}}, \citenamefont {{Smith}}, \citenamefont
  {{Suchyta}}, \citenamefont {{Tarle}}, \citenamefont {{Thomas}}, \citenamefont
  {{To}}, \citenamefont {{Varga}}, \citenamefont {{Weller}}, \citenamefont
  {{Wilkinson}},\ and\ \citenamefont {{DES
  Collaboration}}}]{Zuercher2022despeaks}%
  \BibitemOpen
  \bibfield  {author} {\bibinfo {author} {\bibfnamefont {D.}~\bibnamefont
  {{Z{\"u}rcher}}}, \bibinfo {author} {\bibfnamefont {J.}~\bibnamefont
  {{Fluri}}}, \bibinfo {author} {\bibfnamefont {R.}~\bibnamefont {{Sgier}}},
  \bibinfo {author} {\bibfnamefont {T.}~\bibnamefont {{Kacprzak}}}, \bibinfo
  {author} {\bibfnamefont {M.}~\bibnamefont {{Gatti}}}, \bibinfo {author}
  {\bibfnamefont {C.}~\bibnamefont {{Doux}}}, \bibinfo {author} {\bibfnamefont
  {L.}~\bibnamefont {{Whiteway}}}, \bibinfo {author} {\bibfnamefont
  {A.}~\bibnamefont {{R{\'e}fr{\'e}gier}}}, \bibinfo {author} {\bibfnamefont
  {C.}~\bibnamefont {{Chang}}}, \bibinfo {author} {\bibfnamefont
  {N.}~\bibnamefont {{Jeffrey}}}, \bibinfo {author} {\bibfnamefont
  {B.}~\bibnamefont {{Jain}}}, \bibinfo {author} {\bibfnamefont
  {P.}~\bibnamefont {{Lemos}}}, \bibinfo {author} {\bibfnamefont
  {D.}~\bibnamefont {{Bacon}}}, \bibinfo {author} {\bibfnamefont
  {A.}~\bibnamefont {{Alarcon}}}, \bibinfo {author} {\bibfnamefont
  {A.}~\bibnamefont {{Amon}}}, \bibinfo {author} {\bibfnamefont
  {K.}~\bibnamefont {{Bechtol}}}, \bibinfo {author} {\bibfnamefont
  {M.}~\bibnamefont {{Becker}}}, \bibinfo {author} {\bibfnamefont
  {G.}~\bibnamefont {{Bernstein}}}, \bibinfo {author} {\bibfnamefont
  {A.}~\bibnamefont {{Campos}}}, \bibinfo {author} {\bibfnamefont
  {R.}~\bibnamefont {{Chen}}}, \bibinfo {author} {\bibfnamefont
  {A.}~\bibnamefont {{Choi}}}, \bibinfo {author} {\bibfnamefont
  {C.}~\bibnamefont {{Davis}}}, \bibinfo {author} {\bibfnamefont
  {J.}~\bibnamefont {{Derose}}}, \bibinfo {author} {\bibfnamefont
  {S.}~\bibnamefont {{Dodelson}}}, \bibinfo {author} {\bibfnamefont
  {F.}~\bibnamefont {{Elsner}}}, \bibinfo {author} {\bibfnamefont
  {J.}~\bibnamefont {{Elvin-Poole}}}, \bibinfo {author} {\bibfnamefont
  {S.}~\bibnamefont {{Everett}}}, \bibinfo {author} {\bibfnamefont
  {A.}~\bibnamefont {{Ferte}}}, \bibinfo {author} {\bibfnamefont
  {D.}~\bibnamefont {{Gruen}}}, \bibinfo {author} {\bibfnamefont
  {I.}~\bibnamefont {{Harrison}}}, \bibinfo {author} {\bibfnamefont
  {D.}~\bibnamefont {{Huterer}}}, \bibinfo {author} {\bibfnamefont
  {M.}~\bibnamefont {{Jarvis}}}, \bibinfo {author} {\bibfnamefont {P.~F.}\
  \bibnamefont {{Leget}}}, \bibinfo {author} {\bibfnamefont {N.}~\bibnamefont
  {{Maccrann}}}, \bibinfo {author} {\bibfnamefont {J.}~\bibnamefont
  {{Mccullough}}}, \bibinfo {author} {\bibfnamefont {J.}~\bibnamefont
  {{Muir}}}, \bibinfo {author} {\bibfnamefont {J.}~\bibnamefont {{Myles}}},
  \bibinfo {author} {\bibfnamefont {A.}~\bibnamefont {{Navarro Alsina}}},
  \bibinfo {author} {\bibfnamefont {S.}~\bibnamefont {{Pandey}}}, \bibinfo
  {author} {\bibfnamefont {J.}~\bibnamefont {{Prat}}}, \bibinfo {author}
  {\bibfnamefont {M.}~\bibnamefont {{Raveri}}}, \bibinfo {author}
  {\bibfnamefont {R.~P.}\ \bibnamefont {{Rollins}}}, \bibinfo {author}
  {\bibfnamefont {A.}~\bibnamefont {{Roodman}}}, \bibinfo {author}
  {\bibfnamefont {C.}~\bibnamefont {{Sanchez}}}, \bibinfo {author}
  {\bibfnamefont {L.~F.}\ \bibnamefont {{Secco}}}, \bibinfo {author}
  {\bibfnamefont {E.}~\bibnamefont {{Sheldon}}}, \bibinfo {author}
  {\bibfnamefont {T.}~\bibnamefont {{Shin}}}, \bibinfo {author} {\bibfnamefont
  {M.}~\bibnamefont {{Troxel}}}, \bibinfo {author} {\bibfnamefont
  {I.}~\bibnamefont {{Tutusaus}}}, \bibinfo {author} {\bibfnamefont
  {B.}~\bibnamefont {{Yin}}}, \bibinfo {author} {\bibfnamefont
  {M.}~\bibnamefont {{Aguena}}}, \bibinfo {author} {\bibfnamefont
  {S.}~\bibnamefont {{Allam}}}, \bibinfo {author} {\bibfnamefont
  {F.}~\bibnamefont {{Andrade-Oliveira}}}, \bibinfo {author} {\bibfnamefont
  {J.}~\bibnamefont {{Annis}}}, \bibinfo {author} {\bibfnamefont
  {E.}~\bibnamefont {{Bertin}}}, \bibinfo {author} {\bibfnamefont
  {D.}~\bibnamefont {{Brooks}}}, \bibinfo {author} {\bibfnamefont
  {D.}~\bibnamefont {{Burke}}}, \bibinfo {author} {\bibfnamefont
  {A.}~\bibnamefont {{Carnero Rosell}}}, \bibinfo {author} {\bibfnamefont
  {M.}~\bibnamefont {{Carrasco Kind}}}, \bibinfo {author} {\bibfnamefont
  {J.}~\bibnamefont {{Carretero}}}, \bibinfo {author} {\bibfnamefont
  {F.}~\bibnamefont {{Castander}}}, \bibinfo {author} {\bibfnamefont
  {R.}~\bibnamefont {{Cawthon}}}, \bibinfo {author} {\bibfnamefont
  {C.}~\bibnamefont {{Conselice}}}, \bibinfo {author} {\bibfnamefont
  {M.}~\bibnamefont {{Costanzi}}}, \bibinfo {author} {\bibfnamefont
  {L.}~\bibnamefont {{da Costa}}}, \bibinfo {author} {\bibfnamefont {M.~E.}\
  \bibnamefont {{da Silva Pereira}}}, \bibinfo {author} {\bibfnamefont
  {T.}~\bibnamefont {{Davis}}}, \bibinfo {author} {\bibfnamefont
  {J.}~\bibnamefont {{De Vicente}}}, \bibinfo {author} {\bibfnamefont
  {S.}~\bibnamefont {{Desai}}}, \bibinfo {author} {\bibfnamefont {H.~T.}\
  \bibnamefont {{Diehl}}}, \bibinfo {author} {\bibfnamefont {J.}~\bibnamefont
  {{Dietrich}}}, \bibinfo {author} {\bibfnamefont {P.}~\bibnamefont {{Doel}}},
  \bibinfo {author} {\bibfnamefont {K.}~\bibnamefont {{Eckert}}}, \bibinfo
  {author} {\bibfnamefont {A.}~\bibnamefont {{Evrard}}}, \bibinfo {author}
  {\bibfnamefont {I.}~\bibnamefont {{Ferrero}}}, \bibinfo {author}
  {\bibfnamefont {B.}~\bibnamefont {{Flaugher}}}, \bibinfo {author}
  {\bibfnamefont {P.}~\bibnamefont {{Fosalba}}}, \bibinfo {author}
  {\bibfnamefont {D.}~\bibnamefont {{Friedel}}}, \bibinfo {author}
  {\bibfnamefont {J.}~\bibnamefont {{Frieman}}}, \bibinfo {author}
  {\bibfnamefont {J.}~\bibnamefont {{Garcia-Bellido}}}, \bibinfo {author}
  {\bibfnamefont {E.}~\bibnamefont {{Gaztanaga}}}, \bibinfo {author}
  {\bibfnamefont {D.}~\bibnamefont {{Gerdes}}}, \bibinfo {author}
  {\bibfnamefont {T.}~\bibnamefont {{Giannantonio}}}, \bibinfo {author}
  {\bibfnamefont {R.}~\bibnamefont {{Gruendl}}}, \bibinfo {author}
  {\bibfnamefont {J.}~\bibnamefont {{Gschwend}}}, \bibinfo {author}
  {\bibfnamefont {G.}~\bibnamefont {{Gutierrez}}}, \bibinfo {author}
  {\bibfnamefont {S.}~\bibnamefont {{Hinton}}}, \bibinfo {author}
  {\bibfnamefont {D.~L.}\ \bibnamefont {{Hollowood}}}, \bibinfo {author}
  {\bibfnamefont {K.}~\bibnamefont {{Honscheid}}}, \bibinfo {author}
  {\bibfnamefont {B.}~\bibnamefont {{Hoyle}}}, \bibinfo {author} {\bibfnamefont
  {D.}~\bibnamefont {{James}}}, \bibinfo {author} {\bibfnamefont
  {K.}~\bibnamefont {{Kuehn}}}, \bibinfo {author} {\bibfnamefont
  {N.}~\bibnamefont {{Kuropatkin}}}, \bibinfo {author} {\bibfnamefont
  {O.}~\bibnamefont {{Lahav}}}, \bibinfo {author} {\bibfnamefont
  {C.}~\bibnamefont {{Lidman}}}, \bibinfo {author} {\bibfnamefont
  {M.}~\bibnamefont {{Lima}}}, \bibinfo {author} {\bibfnamefont
  {M.}~\bibnamefont {{Maia}}}, \bibinfo {author} {\bibfnamefont
  {J.}~\bibnamefont {{Marshall}}}, \bibinfo {author} {\bibfnamefont
  {P.}~\bibnamefont {{Melchior}}}, \bibinfo {author} {\bibfnamefont
  {F.}~\bibnamefont {{Menanteau}}}, \bibinfo {author} {\bibfnamefont
  {R.}~\bibnamefont {{Miquel}}}, \bibinfo {author} {\bibfnamefont
  {R.}~\bibnamefont {{Morgan}}}, \bibinfo {author} {\bibfnamefont
  {A.}~\bibnamefont {{Palmese}}}, \bibinfo {author} {\bibfnamefont
  {F.}~\bibnamefont {{Paz-Chinchon}}}, \bibinfo {author} {\bibfnamefont
  {A.}~\bibnamefont {{Pieres}}}, \bibinfo {author} {\bibfnamefont
  {A.}~\bibnamefont {{Plazas Malag{\'o}n}}}, \bibinfo {author} {\bibfnamefont
  {K.}~\bibnamefont {{Reil}}}, \bibinfo {author} {\bibfnamefont
  {M.}~\bibnamefont {{Rodriguez Monroy}}}, \bibinfo {author} {\bibfnamefont
  {K.}~\bibnamefont {{Romer}}}, \bibinfo {author} {\bibfnamefont
  {E.}~\bibnamefont {{Sanchez}}}, \bibinfo {author} {\bibfnamefont
  {V.}~\bibnamefont {{Scarpine}}}, \bibinfo {author} {\bibfnamefont
  {M.}~\bibnamefont {{Schubnell}}}, \bibinfo {author} {\bibfnamefont
  {S.}~\bibnamefont {{Serrano}}}, \bibinfo {author} {\bibfnamefont
  {I.}~\bibnamefont {{Sevilla}}}, \bibinfo {author} {\bibfnamefont
  {M.}~\bibnamefont {{Smith}}}, \bibinfo {author} {\bibfnamefont
  {E.}~\bibnamefont {{Suchyta}}}, \bibinfo {author} {\bibfnamefont
  {G.}~\bibnamefont {{Tarle}}}, \bibinfo {author} {\bibfnamefont
  {D.}~\bibnamefont {{Thomas}}}, \bibinfo {author} {\bibfnamefont
  {C.}~\bibnamefont {{To}}}, \bibinfo {author} {\bibfnamefont {T.~N.}\
  \bibnamefont {{Varga}}}, \bibinfo {author} {\bibfnamefont {J.}~\bibnamefont
  {{Weller}}}, \bibinfo {author} {\bibfnamefont {R.}~\bibnamefont
  {{Wilkinson}}},\ and\ \bibinfo {author} {\bibnamefont {{DES
  Collaboration}}},\ }\bibfield  {title} {\bibinfo {title} {{Dark energy survey
  year 3 results: Cosmology with peaks using an emulator approach}},\ }\href
  {https://doi.org/10.1093/mnras/stac078} {\bibfield  {journal} {\bibinfo
  {journal} {\mnras}\ }\textbf {\bibinfo {volume} {511}},\ \bibinfo {pages}
  {2075} (\bibinfo {year} {2022})},\ \Eprint {https://arxiv.org/abs/2110.10135}
  {arXiv:2110.10135 [astro-ph.CO]} \BibitemShut {NoStop}%
\bibitem [{\citenamefont {{Fluri}}\ \emph {et~al.}(2019)\citenamefont
  {{Fluri}}, \citenamefont {{Kacprzak}}, \citenamefont {{Lucchi}},
  \citenamefont {{Refregier}}, \citenamefont {{Amara}}, \citenamefont
  {{Hofmann}},\ and\ \citenamefont {{Schneider}}}]{Fluri2019kids}%
  \BibitemOpen
  \bibfield  {author} {\bibinfo {author} {\bibfnamefont {J.}~\bibnamefont
  {{Fluri}}}, \bibinfo {author} {\bibfnamefont {T.}~\bibnamefont {{Kacprzak}}},
  \bibinfo {author} {\bibfnamefont {A.}~\bibnamefont {{Lucchi}}}, \bibinfo
  {author} {\bibfnamefont {A.}~\bibnamefont {{Refregier}}}, \bibinfo {author}
  {\bibfnamefont {A.}~\bibnamefont {{Amara}}}, \bibinfo {author} {\bibfnamefont
  {T.}~\bibnamefont {{Hofmann}}},\ and\ \bibinfo {author} {\bibfnamefont
  {A.}~\bibnamefont {{Schneider}}},\ }\bibfield  {title} {\bibinfo {title}
  {{Cosmological constraints with deep learning from KiDS-450 weak lensing
  maps}},\ }\href {https://doi.org/10.1103/PhysRevD.100.063514} {\bibfield
  {journal} {\bibinfo  {journal} {\prd}\ }\textbf {\bibinfo {volume} {100}},\
  \bibinfo {eid} {063514} (\bibinfo {year} {2019})}\BibitemShut {NoStop}%
\bibitem [{\citenamefont {{Fluri}}\ \emph {et~al.}(2022)\citenamefont
  {{Fluri}}, \citenamefont {{Kacprzak}}, \citenamefont {{Lucchi}},
  \citenamefont {{Schneider}}, \citenamefont {{Refregier}},\ and\ \citenamefont
  {{Hofmann}}}]{Fluri2022kids}%
  \BibitemOpen
  \bibfield  {author} {\bibinfo {author} {\bibfnamefont {J.}~\bibnamefont
  {{Fluri}}}, \bibinfo {author} {\bibfnamefont {T.}~\bibnamefont {{Kacprzak}}},
  \bibinfo {author} {\bibfnamefont {A.}~\bibnamefont {{Lucchi}}}, \bibinfo
  {author} {\bibfnamefont {A.}~\bibnamefont {{Schneider}}}, \bibinfo {author}
  {\bibfnamefont {A.}~\bibnamefont {{Refregier}}},\ and\ \bibinfo {author}
  {\bibfnamefont {T.}~\bibnamefont {{Hofmann}}},\ }\bibfield  {title} {\bibinfo
  {title} {{A Full $w$CDM Analysis of KiDS-1000 Weak Lensing Maps using Deep
  Learning}},\ }\href@noop {} {\bibfield  {journal} {\bibinfo  {journal} {arXiv
  e-prints}\ ,\ \bibinfo {eid} {arXiv:2201.07771}} (\bibinfo {year} {2022})},\
  \Eprint {https://arxiv.org/abs/2201.07771} {arXiv:2201.07771 [astro-ph.CO]}
  \BibitemShut {NoStop}%
\bibitem [{\citenamefont {{Gruen}}\ \emph {et~al.}(2018)\citenamefont
  {{Gruen}}, \citenamefont {{Friedrich}}, \citenamefont {{Krause}},
  \citenamefont {{DeRose}}, \citenamefont {{Cawthon}}, \citenamefont {{Davis}},
  \citenamefont {{Elvin-Poole}}, \citenamefont {{Rykoff}}, \citenamefont
  {{Wechsler}}, \citenamefont {{Alarcon}}, \citenamefont {{Bernstein}},
  \citenamefont {{Blazek}}, \citenamefont {{Chang}}, \citenamefont
  {{Clampitt}}, \citenamefont {{Crocce}}, \citenamefont {{De Vicente}},
  \citenamefont {{Gatti}}, \citenamefont {{Gill}}, \citenamefont {{Hartley}},
  \citenamefont {{Hilbert}}, \citenamefont {{Hoyle}}, \citenamefont {{Jain}},
  \citenamefont {{Jarvis}}, \citenamefont {{Lahav}}, \citenamefont
  {{MacCrann}}, \citenamefont {{McClintock}}, \citenamefont {{Prat}},
  \citenamefont {{Rollins}}, \citenamefont {{Ross}}, \citenamefont {{Rozo}},
  \citenamefont {{Samuroff}}, \citenamefont {{S{\'a}nchez}}, \citenamefont
  {{Sheldon}}, \citenamefont {{Troxel}}, \citenamefont {{Zuntz}}, \citenamefont
  {{Abbott}}, \citenamefont {{Abdalla}}, \citenamefont {{Allam}}, \citenamefont
  {{Annis}}, \citenamefont {{Bechtol}}, \citenamefont {{Benoit-L{\'e}vy}},
  \citenamefont {{Bertin}}, \citenamefont {{Bridle}}, \citenamefont {{Brooks}},
  \citenamefont {{Buckley-Geer}}, \citenamefont {{Carnero Rosell}},
  \citenamefont {{Carrasco Kind}}, \citenamefont {{Carretero}}, \citenamefont
  {{Cunha}}, \citenamefont {{D'Andrea}}, \citenamefont {{da Costa}},
  \citenamefont {{Desai}}, \citenamefont {{Diehl}}, \citenamefont {{Dietrich}},
  \citenamefont {{Doel}}, \citenamefont {{Drlica-Wagner}}, \citenamefont
  {{Fernandez}}, \citenamefont {{Flaugher}}, \citenamefont {{Fosalba}},
  \citenamefont {{Frieman}}, \citenamefont {{Garc{\'\i}a-Bellido}},
  \citenamefont {{Gaztanaga}}, \citenamefont {{Giannantonio}}, \citenamefont
  {{Gruendl}}, \citenamefont {{Gschwend}}, \citenamefont {{Gutierrez}},
  \citenamefont {{Honscheid}}, \citenamefont {{James}}, \citenamefont
  {{Jeltema}}, \citenamefont {{Kuehn}}, \citenamefont {{Kuropatkin}},
  \citenamefont {{Lima}}, \citenamefont {{March}}, \citenamefont {{Marshall}},
  \citenamefont {{Martini}}, \citenamefont {{Melchior}}, \citenamefont
  {{Menanteau}}, \citenamefont {{Miquel}}, \citenamefont {{Mohr}},
  \citenamefont {{Plazas}}, \citenamefont {{Roodman}}, \citenamefont
  {{Sanchez}}, \citenamefont {{Scarpine}}, \citenamefont {{Schubnell}},
  \citenamefont {{Sevilla-Noarbe}}, \citenamefont {{Smith}}, \citenamefont
  {{Smith}}, \citenamefont {{Soares-Santos}}, \citenamefont {{Sobreira}},
  \citenamefont {{Swanson}}, \citenamefont {{Tarle}}, \citenamefont {{Thomas}},
  \citenamefont {{Vikram}}, \citenamefont {{Walker}}, \citenamefont {{Weller}},
  \citenamefont {{Zhang}},\ and\ \citenamefont {{DES
  Collaboration}}}]{Gruen2018densitysplit}%
  \BibitemOpen
  \bibfield  {author} {\bibinfo {author} {\bibfnamefont {D.}~\bibnamefont
  {{Gruen}}}, \bibinfo {author} {\bibfnamefont {O.}~\bibnamefont
  {{Friedrich}}}, \bibinfo {author} {\bibfnamefont {E.}~\bibnamefont
  {{Krause}}}, \bibinfo {author} {\bibfnamefont {J.}~\bibnamefont {{DeRose}}},
  \bibinfo {author} {\bibfnamefont {R.}~\bibnamefont {{Cawthon}}}, \bibinfo
  {author} {\bibfnamefont {C.}~\bibnamefont {{Davis}}}, \bibinfo {author}
  {\bibfnamefont {J.}~\bibnamefont {{Elvin-Poole}}}, \bibinfo {author}
  {\bibfnamefont {E.~S.}\ \bibnamefont {{Rykoff}}}, \bibinfo {author}
  {\bibfnamefont {R.~H.}\ \bibnamefont {{Wechsler}}}, \bibinfo {author}
  {\bibfnamefont {A.}~\bibnamefont {{Alarcon}}}, \bibinfo {author}
  {\bibfnamefont {G.~M.}\ \bibnamefont {{Bernstein}}}, \bibinfo {author}
  {\bibfnamefont {J.}~\bibnamefont {{Blazek}}}, \bibinfo {author}
  {\bibfnamefont {C.}~\bibnamefont {{Chang}}}, \bibinfo {author} {\bibfnamefont
  {J.}~\bibnamefont {{Clampitt}}}, \bibinfo {author} {\bibfnamefont
  {M.}~\bibnamefont {{Crocce}}}, \bibinfo {author} {\bibfnamefont
  {J.}~\bibnamefont {{De Vicente}}}, \bibinfo {author} {\bibfnamefont
  {M.}~\bibnamefont {{Gatti}}}, \bibinfo {author} {\bibfnamefont {M.~S.~S.}\
  \bibnamefont {{Gill}}}, \bibinfo {author} {\bibfnamefont {W.~G.}\
  \bibnamefont {{Hartley}}}, \bibinfo {author} {\bibfnamefont {S.}~\bibnamefont
  {{Hilbert}}}, \bibinfo {author} {\bibfnamefont {B.}~\bibnamefont {{Hoyle}}},
  \bibinfo {author} {\bibfnamefont {B.}~\bibnamefont {{Jain}}}, \bibinfo
  {author} {\bibfnamefont {M.}~\bibnamefont {{Jarvis}}}, \bibinfo {author}
  {\bibfnamefont {O.}~\bibnamefont {{Lahav}}}, \bibinfo {author} {\bibfnamefont
  {N.}~\bibnamefont {{MacCrann}}}, \bibinfo {author} {\bibfnamefont
  {T.}~\bibnamefont {{McClintock}}}, \bibinfo {author} {\bibfnamefont
  {J.}~\bibnamefont {{Prat}}}, \bibinfo {author} {\bibfnamefont {R.~P.}\
  \bibnamefont {{Rollins}}}, \bibinfo {author} {\bibfnamefont {A.~J.}\
  \bibnamefont {{Ross}}}, \bibinfo {author} {\bibfnamefont {E.}~\bibnamefont
  {{Rozo}}}, \bibinfo {author} {\bibfnamefont {S.}~\bibnamefont {{Samuroff}}},
  \bibinfo {author} {\bibfnamefont {C.}~\bibnamefont {{S{\'a}nchez}}}, \bibinfo
  {author} {\bibfnamefont {E.}~\bibnamefont {{Sheldon}}}, \bibinfo {author}
  {\bibfnamefont {M.~A.}\ \bibnamefont {{Troxel}}}, \bibinfo {author}
  {\bibfnamefont {J.}~\bibnamefont {{Zuntz}}}, \bibinfo {author} {\bibfnamefont
  {T.~M.~C.}\ \bibnamefont {{Abbott}}}, \bibinfo {author} {\bibfnamefont
  {F.~B.}\ \bibnamefont {{Abdalla}}}, \bibinfo {author} {\bibfnamefont
  {S.}~\bibnamefont {{Allam}}}, \bibinfo {author} {\bibfnamefont
  {J.}~\bibnamefont {{Annis}}}, \bibinfo {author} {\bibfnamefont
  {K.}~\bibnamefont {{Bechtol}}}, \bibinfo {author} {\bibfnamefont
  {A.}~\bibnamefont {{Benoit-L{\'e}vy}}}, \bibinfo {author} {\bibfnamefont
  {E.}~\bibnamefont {{Bertin}}}, \bibinfo {author} {\bibfnamefont {S.~L.}\
  \bibnamefont {{Bridle}}}, \bibinfo {author} {\bibfnamefont {D.}~\bibnamefont
  {{Brooks}}}, \bibinfo {author} {\bibfnamefont {E.}~\bibnamefont
  {{Buckley-Geer}}}, \bibinfo {author} {\bibfnamefont {A.}~\bibnamefont
  {{Carnero Rosell}}}, \bibinfo {author} {\bibfnamefont {M.}~\bibnamefont
  {{Carrasco Kind}}}, \bibinfo {author} {\bibfnamefont {J.}~\bibnamefont
  {{Carretero}}}, \bibinfo {author} {\bibfnamefont {C.~E.}\ \bibnamefont
  {{Cunha}}}, \bibinfo {author} {\bibfnamefont {C.~B.}\ \bibnamefont
  {{D'Andrea}}}, \bibinfo {author} {\bibfnamefont {L.~N.}\ \bibnamefont {{da
  Costa}}}, \bibinfo {author} {\bibfnamefont {S.}~\bibnamefont {{Desai}}},
  \bibinfo {author} {\bibfnamefont {H.~T.}\ \bibnamefont {{Diehl}}}, \bibinfo
  {author} {\bibfnamefont {J.~P.}\ \bibnamefont {{Dietrich}}}, \bibinfo
  {author} {\bibfnamefont {P.}~\bibnamefont {{Doel}}}, \bibinfo {author}
  {\bibfnamefont {A.}~\bibnamefont {{Drlica-Wagner}}}, \bibinfo {author}
  {\bibfnamefont {E.}~\bibnamefont {{Fernandez}}}, \bibinfo {author}
  {\bibfnamefont {B.}~\bibnamefont {{Flaugher}}}, \bibinfo {author}
  {\bibfnamefont {P.}~\bibnamefont {{Fosalba}}}, \bibinfo {author}
  {\bibfnamefont {J.}~\bibnamefont {{Frieman}}}, \bibinfo {author}
  {\bibfnamefont {J.}~\bibnamefont {{Garc{\'\i}a-Bellido}}}, \bibinfo {author}
  {\bibfnamefont {E.}~\bibnamefont {{Gaztanaga}}}, \bibinfo {author}
  {\bibfnamefont {T.}~\bibnamefont {{Giannantonio}}}, \bibinfo {author}
  {\bibfnamefont {R.~A.}\ \bibnamefont {{Gruendl}}}, \bibinfo {author}
  {\bibfnamefont {J.}~\bibnamefont {{Gschwend}}}, \bibinfo {author}
  {\bibfnamefont {G.}~\bibnamefont {{Gutierrez}}}, \bibinfo {author}
  {\bibfnamefont {K.}~\bibnamefont {{Honscheid}}}, \bibinfo {author}
  {\bibfnamefont {D.~J.}\ \bibnamefont {{James}}}, \bibinfo {author}
  {\bibfnamefont {T.}~\bibnamefont {{Jeltema}}}, \bibinfo {author}
  {\bibfnamefont {K.}~\bibnamefont {{Kuehn}}}, \bibinfo {author} {\bibfnamefont
  {N.}~\bibnamefont {{Kuropatkin}}}, \bibinfo {author} {\bibfnamefont
  {M.}~\bibnamefont {{Lima}}}, \bibinfo {author} {\bibfnamefont
  {M.}~\bibnamefont {{March}}}, \bibinfo {author} {\bibfnamefont {J.~L.}\
  \bibnamefont {{Marshall}}}, \bibinfo {author} {\bibfnamefont
  {P.}~\bibnamefont {{Martini}}}, \bibinfo {author} {\bibfnamefont
  {P.}~\bibnamefont {{Melchior}}}, \bibinfo {author} {\bibfnamefont
  {F.}~\bibnamefont {{Menanteau}}}, \bibinfo {author} {\bibfnamefont
  {R.}~\bibnamefont {{Miquel}}}, \bibinfo {author} {\bibfnamefont {J.~J.}\
  \bibnamefont {{Mohr}}}, \bibinfo {author} {\bibfnamefont {A.~A.}\
  \bibnamefont {{Plazas}}}, \bibinfo {author} {\bibfnamefont {A.}~\bibnamefont
  {{Roodman}}}, \bibinfo {author} {\bibfnamefont {E.}~\bibnamefont
  {{Sanchez}}}, \bibinfo {author} {\bibfnamefont {V.}~\bibnamefont
  {{Scarpine}}}, \bibinfo {author} {\bibfnamefont {M.}~\bibnamefont
  {{Schubnell}}}, \bibinfo {author} {\bibfnamefont {I.}~\bibnamefont
  {{Sevilla-Noarbe}}}, \bibinfo {author} {\bibfnamefont {M.}~\bibnamefont
  {{Smith}}}, \bibinfo {author} {\bibfnamefont {R.~C.}\ \bibnamefont
  {{Smith}}}, \bibinfo {author} {\bibfnamefont {M.}~\bibnamefont
  {{Soares-Santos}}}, \bibinfo {author} {\bibfnamefont {F.}~\bibnamefont
  {{Sobreira}}}, \bibinfo {author} {\bibfnamefont {M.~E.~C.}\ \bibnamefont
  {{Swanson}}}, \bibinfo {author} {\bibfnamefont {G.}~\bibnamefont {{Tarle}}},
  \bibinfo {author} {\bibfnamefont {D.}~\bibnamefont {{Thomas}}}, \bibinfo
  {author} {\bibfnamefont {V.}~\bibnamefont {{Vikram}}}, \bibinfo {author}
  {\bibfnamefont {A.~R.}\ \bibnamefont {{Walker}}}, \bibinfo {author}
  {\bibfnamefont {J.}~\bibnamefont {{Weller}}}, \bibinfo {author}
  {\bibfnamefont {Y.}~\bibnamefont {{Zhang}}},\ and\ \bibinfo {author}
  {\bibnamefont {{DES Collaboration}}},\ }\bibfield  {title} {\bibinfo {title}
  {{Density split statistics: Cosmological constraints from counts and lensing
  in cells in DES Y1 and SDSS data}},\ }\href
  {https://doi.org/10.1103/PhysRevD.98.023507} {\bibfield  {journal} {\bibinfo
  {journal} {\prd}\ }\textbf {\bibinfo {volume} {98}},\ \bibinfo {eid} {023507}
  (\bibinfo {year} {2018})}\BibitemShut {NoStop}%
\bibitem [{\citenamefont {{Salvador}}\ \emph {et~al.}(2019)\citenamefont
  {{Salvador}}, \citenamefont {{S{\'a}nchez}}, \citenamefont {{Pagul}},
  \citenamefont {{Garc{\'\i}a-Bellido}}, \citenamefont {{Sanchez}},
  \citenamefont {{Pujol}}, \citenamefont {{Frieman}}, \citenamefont
  {{Gaztanaga}}, \citenamefont {{Ross}}, \citenamefont {{Sevilla-Noarbe}},
  \citenamefont {{Abbott}}, \citenamefont {{Allam}}, \citenamefont {{Annis}},
  \citenamefont {{Avila}}, \citenamefont {{Bertin}}, \citenamefont {{Brooks}},
  \citenamefont {{Burke}}, \citenamefont {{Carnero Rosell}}, \citenamefont
  {{Carrasco Kind}}, \citenamefont {{Carretero}}, \citenamefont {{Castander}},
  \citenamefont {{Cunha}}, \citenamefont {{De Vicente}}, \citenamefont
  {{Diehl}}, \citenamefont {{Doel}}, \citenamefont {{Evrard}}, \citenamefont
  {{Fosalba}}, \citenamefont {{Gruen}}, \citenamefont {{Gruendl}},
  \citenamefont {{Gschwend}}, \citenamefont {{Gutierrez}}, \citenamefont
  {{Hartley}}, \citenamefont {{Hollowood}}, \citenamefont {{James}},
  \citenamefont {{Kuehn}}, \citenamefont {{Kuropatkin}}, \citenamefont
  {{Lahav}}, \citenamefont {{Lima}}, \citenamefont {{March}}, \citenamefont
  {{Marshall}}, \citenamefont {{Menanteau}}, \citenamefont {{Miquel}},
  \citenamefont {{Romer}}, \citenamefont {{Roodman}}, \citenamefont
  {{Scarpine}}, \citenamefont {{Schindler}}, \citenamefont {{Smith}},
  \citenamefont {{Soares-Santos}}, \citenamefont {{Sobreira}}, \citenamefont
  {{Suchyta}}, \citenamefont {{Swanson}}, \citenamefont {{Tarle}},
  \citenamefont {{Thomas}}, \citenamefont {{Vikram}}, \citenamefont
  {{Walker}},\ and\ \citenamefont {{DES Collaboration}}}]{Salvador2019sv}%
  \BibitemOpen
  \bibfield  {author} {\bibinfo {author} {\bibfnamefont {A.~I.}\ \bibnamefont
  {{Salvador}}}, \bibinfo {author} {\bibfnamefont {F.~J.}\ \bibnamefont
  {{S{\'a}nchez}}}, \bibinfo {author} {\bibfnamefont {A.}~\bibnamefont
  {{Pagul}}}, \bibinfo {author} {\bibfnamefont {J.}~\bibnamefont
  {{Garc{\'\i}a-Bellido}}}, \bibinfo {author} {\bibfnamefont {E.}~\bibnamefont
  {{Sanchez}}}, \bibinfo {author} {\bibfnamefont {A.}~\bibnamefont {{Pujol}}},
  \bibinfo {author} {\bibfnamefont {J.}~\bibnamefont {{Frieman}}}, \bibinfo
  {author} {\bibfnamefont {E.}~\bibnamefont {{Gaztanaga}}}, \bibinfo {author}
  {\bibfnamefont {A.~J.}\ \bibnamefont {{Ross}}}, \bibinfo {author}
  {\bibfnamefont {I.}~\bibnamefont {{Sevilla-Noarbe}}}, \bibinfo {author}
  {\bibfnamefont {T.~M.~C.}\ \bibnamefont {{Abbott}}}, \bibinfo {author}
  {\bibfnamefont {S.}~\bibnamefont {{Allam}}}, \bibinfo {author} {\bibfnamefont
  {J.}~\bibnamefont {{Annis}}}, \bibinfo {author} {\bibfnamefont
  {S.}~\bibnamefont {{Avila}}}, \bibinfo {author} {\bibfnamefont
  {E.}~\bibnamefont {{Bertin}}}, \bibinfo {author} {\bibfnamefont
  {D.}~\bibnamefont {{Brooks}}}, \bibinfo {author} {\bibfnamefont {D.~L.}\
  \bibnamefont {{Burke}}}, \bibinfo {author} {\bibfnamefont {A.}~\bibnamefont
  {{Carnero Rosell}}}, \bibinfo {author} {\bibfnamefont {M.}~\bibnamefont
  {{Carrasco Kind}}}, \bibinfo {author} {\bibfnamefont {J.}~\bibnamefont
  {{Carretero}}}, \bibinfo {author} {\bibfnamefont {F.~J.}\ \bibnamefont
  {{Castander}}}, \bibinfo {author} {\bibfnamefont {C.~E.}\ \bibnamefont
  {{Cunha}}}, \bibinfo {author} {\bibfnamefont {J.}~\bibnamefont {{De
  Vicente}}}, \bibinfo {author} {\bibfnamefont {H.~T.}\ \bibnamefont
  {{Diehl}}}, \bibinfo {author} {\bibfnamefont {P.}~\bibnamefont {{Doel}}},
  \bibinfo {author} {\bibfnamefont {A.~E.}\ \bibnamefont {{Evrard}}}, \bibinfo
  {author} {\bibfnamefont {P.}~\bibnamefont {{Fosalba}}}, \bibinfo {author}
  {\bibfnamefont {D.}~\bibnamefont {{Gruen}}}, \bibinfo {author} {\bibfnamefont
  {R.~A.}\ \bibnamefont {{Gruendl}}}, \bibinfo {author} {\bibfnamefont
  {J.}~\bibnamefont {{Gschwend}}}, \bibinfo {author} {\bibfnamefont
  {G.}~\bibnamefont {{Gutierrez}}}, \bibinfo {author} {\bibfnamefont {W.~G.}\
  \bibnamefont {{Hartley}}}, \bibinfo {author} {\bibfnamefont {D.~L.}\
  \bibnamefont {{Hollowood}}}, \bibinfo {author} {\bibfnamefont {D.~J.}\
  \bibnamefont {{James}}}, \bibinfo {author} {\bibfnamefont {K.}~\bibnamefont
  {{Kuehn}}}, \bibinfo {author} {\bibfnamefont {N.}~\bibnamefont
  {{Kuropatkin}}}, \bibinfo {author} {\bibfnamefont {O.}~\bibnamefont
  {{Lahav}}}, \bibinfo {author} {\bibfnamefont {M.}~\bibnamefont {{Lima}}},
  \bibinfo {author} {\bibfnamefont {M.}~\bibnamefont {{March}}}, \bibinfo
  {author} {\bibfnamefont {J.~L.}\ \bibnamefont {{Marshall}}}, \bibinfo
  {author} {\bibfnamefont {F.}~\bibnamefont {{Menanteau}}}, \bibinfo {author}
  {\bibfnamefont {R.}~\bibnamefont {{Miquel}}}, \bibinfo {author}
  {\bibfnamefont {A.~K.}\ \bibnamefont {{Romer}}}, \bibinfo {author}
  {\bibfnamefont {A.}~\bibnamefont {{Roodman}}}, \bibinfo {author}
  {\bibfnamefont {V.}~\bibnamefont {{Scarpine}}}, \bibinfo {author}
  {\bibfnamefont {R.}~\bibnamefont {{Schindler}}}, \bibinfo {author}
  {\bibfnamefont {M.}~\bibnamefont {{Smith}}}, \bibinfo {author} {\bibfnamefont
  {M.}~\bibnamefont {{Soares-Santos}}}, \bibinfo {author} {\bibfnamefont
  {F.}~\bibnamefont {{Sobreira}}}, \bibinfo {author} {\bibfnamefont
  {E.}~\bibnamefont {{Suchyta}}}, \bibinfo {author} {\bibfnamefont {M.~E.~C.}\
  \bibnamefont {{Swanson}}}, \bibinfo {author} {\bibfnamefont {G.}~\bibnamefont
  {{Tarle}}}, \bibinfo {author} {\bibfnamefont {D.}~\bibnamefont {{Thomas}}},
  \bibinfo {author} {\bibfnamefont {V.}~\bibnamefont {{Vikram}}}, \bibinfo
  {author} {\bibfnamefont {A.~R.}\ \bibnamefont {{Walker}}},\ and\ \bibinfo
  {author} {\bibnamefont {{DES Collaboration}}},\ }\bibfield  {title} {\bibinfo
  {title} {{Measuring linear and non-linear galaxy bias using counts-in-cells
  in the Dark Energy Survey Science Verification data}},\ }\href
  {https://doi.org/10.1093/mnras/sty2802} {\bibfield  {journal} {\bibinfo
  {journal} {\mnras}\ }\textbf {\bibinfo {volume} {482}},\ \bibinfo {pages}
  {1435} (\bibinfo {year} {2019})},\ \Eprint {https://arxiv.org/abs/1807.10331}
  {arXiv:1807.10331 [astro-ph.CO]} \BibitemShut {NoStop}%
\bibitem [{\citenamefont {{Hirata}}\ and\ \citenamefont
  {{Seljak}}(2004)}]{Hirata2004intrinsic}%
  \BibitemOpen
  \bibfield  {author} {\bibinfo {author} {\bibfnamefont {C.~M.}\ \bibnamefont
  {{Hirata}}}\ and\ \bibinfo {author} {\bibfnamefont {U.}~\bibnamefont
  {{Seljak}}},\ }\bibfield  {title} {\bibinfo {title} {{Intrinsic
  alignment-lensing interference as a contaminant of cosmic shear}},\ }\href
  {https://doi.org/10.1103/PhysRevD.70.063526} {\bibfield  {journal} {\bibinfo
  {journal} {\prd}\ }\textbf {\bibinfo {volume} {70}},\ \bibinfo {eid} {063526}
  (\bibinfo {year} {2004})}\BibitemShut {NoStop}%
\bibitem [{\citenamefont {{Bridle}}\ and\ \citenamefont
  {{King}}(2007)}]{Bridle2007constraints}%
  \BibitemOpen
  \bibfield  {author} {\bibinfo {author} {\bibfnamefont {S.}~\bibnamefont
  {{Bridle}}}\ and\ \bibinfo {author} {\bibfnamefont {L.}~\bibnamefont
  {{King}}},\ }\bibfield  {title} {\bibinfo {title} {{Dark energy constraints
  from cosmic shear power spectra: impact of intrinsic alignments on
  photometric redshift requirements}},\ }\href
  {https://doi.org/10.1088/1367-2630/9/12/444} {\bibfield  {journal} {\bibinfo
  {journal} {New Journal of Physics}\ }\textbf {\bibinfo {volume} {9}},\
  \bibinfo {pages} {444} (\bibinfo {year} {2007})}\BibitemShut {NoStop}%
\bibitem [{\citenamefont {{Joachimi}}\ \emph {et~al.}(2011)\citenamefont
  {{Joachimi}}, \citenamefont {{Mandelbaum}}, \citenamefont {{Abdalla}},\ and\
  \citenamefont {{Bridle}}}]{Joachimi2011ia}%
  \BibitemOpen
  \bibfield  {author} {\bibinfo {author} {\bibfnamefont {B.}~\bibnamefont
  {{Joachimi}}}, \bibinfo {author} {\bibfnamefont {R.}~\bibnamefont
  {{Mandelbaum}}}, \bibinfo {author} {\bibfnamefont {F.~B.}\ \bibnamefont
  {{Abdalla}}},\ and\ \bibinfo {author} {\bibfnamefont {S.~L.}\ \bibnamefont
  {{Bridle}}},\ }\bibfield  {title} {\bibinfo {title} {{Constraints on
  intrinsic alignment contamination of weak lensing surveys using the MegaZ-LRG
  sample}},\ }\href {https://doi.org/10.1051/0004-6361/201015621} {\bibfield
  {journal} {\bibinfo  {journal} {\aap}\ }\textbf {\bibinfo {volume} {527}},\
  \bibinfo {eid} {A26} (\bibinfo {year} {2011})},\ \Eprint
  {https://arxiv.org/abs/1008.3491} {arXiv:1008.3491 [astro-ph.CO]}
  \BibitemShut {NoStop}%
\bibitem [{\citenamefont {Collaboration}(2020)}]{Planck2018parameters}%
  \BibitemOpen
  \bibfield  {author} {\bibinfo {author} {\bibfnamefont {T.~P.}\ \bibnamefont
  {Collaboration}},\ }\bibfield  {title} {\bibinfo {title} {Planck 2018 results
  vi. cosmological parameters},\ }\href@noop {} {\bibfield  {journal} {\bibinfo
   {journal} {Astronomy and Astrophysics}\ }\textbf {\bibinfo {volume} {641}},\
  \bibinfo {pages} {A6} (\bibinfo {year} {2020})}\BibitemShut {NoStop}%
\bibitem [{\citenamefont {{Rodr{\'\i}guez-Monroy}}\ \emph
  {et~al.}(2022)\citenamefont {{Rodr{\'\i}guez-Monroy}}, \citenamefont
  {{Weaverdyck}}, \citenamefont {{Elvin-Poole}}, \citenamefont {{Crocce}},
  \citenamefont {{Carnero Rosell}}, \citenamefont {{Andrade-Oliveira}},
  \citenamefont {{Avila}}, \citenamefont {{Bechtol}}, \citenamefont
  {{Bernstein}}, \citenamefont {{Blazek}}, \citenamefont {{Camacho}},
  \citenamefont {{Cawthon}}, \citenamefont {{De Vicente}}, \citenamefont
  {{DeRose}}, \citenamefont {{Dodelson}}, \citenamefont {{Everett}},
  \citenamefont {{Fang}}, \citenamefont {{Ferrero}}, \citenamefont
  {{Fert{\'e}}}, \citenamefont {{Friedrich}}, \citenamefont {{Gaztanaga}},
  \citenamefont {{Giannini}}, \citenamefont {{Gruendl}}, \citenamefont
  {{Hartley}}, \citenamefont {{Herner}}, \citenamefont {{Huff}}, \citenamefont
  {{Jarvis}}, \citenamefont {{Krause}}, \citenamefont {{MacCrann}},
  \citenamefont {{Mena-Fern{\'a}ndez}}, \citenamefont {{Muir}}, \citenamefont
  {{Pandey}}, \citenamefont {{Park}}, \citenamefont {{Porredon}}, \citenamefont
  {{Prat}}, \citenamefont {{Rosenfeld}}, \citenamefont {{Ross}}, \citenamefont
  {{Rozo}}, \citenamefont {{Rykoff}}, \citenamefont {{Sanchez}}, \citenamefont
  {{Sanchez Cid}}, \citenamefont {{Sevilla-Noarbe}}, \citenamefont {{Tabbutt}},
  \citenamefont {{To}}, \citenamefont {{Wagoner}}, \citenamefont {{Wechsler}},
  \citenamefont {{Aguena}}, \citenamefont {{Allam}}, \citenamefont {{Amon}},
  \citenamefont {{Annis}}, \citenamefont {{Bacon}}, \citenamefont {{Baxter}},
  \citenamefont {{Bertin}}, \citenamefont {{Bhargava}}, \citenamefont
  {{Brooks}}, \citenamefont {{Burke}}, \citenamefont {{Carrasco Kind}},
  \citenamefont {{Carretero}}, \citenamefont {{Castander}}, \citenamefont
  {{Choi}}, \citenamefont {{Conselice}}, \citenamefont {{Costanzi}},
  \citenamefont {{da Costa}}, \citenamefont {{Pereira}}, \citenamefont
  {{Desai}}, \citenamefont {{Diehl}}, \citenamefont {{Flaugher}}, \citenamefont
  {{Fosalba}}, \citenamefont {{Frieman}}, \citenamefont
  {{Garc{\'\i}a-Bellido}}, \citenamefont {{Giannantonio}}, \citenamefont
  {{Gruen}}, \citenamefont {{Gschwend}}, \citenamefont {{Gutierrez}},
  \citenamefont {{Hinton}}, \citenamefont {{Hollowood}}, \citenamefont
  {{Honscheid}}, \citenamefont {{Huterer}}, \citenamefont {{Jain}},
  \citenamefont {{James}}, \citenamefont {{Kuehn}}, \citenamefont
  {{Kuropatkin}}, \citenamefont {{Lima}}, \citenamefont {{Maia}}, \citenamefont
  {{March}}, \citenamefont {{Marshall}}, \citenamefont {{Melchior}},
  \citenamefont {{Menanteau}}, \citenamefont {{Miller}}, \citenamefont
  {{Miquel}}, \citenamefont {{Mohr}}, \citenamefont {{Morgan}}, \citenamefont
  {{Palmese}}, \citenamefont {{Paz-Chinch{\'o}n}}, \citenamefont {{Pieres}},
  \citenamefont {{Plazas Malag{\'o}n}}, \citenamefont {{Roodman}},
  \citenamefont {{Scarpine}}, \citenamefont {{Serrano}}, \citenamefont
  {{Smith}}, \citenamefont {{Soares-Santos}}, \citenamefont {{Suchyta}},
  \citenamefont {{Tarle}}, \citenamefont {{Thomas}}, \citenamefont {{Varga}},\
  and\ \citenamefont {{DES Collaboration}}}]{RodriguezMonroy2022clustering}%
  \BibitemOpen
  \bibfield  {author} {\bibinfo {author} {\bibfnamefont {M.}~\bibnamefont
  {{Rodr{\'\i}guez-Monroy}}}, \bibinfo {author} {\bibfnamefont
  {N.}~\bibnamefont {{Weaverdyck}}}, \bibinfo {author} {\bibfnamefont
  {J.}~\bibnamefont {{Elvin-Poole}}}, \bibinfo {author} {\bibfnamefont
  {M.}~\bibnamefont {{Crocce}}}, \bibinfo {author} {\bibfnamefont
  {A.}~\bibnamefont {{Carnero Rosell}}}, \bibinfo {author} {\bibfnamefont
  {F.}~\bibnamefont {{Andrade-Oliveira}}}, \bibinfo {author} {\bibfnamefont
  {S.}~\bibnamefont {{Avila}}}, \bibinfo {author} {\bibfnamefont
  {K.}~\bibnamefont {{Bechtol}}}, \bibinfo {author} {\bibfnamefont {G.~M.}\
  \bibnamefont {{Bernstein}}}, \bibinfo {author} {\bibfnamefont
  {J.}~\bibnamefont {{Blazek}}}, \bibinfo {author} {\bibfnamefont
  {H.}~\bibnamefont {{Camacho}}}, \bibinfo {author} {\bibfnamefont
  {R.}~\bibnamefont {{Cawthon}}}, \bibinfo {author} {\bibfnamefont
  {J.}~\bibnamefont {{De Vicente}}}, \bibinfo {author} {\bibfnamefont
  {J.}~\bibnamefont {{DeRose}}}, \bibinfo {author} {\bibfnamefont
  {S.}~\bibnamefont {{Dodelson}}}, \bibinfo {author} {\bibfnamefont
  {S.}~\bibnamefont {{Everett}}}, \bibinfo {author} {\bibfnamefont
  {X.}~\bibnamefont {{Fang}}}, \bibinfo {author} {\bibfnamefont
  {I.}~\bibnamefont {{Ferrero}}}, \bibinfo {author} {\bibfnamefont
  {A.}~\bibnamefont {{Fert{\'e}}}}, \bibinfo {author} {\bibfnamefont
  {O.}~\bibnamefont {{Friedrich}}}, \bibinfo {author} {\bibfnamefont
  {E.}~\bibnamefont {{Gaztanaga}}}, \bibinfo {author} {\bibfnamefont
  {G.}~\bibnamefont {{Giannini}}}, \bibinfo {author} {\bibfnamefont {R.~A.}\
  \bibnamefont {{Gruendl}}}, \bibinfo {author} {\bibfnamefont {W.~G.}\
  \bibnamefont {{Hartley}}}, \bibinfo {author} {\bibfnamefont {K.}~\bibnamefont
  {{Herner}}}, \bibinfo {author} {\bibfnamefont {E.~M.}\ \bibnamefont
  {{Huff}}}, \bibinfo {author} {\bibfnamefont {M.}~\bibnamefont {{Jarvis}}},
  \bibinfo {author} {\bibfnamefont {E.}~\bibnamefont {{Krause}}}, \bibinfo
  {author} {\bibfnamefont {N.}~\bibnamefont {{MacCrann}}}, \bibinfo {author}
  {\bibfnamefont {J.}~\bibnamefont {{Mena-Fern{\'a}ndez}}}, \bibinfo {author}
  {\bibfnamefont {J.}~\bibnamefont {{Muir}}}, \bibinfo {author} {\bibfnamefont
  {S.}~\bibnamefont {{Pandey}}}, \bibinfo {author} {\bibfnamefont
  {Y.}~\bibnamefont {{Park}}}, \bibinfo {author} {\bibfnamefont
  {A.}~\bibnamefont {{Porredon}}}, \bibinfo {author} {\bibfnamefont
  {J.}~\bibnamefont {{Prat}}}, \bibinfo {author} {\bibfnamefont
  {R.}~\bibnamefont {{Rosenfeld}}}, \bibinfo {author} {\bibfnamefont {A.~J.}\
  \bibnamefont {{Ross}}}, \bibinfo {author} {\bibfnamefont {E.}~\bibnamefont
  {{Rozo}}}, \bibinfo {author} {\bibfnamefont {E.~S.}\ \bibnamefont
  {{Rykoff}}}, \bibinfo {author} {\bibfnamefont {E.}~\bibnamefont {{Sanchez}}},
  \bibinfo {author} {\bibfnamefont {D.}~\bibnamefont {{Sanchez Cid}}}, \bibinfo
  {author} {\bibfnamefont {I.}~\bibnamefont {{Sevilla-Noarbe}}}, \bibinfo
  {author} {\bibfnamefont {M.}~\bibnamefont {{Tabbutt}}}, \bibinfo {author}
  {\bibfnamefont {C.}~\bibnamefont {{To}}}, \bibinfo {author} {\bibfnamefont
  {E.~L.}\ \bibnamefont {{Wagoner}}}, \bibinfo {author} {\bibfnamefont {R.~H.}\
  \bibnamefont {{Wechsler}}}, \bibinfo {author} {\bibfnamefont
  {M.}~\bibnamefont {{Aguena}}}, \bibinfo {author} {\bibfnamefont
  {S.}~\bibnamefont {{Allam}}}, \bibinfo {author} {\bibfnamefont
  {A.}~\bibnamefont {{Amon}}}, \bibinfo {author} {\bibfnamefont
  {J.}~\bibnamefont {{Annis}}}, \bibinfo {author} {\bibfnamefont
  {D.}~\bibnamefont {{Bacon}}}, \bibinfo {author} {\bibfnamefont
  {E.}~\bibnamefont {{Baxter}}}, \bibinfo {author} {\bibfnamefont
  {E.}~\bibnamefont {{Bertin}}}, \bibinfo {author} {\bibfnamefont
  {S.}~\bibnamefont {{Bhargava}}}, \bibinfo {author} {\bibfnamefont
  {D.}~\bibnamefont {{Brooks}}}, \bibinfo {author} {\bibfnamefont {D.~L.}\
  \bibnamefont {{Burke}}}, \bibinfo {author} {\bibfnamefont {M.}~\bibnamefont
  {{Carrasco Kind}}}, \bibinfo {author} {\bibfnamefont {J.}~\bibnamefont
  {{Carretero}}}, \bibinfo {author} {\bibfnamefont {F.~J.}\ \bibnamefont
  {{Castander}}}, \bibinfo {author} {\bibfnamefont {A.}~\bibnamefont {{Choi}}},
  \bibinfo {author} {\bibfnamefont {C.}~\bibnamefont {{Conselice}}}, \bibinfo
  {author} {\bibfnamefont {M.}~\bibnamefont {{Costanzi}}}, \bibinfo {author}
  {\bibfnamefont {L.~N.}\ \bibnamefont {{da Costa}}}, \bibinfo {author}
  {\bibfnamefont {M.~E.~S.}\ \bibnamefont {{Pereira}}}, \bibinfo {author}
  {\bibfnamefont {S.}~\bibnamefont {{Desai}}}, \bibinfo {author} {\bibfnamefont
  {H.~T.}\ \bibnamefont {{Diehl}}}, \bibinfo {author} {\bibfnamefont
  {B.}~\bibnamefont {{Flaugher}}}, \bibinfo {author} {\bibfnamefont
  {P.}~\bibnamefont {{Fosalba}}}, \bibinfo {author} {\bibfnamefont
  {J.}~\bibnamefont {{Frieman}}}, \bibinfo {author} {\bibfnamefont
  {J.}~\bibnamefont {{Garc{\'\i}a-Bellido}}}, \bibinfo {author} {\bibfnamefont
  {T.}~\bibnamefont {{Giannantonio}}}, \bibinfo {author} {\bibfnamefont
  {D.}~\bibnamefont {{Gruen}}}, \bibinfo {author} {\bibfnamefont
  {J.}~\bibnamefont {{Gschwend}}}, \bibinfo {author} {\bibfnamefont
  {G.}~\bibnamefont {{Gutierrez}}}, \bibinfo {author} {\bibfnamefont {S.~R.}\
  \bibnamefont {{Hinton}}}, \bibinfo {author} {\bibfnamefont {D.~L.}\
  \bibnamefont {{Hollowood}}}, \bibinfo {author} {\bibfnamefont
  {K.}~\bibnamefont {{Honscheid}}}, \bibinfo {author} {\bibfnamefont
  {D.}~\bibnamefont {{Huterer}}}, \bibinfo {author} {\bibfnamefont
  {B.}~\bibnamefont {{Jain}}}, \bibinfo {author} {\bibfnamefont {D.~J.}\
  \bibnamefont {{James}}}, \bibinfo {author} {\bibfnamefont {K.}~\bibnamefont
  {{Kuehn}}}, \bibinfo {author} {\bibfnamefont {N.}~\bibnamefont
  {{Kuropatkin}}}, \bibinfo {author} {\bibfnamefont {M.}~\bibnamefont
  {{Lima}}}, \bibinfo {author} {\bibfnamefont {M.~A.~G.}\ \bibnamefont
  {{Maia}}}, \bibinfo {author} {\bibfnamefont {M.}~\bibnamefont {{March}}},
  \bibinfo {author} {\bibfnamefont {J.~L.}\ \bibnamefont {{Marshall}}},
  \bibinfo {author} {\bibfnamefont {P.}~\bibnamefont {{Melchior}}}, \bibinfo
  {author} {\bibfnamefont {F.}~\bibnamefont {{Menanteau}}}, \bibinfo {author}
  {\bibfnamefont {C.~J.}\ \bibnamefont {{Miller}}}, \bibinfo {author}
  {\bibfnamefont {R.}~\bibnamefont {{Miquel}}}, \bibinfo {author}
  {\bibfnamefont {J.~J.}\ \bibnamefont {{Mohr}}}, \bibinfo {author}
  {\bibfnamefont {R.}~\bibnamefont {{Morgan}}}, \bibinfo {author}
  {\bibfnamefont {A.}~\bibnamefont {{Palmese}}}, \bibinfo {author}
  {\bibfnamefont {F.}~\bibnamefont {{Paz-Chinch{\'o}n}}}, \bibinfo {author}
  {\bibfnamefont {A.}~\bibnamefont {{Pieres}}}, \bibinfo {author}
  {\bibfnamefont {A.~A.}\ \bibnamefont {{Plazas Malag{\'o}n}}}, \bibinfo
  {author} {\bibfnamefont {A.}~\bibnamefont {{Roodman}}}, \bibinfo {author}
  {\bibfnamefont {V.}~\bibnamefont {{Scarpine}}}, \bibinfo {author}
  {\bibfnamefont {S.}~\bibnamefont {{Serrano}}}, \bibinfo {author}
  {\bibfnamefont {M.}~\bibnamefont {{Smith}}}, \bibinfo {author} {\bibfnamefont
  {M.}~\bibnamefont {{Soares-Santos}}}, \bibinfo {author} {\bibfnamefont
  {E.}~\bibnamefont {{Suchyta}}}, \bibinfo {author} {\bibfnamefont
  {G.}~\bibnamefont {{Tarle}}}, \bibinfo {author} {\bibfnamefont
  {D.}~\bibnamefont {{Thomas}}}, \bibinfo {author} {\bibfnamefont {T.~N.}\
  \bibnamefont {{Varga}}},\ and\ \bibinfo {author} {\bibnamefont {{DES
  Collaboration}}},\ }\bibfield  {title} {\bibinfo {title} {{Dark Energy Survey
  Year 3 results: galaxy clustering and systematics treatment for lens galaxy
  samples}},\ }\href {https://doi.org/10.1093/mnras/stac104} {\bibfield
  {journal} {\bibinfo  {journal} {\mnras}\ }\textbf {\bibinfo {volume} {511}},\
  \bibinfo {pages} {2665} (\bibinfo {year} {2022})},\ \Eprint
  {https://arxiv.org/abs/2105.13540} {arXiv:2105.13540 [astro-ph.CO]}
  \BibitemShut {NoStop}%
\bibitem [{\citenamefont {{Porredon}}\ \emph
  {et~al.}(2021{\natexlab{b}})\citenamefont {{Porredon}}, \citenamefont
  {{Crocce}}, \citenamefont {{Fosalba}}, \citenamefont {{Elvin-Poole}},
  \citenamefont {{Carnero Rosell}}, \citenamefont {{Cawthon}}, \citenamefont
  {{Eifler}}, \citenamefont {{Fang}}, \citenamefont {{Ferrero}}, \citenamefont
  {{Krause}}, \citenamefont {{MacCrann}}, \citenamefont {{Weaverdyck}},
  \citenamefont {{Abbott}}, \citenamefont {{Aguena}}, \citenamefont {{Allam}},
  \citenamefont {{Amon}}, \citenamefont {{Avila}}, \citenamefont {{Bacon}},
  \citenamefont {{Bertin}}, \citenamefont {{Bhargava}}, \citenamefont
  {{Bridle}}, \citenamefont {{Brooks}}, \citenamefont {{Carrasco Kind}},
  \citenamefont {{Carretero}}, \citenamefont {{Castander}}, \citenamefont
  {{Choi}}, \citenamefont {{Costanzi}}, \citenamefont {{da Costa}},
  \citenamefont {{Pereira}}, \citenamefont {{De Vicente}}, \citenamefont
  {{Desai}}, \citenamefont {{Diehl}}, \citenamefont {{Doel}}, \citenamefont
  {{Drlica-Wagner}}, \citenamefont {{Eckert}}, \citenamefont {{Fert{\'e}}},
  \citenamefont {{Flaugher}}, \citenamefont {{Frieman}}, \citenamefont
  {{Garc{\'\i}a-Bellido}}, \citenamefont {{Gaztanaga}}, \citenamefont
  {{Gerdes}}, \citenamefont {{Giannantonio}}, \citenamefont {{Gruen}},
  \citenamefont {{Gruendl}}, \citenamefont {{Gschwend}}, \citenamefont
  {{Gutierrez}}, \citenamefont {{Hartley}}, \citenamefont {{Hinton}},
  \citenamefont {{Hollowood}}, \citenamefont {{Honscheid}}, \citenamefont
  {{Hoyle}}, \citenamefont {{James}}, \citenamefont {{Jarvis}}, \citenamefont
  {{Kuehn}}, \citenamefont {{Kuropatkin}}, \citenamefont {{Maia}},
  \citenamefont {{Marshall}}, \citenamefont {{Menanteau}}, \citenamefont
  {{Miquel}}, \citenamefont {{Morgan}}, \citenamefont {{Palmese}},
  \citenamefont {{Pandey}}, \citenamefont {{Paz-Chinch{\'o}n}}, \citenamefont
  {{Plazas}}, \citenamefont {{Rodriguez-Monroy}}, \citenamefont {{Roodman}},
  \citenamefont {{Samuroff}}, \citenamefont {{Sanchez}}, \citenamefont
  {{Scarpine}}, \citenamefont {{Serrano}}, \citenamefont {{Sevilla-Noarbe}},
  \citenamefont {{Smith}}, \citenamefont {{Soares-Santos}}, \citenamefont
  {{Suchyta}}, \citenamefont {{Swanson}}, \citenamefont {{Tarle}},
  \citenamefont {{To}}, \citenamefont {{Varga}}, \citenamefont {{Weller}},
  \citenamefont {{Wilkinson}},\ and\ \citenamefont {{DES
  Collaboration}}}]{Porredon2021optimizing}%
  \BibitemOpen
  \bibfield  {author} {\bibinfo {author} {\bibfnamefont {A.}~\bibnamefont
  {{Porredon}}}, \bibinfo {author} {\bibfnamefont {M.}~\bibnamefont
  {{Crocce}}}, \bibinfo {author} {\bibfnamefont {P.}~\bibnamefont {{Fosalba}}},
  \bibinfo {author} {\bibfnamefont {J.}~\bibnamefont {{Elvin-Poole}}}, \bibinfo
  {author} {\bibfnamefont {A.}~\bibnamefont {{Carnero Rosell}}}, \bibinfo
  {author} {\bibfnamefont {R.}~\bibnamefont {{Cawthon}}}, \bibinfo {author}
  {\bibfnamefont {T.~F.}\ \bibnamefont {{Eifler}}}, \bibinfo {author}
  {\bibfnamefont {X.}~\bibnamefont {{Fang}}}, \bibinfo {author} {\bibfnamefont
  {I.}~\bibnamefont {{Ferrero}}}, \bibinfo {author} {\bibfnamefont
  {E.}~\bibnamefont {{Krause}}}, \bibinfo {author} {\bibfnamefont
  {N.}~\bibnamefont {{MacCrann}}}, \bibinfo {author} {\bibfnamefont
  {N.}~\bibnamefont {{Weaverdyck}}}, \bibinfo {author} {\bibfnamefont
  {T.~M.~C.}\ \bibnamefont {{Abbott}}}, \bibinfo {author} {\bibfnamefont
  {M.}~\bibnamefont {{Aguena}}}, \bibinfo {author} {\bibfnamefont
  {S.}~\bibnamefont {{Allam}}}, \bibinfo {author} {\bibfnamefont
  {A.}~\bibnamefont {{Amon}}}, \bibinfo {author} {\bibfnamefont
  {S.}~\bibnamefont {{Avila}}}, \bibinfo {author} {\bibfnamefont
  {D.}~\bibnamefont {{Bacon}}}, \bibinfo {author} {\bibfnamefont
  {E.}~\bibnamefont {{Bertin}}}, \bibinfo {author} {\bibfnamefont
  {S.}~\bibnamefont {{Bhargava}}}, \bibinfo {author} {\bibfnamefont {S.~L.}\
  \bibnamefont {{Bridle}}}, \bibinfo {author} {\bibfnamefont {D.}~\bibnamefont
  {{Brooks}}}, \bibinfo {author} {\bibfnamefont {M.}~\bibnamefont {{Carrasco
  Kind}}}, \bibinfo {author} {\bibfnamefont {J.}~\bibnamefont {{Carretero}}},
  \bibinfo {author} {\bibfnamefont {F.~J.}\ \bibnamefont {{Castander}}},
  \bibinfo {author} {\bibfnamefont {A.}~\bibnamefont {{Choi}}}, \bibinfo
  {author} {\bibfnamefont {M.}~\bibnamefont {{Costanzi}}}, \bibinfo {author}
  {\bibfnamefont {L.~N.}\ \bibnamefont {{da Costa}}}, \bibinfo {author}
  {\bibfnamefont {M.~E.~S.}\ \bibnamefont {{Pereira}}}, \bibinfo {author}
  {\bibfnamefont {J.}~\bibnamefont {{De Vicente}}}, \bibinfo {author}
  {\bibfnamefont {S.}~\bibnamefont {{Desai}}}, \bibinfo {author} {\bibfnamefont
  {H.~T.}\ \bibnamefont {{Diehl}}}, \bibinfo {author} {\bibfnamefont
  {P.}~\bibnamefont {{Doel}}}, \bibinfo {author} {\bibfnamefont
  {A.}~\bibnamefont {{Drlica-Wagner}}}, \bibinfo {author} {\bibfnamefont
  {K.}~\bibnamefont {{Eckert}}}, \bibinfo {author} {\bibfnamefont
  {A.}~\bibnamefont {{Fert{\'e}}}}, \bibinfo {author} {\bibfnamefont
  {B.}~\bibnamefont {{Flaugher}}}, \bibinfo {author} {\bibfnamefont
  {J.}~\bibnamefont {{Frieman}}}, \bibinfo {author} {\bibfnamefont
  {J.}~\bibnamefont {{Garc{\'\i}a-Bellido}}}, \bibinfo {author} {\bibfnamefont
  {E.}~\bibnamefont {{Gaztanaga}}}, \bibinfo {author} {\bibfnamefont {D.~W.}\
  \bibnamefont {{Gerdes}}}, \bibinfo {author} {\bibfnamefont {T.}~\bibnamefont
  {{Giannantonio}}}, \bibinfo {author} {\bibfnamefont {D.}~\bibnamefont
  {{Gruen}}}, \bibinfo {author} {\bibfnamefont {R.~A.}\ \bibnamefont
  {{Gruendl}}}, \bibinfo {author} {\bibfnamefont {J.}~\bibnamefont
  {{Gschwend}}}, \bibinfo {author} {\bibfnamefont {G.}~\bibnamefont
  {{Gutierrez}}}, \bibinfo {author} {\bibfnamefont {W.~G.}\ \bibnamefont
  {{Hartley}}}, \bibinfo {author} {\bibfnamefont {S.~R.}\ \bibnamefont
  {{Hinton}}}, \bibinfo {author} {\bibfnamefont {D.~L.}\ \bibnamefont
  {{Hollowood}}}, \bibinfo {author} {\bibfnamefont {K.}~\bibnamefont
  {{Honscheid}}}, \bibinfo {author} {\bibfnamefont {B.}~\bibnamefont
  {{Hoyle}}}, \bibinfo {author} {\bibfnamefont {D.~J.}\ \bibnamefont
  {{James}}}, \bibinfo {author} {\bibfnamefont {M.}~\bibnamefont {{Jarvis}}},
  \bibinfo {author} {\bibfnamefont {K.}~\bibnamefont {{Kuehn}}}, \bibinfo
  {author} {\bibfnamefont {N.}~\bibnamefont {{Kuropatkin}}}, \bibinfo {author}
  {\bibfnamefont {M.~A.~G.}\ \bibnamefont {{Maia}}}, \bibinfo {author}
  {\bibfnamefont {J.~L.}\ \bibnamefont {{Marshall}}}, \bibinfo {author}
  {\bibfnamefont {F.}~\bibnamefont {{Menanteau}}}, \bibinfo {author}
  {\bibfnamefont {R.}~\bibnamefont {{Miquel}}}, \bibinfo {author}
  {\bibfnamefont {R.}~\bibnamefont {{Morgan}}}, \bibinfo {author}
  {\bibfnamefont {A.}~\bibnamefont {{Palmese}}}, \bibinfo {author}
  {\bibfnamefont {S.}~\bibnamefont {{Pandey}}}, \bibinfo {author}
  {\bibfnamefont {F.}~\bibnamefont {{Paz-Chinch{\'o}n}}}, \bibinfo {author}
  {\bibfnamefont {A.~A.}\ \bibnamefont {{Plazas}}}, \bibinfo {author}
  {\bibfnamefont {M.}~\bibnamefont {{Rodriguez-Monroy}}}, \bibinfo {author}
  {\bibfnamefont {A.}~\bibnamefont {{Roodman}}}, \bibinfo {author}
  {\bibfnamefont {S.}~\bibnamefont {{Samuroff}}}, \bibinfo {author}
  {\bibfnamefont {E.}~\bibnamefont {{Sanchez}}}, \bibinfo {author}
  {\bibfnamefont {V.}~\bibnamefont {{Scarpine}}}, \bibinfo {author}
  {\bibfnamefont {S.}~\bibnamefont {{Serrano}}}, \bibinfo {author}
  {\bibfnamefont {I.}~\bibnamefont {{Sevilla-Noarbe}}}, \bibinfo {author}
  {\bibfnamefont {M.}~\bibnamefont {{Smith}}}, \bibinfo {author} {\bibfnamefont
  {M.}~\bibnamefont {{Soares-Santos}}}, \bibinfo {author} {\bibfnamefont
  {E.}~\bibnamefont {{Suchyta}}}, \bibinfo {author} {\bibfnamefont {M.~E.~C.}\
  \bibnamefont {{Swanson}}}, \bibinfo {author} {\bibfnamefont {G.}~\bibnamefont
  {{Tarle}}}, \bibinfo {author} {\bibfnamefont {C.}~\bibnamefont {{To}}},
  \bibinfo {author} {\bibfnamefont {T.~N.}\ \bibnamefont {{Varga}}}, \bibinfo
  {author} {\bibfnamefont {J.}~\bibnamefont {{Weller}}}, \bibinfo {author}
  {\bibfnamefont {R.~D.}\ \bibnamefont {{Wilkinson}}},\ and\ \bibinfo {author}
  {\bibnamefont {{DES Collaboration}}},\ }\bibfield  {title} {\bibinfo {title}
  {{Dark Energy Survey Year 3 results: Optimizing the lens sample in a combined
  galaxy clustering and galaxy-galaxy lensing analysis}},\ }\href
  {https://doi.org/10.1103/PhysRevD.103.043503} {\bibfield  {journal} {\bibinfo
   {journal} {\prd}\ }\textbf {\bibinfo {volume} {103}},\ \bibinfo {eid}
  {043503} (\bibinfo {year} {2021}{\natexlab{b}})},\ \Eprint
  {https://arxiv.org/abs/2011.03411} {arXiv:2011.03411 [astro-ph.CO]}
  \BibitemShut {NoStop}%
\bibitem [{\citenamefont {{Potter}}\ \emph {et~al.}(2017)\citenamefont
  {{Potter}}, \citenamefont {{Stadel}},\ and\ \citenamefont
  {{Teyssier}}}]{potter2016pkdgrav3}%
  \BibitemOpen
  \bibfield  {author} {\bibinfo {author} {\bibfnamefont {D.}~\bibnamefont
  {{Potter}}}, \bibinfo {author} {\bibfnamefont {J.}~\bibnamefont {{Stadel}}},\
  and\ \bibinfo {author} {\bibfnamefont {R.}~\bibnamefont {{Teyssier}}},\
  }\bibfield  {title} {\bibinfo {title} {{PKDGRAV3: beyond trillion particle
  cosmological simulations for the next era of galaxy surveys}},\ }\href
  {https://doi.org/10.1186/s40668-017-0021-1} {\bibfield  {journal} {\bibinfo
  {journal} {Computational Astrophysics and Cosmology}\ }\textbf {\bibinfo
  {volume} {4}},\ \bibinfo {eid} {2} (\bibinfo {year} {2017})},\ \Eprint
  {https://arxiv.org/abs/1609.08621} {arXiv:1609.08621 [astro-ph.IM]}
  \BibitemShut {NoStop}%
\bibitem [{\citenamefont {Hahn}\ and\ \citenamefont {Abel}(2011)}]{MUSIC2011}%
  \BibitemOpen
  \bibfield  {author} {\bibinfo {author} {\bibfnamefont {O.}~\bibnamefont
  {Hahn}}\ and\ \bibinfo {author} {\bibfnamefont {T.}~\bibnamefont {Abel}},\
  }\bibfield  {title} {\bibinfo {title} {{Multi-scale initial conditions for
  cosmological simulations}},\ }\href
  {https://doi.org/10.1111/j.1365-2966.2011.18820.x} {\bibfield  {journal}
  {\bibinfo  {journal} {Monthly Notices of the Royal Astronomical Society}\
  }\textbf {\bibinfo {volume} {415}},\ \bibinfo {pages} {2101} (\bibinfo {year}
  {2011})},\ \Eprint
  {https://arxiv.org/abs/http://oup.prod.sis.lan/mnras/article-pdf/415/3/2101/5967972/mnras0415-2101.pdf}
  {http://oup.prod.sis.lan/mnras/article-pdf/415/3/2101/5967972/mnras0415-2101.pdf}
  \BibitemShut {NoStop}%
\bibitem [{\citenamefont {Sgier}\ \emph {et~al.}(2019)\citenamefont {Sgier},
  \citenamefont {R{\'e}fr{\'e}gier}, \citenamefont {Amara},\ and\ \citenamefont
  {Nicola}}]{Sgier2018}%
  \BibitemOpen
  \bibfield  {author} {\bibinfo {author} {\bibfnamefont {R.}~\bibnamefont
  {Sgier}}, \bibinfo {author} {\bibfnamefont {A.}~\bibnamefont
  {R{\'e}fr{\'e}gier}}, \bibinfo {author} {\bibfnamefont {A.}~\bibnamefont
  {Amara}},\ and\ \bibinfo {author} {\bibfnamefont {A.}~\bibnamefont
  {Nicola}},\ }\bibfield  {title} {\bibinfo {title} {Fast generation of
  covariance matrices for weak lensing},\ }\href@noop {} {\bibfield  {journal}
  {\bibinfo  {journal} {Journal of Cosmology and Astroparticle Physics}\
  }\textbf {\bibinfo {volume} {2019}}\bibinfo  {number} { (01)},\ \bibinfo
  {pages} {044}}\BibitemShut {NoStop}%
\bibitem [{\citenamefont {{Z{\"u}rcher}}\ \emph {et~al.}(2021)\citenamefont
  {{Z{\"u}rcher}}, \citenamefont {{Fluri}}, \citenamefont {{Sgier}},
  \citenamefont {{Kacprzak}},\ and\ \citenamefont
  {{Refregier}}}]{zuercher2020forecast}%
  \BibitemOpen
\bibfield  {number} {  }\bibfield  {author} {\bibinfo {author} {\bibfnamefont
  {D.}~\bibnamefont {{Z{\"u}rcher}}}, \bibinfo {author} {\bibfnamefont
  {J.}~\bibnamefont {{Fluri}}}, \bibinfo {author} {\bibfnamefont
  {R.}~\bibnamefont {{Sgier}}}, \bibinfo {author} {\bibfnamefont
  {T.}~\bibnamefont {{Kacprzak}}},\ and\ \bibinfo {author} {\bibfnamefont
  {A.}~\bibnamefont {{Refregier}}},\ }\bibfield  {title} {\bibinfo {title}
  {{Cosmological forecast for non-Gaussian statistics in large-scale weak
  lensing surveys}},\ }\href {https://doi.org/10.1088/1475-7516/2021/01/028}
  {\bibfield  {journal} {\bibinfo  {journal} {\jcap}\ }\textbf {\bibinfo
  {volume} {2021}},\ \bibinfo {eid} {028} (\bibinfo {year} {2021})}\BibitemShut
  {NoStop}%
\bibitem [{\citenamefont {{Friedrich}}\ \emph {et~al.}(2018)\citenamefont
  {{Friedrich}}, \citenamefont {{Gruen}}, \citenamefont {{DeRose}},
  \citenamefont {{Kirk}}, \citenamefont {{Krause}}, \citenamefont
  {{McClintock}}, \citenamefont {{Rykoff}}, \citenamefont {{Seitz}},
  \citenamefont {{Wechsler}}, \citenamefont {{Bernstein}}, \citenamefont
  {{Blazek}}, \citenamefont {{Chang}}, \citenamefont {{Hilbert}}, \citenamefont
  {{Jain}}, \citenamefont {{Kovacs}}, \citenamefont {{Lahav}}, \citenamefont
  {{Abdalla}}, \citenamefont {{Allam}}, \citenamefont {{Annis}}, \citenamefont
  {{Bechtol}}, \citenamefont {{Benoit-L{\'e}vy}}, \citenamefont {{Bertin}},
  \citenamefont {{Brooks}}, \citenamefont {{Carnero Rosell}}, \citenamefont
  {{Carrasco Kind}}, \citenamefont {{Carretero}}, \citenamefont {{Cunha}},
  \citenamefont {{D'Andrea}}, \citenamefont {{da Costa}}, \citenamefont
  {{Davis}}, \citenamefont {{Desai}}, \citenamefont {{Diehl}}, \citenamefont
  {{Dietrich}}, \citenamefont {{Drlica-Wagner}}, \citenamefont {{Eifler}},
  \citenamefont {{Fosalba}}, \citenamefont {{Frieman}}, \citenamefont
  {{Garc{\'\i}a-Bellido}}, \citenamefont {{Gaztanaga}}, \citenamefont
  {{Gerdes}}, \citenamefont {{Giannantonio}}, \citenamefont {{Gruendl}},
  \citenamefont {{Gschwend}}, \citenamefont {{Gutierrez}}, \citenamefont
  {{Honscheid}}, \citenamefont {{James}}, \citenamefont {{Jarvis}},
  \citenamefont {{Kuehn}}, \citenamefont {{Kuropatkin}}, \citenamefont
  {{Lima}}, \citenamefont {{March}}, \citenamefont {{Marshall}}, \citenamefont
  {{Melchior}}, \citenamefont {{Menanteau}}, \citenamefont {{Miquel}},
  \citenamefont {{Mohr}}, \citenamefont {{Nord}}, \citenamefont {{Plazas}},
  \citenamefont {{Sanchez}}, \citenamefont {{Scarpine}}, \citenamefont
  {{Schindler}}, \citenamefont {{Schubnell}}, \citenamefont {{Sevilla-Noarbe}},
  \citenamefont {{Sheldon}}, \citenamefont {{Smith}}, \citenamefont
  {{Soares-Santos}}, \citenamefont {{Sobreira}}, \citenamefont {{Suchyta}},
  \citenamefont {{Swanson}}, \citenamefont {{Tarle}}, \citenamefont {{Thomas}},
  \citenamefont {{Troxel}}, \citenamefont {{Vikram}}, \citenamefont
  {{Weller}},\ and\ \citenamefont {{DES
  Collaboration}}}]{Friedrich2018density}%
  \BibitemOpen
  \bibfield  {author} {\bibinfo {author} {\bibfnamefont {O.}~\bibnamefont
  {{Friedrich}}}, \bibinfo {author} {\bibfnamefont {D.}~\bibnamefont
  {{Gruen}}}, \bibinfo {author} {\bibfnamefont {J.}~\bibnamefont {{DeRose}}},
  \bibinfo {author} {\bibfnamefont {D.}~\bibnamefont {{Kirk}}}, \bibinfo
  {author} {\bibfnamefont {E.}~\bibnamefont {{Krause}}}, \bibinfo {author}
  {\bibfnamefont {T.}~\bibnamefont {{McClintock}}}, \bibinfo {author}
  {\bibfnamefont {E.~S.}\ \bibnamefont {{Rykoff}}}, \bibinfo {author}
  {\bibfnamefont {S.}~\bibnamefont {{Seitz}}}, \bibinfo {author} {\bibfnamefont
  {R.~H.}\ \bibnamefont {{Wechsler}}}, \bibinfo {author} {\bibfnamefont
  {G.~M.}\ \bibnamefont {{Bernstein}}}, \bibinfo {author} {\bibfnamefont
  {J.}~\bibnamefont {{Blazek}}}, \bibinfo {author} {\bibfnamefont
  {C.}~\bibnamefont {{Chang}}}, \bibinfo {author} {\bibfnamefont
  {S.}~\bibnamefont {{Hilbert}}}, \bibinfo {author} {\bibfnamefont
  {B.}~\bibnamefont {{Jain}}}, \bibinfo {author} {\bibfnamefont
  {A.}~\bibnamefont {{Kovacs}}}, \bibinfo {author} {\bibfnamefont
  {O.}~\bibnamefont {{Lahav}}}, \bibinfo {author} {\bibfnamefont {F.~B.}\
  \bibnamefont {{Abdalla}}}, \bibinfo {author} {\bibfnamefont {S.}~\bibnamefont
  {{Allam}}}, \bibinfo {author} {\bibfnamefont {J.}~\bibnamefont {{Annis}}},
  \bibinfo {author} {\bibfnamefont {K.}~\bibnamefont {{Bechtol}}}, \bibinfo
  {author} {\bibfnamefont {A.}~\bibnamefont {{Benoit-L{\'e}vy}}}, \bibinfo
  {author} {\bibfnamefont {E.}~\bibnamefont {{Bertin}}}, \bibinfo {author}
  {\bibfnamefont {D.}~\bibnamefont {{Brooks}}}, \bibinfo {author}
  {\bibfnamefont {A.}~\bibnamefont {{Carnero Rosell}}}, \bibinfo {author}
  {\bibfnamefont {M.}~\bibnamefont {{Carrasco Kind}}}, \bibinfo {author}
  {\bibfnamefont {J.}~\bibnamefont {{Carretero}}}, \bibinfo {author}
  {\bibfnamefont {C.~E.}\ \bibnamefont {{Cunha}}}, \bibinfo {author}
  {\bibfnamefont {C.~B.}\ \bibnamefont {{D'Andrea}}}, \bibinfo {author}
  {\bibfnamefont {L.~N.}\ \bibnamefont {{da Costa}}}, \bibinfo {author}
  {\bibfnamefont {C.}~\bibnamefont {{Davis}}}, \bibinfo {author} {\bibfnamefont
  {S.}~\bibnamefont {{Desai}}}, \bibinfo {author} {\bibfnamefont {H.~T.}\
  \bibnamefont {{Diehl}}}, \bibinfo {author} {\bibfnamefont {J.~P.}\
  \bibnamefont {{Dietrich}}}, \bibinfo {author} {\bibfnamefont
  {A.}~\bibnamefont {{Drlica-Wagner}}}, \bibinfo {author} {\bibfnamefont
  {T.~F.}\ \bibnamefont {{Eifler}}}, \bibinfo {author} {\bibfnamefont
  {P.}~\bibnamefont {{Fosalba}}}, \bibinfo {author} {\bibfnamefont
  {J.}~\bibnamefont {{Frieman}}}, \bibinfo {author} {\bibfnamefont
  {J.}~\bibnamefont {{Garc{\'\i}a-Bellido}}}, \bibinfo {author} {\bibfnamefont
  {E.}~\bibnamefont {{Gaztanaga}}}, \bibinfo {author} {\bibfnamefont {D.~W.}\
  \bibnamefont {{Gerdes}}}, \bibinfo {author} {\bibfnamefont {T.}~\bibnamefont
  {{Giannantonio}}}, \bibinfo {author} {\bibfnamefont {R.~A.}\ \bibnamefont
  {{Gruendl}}}, \bibinfo {author} {\bibfnamefont {J.}~\bibnamefont
  {{Gschwend}}}, \bibinfo {author} {\bibfnamefont {G.}~\bibnamefont
  {{Gutierrez}}}, \bibinfo {author} {\bibfnamefont {K.}~\bibnamefont
  {{Honscheid}}}, \bibinfo {author} {\bibfnamefont {D.~J.}\ \bibnamefont
  {{James}}}, \bibinfo {author} {\bibfnamefont {M.}~\bibnamefont {{Jarvis}}},
  \bibinfo {author} {\bibfnamefont {K.}~\bibnamefont {{Kuehn}}}, \bibinfo
  {author} {\bibfnamefont {N.}~\bibnamefont {{Kuropatkin}}}, \bibinfo {author}
  {\bibfnamefont {M.}~\bibnamefont {{Lima}}}, \bibinfo {author} {\bibfnamefont
  {M.}~\bibnamefont {{March}}}, \bibinfo {author} {\bibfnamefont {J.~L.}\
  \bibnamefont {{Marshall}}}, \bibinfo {author} {\bibfnamefont
  {P.}~\bibnamefont {{Melchior}}}, \bibinfo {author} {\bibfnamefont
  {F.}~\bibnamefont {{Menanteau}}}, \bibinfo {author} {\bibfnamefont
  {R.}~\bibnamefont {{Miquel}}}, \bibinfo {author} {\bibfnamefont {J.~J.}\
  \bibnamefont {{Mohr}}}, \bibinfo {author} {\bibfnamefont {B.}~\bibnamefont
  {{Nord}}}, \bibinfo {author} {\bibfnamefont {A.~A.}\ \bibnamefont
  {{Plazas}}}, \bibinfo {author} {\bibfnamefont {E.}~\bibnamefont {{Sanchez}}},
  \bibinfo {author} {\bibfnamefont {V.}~\bibnamefont {{Scarpine}}}, \bibinfo
  {author} {\bibfnamefont {R.}~\bibnamefont {{Schindler}}}, \bibinfo {author}
  {\bibfnamefont {M.}~\bibnamefont {{Schubnell}}}, \bibinfo {author}
  {\bibfnamefont {I.}~\bibnamefont {{Sevilla-Noarbe}}}, \bibinfo {author}
  {\bibfnamefont {E.}~\bibnamefont {{Sheldon}}}, \bibinfo {author}
  {\bibfnamefont {M.}~\bibnamefont {{Smith}}}, \bibinfo {author} {\bibfnamefont
  {M.}~\bibnamefont {{Soares-Santos}}}, \bibinfo {author} {\bibfnamefont
  {F.}~\bibnamefont {{Sobreira}}}, \bibinfo {author} {\bibfnamefont
  {E.}~\bibnamefont {{Suchyta}}}, \bibinfo {author} {\bibfnamefont {M.~E.~C.}\
  \bibnamefont {{Swanson}}}, \bibinfo {author} {\bibfnamefont {G.}~\bibnamefont
  {{Tarle}}}, \bibinfo {author} {\bibfnamefont {D.}~\bibnamefont {{Thomas}}},
  \bibinfo {author} {\bibfnamefont {M.~A.}\ \bibnamefont {{Troxel}}}, \bibinfo
  {author} {\bibfnamefont {V.}~\bibnamefont {{Vikram}}}, \bibinfo {author}
  {\bibfnamefont {J.}~\bibnamefont {{Weller}}},\ and\ \bibinfo {author}
  {\bibnamefont {{DES Collaboration}}},\ }\bibfield  {title} {\bibinfo {title}
  {{Density split statistics: Joint model of counts and lensing in cells}},\
  }\href {https://doi.org/10.1103/PhysRevD.98.023508} {\bibfield  {journal}
  {\bibinfo  {journal} {\prd}\ }\textbf {\bibinfo {volume} {98}},\ \bibinfo
  {eid} {023508} (\bibinfo {year} {2018})}\BibitemShut {NoStop}%
\bibitem [{\citenamefont {Makitalo}\ and\ \citenamefont
  {Foi}(2011)}]{Makitalo2011anscombe}%
  \BibitemOpen
  \bibfield  {author} {\bibinfo {author} {\bibfnamefont {M.}~\bibnamefont
  {Makitalo}}\ and\ \bibinfo {author} {\bibfnamefont {A.}~\bibnamefont {Foi}},\
  }\bibfield  {title} {\bibinfo {title} {A closed-form approximation of the
  exact unbiased inverse of the anscombe variance-stabilizing transformation},\
  }\href {https://doi.org/10.1109/TIP.2011.2121085} {\bibfield  {journal}
  {\bibinfo  {journal} {IEEE Transactions on Image Processing}\ }\textbf
  {\bibinfo {volume} {20}},\ \bibinfo {pages} {2697} (\bibinfo {year}
  {2011})}\BibitemShut {NoStop}%
\bibitem [{\citenamefont {{Schneider}}\ \emph {et~al.}(2020)\citenamefont
  {{Schneider}}, \citenamefont {{Stoira}}, \citenamefont {{Refregier}},
  \citenamefont {{Weiss}}, \citenamefont {{Knabenhans}}, \citenamefont
  {{Stadel}},\ and\ \citenamefont {{Teyssier}}}]{Schneider2020baryons1}%
  \BibitemOpen
  \bibfield  {author} {\bibinfo {author} {\bibfnamefont {A.}~\bibnamefont
  {{Schneider}}}, \bibinfo {author} {\bibfnamefont {N.}~\bibnamefont
  {{Stoira}}}, \bibinfo {author} {\bibfnamefont {A.}~\bibnamefont
  {{Refregier}}}, \bibinfo {author} {\bibfnamefont {A.~J.}\ \bibnamefont
  {{Weiss}}}, \bibinfo {author} {\bibfnamefont {M.}~\bibnamefont
  {{Knabenhans}}}, \bibinfo {author} {\bibfnamefont {J.}~\bibnamefont
  {{Stadel}}},\ and\ \bibinfo {author} {\bibfnamefont {R.}~\bibnamefont
  {{Teyssier}}},\ }\bibfield  {title} {\bibinfo {title} {{Baryonic effects for
  weak lensing. Part I. Power spectrum and covariance matrix}},\ }\href
  {https://doi.org/10.1088/1475-7516/2020/04/019} {\bibfield  {journal}
  {\bibinfo  {journal} {\jcap}\ }\textbf {\bibinfo {volume} {2020}},\ \bibinfo
  {eid} {019} (\bibinfo {year} {2020})}\BibitemShut {NoStop}%
\bibitem [{\citenamefont {{Weiss}}\ \emph {et~al.}(2019)\citenamefont
  {{Weiss}}, \citenamefont {{Schneider}}, \citenamefont {{Sgier}},
  \citenamefont {{Kacprzak}}, \citenamefont {{Amara}},\ and\ \citenamefont
  {{Refregier}}}]{Weiss2019peakbaryons}%
  \BibitemOpen
  \bibfield  {author} {\bibinfo {author} {\bibfnamefont {A.~J.}\ \bibnamefont
  {{Weiss}}}, \bibinfo {author} {\bibfnamefont {A.}~\bibnamefont
  {{Schneider}}}, \bibinfo {author} {\bibfnamefont {R.}~\bibnamefont
  {{Sgier}}}, \bibinfo {author} {\bibfnamefont {T.}~\bibnamefont {{Kacprzak}}},
  \bibinfo {author} {\bibfnamefont {A.}~\bibnamefont {{Amara}}},\ and\ \bibinfo
  {author} {\bibfnamefont {A.}~\bibnamefont {{Refregier}}},\ }\href@noop {}
  {\bibfield  {journal} {\bibinfo  {journal} {\jcap}\ }\textbf {\bibinfo
  {volume} {2019}},\ \bibinfo {eid} {011} (\bibinfo {year} {2019})}\BibitemShut
  {NoStop}%
\bibitem [{\citenamefont {{He}}\ \emph {et~al.}(2016)\citenamefont {{He}},
  \citenamefont {{Zhang}}, \citenamefont {{Ren}},\ and\ \citenamefont
  {{Sun}}}]{He2015resnet}%
  \BibitemOpen
  \bibfield  {author} {\bibinfo {author} {\bibfnamefont {K.}~\bibnamefont
  {{He}}}, \bibinfo {author} {\bibfnamefont {X.}~\bibnamefont {{Zhang}}},
  \bibinfo {author} {\bibfnamefont {S.}~\bibnamefont {{Ren}}},\ and\ \bibinfo
  {author} {\bibfnamefont {J.}~\bibnamefont {{Sun}}},\ }\bibfield  {title}
  {\bibinfo {title} {Deep residual learning for image recognition},\ }\href
  {https://doi.org/10.1109/CVPR.2016.90} {\bibfield  {journal} {\bibinfo
  {journal} {IEEE Conf. on Computer Vision and Pattern Recognition (CVPR)}\ ,\
  \bibinfo {pages} {770}} (\bibinfo {year} {2016})}\BibitemShut {NoStop}%
\bibitem [{\citenamefont {Kingma}\ and\ \citenamefont {Ba}(2014)}]{Kingma2014}%
  \BibitemOpen
  \bibfield  {author} {\bibinfo {author} {\bibfnamefont {D.~P.}\ \bibnamefont
  {Kingma}}\ and\ \bibinfo {author} {\bibfnamefont {J.~L.}\ \bibnamefont
  {Ba}},\ }\href
  {https://doi.org/http://doi.acm.org.ezproxy.lib.ucf.edu/10.1145/1830483.1830503}
  {\bibinfo {title} {{Adam: A Method for Stochastic Optimization}}} (\bibinfo
  {year} {2014})\BibitemShut {NoStop}%
\bibitem [{\citenamefont {Seetharaman}\ \emph {et~al.}(2020)\citenamefont
  {Seetharaman}, \citenamefont {Wichern}, \citenamefont {Pardo},\ and\
  \citenamefont {Le~Roux}}]{Seetharaman2020autoclip}%
  \BibitemOpen
  \bibfield  {author} {\bibinfo {author} {\bibfnamefont {P.}~\bibnamefont
  {Seetharaman}}, \bibinfo {author} {\bibfnamefont {G.}~\bibnamefont
  {Wichern}}, \bibinfo {author} {\bibfnamefont {B.}~\bibnamefont {Pardo}},\
  and\ \bibinfo {author} {\bibfnamefont {J.}~\bibnamefont {Le~Roux}},\
  }\bibfield  {title} {\bibinfo {title} {Autoclip: Adaptive gradient clipping
  for source separation networks},\ }in\ \href@noop {} {\emph {\bibinfo
  {booktitle} {2020 IEEE 30th International Workshop on Machine Learning for
  Signal Processing (MLSP)}}}\ (\bibinfo {organization} {IEEE},\ \bibinfo
  {year} {2020})\ pp.\ \bibinfo {pages} {1--6}\BibitemShut {NoStop}%
\bibitem [{\citenamefont {Abadi}\ \emph {et~al.}(2015)\citenamefont {Abadi},
  \citenamefont {Agarwal}, \citenamefont {Barham}, \citenamefont {Brevdo},
  \citenamefont {Chen}, \citenamefont {Citro}, \citenamefont {Corrado},
  \citenamefont {Davis}, \citenamefont {Dean}, \citenamefont {Devin},
  \citenamefont {Ghemawat}, \citenamefont {Goodfellow}, \citenamefont {Harp},
  \citenamefont {Irving}, \citenamefont {Isard}, \citenamefont {Jia},
  \citenamefont {Jozefowicz}, \citenamefont {Kaiser}, \citenamefont {Kudlur},
  \citenamefont {Levenberg}, \citenamefont {Man\'{e}}, \citenamefont {Monga},
  \citenamefont {Moore}, \citenamefont {Murray}, \citenamefont {Olah},
  \citenamefont {Schuster}, \citenamefont {Shlens}, \citenamefont {Steiner},
  \citenamefont {Sutskever}, \citenamefont {Talwar}, \citenamefont {Tucker},
  \citenamefont {Vanhoucke}, \citenamefont {Vasudevan}, \citenamefont
  {Vi\'{e}gas}, \citenamefont {Vinyals}, \citenamefont {Warden}, \citenamefont
  {Wattenberg}, \citenamefont {Wicke}, \citenamefont {Yu},\ and\ \citenamefont
  {Zheng}}]{tensorflow2015-whitepaper}%
  \BibitemOpen
  \bibfield  {author} {\bibinfo {author} {\bibfnamefont {M.}~\bibnamefont
  {Abadi}}, \bibinfo {author} {\bibfnamefont {A.}~\bibnamefont {Agarwal}},
  \bibinfo {author} {\bibfnamefont {P.}~\bibnamefont {Barham}}, \bibinfo
  {author} {\bibfnamefont {E.}~\bibnamefont {Brevdo}}, \bibinfo {author}
  {\bibfnamefont {Z.}~\bibnamefont {Chen}}, \bibinfo {author} {\bibfnamefont
  {C.}~\bibnamefont {Citro}}, \bibinfo {author} {\bibfnamefont {G.~S.}\
  \bibnamefont {Corrado}}, \bibinfo {author} {\bibfnamefont {A.}~\bibnamefont
  {Davis}}, \bibinfo {author} {\bibfnamefont {J.}~\bibnamefont {Dean}},
  \bibinfo {author} {\bibfnamefont {M.}~\bibnamefont {Devin}}, \bibinfo
  {author} {\bibfnamefont {S.}~\bibnamefont {Ghemawat}}, \bibinfo {author}
  {\bibfnamefont {I.}~\bibnamefont {Goodfellow}}, \bibinfo {author}
  {\bibfnamefont {A.}~\bibnamefont {Harp}}, \bibinfo {author} {\bibfnamefont
  {G.}~\bibnamefont {Irving}}, \bibinfo {author} {\bibfnamefont
  {M.}~\bibnamefont {Isard}}, \bibinfo {author} {\bibfnamefont
  {Y.}~\bibnamefont {Jia}}, \bibinfo {author} {\bibfnamefont {R.}~\bibnamefont
  {Jozefowicz}}, \bibinfo {author} {\bibfnamefont {L.}~\bibnamefont {Kaiser}},
  \bibinfo {author} {\bibfnamefont {M.}~\bibnamefont {Kudlur}}, \bibinfo
  {author} {\bibfnamefont {J.}~\bibnamefont {Levenberg}}, \bibinfo {author}
  {\bibfnamefont {D.}~\bibnamefont {Man\'{e}}}, \bibinfo {author}
  {\bibfnamefont {R.}~\bibnamefont {Monga}}, \bibinfo {author} {\bibfnamefont
  {S.}~\bibnamefont {Moore}}, \bibinfo {author} {\bibfnamefont
  {D.}~\bibnamefont {Murray}}, \bibinfo {author} {\bibfnamefont
  {C.}~\bibnamefont {Olah}}, \bibinfo {author} {\bibfnamefont {M.}~\bibnamefont
  {Schuster}}, \bibinfo {author} {\bibfnamefont {J.}~\bibnamefont {Shlens}},
  \bibinfo {author} {\bibfnamefont {B.}~\bibnamefont {Steiner}}, \bibinfo
  {author} {\bibfnamefont {I.}~\bibnamefont {Sutskever}}, \bibinfo {author}
  {\bibfnamefont {K.}~\bibnamefont {Talwar}}, \bibinfo {author} {\bibfnamefont
  {P.}~\bibnamefont {Tucker}}, \bibinfo {author} {\bibfnamefont
  {V.}~\bibnamefont {Vanhoucke}}, \bibinfo {author} {\bibfnamefont
  {V.}~\bibnamefont {Vasudevan}}, \bibinfo {author} {\bibfnamefont
  {F.}~\bibnamefont {Vi\'{e}gas}}, \bibinfo {author} {\bibfnamefont
  {O.}~\bibnamefont {Vinyals}}, \bibinfo {author} {\bibfnamefont
  {P.}~\bibnamefont {Warden}}, \bibinfo {author} {\bibfnamefont
  {M.}~\bibnamefont {Wattenberg}}, \bibinfo {author} {\bibfnamefont
  {M.}~\bibnamefont {Wicke}}, \bibinfo {author} {\bibfnamefont
  {Y.}~\bibnamefont {Yu}},\ and\ \bibinfo {author} {\bibfnamefont
  {X.}~\bibnamefont {Zheng}},\ }\href {https://www.tensorflow.org/} {\bibinfo
  {title} {{TensorFlow}: Large-scale machine learning on heterogeneous
  systems}} (\bibinfo {year} {2015}),\ \bibinfo {note} {software available from
  tensorflow.org}\BibitemShut {NoStop}%
\bibitem [{\citenamefont {{Foreman-Mackey}}\ \emph {et~al.}(2013)\citenamefont
  {{Foreman-Mackey}}, \citenamefont {{Hogg}}, \citenamefont {{Lang}},\ and\
  \citenamefont {{Goodman}}}]{ForemanMackey2013emcee}%
  \BibitemOpen
  \bibfield  {author} {\bibinfo {author} {\bibfnamefont {D.}~\bibnamefont
  {{Foreman-Mackey}}}, \bibinfo {author} {\bibfnamefont {D.~W.}\ \bibnamefont
  {{Hogg}}}, \bibinfo {author} {\bibfnamefont {D.}~\bibnamefont {{Lang}}},\
  and\ \bibinfo {author} {\bibfnamefont {J.}~\bibnamefont {{Goodman}}},\
  }\bibfield  {title} {\bibinfo {title} {{emcee: The MCMC Hammer}},\ }\href
  {https://doi.org/10.1086/670067} {\bibfield  {journal} {\bibinfo  {journal}
  {\pasp}\ }\textbf {\bibinfo {volume} {125}},\ \bibinfo {pages} {306}
  (\bibinfo {year} {2013})},\ \Eprint {https://arxiv.org/abs/1202.3665}
  {arXiv:1202.3665 [astro-ph.IM]} \BibitemShut {NoStop}%
\bibitem [{\citenamefont {{Kratochvil}}\ \emph {et~al.}(2012)\citenamefont
  {{Kratochvil}}, \citenamefont {{Lim}}, \citenamefont {{Wang}}, \citenamefont
  {{Haiman}}, \citenamefont {{May}},\ and\ \citenamefont
  {{Huffenberger}}}]{kratochvil2011probing}%
  \BibitemOpen
  \bibfield  {author} {\bibinfo {author} {\bibfnamefont {J.~M.}\ \bibnamefont
  {{Kratochvil}}}, \bibinfo {author} {\bibfnamefont {E.~A.}\ \bibnamefont
  {{Lim}}}, \bibinfo {author} {\bibfnamefont {S.}~\bibnamefont {{Wang}}},
  \bibinfo {author} {\bibfnamefont {Z.}~\bibnamefont {{Haiman}}}, \bibinfo
  {author} {\bibfnamefont {M.}~\bibnamefont {{May}}},\ and\ \bibinfo {author}
  {\bibfnamefont {K.}~\bibnamefont {{Huffenberger}}},\ }\bibfield  {title}
  {\bibinfo {title} {{Probing cosmology with weak lensing Minkowski
  functionals}},\ }\href {https://doi.org/10.1103/PhysRevD.85.103513}
  {\bibfield  {journal} {\bibinfo  {journal} {\prd}\ }\textbf {\bibinfo
  {volume} {85}},\ \bibinfo {eid} {103513} (\bibinfo {year}
  {2012})}\BibitemShut {NoStop}%
\bibitem [{\citenamefont {Harnois-D{\'e}raps}\ \emph
  {et~al.}(2021)\citenamefont {Harnois-D{\'e}raps}, \citenamefont {Martinet},
  \citenamefont {Castro}, \citenamefont {Dolag}, \citenamefont {Giblin},
  \citenamefont {Heymans}, \citenamefont {Hildebrandt},\ and\ \citenamefont
  {Xia}}]{HarnoisDeraps2020peaks}%
  \BibitemOpen
  \bibfield  {author} {\bibinfo {author} {\bibfnamefont {J.}~\bibnamefont
  {Harnois-D{\'e}raps}}, \bibinfo {author} {\bibfnamefont {N.}~\bibnamefont
  {Martinet}}, \bibinfo {author} {\bibfnamefont {T.}~\bibnamefont {Castro}},
  \bibinfo {author} {\bibfnamefont {K.}~\bibnamefont {Dolag}}, \bibinfo
  {author} {\bibfnamefont {B.}~\bibnamefont {Giblin}}, \bibinfo {author}
  {\bibfnamefont {C.}~\bibnamefont {Heymans}}, \bibinfo {author} {\bibfnamefont
  {H.}~\bibnamefont {Hildebrandt}},\ and\ \bibinfo {author} {\bibfnamefont
  {Q.}~\bibnamefont {Xia}},\ }\bibfield  {title} {\bibinfo {title} {Cosmic
  shear cosmology beyond two-point statistics: a combined peak count and
  correlation function analysis of des-y1},\ }\href@noop {} {\bibfield
  {journal} {\bibinfo  {journal} {Monthly Notices of the Royal Astronomical
  Society}\ }\textbf {\bibinfo {volume} {506}},\ \bibinfo {pages} {1623}
  (\bibinfo {year} {2021})}\BibitemShut {NoStop}%
\bibitem [{\citenamefont {Matilla}\ \emph {et~al.}(2020)\citenamefont
  {Matilla}, \citenamefont {Sharma}, \citenamefont {Hsu},\ and\ \citenamefont
  {Haiman}}]{ZorillaMatilla2020interpreting}%
  \BibitemOpen
  \bibfield  {author} {\bibinfo {author} {\bibfnamefont {J.~M.~Z.}\
  \bibnamefont {Matilla}}, \bibinfo {author} {\bibfnamefont {M.}~\bibnamefont
  {Sharma}}, \bibinfo {author} {\bibfnamefont {D.}~\bibnamefont {Hsu}},\ and\
  \bibinfo {author} {\bibfnamefont {Z.}~\bibnamefont {Haiman}},\ }\bibfield
  {title} {\bibinfo {title} {Interpreting deep learning models for weak
  lensing},\ }\href {https://doi.org/10.1103/PhysRevD.102.123506} {\bibfield
  {journal} {\bibinfo  {journal} {Phys. Rev. D}\ }\textbf {\bibinfo {volume}
  {102}},\ \bibinfo {pages} {123506} (\bibinfo {year} {2020})}\BibitemShut
  {NoStop}%
\bibitem [{\citenamefont {Olah}\ \emph {et~al.}(2018)\citenamefont {Olah},
  \citenamefont {Satyanarayan}, \citenamefont {Johnson}, \citenamefont
  {Carter}, \citenamefont {Schubert}, \citenamefont {Ye},\ and\ \citenamefont
  {Mordvintsev}}]{olah2018building}%
  \BibitemOpen
  \bibfield  {author} {\bibinfo {author} {\bibfnamefont {C.}~\bibnamefont
  {Olah}}, \bibinfo {author} {\bibfnamefont {A.}~\bibnamefont {Satyanarayan}},
  \bibinfo {author} {\bibfnamefont {I.}~\bibnamefont {Johnson}}, \bibinfo
  {author} {\bibfnamefont {S.}~\bibnamefont {Carter}}, \bibinfo {author}
  {\bibfnamefont {L.}~\bibnamefont {Schubert}}, \bibinfo {author}
  {\bibfnamefont {K.}~\bibnamefont {Ye}},\ and\ \bibinfo {author}
  {\bibfnamefont {A.}~\bibnamefont {Mordvintsev}},\ }\bibfield  {title}
  {\bibinfo {title} {The building blocks of interpretability},\ }\bibfield
  {journal} {\bibinfo  {journal} {Distill}\ }\href
  {https://doi.org/10.23915/distill.00010} {10.23915/distill.00010} (\bibinfo
  {year} {2018}),\ \bibinfo {note}
  {https://distill.pub/2018/building-blocks}\BibitemShut {NoStop}%
\bibitem [{\citenamefont {Samek}\ \emph {et~al.}(2021)\citenamefont {Samek},
  \citenamefont {Montavon}, \citenamefont {Lapuschkin}, \citenamefont
  {Anders},\ and\ \citenamefont {Müller}}]{Samek2021explaining}%
  \BibitemOpen
  \bibfield  {author} {\bibinfo {author} {\bibfnamefont {W.}~\bibnamefont
  {Samek}}, \bibinfo {author} {\bibfnamefont {G.}~\bibnamefont {Montavon}},
  \bibinfo {author} {\bibfnamefont {S.}~\bibnamefont {Lapuschkin}}, \bibinfo
  {author} {\bibfnamefont {C.~J.}\ \bibnamefont {Anders}},\ and\ \bibinfo
  {author} {\bibfnamefont {K.-R.}\ \bibnamefont {Müller}},\ }\bibfield
  {title} {\bibinfo {title} {Explaining deep neural networks and beyond: A
  review of methods and applications},\ }\href
  {https://doi.org/10.1109/JPROC.2021.3060483} {\bibfield  {journal} {\bibinfo
  {journal} {Proceedings of the IEEE}\ }\textbf {\bibinfo {volume} {109}},\
  \bibinfo {pages} {247} (\bibinfo {year} {2021})}\BibitemShut {NoStop}%
\bibitem [{\citenamefont {Berner}\ \emph {et~al.}(2021)\citenamefont {Berner},
  \citenamefont {Refregier}, \citenamefont {Sgier}, \citenamefont {Kacprzak},
  \citenamefont {Tortorelli},\ and\ \citenamefont {Monaco}}]{Berner2021rapid}%
  \BibitemOpen
  \bibfield  {author} {\bibinfo {author} {\bibfnamefont {P.}~\bibnamefont
  {Berner}}, \bibinfo {author} {\bibfnamefont {A.}~\bibnamefont {Refregier}},
  \bibinfo {author} {\bibfnamefont {R.}~\bibnamefont {Sgier}}, \bibinfo
  {author} {\bibfnamefont {T.}~\bibnamefont {Kacprzak}}, \bibinfo {author}
  {\bibfnamefont {L.}~\bibnamefont {Tortorelli}},\ and\ \bibinfo {author}
  {\bibfnamefont {P.}~\bibnamefont {Monaco}},\ }\bibfield  {title} {\bibinfo
  {title} {Rapid simulations of halo and subhalo clustering},\ }\href@noop {}
  {\bibfield  {journal} {\bibinfo  {journal} {arXiv preprint arXiv:2112.08389}\
  } (\bibinfo {year} {2021})}\BibitemShut {NoStop}%
\bibitem [{\citenamefont {{Taffoni}}\ \emph {et~al.}(2002)\citenamefont
  {{Taffoni}}, \citenamefont {{Monaco}},\ and\ \citenamefont
  {{Theuns}}}]{Taffoni2002pinnochio}%
  \BibitemOpen
  \bibfield  {author} {\bibinfo {author} {\bibfnamefont {G.}~\bibnamefont
  {{Taffoni}}}, \bibinfo {author} {\bibfnamefont {P.}~\bibnamefont
  {{Monaco}}},\ and\ \bibinfo {author} {\bibfnamefont {T.}~\bibnamefont
  {{Theuns}}},\ }\bibfield  {title} {\bibinfo {title} {{PINOCCHIO and the
  hierarchical build-up of dark matter haloes}},\ }\href
  {https://doi.org/10.1046/j.1365-8711.2002.05441.x} {\bibfield  {journal}
  {\bibinfo  {journal} {\mnras}\ }\textbf {\bibinfo {volume} {333}},\ \bibinfo
  {pages} {623} (\bibinfo {year} {2002})},\ \Eprint
  {https://arxiv.org/abs/astro-ph/0109324} {arXiv:astro-ph/0109324 [astro-ph]}
  \BibitemShut {NoStop}%
\bibitem [{\citenamefont {{DeRose}}\ \emph {et~al.}(2019)\citenamefont
  {{DeRose}}, \citenamefont {{Wechsler}}, \citenamefont {{Becker}},
  \citenamefont {{Busha}}, \citenamefont {{Rykoff}}, \citenamefont
  {{MacCrann}}, \citenamefont {{Erickson}}, \citenamefont {{Evrard}},
  \citenamefont {{Kravtsov}}, \citenamefont {{Gruen}}, \citenamefont {{Allam}},
  \citenamefont {{Avila}}, \citenamefont {{Bridle}}, \citenamefont {{Brooks}},
  \citenamefont {{Buckley-Geer}}, \citenamefont {{Carnero Rosell}},
  \citenamefont {{Carrasco Kind}}, \citenamefont {{Carretero}}, \citenamefont
  {{Castander}}, \citenamefont {{Cawthon}}, \citenamefont {{Crocce}},
  \citenamefont {{da Costa}}, \citenamefont {{Davis}}, \citenamefont {{De
  Vicente}}, \citenamefont {{Dietrich}}, \citenamefont {{Doel}}, \citenamefont
  {{Drlica-Wagner}}, \citenamefont {{Fosalba}}, \citenamefont {{Frieman}},
  \citenamefont {{Garcia-Bellido}}, \citenamefont {{Gutierrez}}, \citenamefont
  {{Hartley}}, \citenamefont {{Hollowood}}, \citenamefont {{Hoyle}},
  \citenamefont {{James}}, \citenamefont {{Krause}}, \citenamefont {{Kuehn}},
  \citenamefont {{Kuropatkin}}, \citenamefont {{Lima}}, \citenamefont {{Maia}},
  \citenamefont {{Menanteau}}, \citenamefont {{Miller}}, \citenamefont
  {{Miquel}}, \citenamefont {{Ogando}}, \citenamefont {{Plazas Malag{\'o}n}},
  \citenamefont {{Romer}}, \citenamefont {{Sanchez}}, \citenamefont
  {{Schindler}}, \citenamefont {{Serrano}}, \citenamefont {{Sevilla-Noarbe}},
  \citenamefont {{Smith}}, \citenamefont {{Suchyta}}, \citenamefont
  {{Swanson}}, \citenamefont {{Tarle}},\ and\ \citenamefont
  {{Vikram}}}]{DeRose2019buzzard}%
  \BibitemOpen
  \bibfield  {author} {\bibinfo {author} {\bibfnamefont {J.}~\bibnamefont
  {{DeRose}}}, \bibinfo {author} {\bibfnamefont {R.~H.}\ \bibnamefont
  {{Wechsler}}}, \bibinfo {author} {\bibfnamefont {M.~R.}\ \bibnamefont
  {{Becker}}}, \bibinfo {author} {\bibfnamefont {M.~T.}\ \bibnamefont
  {{Busha}}}, \bibinfo {author} {\bibfnamefont {E.~S.}\ \bibnamefont
  {{Rykoff}}}, \bibinfo {author} {\bibfnamefont {N.}~\bibnamefont
  {{MacCrann}}}, \bibinfo {author} {\bibfnamefont {B.}~\bibnamefont
  {{Erickson}}}, \bibinfo {author} {\bibfnamefont {A.~E.}\ \bibnamefont
  {{Evrard}}}, \bibinfo {author} {\bibfnamefont {A.}~\bibnamefont
  {{Kravtsov}}}, \bibinfo {author} {\bibfnamefont {D.}~\bibnamefont {{Gruen}}},
  \bibinfo {author} {\bibfnamefont {S.}~\bibnamefont {{Allam}}}, \bibinfo
  {author} {\bibfnamefont {S.}~\bibnamefont {{Avila}}}, \bibinfo {author}
  {\bibfnamefont {S.}~\bibnamefont {{Bridle}}}, \bibinfo {author}
  {\bibfnamefont {D.}~\bibnamefont {{Brooks}}}, \bibinfo {author}
  {\bibfnamefont {E.}~\bibnamefont {{Buckley-Geer}}}, \bibinfo {author}
  {\bibfnamefont {A.}~\bibnamefont {{Carnero Rosell}}}, \bibinfo {author}
  {\bibfnamefont {M.}~\bibnamefont {{Carrasco Kind}}}, \bibinfo {author}
  {\bibfnamefont {J.}~\bibnamefont {{Carretero}}}, \bibinfo {author}
  {\bibfnamefont {F.~J.}\ \bibnamefont {{Castander}}}, \bibinfo {author}
  {\bibfnamefont {R.}~\bibnamefont {{Cawthon}}}, \bibinfo {author}
  {\bibfnamefont {M.}~\bibnamefont {{Crocce}}}, \bibinfo {author}
  {\bibfnamefont {L.~N.}\ \bibnamefont {{da Costa}}}, \bibinfo {author}
  {\bibfnamefont {C.}~\bibnamefont {{Davis}}}, \bibinfo {author} {\bibfnamefont
  {J.}~\bibnamefont {{De Vicente}}}, \bibinfo {author} {\bibfnamefont {J.~P.}\
  \bibnamefont {{Dietrich}}}, \bibinfo {author} {\bibfnamefont
  {P.}~\bibnamefont {{Doel}}}, \bibinfo {author} {\bibfnamefont
  {A.}~\bibnamefont {{Drlica-Wagner}}}, \bibinfo {author} {\bibfnamefont
  {P.}~\bibnamefont {{Fosalba}}}, \bibinfo {author} {\bibfnamefont
  {J.}~\bibnamefont {{Frieman}}}, \bibinfo {author} {\bibfnamefont
  {J.}~\bibnamefont {{Garcia-Bellido}}}, \bibinfo {author} {\bibfnamefont
  {G.}~\bibnamefont {{Gutierrez}}}, \bibinfo {author} {\bibfnamefont {W.~G.}\
  \bibnamefont {{Hartley}}}, \bibinfo {author} {\bibfnamefont {D.~L.}\
  \bibnamefont {{Hollowood}}}, \bibinfo {author} {\bibfnamefont
  {B.}~\bibnamefont {{Hoyle}}}, \bibinfo {author} {\bibfnamefont {D.~J.}\
  \bibnamefont {{James}}}, \bibinfo {author} {\bibfnamefont {E.}~\bibnamefont
  {{Krause}}}, \bibinfo {author} {\bibfnamefont {K.}~\bibnamefont {{Kuehn}}},
  \bibinfo {author} {\bibfnamefont {N.}~\bibnamefont {{Kuropatkin}}}, \bibinfo
  {author} {\bibfnamefont {M.}~\bibnamefont {{Lima}}}, \bibinfo {author}
  {\bibfnamefont {M.~A.~G.}\ \bibnamefont {{Maia}}}, \bibinfo {author}
  {\bibfnamefont {F.}~\bibnamefont {{Menanteau}}}, \bibinfo {author}
  {\bibfnamefont {C.~J.}\ \bibnamefont {{Miller}}}, \bibinfo {author}
  {\bibfnamefont {R.}~\bibnamefont {{Miquel}}}, \bibinfo {author}
  {\bibfnamefont {R.~L.~C.}\ \bibnamefont {{Ogando}}}, \bibinfo {author}
  {\bibfnamefont {A.}~\bibnamefont {{Plazas Malag{\'o}n}}}, \bibinfo {author}
  {\bibfnamefont {A.~K.}\ \bibnamefont {{Romer}}}, \bibinfo {author}
  {\bibfnamefont {E.}~\bibnamefont {{Sanchez}}}, \bibinfo {author}
  {\bibfnamefont {R.}~\bibnamefont {{Schindler}}}, \bibinfo {author}
  {\bibfnamefont {S.}~\bibnamefont {{Serrano}}}, \bibinfo {author}
  {\bibfnamefont {I.}~\bibnamefont {{Sevilla-Noarbe}}}, \bibinfo {author}
  {\bibfnamefont {M.}~\bibnamefont {{Smith}}}, \bibinfo {author} {\bibfnamefont
  {E.}~\bibnamefont {{Suchyta}}}, \bibinfo {author} {\bibfnamefont {M.~E.~C.}\
  \bibnamefont {{Swanson}}}, \bibinfo {author} {\bibfnamefont {G.}~\bibnamefont
  {{Tarle}}},\ and\ \bibinfo {author} {\bibfnamefont {V.}~\bibnamefont
  {{Vikram}}},\ }\bibfield  {title} {\bibinfo {title} {{The Buzzard Flock: Dark
  Energy Survey Synthetic Sky Catalogs}},\ }\href
  {https://arxiv.org/abs/1901.02401} {\bibfield  {journal} {\bibinfo  {journal}
  {arXiv e-prints}\ ,\ \bibinfo {eid} {arXiv:1901.02401}} (\bibinfo {year}
  {2019})}\BibitemShut {NoStop}%
\bibitem [{\citenamefont {{Fosalba}}\ \emph {et~al.}(2015)\citenamefont
  {{Fosalba}}, \citenamefont {{Gazta{\~n}aga}}, \citenamefont {{Castander}},\
  and\ \citenamefont {{Crocce}}}]{Fosalba2015mice}%
  \BibitemOpen
  \bibfield  {author} {\bibinfo {author} {\bibfnamefont {P.}~\bibnamefont
  {{Fosalba}}}, \bibinfo {author} {\bibfnamefont {E.}~\bibnamefont
  {{Gazta{\~n}aga}}}, \bibinfo {author} {\bibfnamefont {F.~J.}\ \bibnamefont
  {{Castander}}},\ and\ \bibinfo {author} {\bibfnamefont {M.}~\bibnamefont
  {{Crocce}}},\ }\bibfield  {title} {\bibinfo {title} {{The MICE Grand
  Challenge light-cone simulation - III. Galaxy lensing mocks from all-sky
  lensing maps}},\ }\href {https://doi.org/10.1093/mnras/stu2464} {\bibfield
  {journal} {\bibinfo  {journal} {\mnras}\ }\textbf {\bibinfo {volume} {447}},\
  \bibinfo {pages} {1319} (\bibinfo {year} {2015})},\ \Eprint
  {https://arxiv.org/abs/1312.2947} {arXiv:1312.2947 [astro-ph.CO]}
  \BibitemShut {NoStop}%
\bibitem [{\citenamefont {{Friedrich}}\ \emph {et~al.}(2022)\citenamefont
  {{Friedrich}}, \citenamefont {{Halder}}, \citenamefont {{Boyle}},
  \citenamefont {{Uhlemann}}, \citenamefont {{Britt}}, \citenamefont {{Codis}},
  \citenamefont {{Gruen}},\ and\ \citenamefont {{Hahn}}}]{Friedrich2022pdf}%
  \BibitemOpen
  \bibfield  {author} {\bibinfo {author} {\bibfnamefont {O.}~\bibnamefont
  {{Friedrich}}}, \bibinfo {author} {\bibfnamefont {A.}~\bibnamefont
  {{Halder}}}, \bibinfo {author} {\bibfnamefont {A.}~\bibnamefont {{Boyle}}},
  \bibinfo {author} {\bibfnamefont {C.}~\bibnamefont {{Uhlemann}}}, \bibinfo
  {author} {\bibfnamefont {D.}~\bibnamefont {{Britt}}}, \bibinfo {author}
  {\bibfnamefont {S.}~\bibnamefont {{Codis}}}, \bibinfo {author} {\bibfnamefont
  {D.}~\bibnamefont {{Gruen}}},\ and\ \bibinfo {author} {\bibfnamefont
  {C.}~\bibnamefont {{Hahn}}},\ }\bibfield  {title} {\bibinfo {title} {{The PDF
  perspective on the tracer-matter connection: Lagrangian bias and
  non-Poissonian shot noise}},\ }\href {https://doi.org/10.1093/mnras/stab3703}
  {\bibfield  {journal} {\bibinfo  {journal} {\mnras}\ }\textbf {\bibinfo
  {volume} {510}},\ \bibinfo {pages} {5069} (\bibinfo {year} {2022})},\ \Eprint
  {https://arxiv.org/abs/2107.02300} {arXiv:2107.02300 [astro-ph.CO]}
  \BibitemShut {NoStop}%
\bibitem [{\citenamefont {Harnois-Déraps}\ \emph {et~al.}(2018)\citenamefont
  {Harnois-Déraps}, \citenamefont {Amon}, \citenamefont {Choi}, \citenamefont
  {Demchenko}, \citenamefont {Heymans}, \citenamefont {Kannawadi},
  \citenamefont {Nakajima}, \citenamefont {Sirks}, \citenamefont
  {van Waerbeke}, \citenamefont {Cai}, \citenamefont {Giblin}, \citenamefont
  {Hildebrandt}, \citenamefont {Hoekstra}, \citenamefont {Miller},\ and\
  \citenamefont {Tröster}}]{HarnoisDeraps2018simulations}%
  \BibitemOpen
  \bibfield  {author} {\bibinfo {author} {\bibfnamefont {J.}~\bibnamefont
  {Harnois-Déraps}}, \bibinfo {author} {\bibfnamefont {A.}~\bibnamefont
  {Amon}}, \bibinfo {author} {\bibfnamefont {A.}~\bibnamefont {Choi}}, \bibinfo
  {author} {\bibfnamefont {V.}~\bibnamefont {Demchenko}}, \bibinfo {author}
  {\bibfnamefont {C.}~\bibnamefont {Heymans}}, \bibinfo {author} {\bibfnamefont
  {A.}~\bibnamefont {Kannawadi}}, \bibinfo {author} {\bibfnamefont
  {R.}~\bibnamefont {Nakajima}}, \bibinfo {author} {\bibfnamefont
  {E.}~\bibnamefont {Sirks}}, \bibinfo {author} {\bibfnamefont
  {L.}~\bibnamefont {van Waerbeke}}, \bibinfo {author} {\bibfnamefont {Y.-C.}\
  \bibnamefont {Cai}}, \bibinfo {author} {\bibfnamefont {B.}~\bibnamefont
  {Giblin}}, \bibinfo {author} {\bibfnamefont {H.}~\bibnamefont {Hildebrandt}},
  \bibinfo {author} {\bibfnamefont {H.}~\bibnamefont {Hoekstra}}, \bibinfo
  {author} {\bibfnamefont {L.}~\bibnamefont {Miller}},\ and\ \bibinfo {author}
  {\bibfnamefont {T.}~\bibnamefont {Tröster}},\ }\bibfield  {title} {\bibinfo
  {title} {{Cosmological simulations for combined-probe analyses: covariance
  and neighbour-exclusion bias}},\ }\href
  {https://doi.org/10.1093/mnras/sty2319} {\bibfield  {journal} {\bibinfo
  {journal} {Monthly Notices of the Royal Astronomical Society}\ }\textbf
  {\bibinfo {volume} {481}},\ \bibinfo {pages} {1337} (\bibinfo {year}
  {2018})},\ \Eprint
  {https://arxiv.org/abs/https://academic.oup.com/mnras/article-pdf/481/1/1337/25720036/sty2319.pdf}
  {https://academic.oup.com/mnras/article-pdf/481/1/1337/25720036/sty2319.pdf}
  \BibitemShut {NoStop}%
\bibitem [{\citenamefont {Schneider}\ \emph {et~al.}(2020)\citenamefont
  {Schneider}, \citenamefont {Refregier}, \citenamefont {Grandis},
  \citenamefont {Eckert}, \citenamefont {Stoira}, \citenamefont {Kacprzak},
  \citenamefont {Knabenhans}, \citenamefont {Stadel},\ and\ \citenamefont
  {Teyssier}}]{Schneider2020xray}%
  \BibitemOpen
  \bibfield  {author} {\bibinfo {author} {\bibfnamefont {A.}~\bibnamefont
  {Schneider}}, \bibinfo {author} {\bibfnamefont {A.}~\bibnamefont
  {Refregier}}, \bibinfo {author} {\bibfnamefont {S.}~\bibnamefont {Grandis}},
  \bibinfo {author} {\bibfnamefont {D.}~\bibnamefont {Eckert}}, \bibinfo
  {author} {\bibfnamefont {N.}~\bibnamefont {Stoira}}, \bibinfo {author}
  {\bibfnamefont {T.}~\bibnamefont {Kacprzak}}, \bibinfo {author}
  {\bibfnamefont {M.}~\bibnamefont {Knabenhans}}, \bibinfo {author}
  {\bibfnamefont {J.}~\bibnamefont {Stadel}},\ and\ \bibinfo {author}
  {\bibfnamefont {R.}~\bibnamefont {Teyssier}},\ }\bibfield  {title} {\bibinfo
  {title} {Baryonic effects for weak lensing. part {II}. combination with x-ray
  data and extended cosmologies},\ }\href
  {https://doi.org/10.1088/1475-7516/2020/04/020} {\bibfield  {journal}
  {\bibinfo  {journal} {Journal of Cosmology and Astroparticle Physics}\
  }\textbf {\bibinfo {volume} {2020}}\bibinfo  {number} { (04)},\ \bibinfo
  {pages} {020}}\BibitemShut {NoStop}%
\bibitem [{\citenamefont {{Samuroff}}\ \emph {et~al.}(2019)\citenamefont
  {{Samuroff}}, \citenamefont {{Blazek}}, \citenamefont {{Troxel}},
  \citenamefont {{MacCrann}}, \citenamefont {{Krause}}, \citenamefont
  {{Leonard}}, \citenamefont {{Prat}}, \citenamefont {{Gruen}}, \citenamefont
  {{Dodelson}}, \citenamefont {{Eifler}}, \citenamefont {{Gatti}},
  \citenamefont {{Hartley}}, \citenamefont {{Hoyle}}, \citenamefont {{Larsen}},
  \citenamefont {{Zuntz}}, \citenamefont {{Abbott}}, \citenamefont {{Allam}},
  \citenamefont {{Annis}}, \citenamefont {{Bernstein}}, \citenamefont
  {{Bertin}}, \citenamefont {{Bridle}}, \citenamefont {{Brooks}}, \citenamefont
  {{Carnero Rosell}}, \citenamefont {{Carrasco Kind}}, \citenamefont
  {{Carretero}}, \citenamefont {{Castander}}, \citenamefont {{Cunha}},
  \citenamefont {{da Costa}}, \citenamefont {{Davis}}, \citenamefont {{De
  Vicente}}, \citenamefont {{DePoy}}, \citenamefont {{Desai}}, \citenamefont
  {{Diehl}}, \citenamefont {{Dietrich}}, \citenamefont {{Doel}}, \citenamefont
  {{Flaugher}}, \citenamefont {{Fosalba}}, \citenamefont {{Frieman}},
  \citenamefont {{Garc{\'\i}a-Bellido}}, \citenamefont {{Gaztanaga}},
  \citenamefont {{Gerdes}}, \citenamefont {{Gruendl}}, \citenamefont
  {{Gschwend}}, \citenamefont {{Gutierrez}}, \citenamefont {{Hollowood}},
  \citenamefont {{Honscheid}}, \citenamefont {{James}}, \citenamefont
  {{Kuehn}}, \citenamefont {{Kuropatkin}}, \citenamefont {{Lima}},
  \citenamefont {{Maia}}, \citenamefont {{March}}, \citenamefont {{Marshall}},
  \citenamefont {{Martini}}, \citenamefont {{Melchior}}, \citenamefont
  {{Menanteau}}, \citenamefont {{Miller}}, \citenamefont {{Miquel}},
  \citenamefont {{Ogando}}, \citenamefont {{Plazas}}, \citenamefont
  {{Sanchez}}, \citenamefont {{Scarpine}}, \citenamefont {{Schindler}},
  \citenamefont {{Schubnell}}, \citenamefont {{Serrano}}, \citenamefont
  {{Sevilla-Noarbe}}, \citenamefont {{Sheldon}}, \citenamefont {{Smith}},
  \citenamefont {{Sobreira}}, \citenamefont {{Suchyta}}, \citenamefont
  {{Tarle}}, \citenamefont {{Thomas}}, \citenamefont {{Vikram}},\ and\
  \citenamefont {{DES Collaboration}}}]{Samuroff2019desia}%
  \BibitemOpen
\bibfield  {number} {  }\bibfield  {author} {\bibinfo {author} {\bibfnamefont
  {S.}~\bibnamefont {{Samuroff}}}, \bibinfo {author} {\bibfnamefont
  {J.}~\bibnamefont {{Blazek}}}, \bibinfo {author} {\bibfnamefont {M.~A.}\
  \bibnamefont {{Troxel}}}, \bibinfo {author} {\bibfnamefont {N.}~\bibnamefont
  {{MacCrann}}}, \bibinfo {author} {\bibfnamefont {E.}~\bibnamefont
  {{Krause}}}, \bibinfo {author} {\bibfnamefont {C.~D.}\ \bibnamefont
  {{Leonard}}}, \bibinfo {author} {\bibfnamefont {J.}~\bibnamefont {{Prat}}},
  \bibinfo {author} {\bibfnamefont {D.}~\bibnamefont {{Gruen}}}, \bibinfo
  {author} {\bibfnamefont {S.}~\bibnamefont {{Dodelson}}}, \bibinfo {author}
  {\bibfnamefont {T.~F.}\ \bibnamefont {{Eifler}}}, \bibinfo {author}
  {\bibfnamefont {M.}~\bibnamefont {{Gatti}}}, \bibinfo {author} {\bibfnamefont
  {W.~G.}\ \bibnamefont {{Hartley}}}, \bibinfo {author} {\bibfnamefont
  {B.}~\bibnamefont {{Hoyle}}}, \bibinfo {author} {\bibfnamefont
  {P.}~\bibnamefont {{Larsen}}}, \bibinfo {author} {\bibfnamefont
  {J.}~\bibnamefont {{Zuntz}}}, \bibinfo {author} {\bibfnamefont {T.~M.~C.}\
  \bibnamefont {{Abbott}}}, \bibinfo {author} {\bibfnamefont {S.}~\bibnamefont
  {{Allam}}}, \bibinfo {author} {\bibfnamefont {J.}~\bibnamefont {{Annis}}},
  \bibinfo {author} {\bibfnamefont {G.~M.}\ \bibnamefont {{Bernstein}}},
  \bibinfo {author} {\bibfnamefont {E.}~\bibnamefont {{Bertin}}}, \bibinfo
  {author} {\bibfnamefont {S.~L.}\ \bibnamefont {{Bridle}}}, \bibinfo {author}
  {\bibfnamefont {D.}~\bibnamefont {{Brooks}}}, \bibinfo {author}
  {\bibfnamefont {A.}~\bibnamefont {{Carnero Rosell}}}, \bibinfo {author}
  {\bibfnamefont {M.}~\bibnamefont {{Carrasco Kind}}}, \bibinfo {author}
  {\bibfnamefont {J.}~\bibnamefont {{Carretero}}}, \bibinfo {author}
  {\bibfnamefont {F.~J.}\ \bibnamefont {{Castander}}}, \bibinfo {author}
  {\bibfnamefont {C.~E.}\ \bibnamefont {{Cunha}}}, \bibinfo {author}
  {\bibfnamefont {L.~N.}\ \bibnamefont {{da Costa}}}, \bibinfo {author}
  {\bibfnamefont {C.}~\bibnamefont {{Davis}}}, \bibinfo {author} {\bibfnamefont
  {J.}~\bibnamefont {{De Vicente}}}, \bibinfo {author} {\bibfnamefont {D.~L.}\
  \bibnamefont {{DePoy}}}, \bibinfo {author} {\bibfnamefont {S.}~\bibnamefont
  {{Desai}}}, \bibinfo {author} {\bibfnamefont {H.~T.}\ \bibnamefont
  {{Diehl}}}, \bibinfo {author} {\bibfnamefont {J.~P.}\ \bibnamefont
  {{Dietrich}}}, \bibinfo {author} {\bibfnamefont {P.}~\bibnamefont {{Doel}}},
  \bibinfo {author} {\bibfnamefont {B.}~\bibnamefont {{Flaugher}}}, \bibinfo
  {author} {\bibfnamefont {P.}~\bibnamefont {{Fosalba}}}, \bibinfo {author}
  {\bibfnamefont {J.}~\bibnamefont {{Frieman}}}, \bibinfo {author}
  {\bibfnamefont {J.}~\bibnamefont {{Garc{\'\i}a-Bellido}}}, \bibinfo {author}
  {\bibfnamefont {E.}~\bibnamefont {{Gaztanaga}}}, \bibinfo {author}
  {\bibfnamefont {D.~W.}\ \bibnamefont {{Gerdes}}}, \bibinfo {author}
  {\bibfnamefont {R.~A.}\ \bibnamefont {{Gruendl}}}, \bibinfo {author}
  {\bibfnamefont {J.}~\bibnamefont {{Gschwend}}}, \bibinfo {author}
  {\bibfnamefont {G.}~\bibnamefont {{Gutierrez}}}, \bibinfo {author}
  {\bibfnamefont {D.~L.}\ \bibnamefont {{Hollowood}}}, \bibinfo {author}
  {\bibfnamefont {K.}~\bibnamefont {{Honscheid}}}, \bibinfo {author}
  {\bibfnamefont {D.~J.}\ \bibnamefont {{James}}}, \bibinfo {author}
  {\bibfnamefont {K.}~\bibnamefont {{Kuehn}}}, \bibinfo {author} {\bibfnamefont
  {N.}~\bibnamefont {{Kuropatkin}}}, \bibinfo {author} {\bibfnamefont
  {M.}~\bibnamefont {{Lima}}}, \bibinfo {author} {\bibfnamefont {M.~A.~G.}\
  \bibnamefont {{Maia}}}, \bibinfo {author} {\bibfnamefont {M.}~\bibnamefont
  {{March}}}, \bibinfo {author} {\bibfnamefont {J.~L.}\ \bibnamefont
  {{Marshall}}}, \bibinfo {author} {\bibfnamefont {P.}~\bibnamefont
  {{Martini}}}, \bibinfo {author} {\bibfnamefont {P.}~\bibnamefont
  {{Melchior}}}, \bibinfo {author} {\bibfnamefont {F.}~\bibnamefont
  {{Menanteau}}}, \bibinfo {author} {\bibfnamefont {C.~J.}\ \bibnamefont
  {{Miller}}}, \bibinfo {author} {\bibfnamefont {R.}~\bibnamefont {{Miquel}}},
  \bibinfo {author} {\bibfnamefont {R.~L.~C.}\ \bibnamefont {{Ogando}}},
  \bibinfo {author} {\bibfnamefont {A.~A.}\ \bibnamefont {{Plazas}}}, \bibinfo
  {author} {\bibfnamefont {E.}~\bibnamefont {{Sanchez}}}, \bibinfo {author}
  {\bibfnamefont {V.}~\bibnamefont {{Scarpine}}}, \bibinfo {author}
  {\bibfnamefont {R.}~\bibnamefont {{Schindler}}}, \bibinfo {author}
  {\bibfnamefont {M.}~\bibnamefont {{Schubnell}}}, \bibinfo {author}
  {\bibfnamefont {S.}~\bibnamefont {{Serrano}}}, \bibinfo {author}
  {\bibfnamefont {I.}~\bibnamefont {{Sevilla-Noarbe}}}, \bibinfo {author}
  {\bibfnamefont {E.}~\bibnamefont {{Sheldon}}}, \bibinfo {author}
  {\bibfnamefont {M.}~\bibnamefont {{Smith}}}, \bibinfo {author} {\bibfnamefont
  {F.}~\bibnamefont {{Sobreira}}}, \bibinfo {author} {\bibfnamefont
  {E.}~\bibnamefont {{Suchyta}}}, \bibinfo {author} {\bibfnamefont
  {G.}~\bibnamefont {{Tarle}}}, \bibinfo {author} {\bibfnamefont
  {D.}~\bibnamefont {{Thomas}}}, \bibinfo {author} {\bibfnamefont
  {V.}~\bibnamefont {{Vikram}}},\ and\ \bibinfo {author} {\bibnamefont {{DES
  Collaboration}}},\ }\bibfield  {title} {\bibinfo {title} {{Dark Energy Survey
  Year 1 results: constraints on intrinsic alignments and their colour
  dependence from galaxy clustering and weak lensing}},\ }\href
  {https://doi.org/10.1093/mnras/stz2197} {\bibfield  {journal} {\bibinfo
  {journal} {\mnras}\ }\textbf {\bibinfo {volume} {489}},\ \bibinfo {pages}
  {5453} (\bibinfo {year} {2019})}\BibitemShut {NoStop}%
\bibitem [{\citenamefont {et~al.}({\natexlab{b}})}]{Kacprzak2022cosmogrid}%
  \BibitemOpen
  \bibfield  {author} {\bibinfo {author} {\bibfnamefont {K.}~\bibnamefont
  {et~al.}},\ }\bibfield  {title} {\bibinfo {title} {Cosmogrid: a fully
  numerical $w$cdm theory prediction for large scale structure with beyond
  gaussian statistics},\ }\href@noop {} {\bibfield  {journal} {\bibinfo
  {journal} {in prep}\ } ({\natexlab{b}})}\BibitemShut {NoStop}%
\bibitem [{\citenamefont {Pedregosa}\ \emph {et~al.}(2011)\citenamefont
  {Pedregosa}, \citenamefont {Varoquaux}, \citenamefont {Gramfort},
  \citenamefont {Michel}, \citenamefont {Thirion}, \citenamefont {Grisel},
  \citenamefont {Blondel}, \citenamefont {Prettenhofer}, \citenamefont {Weiss},
  \citenamefont {Dubourg}, \citenamefont {Vanderplas}, \citenamefont {Passos},
  \citenamefont {Cournapeau}, \citenamefont {Brucher}, \citenamefont {Perrot},\
  and\ \citenamefont {Duchesnay}}]{scikit-learn}%
  \BibitemOpen
  \bibfield  {author} {\bibinfo {author} {\bibfnamefont {F.}~\bibnamefont
  {Pedregosa}}, \bibinfo {author} {\bibfnamefont {G.}~\bibnamefont
  {Varoquaux}}, \bibinfo {author} {\bibfnamefont {A.}~\bibnamefont {Gramfort}},
  \bibinfo {author} {\bibfnamefont {V.}~\bibnamefont {Michel}}, \bibinfo
  {author} {\bibfnamefont {B.}~\bibnamefont {Thirion}}, \bibinfo {author}
  {\bibfnamefont {O.}~\bibnamefont {Grisel}}, \bibinfo {author} {\bibfnamefont
  {M.}~\bibnamefont {Blondel}}, \bibinfo {author} {\bibfnamefont
  {P.}~\bibnamefont {Prettenhofer}}, \bibinfo {author} {\bibfnamefont
  {R.}~\bibnamefont {Weiss}}, \bibinfo {author} {\bibfnamefont
  {V.}~\bibnamefont {Dubourg}}, \bibinfo {author} {\bibfnamefont
  {J.}~\bibnamefont {Vanderplas}}, \bibinfo {author} {\bibfnamefont
  {A.}~\bibnamefont {Passos}}, \bibinfo {author} {\bibfnamefont
  {D.}~\bibnamefont {Cournapeau}}, \bibinfo {author} {\bibfnamefont
  {M.}~\bibnamefont {Brucher}}, \bibinfo {author} {\bibfnamefont
  {M.}~\bibnamefont {Perrot}},\ and\ \bibinfo {author} {\bibfnamefont
  {E.}~\bibnamefont {Duchesnay}},\ }\bibfield  {title} {\bibinfo {title}
  {Scikit-learn: Machine learning in {P}ython},\ }\href@noop {} {\bibfield
  {journal} {\bibinfo  {journal} {Journal of Machine Learning Research}\
  }\textbf {\bibinfo {volume} {12}},\ \bibinfo {pages} {2825} (\bibinfo {year}
  {2011})}\BibitemShut {NoStop}%
\bibitem [{\citenamefont {Liu}\ \emph {et~al.}(2021)\citenamefont {Liu},
  \citenamefont {Xu}, \citenamefont {Jiang},\ and\ \citenamefont
  {Wong}}]{Liu2021density}%
  \BibitemOpen
  \bibfield  {author} {\bibinfo {author} {\bibfnamefont {Q.}~\bibnamefont
  {Liu}}, \bibinfo {author} {\bibfnamefont {J.}~\bibnamefont {Xu}}, \bibinfo
  {author} {\bibfnamefont {R.}~\bibnamefont {Jiang}},\ and\ \bibinfo {author}
  {\bibfnamefont {W.~H.}\ \bibnamefont {Wong}},\ }\bibfield  {title} {\bibinfo
  {title} {Density estimation using deep generative neural networks},\ }\href
  {https://doi.org/10.1073/pnas.2101344118} {\bibfield  {journal} {\bibinfo
  {journal} {Proceedings of the National Academy of Sciences}\ }\textbf
  {\bibinfo {volume} {118}},\ \bibinfo {pages} {e2101344118} (\bibinfo {year}
  {2021})},\ \Eprint
  {https://arxiv.org/abs/https://www.pnas.org/doi/pdf/10.1073/pnas.2101344118}
  {https://www.pnas.org/doi/pdf/10.1073/pnas.2101344118} \BibitemShut {NoStop}%
\bibitem [{\citenamefont {Dutordoir}\ \emph {et~al.}(2018)\citenamefont
  {Dutordoir}, \citenamefont {Salimbeni}, \citenamefont {Deisenroth},\ and\
  \citenamefont {Hensman}}]{Dutordoir2018gpdensity}%
  \BibitemOpen
  \bibfield  {author} {\bibinfo {author} {\bibfnamefont {V.}~\bibnamefont
  {Dutordoir}}, \bibinfo {author} {\bibfnamefont {H.}~\bibnamefont
  {Salimbeni}}, \bibinfo {author} {\bibfnamefont {M.~P.}\ \bibnamefont
  {Deisenroth}},\ and\ \bibinfo {author} {\bibfnamefont {J.}~\bibnamefont
  {Hensman}},\ }\bibfield  {title} {\bibinfo {title} {Gaussian process
  conditional density estimation},\ }in\ \href@noop {} {\emph {\bibinfo
  {booktitle} {Proceedings of the 32nd International Conference on Neural
  Information Processing Systems}}},\ \bibinfo {series and number} {NIPS'18}\
  (\bibinfo  {publisher} {Curran Associates Inc.},\ \bibinfo {address} {Red
  Hook, NY, USA},\ \bibinfo {year} {2018})\ p.\ \bibinfo {pages}
  {2391–2401}\BibitemShut {NoStop}%
\bibitem [{\citenamefont {Papamakarios}\ \emph {et~al.}(2017)\citenamefont
  {Papamakarios}, \citenamefont {Pavlakou},\ and\ \citenamefont
  {Murray}}]{papamakarios2017masked}%
  \BibitemOpen
  \bibfield  {author} {\bibinfo {author} {\bibfnamefont {G.}~\bibnamefont
  {Papamakarios}}, \bibinfo {author} {\bibfnamefont {T.}~\bibnamefont
  {Pavlakou}},\ and\ \bibinfo {author} {\bibfnamefont {I.}~\bibnamefont
  {Murray}},\ }\bibfield  {title} {\bibinfo {title} {Masked autoregressive flow
  for density estimation},\ }\href@noop {} {\bibfield  {journal} {\bibinfo
  {journal} {Advances in neural information processing systems}\ }\textbf
  {\bibinfo {volume} {30}} (\bibinfo {year} {2017})}\BibitemShut {NoStop}%
\bibitem [{\citenamefont {Heavens}\ \emph {et~al.}(2017)\citenamefont
  {Heavens}, \citenamefont {Sellentin}, \citenamefont {de~Mijolla},\ and\
  \citenamefont {Vianello}}]{Heavens2017moped}%
  \BibitemOpen
  \bibfield  {author} {\bibinfo {author} {\bibfnamefont {A.}~\bibnamefont
  {Heavens}}, \bibinfo {author} {\bibfnamefont {E.}~\bibnamefont {Sellentin}},
  \bibinfo {author} {\bibfnamefont {D.}~\bibnamefont {de~Mijolla}},\ and\
  \bibinfo {author} {\bibfnamefont {A.}~\bibnamefont {Vianello}},\ }\bibfield
  {title} {\bibinfo {title} {{Massive data compression for parameter-dependent
  covariance matrices}},\ }\href {https://doi.org/10.1093/mnras/stx2326}
  {\bibfield  {journal} {\bibinfo  {journal} {Mon. Not. Roy. Astron. Soc.}\
  }\textbf {\bibinfo {volume} {472}},\ \bibinfo {pages} {4244} (\bibinfo {year}
  {2017})},\ \Eprint {https://arxiv.org/abs/1707.06529} {arXiv:1707.06529
  [astro-ph.CO]} \BibitemShut {NoStop}%
\bibitem [{\citenamefont {hsin Chen}\ \emph {et~al.}(2015)\citenamefont {hsin
  Chen}, \citenamefont {Moreno}, \citenamefont {Sainath}, \citenamefont
  {Visontai}, \citenamefont {Alvarez},\ and\ \citenamefont
  {Parada}}]{Chen2015locally}%
  \BibitemOpen
  \bibfield  {author} {\bibinfo {author} {\bibfnamefont {Y.}~\bibnamefont {hsin
  Chen}}, \bibinfo {author} {\bibfnamefont {I.~L.}\ \bibnamefont {Moreno}},
  \bibinfo {author} {\bibfnamefont {T.}~\bibnamefont {Sainath}}, \bibinfo
  {author} {\bibfnamefont {M.}~\bibnamefont {Visontai}}, \bibinfo {author}
  {\bibfnamefont {R.}~\bibnamefont {Alvarez}},\ and\ \bibinfo {author}
  {\bibfnamefont {C.}~\bibnamefont {Parada}},\ }\bibfield  {title} {\bibinfo
  {title} {Locally-connected and convolutional neural networks for small
  footprint speaker recognition},\ }in\ \href@noop {} {\emph {\bibinfo
  {booktitle} {Interspeech}}}\ (\bibinfo {year} {2015})\BibitemShut {NoStop}%
\bibitem [{\citenamefont {Kaiser}\ \emph {et~al.}(2018)\citenamefont {Kaiser},
  \citenamefont {Gomez},\ and\ \citenamefont {Chollet}}]{Kaiser2018depthwise}%
  \BibitemOpen
  \bibfield  {author} {\bibinfo {author} {\bibfnamefont {L.}~\bibnamefont
  {Kaiser}}, \bibinfo {author} {\bibfnamefont {A.~N.}\ \bibnamefont {Gomez}},\
  and\ \bibinfo {author} {\bibfnamefont {F.}~\bibnamefont {Chollet}},\
  }\bibfield  {title} {\bibinfo {title} {Depthwise separable convolutions for
  neural machine translation},\ }in\ \href
  {https://openreview.net/forum?id=S1jBcueAb} {\emph {\bibinfo {booktitle}
  {International Conference on Learning Representations}}}\ (\bibinfo {year}
  {2018})\BibitemShut {NoStop}%
\end{thebibliography}%

\appendix


\section{Redshift bins}
\label{sec:reshift_bins}

\begin{figure}
    \includegraphics[width=0.35\textwidth]{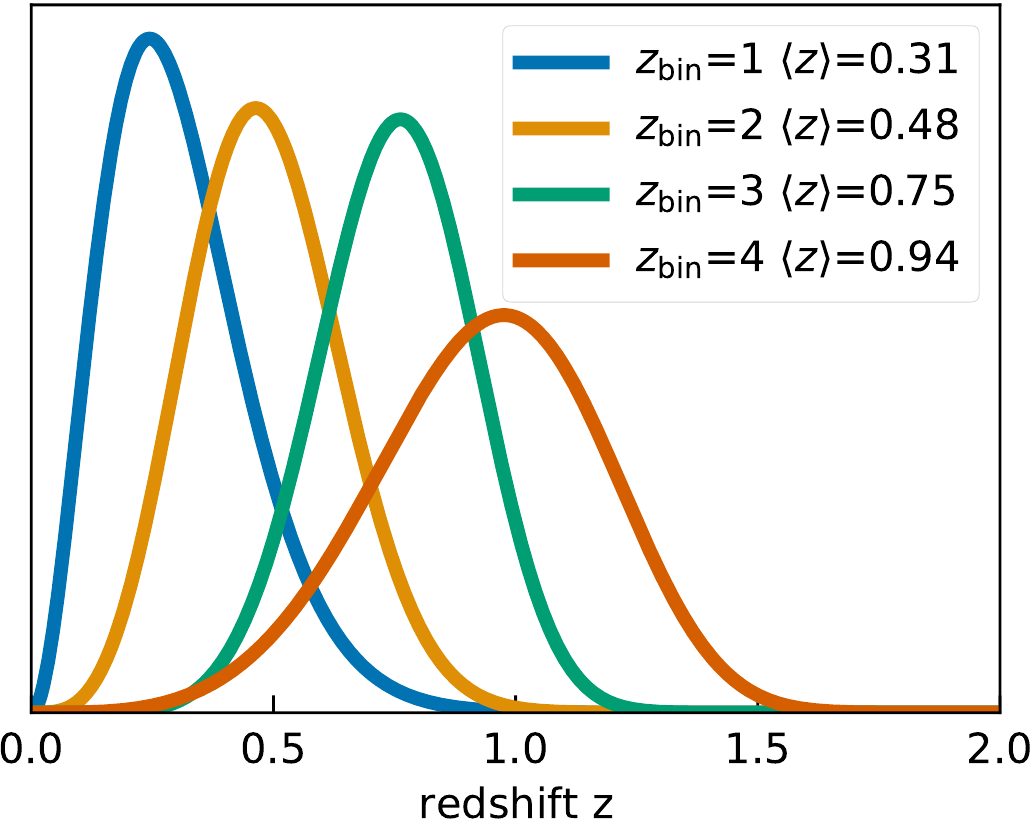} 
    \caption{Redshift bins used in this work.}
    \label{fig:nz}
\end{figure}

In this work we use a generic Stage-III simulated survey with four redshift bins.
These bins are the same as used in \cite{Fischbacher2022ia} and have the 
mean redshifts of $\langle z \rangle=0.31, 0.48, 0.75, 0.94$.
The shape of the bins is shown in Figure~\ref{fig:nz}.
The last bin is particularly wide, as is the case for photometric surveys; the uncertainty on the redshift for distant galaxies is high, which causes the bin to be broader.

\begin{figure}
    \includegraphics[width=0.4\textwidth]{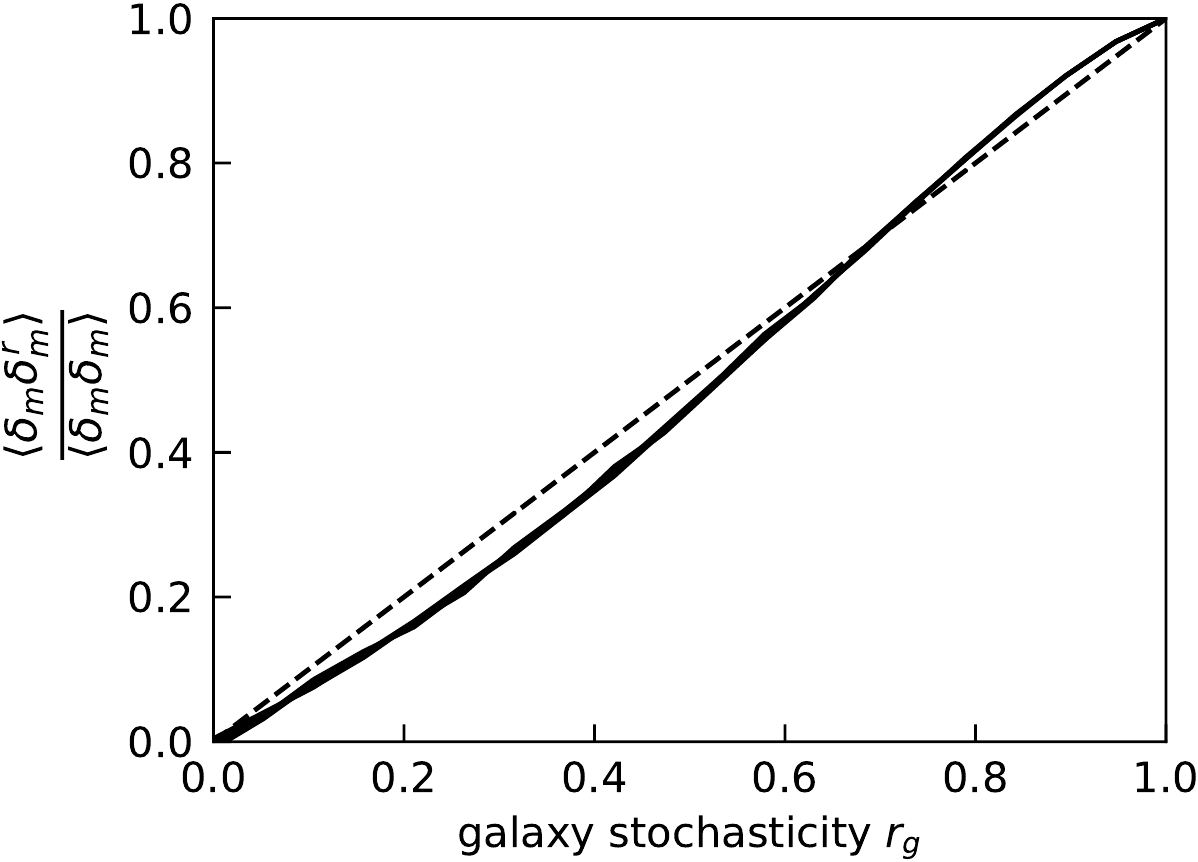} 
    \caption{Stochasticity parameter \rstoch\ used in the model and the corresponding cross correlation between the galaxy density contrast $\delta_g$ and the underlying dark matter density field $\delta_m$.}
    \label{fig:stoch}
\end{figure}


\section{Maps of galaxy clustering with stochasticity}
\label{sec:stochasticity}

Galaxy stochasticity describes the degree of correlation between the galaxy density contrast $\delta_g$ and the underlying dark matter density field $\delta_m$.
The parameter $r$ is the correlation coefficient between these fields, so that $r = \langle \delta_g \delta_m \rangle/ \langle \delta_m \delta_m \rangle$.
To create maps with varying degrees of correlation, we add random uniform noise to phases of the $\delta_m$ field, as described in Equation~\ref{eqn:stochasticity}.
This equation uses a proxy parameter $r_s$, which is part of the model.
We find the factor $(1-r_s)^{2/3}$ empirically;
we calculate the cross-correlations $\langle \delta_g \delta_m \rangle/ \langle \delta_m \delta_m \rangle$ and find that this expression brings the relation $r_s/r \approx 1$, with deviations of 10-20\%.
Figure~\ref{fig:stoch} shows this correlation as a function of the parameter $r_s$, for four redshift bins.
Each line is an average cross correlation from 36 different $5\times5$ deg maps.
The function in Equation~\ref{eqn:stochasticity} could probably be improved, but it is sufficient for the purpose of this paper.



\section{Likelihood modelling}
\label{sec:mdn_tests}


As introduced in Section~\ref{sec:likelihood}, we model the likelihood of the predicted summary $\theta_p$ given true parameter value $\theta_t$ using a Mixture Density Network.
This network takes in values of $\theta_t$ and outputs the parameter of a Gaussian Mixture Model: means of components $\mu_j$, their covariance matrices $\Psi_j$, and their relative weights $w_j$.
This is a simple network with 3 layers and 256 hidden units each, and a Relu activation.
The loss is the negative log likelihood of the samples
\begin{equation*}
L = -\log \mathcal{N}\big[\theta_p | \mu_{1,\dots,K}(\theta_t), \Psi_{1,\dots,K}(\theta_t), w_{1,\dots,K}(\theta_t) \big], 
 \end{equation*}
where $\mathcal{N}$ is the normal distribution PDF.
We use $K=4$ Gaussians in our model.
As described in Section~\ref{sec:likelihood}, we use a training set of 7615200 samples from $p(\theta_p|\theta_t)$, for each CNN and PSD network. 
We train these networks using the \textsc{Adam} optimizer \cite{Kingma2014} with the initial learning rate of 0.001. 
We leave 25\% of the training set as the validation set.
Before passing them the neural networks, we rescale $\theta_t$ using the Robust Scaler and $\theta_p$ with \mbox{MinMax} Scaler \cite{scikit-learn} in the range $[10^{-5}, 1-10^{-5}]$, followed by a Normal Percentile Function (PPF).
These invertible transformations make it easier for the GMM to model this data.
We use early stopping and pick the model with the best validation loss.

\begin{figure}
    \includegraphics[width=0.49\textwidth]{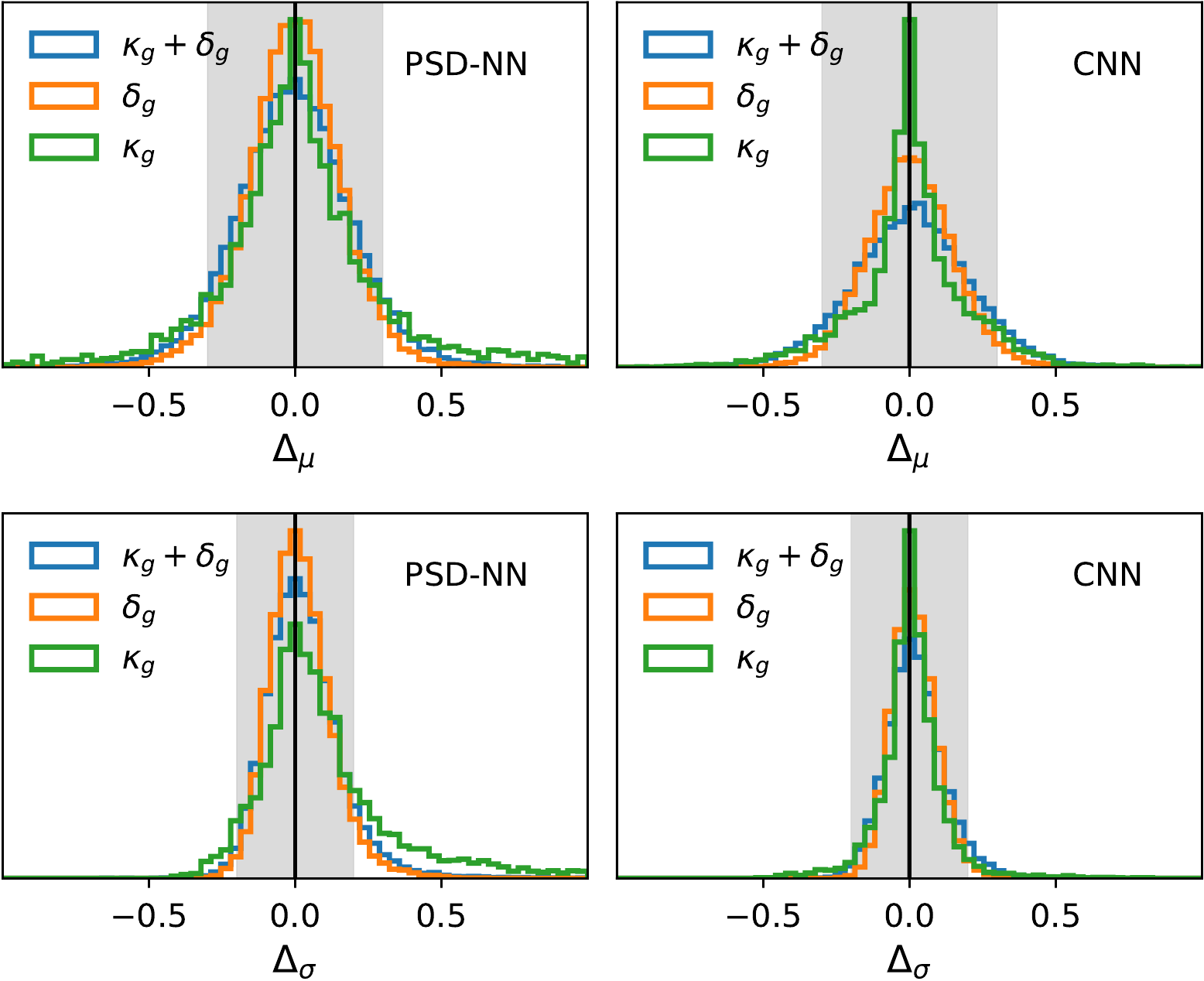} 
    \caption{
    Verification of the accuracy of the conditional density estimation of the likelihood $p(\theta_p|\theta_t)$ of predicted summary $\theta_p$ given true parameter value $\theta_t$, modeled using a Mixture Density Network (MDN).
    The upper panels show histograms of $\Delta_\mu$: fractional differences between the summaries predicted by the CNN and PSD-NN models $\theta_p$ and samples $\theta_s$ from the estimated density $p(\theta_p|\theta_t)$, compared to the standard deviation of $\theta_p$ (see Equation~\ref{eqn:likemodel_mu}).
    The shaded regions correspond $<0.3\sigma$, which is the level of error that will not have significant impact on the results.
    The lower panels show the fractional difference $\Delta_\sigma$ between standard deviations of $\theta_p$ and $\theta_s$, as defined in Equation~\ref{eqn:likemodel_sig}, with shaded regions corresponding $<0.2\sigma$.
    The histograms consist of 5700 parameter combinations from the prior space defined in Table~\ref{tab:params}, concatenating all model parameters.
    }
    \label{fig:likemodel_check}
\end{figure}

To validate the precision of the likelihood model, we perform the following test.
First, we create a validation set of $\theta_p^{\mathrm{val}}$ at fixed $\theta_t^{\mathrm{val}}$, which contains 5700 parameter combinations.
For each $\theta_t^{\mathrm{val}}$, we sample new predictions $\theta_s^{\mathrm{val}}$ from the MDN model.
We compare these predictions with the corresponding summaries in the training set $\theta_p^{\mathrm{val}}$.
For each $\theta_t$, we calculate the significance of the difference of means of these samples $\Delta_\mu$, as shown in Equation~\ref{eqn:likemodel_mu}.
We also compare the scatter in the summaries in the samples predicted by the CNNs/PSDs and the MDN density estimator.
To do this, we calculate the standard deviations for each parameter in the summary vector, and compare their fractional differences $\Delta_\sigma$, shown in Equation~\ref{eqn:likemodel_sig}:
\begin{align}
\Delta_\mu &= ( \mathrm{mean} \ \theta_s^{\mathrm{val}} -  \theta_p^{\mathrm{val}})/ (\mathrm{std} \  \theta_p^{\mathrm{val}}) \label{eqn:likemodel_mu}\\
\Delta_\sigma &= ( \mathrm{std} \ \theta_s^{\mathrm{val}} - \mathrm{std} \  \theta_p^{\mathrm{val}})/ (\mathrm{std} \  \theta_p^{\mathrm{val}}). \label{eqn:likemodel_sig}
\end{align}
We then plot the distribution of $\Delta_\mu$ from all 5700 parameter sets and all 6 models, as shown in upper panel of Figure~\ref{fig:likemodel_check}.
To make this histogram, we use fractional differences for all parameters inside the $\theta$ vector.
The MDN is unbiased and much smaller than the uncertainty of the summary, on the level of $<0.3\sigma$.
The fact that the MDN is unbiased and with low scatter indicates that it estimates the conditional density of $p(\theta_p|\theta_t)$ sufficiently well.
We plot the histogram of $\Delta_\sigma$ in the bottom panels of Figure~\ref{fig:likemodel_check}.
The fractional difference distribution is, again, centered around zero, with differences on the level of $0.1\sigma$
The MDN modelling for $\kappa_g$ is slightly worse than for other probes, most likely due to its comparatively worse overall constraining power, with posterior probability mass often hitting the prior boundaries.
Given that both the central values of the summaries from the MDN model and their uncertainty are unbiased and with relatively low scatter, we consider this method to be sufficiency precise for the purpose of this paper. 

Finally, we test if a more complicated mixture model would be needed.
We run the fiducial combined probes analysis with number of Gaussians increased to $K=8$ and the number of neural network hidden units to 512.
We find no fundamental differences in the results, which suggests that there is no need to use a larger model. 
More advanced density estimation methods can be used for this step, such as \cite{Liu2021density,Dutordoir2018gpdensity,papamakarios2017masked}; we leave these improvements to future work.

\section{Neural networks for PSD}
\label{sec:psd_nets}

\begin{figure}
    \includegraphics[width=0.45\textwidth]{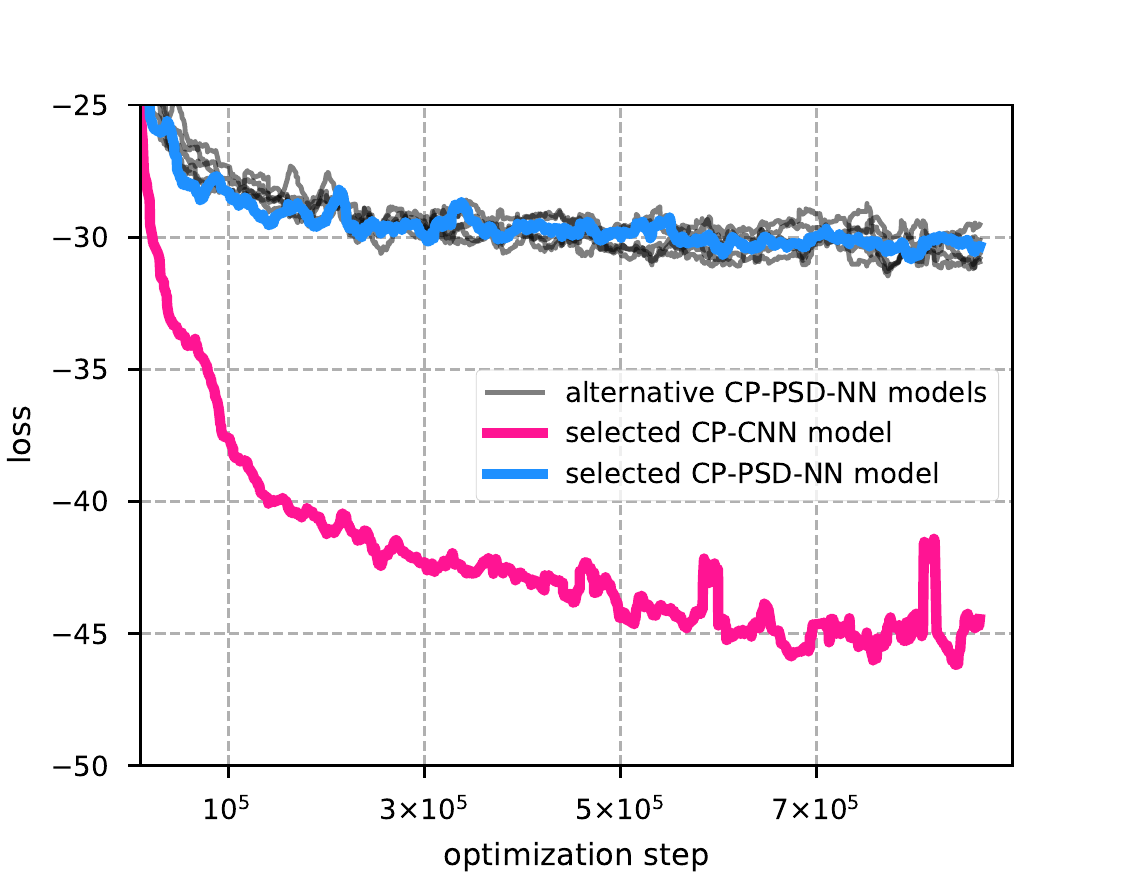} \\
     \caption{
     Training loss during the optimization process for the main \mbox{CP-CNN} (pink), the main \mbox{CP-PSD-NN} (blue), and all the alternative \mbox{CP-PSD-NN} networks.
     In this work we use likelihood loss, as defined in Equation~\ref{eqn:likelihoodloss}, which can have negative values.
     }
    \label{fig:psd_opt}
\end{figure}

In this work we calculate constraints from power spectra by compressing the PSD vectors into summaries using a simple neural network.
This requires choosing an architecture and training of the PSD-NN networks.
These choices can affect the size of the final constraint.
For example, a NN with a small number of neurons would not be able to capture all variation in the PSD vectors in the training set.
A sufficiently large network would, however, be able to extract almost all the information from PSDs, as long as the number of network outputs is greater or equal to the number of model parameters \cite{Heavens2017moped}.

In order to have a fair comparison between the PSD and CNN results, we make sure that the PSD-NN extract close to full information from the PSD vectors.
For a single case of \mbox{CP-PSD-NN}, we perform a convergence test: we increase the number of trainable parameters in the networks to see if this makes a difference.
We also test different network architectures to make sure that the results are not significantly affected by this choice.
To test this, we create 5 alternatives to the main model described in Section~\ref{sec:methods}.
We use three different architectures, and for each of them we try a standard and an extra-large version. 
The architectures are:

\begin{enumerate}
    \item 
    Classic NN (as in main model) with flattened PSD vector connected to 2 dense hidden layers with 1024 units, which then map to the outputs.
    Relu activation was used. The total number of trainable parameters was 1,823,779.
    The XL version had 3 hidden layers and 2,873,379 parameters.
    
    \item 
    Locally-connected convolutional neural net, which treated the PSD vectors as 1D channels. 
    The channels were passed through locally connected, 1D convolutional layers \cite{Chen2015locally}.
    This architecture has 128 filters with size of 3 and separated by stride of 2, for each part of the PSD vector.
    Here the filters are shared across channels.
    We use two such layers, followed by 4 residual layers, which are then fully connected to the output.
    The total number of trainable parameters was 734,883.
    The XL version had 256 filters, 8 residual layers, and 4,224,291 parameters.

    \item
    Separable convolutional neural net \cite{Kaiser2018depthwise}, which also used PSD vectors as 1D channels.
    This architecture uses a depth-wise 1D convolution that acts separately on channels, followed by a point-wise convolution that mixes channels.
    We used 2 such layers, each with 128 filters, with kernel size of 3 and a stride of 2.
    They were followed by 4 residual layers, which are later flattened and fully-connected to the output.
    This network has 438,415 trainable parameters.
    The XL version had 256 filters and 8 residual layers, which totaled to 3,270,799 parameters.
\end{enumerate}

We train these networks for 885k steps with batch size of 32 and learning rate of 0.0025.
The loss function progress during training is shown in Figure~\ref{fig:psd_opt}.
The alternative networks are shown in grey, while the main \mbox{CP-PSD-NN} in blue.
For comparison, the main deep learning result, \mbox{CP-CNN}, is shown in pink.
We notice that the loss has not decreased much for the last 100k steps, which indicates that all networks are almost perfectly converged.
There is some small variation in the values of the loss function at the end of training among the PSD networks.
Importantly, the XL networks do not achieve significantly lower loss at the end of training than the main \mbox{CP-PSD-NN}.
All the PSD networks, however, have much higher final loss than deep learning method \mbox{CP-CNN}.
This suggests that training the PSD networks for longer, or using larger models, would not improve the loss function significantly.
We conclude that the main \mbox{PSD-NN} used in this work effectively captures almost all the information contained in the PSD vectors, and can be used in fair comparison with the deep learning method.
This test was performed only for the combined probes, which is the most complex case; the conclusions of this test should also hold if repeated for the individual probes.

\section{Data transformations}
\label{sec:numerics}

To ensure that the results are numerically stable, we apply a number of transformation to the maps before passing them to the neural networks.
All these operations are lossless and identical for both CNN and PSD.
We bring the dynamic range of the maps close to the range between -1 and 1, which generally helps to improve the training: 
We scale the $\kappa_g$ map by a factor of $(0.005^2 + \sigma_{\mathrm{pix}}^2)^{-0.5}$, where $\sigma^{\mathrm{pix}}$ is the standard deviation of shape noise in a pixel.
For $\delta_g$, we use the transformation \mbox{$\delta_g \leftarrow \delta_g/N^{\mathrm{pix}}_g - 1$}, where $N^{\mathrm{pix}}_g$ is the average number of galaxies in a pixel calculated.
We also make sure that the number of galaxies in the $\delta_g$ map is always $\geq0$ to prevent negative Poisson $\lambda$ parameters, that can very rarely occur.
We do this by rescaling the density contrast $\delta_m$ such that the lowest pixel value is $-1$, but the total number of galaxies is preserved.
To do this, we first clip $\delta_m$ to the value of -1 and create $\delta_m^{\mathrm{clip}}$, and the final map is \mbox{$\delta_m \leftarrow \delta_m^{\mathrm{clip}} \cdot \sum \delta_m / (\sum \delta_m^{\mathrm{clip}})$}.
We also transform the output parameters $\theta$ to by using \Seight\ instead of \sigeight, and multiplying the \AIA\ by a factor of 0.1. 
This operation is reversed at prediction step.

\section{Approximate Poisson noise}
\label{sec:anscombe}
As currently there is no Poisson noise generator available on GPU in \textsc{Tensorflow}, we use the inverse Anscombe transform \cite{Makitalo2011anscombe} to approximate it.
We sample random Poisson number $z$ with mean $\lambda$ using this equation
\small
\begin{align}
    z & \sim \mathrm{Normal} \left[ 2{\sqrt {\lambda+{\tfrac {3}{8}}}}-{\tfrac {1}{4} \lambda^{-1/2}}, 1 \right] \\
    z & \leftarrow \begin{cases} 
      0 & z<1 \\
      {\frac {1}{4}}z^{2}-{\frac {1}{8}}+{\frac {1}{4}}{\sqrt {\frac {3}{2}}}z^{-1}-{\frac {11}{8}}z^{-2}+{\frac {5}{8}}{\sqrt {\frac {3}{2}}}z^{-3}   & z \geq 1 \\ 
   \end{cases} \\
   z &\leftarrow \mathrm{round}[\mathrm{max} [z, 0]]
\end{align}
\normalsize
We verify that this approximation is very close to Poisson and should be sufficiently precise for the purpose of this work.


\end{document}